\documentclass[twocolumn]{aastex62}
\usepackage{graphicx}
\usepackage{esint}
\usepackage{epstopdf}
\usepackage{longtable}
\usepackage{natbib}

\begin{document}

\title{Galactic interstellar sulfur isotopes: A radial $^{32}$S$/$$^{34}$S gradient?}

\correspondingauthor{Jiangshui Zhang}
\email{jszhang@gzhu.edu.cn}

\author{H.Z. Yu}
\affil{Center for Astrophysics, Guangzhou University, 510006 Guangzhou, PR China}

\author{J.S. Zhang}
\affil{Center for Astrophysics, Guangzhou University, 510006 Guangzhou, PR China}

\author{C. Henkel }
\affiliation{Max-Planck-Institut f\"{u}r Radioastronomie, Auf dem H\"{u}gel 69, 53121 Bonn, Germany}
\affiliation{Astronomy Department, King Abdulaziz University, P.O. Box 80203, 21589 Jeddah, Saudi Arabia}

\author{Y.T. Yan}
\affiliation{Max-Planck-Institut f\"{u}r Radioastronomie, Auf dem H\"{u}gel 69, 53121 Bonn, Germany}
\affiliation{Center for Astrophysics, Guangzhou University, 510006 Guangzhou, PR China}

\author{W. Liu}
\affil{Center for Astrophysics, Guangzhou University, 510006 Guangzhou, PR China}

\author{X. D. Tang}
\affiliation{Xinjiang Astronomical Observatory, Chinese Academy of Sciences, 830011 Urumqi, PR China}
\affiliation{Key Laboratory of Radio Astronomy, Chinese Academy of Sciences, PR China}

\author{N. Langer}
\affiliation{Max-Planck-Institut f\"{u}r Radioastronomie, Auf dem H\"{u}gel 69, 53121 Bonn, Germany}
\affiliation{Argelander-Insitut f\"{u}r Astronomie, Universit\"{a}t Bonn, Auf dem H\"{u}gel 71, 53121 Bonn, Germany}

\author{T.C. Luan}
\affil{Center for Astrophysics, Guangzhou University, 510006 Guangzhou, PR China}

\author{J.L. Chen}
\affil{Center for Astrophysics, Guangzhou University, 510006 Guangzhou, PR China}

\author{Y.X. Wang}
\affil{Center for Astrophysics, Guangzhou University, 510006 Guangzhou, PR China}

\author{G.G. Deng}
\affil{Center for Astrophysics, Guangzhou University, 510006 Guangzhou, PR China}

\author{Y.P. Zhou}
\affil{Center for Astrophysics, Guangzhou University, 510006 Guangzhou, PR China}

\begin{abstract}
We present observations of $^{12}$C$^{32}$S, $^{12}$C$^{34}$S, $^{13}$C$^{32}$S and $^{12}$C$^{33}$S J=2$-$1 lines toward a large sample of massive star forming regions by using the Arizona Radio Observatory 12-m telescope and the IRAM\,30-m. Taking new measurements of the carbon $^{12}$C/$^{13}$C ratio, the $^{32}$S$/$$^{34}$S isotope ratio was determined from the integrated $^{13}$C$^{32}$S/$^{12}$C$^{34}$S line intensity ratios for our sample. Our analysis shows a $^{32}$S$/$$^{34}$S gradient from the inner Galaxy out to a galactocentric distance of 12\,kpc. An unweighted least-squares fit to our data yields $^{32}$S$/$$^{34}$S = (1.56 $\pm$ 0.17)$\rm D_{\rm GC}$ + (6.75 $\pm$ 1.22) with a correlation coefficient of 0.77. Errors represent 1$\sigma$ standard deviations. Testing this result by (a) excluding the Galactic center region, (b) excluding all sources with C$^{34}$S opacities $>$ 0.25, (c) combining our data and old data from previous study, and (d) using different sets of carbon isotope ratios leads to the conclusion that the observed $^{32}$S$/$$^{34}$S gradient is not an artefact but persists irrespective of the choice of the sample and carbon isotope data. A gradient with rising $^{32}$S$/$$^{34}$S values as a function of galactocentric radius implies that the solar system ratio should be larger than that of the local interstellar medium. With the new carbon isotope ratios we obtain indeed a local $^{32}$S$/$$^{34}$S isotope ratio about 10$\%$ below the solar system one, as expected in case of decreasing $^{32}$S$/$$^{34}$S ratios with time and increased amounts of stellar processing. However, taking older carbon isotope ratios based on a lesser amount of data, such a decrease is not seen. No systematic variation of $^{34}$S$/$$^{33}$S ratios along galactocentric distance was found. The average value is 5.9 $\pm$ 1.5, the error denoting the standard deviation of an individual measurement.

\end{abstract}


\section{Introduction} \label{sec:intro}
Determining abundance gradients across the Milky Way Galaxy is a powerful tool to trace its chemical evolution \citep{2005ApJ...634.1126M}. Residing in a Galaxy which formed from inside out, stellar processing is least advanced in its outskirts and most advanced in its inner regions, especially in the Central Molecular Zone \citep{2001ApJ...554.1044C}. Measuring isotope ratios as a function of distance from the Galactic center ($D_{\rm GC}$), it is possible to trace back the star formation history and/or initial mass function (IMF) along the Galactic plane with different $D_{\rm GC}$ \citep{1994ARA&A..32..191W, 2018Natur.558..260Z}. Isotopic ratios, such as $^{12}$C$/$$^{13}$C, $^{14}$N$/$$^{15}$N, $^{18}$O$/$$^{17}$O and $^{32}$S$/$$^{34}$S, are linked to the evolution of stars with different masses. The isotopic abundances of carbon (C), nitrogen (N) and oxygen (O) can provide specific information on hydrogen burning and helium burning, including the carbon-nitrogen-oyxgen (CNO) cycle \citep{1992A&ARv...4....1W}. Sulfur is a powerful tool to trace the evolution of massive stars in their late stage of evolution. Sulfur provides, unlike C, N, or O, a total of four stable isotopes, $^{32}$S, $^{34}$S, $^{33}$S, $^{36}$S. Abundance ratios are 95.02 : 4.21 : 0.75 : 0.021 in the solar system \citep{1989GeCoA..53..197A}.

CS (carbon monosulfide) is most ubiquitous among the sulfur bearing molecules in the interstellar medium (ISM), which makes it the most prominent species to study sulfur isotopic ratios. For the interstellar medium, \citet{1996A&A...305..960C} measured abundance ratios of 24.4 $\pm$ 5.0 and 6.27 $\pm$ 1.01 (the errors are the standard deviation of the mean) for $^{32}$S$/$$^{34}$S and $^{34}$S$/$$^{33}$S, respectively. These results, mainly referring to sources with galactocentric distances close to or slightly inside the solar circle, are consistent with the solar system values mentioned above. \citet{1996A&A...313L...1M} measured $^{34}$S$/$$^{36}$S abundance ratios for the first time in the interstellar medium, obtaining a ratio of 115 $\pm$ 17. This abundance ratio is approximately half the solar system value. A similarly low value was also found in the late-type giant IRC+10216 \citep{2004A&A...426..219M}.

The gradient of Galactic interstellar sulfur isotope ratios proposed by \citet{1996A&A...305..960C} yielded $^{32}$S$/$$^{34}$S = (3.3 $\pm$ 0.5)$D_{\rm GC}$ + (4.1 $\pm$ 3.1), with $D_{\rm GC}$ here and elsewhere in units of kpc. No corresponding systematic variation of $^{34}$S$/$$^{33}$S and $^{34}$S$/$$^{36}$S was found within the Galactic disk. However, the $^{32}$S$/$$^{34}$S gradient proposed by \citet{1996A&A...305..960C} suffers from small number statistics, in particular for sources in the Galactic center and far outer Galaxy and a lack of sources from the northern sky. In addition, the distances in previous studies are mainly derived from kinematic methods, which may cause large uncertainties, not only for the distance to the Sun, but also for the distance to the center of the Milky Way \citep[e.g.,][]{2006Sci...311..54X}. Furthermore, a more accurate value of $^{12}$C/$^{13}$C for each source is desirable to determine the $^{32}$S$/$$^{34}$S isotope ratio from $^{13}$C$^{32}$S/$^{12}$C$^{34}$S.

Therefore, we are performing observations on CS $J$=2$-$1 and some of its rare isotopologues toward a large sample of massive star forming regions with accurate distance, to measure  $^{32}$S$/$$^{34}$S, $^{34}$S$/$$^{33}$S and later also $^{34}$S$/$$^{36}$S ratios and check for possible gradients. Here observations of $^{12}$C$^{32}$S (hereafter CS), $^{12}$C$^{34}$S (hereafter C$^{34}$S), $^{13}$C$^{32}$S (hereafter $^{13}$CS) and $^{12}$C$^{33}$S (hereafter C$^{33}$S) $J$$=$2$-$1 are presented toward our large sample. In Sect. 2, the sources selected for our sample and our observations are introduced. In Section 3, the main results of our measurements are illustrated. In Sect. 4, we discuss potential effects that could seriously contaminate derived isotope ratios. A short summary is presented in Sect. 5.

\section{Source Selection and Observations} \label{sec:Observation}
\subsection{Sample Selection and Distance }
During the last decade, masers have been measured with high angular resolution in more than 100 massive star forming regions to determine trigonometric parallaxes and thus distances as well as proper motions \citep{2014ApJ...783..130R, 2019ApJ...885..131R}. Among them, 95 sources in the ARO 12-m sky were selected as targets for our measurement of sulfur isotope ratios. In contrast to the southern sources observed by \citet{1996A&A...305..960C}, most of these sources are part of the northern sky. The galactocentric distance of our targets was calculated from the heliocentric distance \citep{2009ApJ...699.1153R}:
\begin{equation}
D_{\rm GC} = \sqrt{(R_0\cos(l)-d)^{2} + R_{0}^{2}\sin^2(l)}.
\end{equation}
R$_0$ is the distance of the Sun as seen from the Galactic center (R$_0$ = 8.122 $\pm$ 0.031 kpc as obtained by the \citet{2018A&A...615L..15G}), $l$ is the Galactic longitude and $d$ is the distance of the object from the Sun as determined from the parallax method \citep{2014ApJ...783..130R}. The 1$\sigma$ uncertainty in R$_0$ is with $\sim$ 0.4$\%$ too small to play a significant role in our study and will therefore be ignored in the following.

\subsection{Observations and Data Reduction}
The observations were performed in November and December 2018 using the Arizona Radio Observatory 12m telescope (ARO 12-m)\footnote{The ARO 12-m telescope is currently operated by the Arizona Radio Observatory, at Steward Observatory (University of Arizona), with partial support from the Research Corporation.} toward 95 massive star forming regions. 61 of these sources were successfully detected in all four CS isotopologues, i.e., CS, C$^{34}$S, $^{13}$CS and C$^{33}$S. A 3 mm Sideband Separating (SBS) Receiver was employed covering the ALMA band 3 (84 $-$ 116 GHz). Our observations were taken in the standard position switching mode with an off position 30$\arcmin$ west in azimuth. The center frequencies were set at 97.980986, 96.412982, 92.494299 and 97.172086 GHz for the CS, C$^{34}$S, $^{13}$CS and C$^{33}$S $J$$=$2$-$1 lines \citep{1996A&A...305..960C}, respectively; the beam size was $\sim$ 63$\arcsec$. The temperature scale was determined using the chopper wheel method in units of $T_{\rm A}^{*}$ \citep{2018ApJ...862...63C}. $T_{A}^{*}$ was converted to main beam brightness temperature by the equation $T_{\rm mb}$ = $T_{A}^{*}$/$\eta_{b}$, where $\eta_{b}$ is the main beam efficiency, ranging from 0.82 to 0.88, determined from continuum observations of Jupiter. An ARO Wideband Spectrometer (AROWS) backend was used for these measurements with channel widths of 156.25\,kHz and 78.125\,kHz, corresponding to $\sim$ 0.5 and 0.25 km\,s$^{-1}$ and associated bandwidths of 1000 and 500 MHz per polarization, respectively.

Six sources among our sample were also observed by the IRAM 30-m\footnote{The IRAM 30-m is supported by INSU/CNRS (France), MPG (Germany), and IGN (Spain).} with the EMIR heterodyne receiver during June 2016, which can be used to check in how far different beam sizes affect determined abundance ratios. We applied the $\lambda$ $\sim$ 3\,mm E090 band to observe the $J$$=$2$-$1 lines of the CS, C$^{34}$S, $^{13}$CS and C$^{33}$S isotopologues.  A fast Fourier Transform Spectrometer (FTS) backend in the wide band mode was employed in our observations, with a spectral resolution of 400 kHz ($\sim$ 0.6 km\,s$^{-1}$ at 96 GHz). The accuracy of calibration was $\pm$ 15$\%$. Because of the large bandwidth available for the FTS (16 GHz, containing four IF subbands with 4 GHz each in dual polarization), all CS isotopologues could be observed simultaneously. A standard position switching mode was adopted in our observations with an HPBW of $\sim$ 25$\arcsec$. The off-position was set at 30$\arcmin$ in azimuth. The observed antenna temperature scale $T_{A}^{*}$, calibrated like the ARO-12m data with a chopper wheel, was converted into main beam brightness temperature $T_{\rm mb}$ = $T_{A}^{*}$/$\eta_b$, where $\eta_b$ is the main beam efficiency, $\sim$ 0.81.

The data reduction was conducted with CLASS from the GILDAS software package\footnote{http://www.iram.fr/IRAMFR/GILDAS.}. The data reduction process includes linear baseline subtractions and multiple Gaussian fitting. The spectra were averaged from multiple observations including two polarizations for each transition line.

\section{Analysis and results} \label{sec:Analysis results}
As already mentioned in Sect. 2.2, we have detected all four CS isotopologues toward 61 of 95 sources from our ARO 12-m survey. Among those 61 sources, two sources (G049.48-00.36 and G043.16+00.01) show two clearly separated velocity components in the spectra of all four CS isotopologues, a dominant one and a significantly weaker one. Within the margins of error, the integrated line intensity ratios between the two velocity components are the same for the rare isotopologues. Therefore we used also for these two sources integrated intensities covering the entire velocity range exhibiting line emission.

All these spectra are shown in Fig 1. For all spectra, we subtracted baselines and then performed Gaussian fits. The resulting fitting parameters are shown in Tables 1 and 2. For the remaining 34 sources with detections of less than four isotopologues, spectra are shown in Fig. 2.

CS is the most abundant isotopologue and is often slightly saturated in star forming regions \citep[e.g.,][]{1980APJ...235.437L}. Thus, $^{32}$S$/$$^{34}$S abundance ratios cannot be directly obtained from CS/C$^{34}$S line intensity ratios. However, with the assumption that C$^{34}$S and $^{13}$CS are unsaturated, the $^{32}$S$/$$^{34}$S ratio can be determined from the integrated intensity ratio $I$($^{13}$CS)/$I$(C$^{34}$S) when we have reliable $^{12}$C/$^{13}$C ratios:
\begin{equation}
\frac{^{32}\rm S}{^{34}\rm S} \sim \frac{^{12}\rm C}{^{13}\rm C} \frac{ I(^{13}\rm C^{32}\rm S)}{ I(^{12}\rm C^{34}\rm S)}.
\end{equation}
\citet{1980APJ...235.437L} determined CS $J$ = 2$-$1 optical depths up to three in massive star forming regions. From their result, as long as $^{12}$C/$^{13}$C and $^{32}$S/$^{34}$S are $\gg$ 3 (see below), all rare CS isotopologues can be considered as optically thin (see Sect. 4.4 for a more detailed discussion).

The latest results analyzing $^{12}$C/$^{13}$C ratios in the Galactic disk and center region were reported by \citet{2019ApJ...877..154Y}. These are based on systematic observations of K$-$doublet lines of H$_2$CO and H$_2$$^{13}$CO at C ($\sim$5 GHz) and Ku ($\sim$15 GHz) bands toward a large sample of Galactic molecular clouds. Through RADEX non$-$LTE modelling accounting for radiative transfer effects and taking reliable distance values from trigonometric parallax measurements for a significant percentage (32 $\%$) of these sources, they obtained a linear fit result of
\begin{equation}
\frac{^{12}\rm C}{^{13}\rm C} = (5.08 \pm 1.10) (D_{\rm GC}) + (11.86 \pm 6.60).
\end{equation}
Thus, equipped with accurate trigonometric parallax distances, this new relation can be used here to adopt proper $^{12}$C/$^{13}$C values and on this basis, with equation (2), also sulfur isotope ratios for our sources (see Table 3).

The $^{34}$S$/$$^{33}$S ratio can be determined directly by line intensity ratios of C$^{34}$S and C$^{33}$S with the assumption (see Sect. 4.4 for more details) that both C$^{34}$S and C$^{33}$S are optically thin:
\begin{equation}
\frac{^{34}\rm S}{^{33}\rm S} \sim \frac{ I(^{12}\rm C^{34}\rm S)}{ I(^{12}\rm C^{33}\rm S)}.
\end{equation}
Relevant parameters for all 61 sources are listed in Table 3.

\section{Discussion} \label{sec:Discussion}
Although the results from our sulfur isotope ratio determinations are more reliable than previous results, due to a larger sample, trigonometric parallax distances and newly measured carbon isotope ratios, some not yet sufficiently addressed complications remain that are discussed below in Sects. 4.1 $-$ 4.4. We then present our new data which are also being compared to previous data sets in Sect. 4.5.1, while data in the more local environment of the Galactic disk are discussed in Sect. 4.5.2. Conclusions with respect to nucleosynthesis and the chemical evolution of the Galaxy are outlined in Sect. 4.5.3.

\subsection{Beam Size Effect}
As mentioned in Sect.2, 6 sources (G035.19-00.74, G049.48-00.36, G133.94+01.06, G031.28+00.06, G043.89-00.78 and G059.78+00.06) were observed by both the ARO 12-m and IRAM 30-m telescopes, which can be used to check beam size effects potentially influencing our abundance ratios.  A comparison between the spectra from the ARO 12-m and the IRAM 30-m is shown in Fig 3. Measured line parameters are given in Tables 1 and 2. If the scale of our targets is smaller than the beam size, filling factors are less than unity. So measured line intensities are diluted. In this case, the intensity of detected lines has to be adjusted for the beam dilution effect $\theta_{s}^{2}$/($\theta_{s}^{2}$ + $\theta_{beam}^{2}$) ($\eta_{\rm BD}$), where $\theta_{s}$ and $\theta_{beam}$ are source size and beam size, respectively \citep{ 2017A&A...606A..74Z}. Thus, brightness temperatures ($T_{\rm B}$) of the sources can be derived from the main beam brightness temperature ($T_{\rm mb}$) dilution:
\begin{equation}
 T_B = \frac{ T_{\rm mb}}{\eta_{BD}} = T_{\rm mb}\frac{\theta_{s}^{2} + \theta_{beam}^{2}}{\theta_{s}^{2}}.
\end{equation}
There are two $T$$\rm _{mb}$ values, measured by the IRAM 30-m and the ARO 12-m with beam sizes of $\sim$ 25$\arcsec$ and $\sim$ 63$\arcsec$. Therefore the source size can be estimated by Eq. 5. We employed CS, C$^{34}$S, $^{13}$CS and C$^{33}$S line data to estimate the size of those 6 sources and results are listed in Table 4. Source sizes turn out to be in the range 24$\arcsec$ to 126$\arcsec$ and are quite different between estimations from different isotopes. To the contrary, the sulfur isotope abundance ratios do not show significant differences between the ARO 12-m and the IRAM 30-m measurements. Within the error ranges, the average abundance ratios determined from the ARO 12-m (17.6 $\pm$ 2.7 and 4.9 $\pm$ 0.7 for $\rm ^{32}S/ ^{34}S$ and $\rm ^{34}S/^{33}S$ , respectively) are consistent with those from the IRAM 30-m observations (18.1 $\pm$ 4.1 and 5.5 $\pm$ 0.8; the errors given in this subsection represent standard deviations for individual sources). In addition, two sources (Sgr B2, G111.54+00.77 (NGC 7538)) among our sample were also observed by the 7-m antenna of the Crawford Hill Observatory \citep{1980ApJ...240...65F}. The $I(^{13}\rm CS)$/$I(\rm C^{34}S)$ line intensity ratios measured by \citet{1980ApJ...240...65F} yield 0.62 $\pm$ 0.04 and 0.43 $\pm$ 0.3 with a beam size of $\sim$ 2.2$\arcmin$, which is also consistent with our results from the ARO 12-m (0.60 $\pm$ 0.03 and 0.38 $\pm$ 0.04). Thus, our measurements of sulfur isotope ratios are not seriously affected by beam size effects.

\subsection{Hyperfine Structure}
$^{13}$CS and C$^{33}$S J$=$2$-$1 lines are split by hyperfine interactions and consist of three and eight hyperfine components encompassing frequency ranges of 0.09 and 9 MHz, respectively \citep{1981Bogey...81...256B}. For $^{13}$CS, the corresponding velocity range is only 0.3\,km\,s$^{-1}$. The central and strongest group of hyperfine components of C$^{33}$S $J$=2$-$1, including 93$\%$ of the emission under conditions of Local Thermodynamic Equilibrium (LTE) in the optically thin case, is only spread over $\pm$ 0.030\,MHz \citep{1981Bogey...81...256B}. Therefore, the corresponding velocity range is as small as $\pm$ 0.09\,km\,s$^{-1}$. Hyperfine splitting slightly broadens the lines in the case of $^{13}$CS and C$^{33}$S $J$$=$2$-$1. For our sources, line widths are typically in the range of 3 to 7\,km\,s$^{-1}$, so that significant line broadening by hyperfine splitting can be neglected for most cases (see also Sect. 4.4). However, in a few cases (G005.88, G012.68, G012.80, G012.88, G013.87, G014.33, G014.63, G028.86, G029.95, G031.28, G045.07, G079.87, G081.75), we detected the weak C$^{33}$S $J$=2$-$1 hyperfine component at rest frequency 97.174996 GHz. In these cases, the line intensities from the main and satellite hyperfine features were added.

\subsection{Fractionation}
Observed isotope ratios cannot reveal the stellar nucleosynthesis and chemical evolution in case the ratios are significantly affected by fractionation \citep{1996A&A...305..960C}. In order to investigate the effect of fractionation on sulfur isotopes, $^{32}$S$/$$^{34}$S and $^{34}$S$/$$^{33}$S ratios are plotted against gas kinetic temperatures, $T_{\rm k}$, for different radial galactocentric bins (Fig. 4). Kinetic temperatures for our sources were derived by \citet{1993A&AS...98...51H}, \citet{1996A&A...306..267S}, \citet{1996A&A...308..573M}, \citet{1998A&A...336..991T}, \citet{2005ApJ...634.1126M}, \citet{2010ApJ...717.1157D}, \citet{2010MNRAS.402.2682H}, \citet{2011ApJ...741..110D}, \citet{2011MNRAS.418.1689U}, \citet{2012A&A...544A.146W}, \citet{2014ApJ...785...55O} and \citet{2016ApJ...822...59S}, which were estimated from the para-NH$_3$ (1, 1) and (2, 2) transitions. No correlations are found between sulfur isotope ratios and $T_{\rm k}$, which indicates that fractionation is negligible for our targets. Sulfur isotope ratios are spread over sources with a wide range of temperatures, around $T_{\rm k}$ $\sim$ 20 $-$ 40\,K. The energies of the ground vibrational states between CS and C$^{34}$S (C$^{33}$S and C$^{34}$S) only differ by 7.5 (4) K. This is not sufficient to yield significant fractionation in massive star forming regions with $T_{\rm k}$ $\gg$ 10 K \citep{1996A&A...305..960C}.

\subsection{Line Saturation}
We derive the $^{32}$S$/$$^{34}$S and $^{34}$S$/$$^{33}$S isotopic ratios with the assumption that $^{13}$CS, C$^{33}$S and C$^{34}$S are unsaturated and show equal excitation temperature. To quantify potential line saturation effects related to CS and its rarer isotopologues, some details need to be further discussed.

With the assumption that CS and $^{13}$CS share the same beam filling factor and excitation temperature, the maximum optical depth of the main isotopologue CS ($\tau(^{12}\rm C\rm S)$) can be estimated from:
\begin{equation}
\frac{T_{\rm mb}(^{12}\rm C\rm S)}{T_{\rm mb}(^{13}\rm C\rm S)} \sim \frac{1 - e^{-\tau(^{12}\rm C\rm S)}}{1 - e^{-\tau(^{12}\rm C\rm S)/R}},\,R = \frac{^{12}\rm C}{^{13}\rm C}.
\end{equation}
Likewise, the maximum optical depth of $^{13}$CS ($\tau(^{13}\rm C\rm S)$) can be estimated from:
\begin{equation}
\frac{T_{\rm {mb}}(^{12}\rm C\rm S)}{T_{\rm mb}(^{13}\rm C\rm S)} \sim \frac{1 - e^{-\tau(^{13}\rm C\rm S)R}}{1 - e^{-\tau(^{13}\rm C\rm S)}},\,R = \frac{^{12}\rm C}{^{13}\rm C}.
\end{equation}
As shown in Table 3, the peak $\tau(^{12}\rm CS)$ and $\tau(^{13}\rm CS)$ for our targets range from 0.3 to 8.7 and 0.004 to 0.27, respectively. Thus $^{13}$CS is optically thin.

What remains to be analyzed is C$^{34}$S with line intensities that are weaker than those of CS but stronger than those of $^{13}$CS. The line intensity ratios of $I$(C$^{34}$S)/$I$($^{13}$CS) and $I$(C$^{34}$S)/$I$(C$^{33}$S) cannot reflect the actual abundance ratios provided that C$^{34}$S is saturated. In this case, the abundance ratios of $^{32}$S$/$$^{34}$S and $^{34}$S$/$$^{33}$S will be overestimated and underestimated, respectively (see eqs. (2) and (4)).

In the following we use three ways to check for possible C$^{34}$S line saturation. Firstly, we compare the line widths of C$^{34}$S with those of other rare CS isotopologues \citep{1996A&A...305..960C}. A comparison among our sources shows: $\Delta$($\Delta$$v_{1/2}$(C$^{34}$S, CS)) = 1.05 $\pm$ 0.80 $\rm km$ $\rm  s^{-1}$, $\Delta$($\Delta$$v_{1/2}$(C$^{34}$S, $^{13}$CS)) = 0.39 $\pm$ 0.40 $\rm km$ $\rm  s^{-1}$ and $\Delta$($\Delta$$v_{1/2}$(C$^{34}$S, C$^{33}$S)) = $-$0.26 $\pm$ 0.86 $\rm km$ $\rm  s^{-1}$ (the errors represent standard deviations for individual sources). Most of the CS lines (more than 90$\%$) tend to have higher line widths than other isotopologues in spite of not being affected by hyperfine splitting. This indicates that CS is often saturated. However, the line widths of C$^{34}$S are mostly similar to those of C$^{33}$S and $^{13}$CS. As already indicated in Sect. 4.2, hyperfine structure slightly broadens the lines of $^{13}$CS and C$^{33}$S, but this is negligible for our sample because observed line widths of several km\,s$^{-1}$ are too large to lead to a significant effect.

While this analysis shows that saturated C$^{34}$S $J$$=$2$-$1 lines cannot be common, there is an even stricter way to quantify our C$^{34}$S opacities, namely to look directly for the correlation between line intensity ratios $I$(C$^{34}$S)/$I$($^{13}$CS) and $I$(CS)/$I$(C$^{34}$S) \citep{1980ApJ...240...65F}. The $I$(C$^{34}$S)/$I$($^{13}$CS) line intensity ratios with smaller $I$(CS)/$I$(C$^{34}$S) would be reduced provided that the C$^{34}$S $J$$=$2$-$1 line is optically thick. As shown in Fig. 5(a), there is no obvious correlation between $I$(C$^{34}$S)/$I$($^{13}$CS) and $I$(CS)/$I$(C$^{34}$S). Also no correlation is found between $I$(CS)/$I$(C$^{34}$S) and $I$(C$^{34}$S)/$I$(C$^{33}$S) in Fig. 5(b). It implies that saturation of C$^{34}$S is not significant for our sample.

Nevertheless, C$^{34}$S $J$$=$2$-$1 saturation effects cannot be ruled out in a few particularly opaque sources. Therefore, the most critical sources have to be addressed individually. Table 3 indicates that the peak $\tau(^{13}\rm CS)$ for G012.88+00.48 is 0.27. While this implies that $^{13}$CS and the even weaker C$^{33}$S lines (by typical factors of $\sim$ 2 to 4 relative to $^{13}$CS, see Table 1) must be optically thin, higher line intensities show that the optical depth of C$^{34}$S must be $>$ 0.27.  Assuming identical beam filling factors and excitation temperatures, the C$^{34}$S and $^{13}$CS line intensity ratio can be expressed as:
\begin{equation}
\frac{T_{\rm mb}(\rm C\rm ^{34}S)}{T_{\rm mb}(^{13}\rm C\rm S)} = \frac{1 - e^{-\tau(\rm C\rm ^{34}S)}}{1 - e^{-\tau(^{13}\rm C\rm S)}}.
\end{equation}
Thus, with $\tau$($^{13}$CS) being known (see eq. (7)), the optical depth of C$^{34}$S, $\tau$(C$^{34}$S), can be derived. As shown in Table 3, $\tau$(C$^{34}$S) is with 0.63 still well below unity, which demonstrates that all of our targets are not optically thick in the C$^{34}$S $J$$=$2$-$1 line. However, in a few rare instances, intermediate opacities are apparently reached, which will be further discussed in Sect. 4.5.1.

Another effect to be considered is that line saturation can enhance the excitation temperature ($T_{\rm ex}$) with respect to the optically thin limit, which can be estimated when calculating the statistical equilibrium populations of the $J$ = 1 and 2 states \citep{1996A&A...305..960C}. In order to analyze what kind of $T$$_{\rm ex}$ enhancements could affect the CS opacities, we performed calculations by using the RADEX non-LTE (Local Thermodynamic Equilibrium) model \citep{2007A&A...468..627V} approximation for a uniform sphere. The collision rates for CS were taken from \citet{2006A&A...451.1125L}. Populations of the 31 first rotational levels of the CS molecule were calculated for collisions with H$_2$ for temperatures ranging from 10 K to 300 K. In Fig. 6, we plot the excitation temperature against the CS $J$=2$-$1 optical depth for different kinetic temperatures ($T_{\rm k}$) and particle densities $n$($\rm H_{2}$). Fig. 6 indicates that $T$$_{\rm ex}$ is only slightly increased with increasing optical depth for CS opacities below unity, while there is a dramatic rise in $T$$_{\rm ex}$ for opacities well beyond this value. Our sulfur isotope ratios are derived from lines with opacities below unity (C$^{34}$S, $^{13}$CS and C$^{33}$S). Since we focus exclusively on the rare isotopologues for the isotope ratios, the enhancement of $T$$_{\rm ex}$ is not significant.

\subsection{Galactic Sulfur Isotope Gradient: Nucleosynthesis and Chemical Evolution}
\subsubsection{$^{32}$\rm S$/$$^{34}$\rm S and $^{34}$\rm S$/$$^{33}$\rm S across the Galaxy}
The $^{32}$S$/$$^{34}$S isotope ratios of our sample of 61 sources are plotted as a function of galactocentric distance in Fig. 7(a). For G049.48-00.36 and G043.16+00.01 with two velocity components, we take as mentioned in Sect. 3 their average ratios for our analysis.  A strong positive gradient in $^{32}$S$/$$^{34}$S along $D_{\rm GC}$ can be found, qualitatively confirming the finding of \citet{1996A&A...305..960C}. An unweighted least-squares linear fit gives
\begin{equation}
 ^{32}\rm S / ^{34}\rm S = (1.56 \pm 0.17)\mbox{D}_{GC} + (6.75 \pm 1.22)
\end{equation}
with a correlation coefficient of 0.77, the errors representing 1 $\sigma$ standard deviations. For comparison, the result of \citet{1996A&A...305..960C} is also presented. We find that their ratios tend to be larger than ours, showing a steeper gradient. Considering the previous $^{12}$C/$^{13}$C ratios and distance model in the 1980s they have taken and for self-consistency, here we adopt again the most recent $^{12}$C/$^{13}$C values \citep{2019ApJ...877..154Y} and distance model \citep{2016ApJ...823..77R, 2019ApJ...885..131R} for their data to obtain new $^{32}$S$/$$^{34}$S ratios. Both our results and their modified results are plotted together in Fig. 8. We then find that their $^{32}$S$/$$^{34}$S results are consistent with ours, even though their data are southern sources and thus lack accurate distances measured with the parallax method.  An unweighted least-squares fit to all 81 data points then yields
\begin{equation}
 ^{32} \mbox{S} / ^{34} \mbox{S} = (1.50 \pm 0.17)D_{\rm GC} + (7.90 \pm 1.17)
\end{equation}
with a correlation coefficient of 0.71 (Fig. 8).

The sample covers sources both in the southern \citep{1996A&A...305..960C} and northern (this sample) hemisphere, mostly at galactocentric distances between 2 and 10 kpc. More data from the outskirts of the Galactic disk would still be desirable. Furthermore, we only have one source belonging to the Galactic center region, which is clearly not sufficient to characterize this important region inside the Galactic disk. Finally, while being in qualitative agreement with \citet{1996A&A...305..960C}, we should nevertheless test the reliability of our $^{32}$S$/$$^{34}$S gradient beyond the derivation of the above mentioned Pearson correlation coefficient. Since our Galactic center source is an outlier with respect to its galactocentric distance (Fig. 8), it may have a strong influence onto the overall slope of our derived linear fit. Therefore an analysis of the disk sources alone, excluding the Galactic center, is mandatory. Is the presence of a gradient then still convincing?

As shown in Fig 8, we also present unweighted least-squares fits without the Galactic center value for our data and obtain
\begin{equation}
^{32} \mbox{S} / ^{34} \mbox{S} = (1.57 \pm 0.19)D_{\rm GC} + (6.68 \pm 1.35).
\end{equation}
The combination of our data with those of \citet{1996A&A...305..960C} yields
\begin{equation}
^{32} \mbox{S} / ^{34} \mbox{S} = (1.48 \pm 0.18)D_{\rm GC} + (8.05 \pm 1.28).
\end{equation}
Correlation coefficients are 0.74 and 0.68, respectively. A comparison of eq.(11) with eq.(9) and of eq.(12) with eq.(10) demonstrates, that the exclusion of the Galactic center value is not significantly affecting neither the fitted slopes and intercepts, nor the correlation coefficients.

In view of the caveats related to C$^{34}$S $J$$=$2$-$1 line saturation (see Sect. 4.4), we also selected a subsample from our data and those of \citet{1996A&A...305..960C} with particularly low opacities. Here, 47 sources (this work) and 7 sources \citep{1996A&A...305..960C} with $\tau$(C$^{34}$S) $<$ 0.25 are selected (see Fig. 9). An unweighted least-squares fit to the subsample yields:
\begin{equation}
 ^{32}\rm S / ^{34}\rm S = (1.61 \pm 0.22)D_{GC} + (6.64 \pm 1.69).
\end{equation}
with a correlation coefficient of 0.71. Taking out the Galactic center value, we get
\begin{equation}
 ^{32}\rm S / ^{34}\rm S = (1.64 \pm 0.26)D_{GC} + (6.48 \pm 1.99).
\end{equation}
with a correlation coefficient of 0.67. There is no notable difference compared with the fitting of all the data, which again indicates that line saturation of C$^{34}$S does not impact the $^{32}$S$/$$^{34}$S gradient significantly.

To check another potential bias in our analysis, $^{32}$S$/$$^{34}$S and $^{34}$S$/$$^{33}$S abundance ratios are also plotted as a function of distance from the Sun (Fig. 10). No systematic variation between isotope ratios and distance is found, which indicates that our sulfur isotope ratios are not significantly affected by a distance bias. This complements our result related to the application of different beam sizes (Sect. 4.1).

To further check the $^{32}$S$/$$^{34}$S gradient with a different set of $^{12}$C/$^{13}$C values, we also take the carbon isotope ratios from \citet{2005ApJ...634.1126M}, based on old data obtained from CN, C$^{18}$O and H$_2$CO. We also take their old distances to compare the results based on the different molecular species (see Fig 11). Unweighted least-squares fits in combination with our data and those of \citet{1996A&A...305..960C} yield
\begin{equation}
^{32} \mbox{S} / ^{34} \mbox{S} = (1.80 \pm 0.19)D_{\rm GC} + (8.50 \pm 1.35),
\end{equation}
\begin{equation}
^{32} \mbox{S} / ^{34} \mbox{S} = (1.52 \pm 0.20)D_{\rm GC} + (11.53 \pm 1.40)
\end{equation}
and
\begin{equation}
^{32} \mbox{S} / ^{34} \mbox{S} = (2.25 \pm 0.25)D_{\rm GC} + (11.98 \pm 1.74).
\end{equation}
for $^{12}$C/$^{13}$C values from CN, C$^{18}$O and H$_2$CO, respectively. Correlation coefficients are 0.72, 0.65 and 0.71.

We also performed unweighted least-squares fits to these data excluding the Galactic center region (see Fig 11). This yields
\begin{equation}
^{32} \mbox{S} / ^{34} \mbox{S} = (1.77 \pm 0.21)D_{\rm GC} + (8.71 \pm 1.47),
\end{equation}
\begin{equation}
^{32} \mbox{S} / ^{34} \mbox{S} = (1.51 \pm 0.22)D_{\rm GC} + (11.57 \pm 1.52)
\end{equation}
and
\begin{equation}
^{32} \mbox{S} / ^{34} \mbox{S} = (2.22 \pm 0.27)D_{\rm GC} + (12.19 \pm 1.92)
\end{equation}
Correlation coefficients are 0.69, 0.62 and 0.68, respectively.

In summary, the fitting results we illustrated above show that the $^{32}$S$/$$^{34}$S gradient as a function of galactocentric radius also exhibits high correlation coefficients without the Galactic center value. Furthermore, the gradient persists using different carbon isotope data sets. It is therefore credible (for additional evidence, see also Sect. 4.5.2).

For the $^{34}$S$/$$^{33}$S isotope ratios, our results and those of \citet{1996A&A...305..960C} are plotted together against galacocentric distance in Figure 7(b).  The scatter is large and no obvious hint for a $^{34}$S$/$$^{33}$S gradient can be found from the two individual data sets as well as from their combination. The average $^{34}$S$/$$^{33}$S values of our sample, including 61 sources, is 5.9 $\pm$ 1.5 (the error represents the standard deviations for individual targets), which agrees with the value obtained by \citet{1996A&A...305..960C}, 6.27 $\pm$ 1.01.

\subsubsection{The Local ISM}
With the $^{32}$S$/$$^{34}$S isotope ratio apparently being smaller in the older and more processed inner regions of the Galaxy \citep[e.g.,][]{2001ApJ...554.1044C}, we should be able to measure a related  effect also in the more local environment of the Galactic disk. The solar system, having an age of about 4.6\,Gyr, formed at a time when the local Galaxy was less enriched by stellar nucleosynthesis and should thus be characterized by a higher $^{32}$S$/$$^{34}$S isotope ratio than the local interstellar medium, which has also been enriched by stellar ejecta during the past billions of years.

Considering the gradients presented in eqs.(9) to (14), based on our data and the carbon isotope ratios of \citet{2019ApJ...877..154Y}, we obtain with $D$$_{\rm GC}$ $=$ 8.122 kpc consistent $^{32}$S$/$$^{34}$S values in the small range between 19.4 and 20.1. These are lower than the solar system ratio of 22.6 (Sect.\,1) by about 10$\%$ and are supporting our expectation of a local $^{32}$S$/$$^{34}$S ratio decreasing with time. However, taking instead the carbon isotope ratios from \citet{2005ApJ...634.1126M}, we end up with $^{32}$S$/$$^{34}$S values of 23 to 31, i.e. values that are larger than the solar system value. The gradient reported by \citet{1996A&A...305..960C} yields 30.9.

Considering only sources near the solar circle, we also compare the solar system value with targets at galactocentric distances between 6.5 and 9.5 kpc. For the latter sample, we have 24 sources from this work and 10 sources from \citet{1996A&A...305..960C}. The average distances are 7.6 kpc and 7.9 kpc for our sample and the combination of our data with those of \citet{1996A&A...305..960C}, respectively. Furthermore, the sample with $\tau$(C$^{34}$S) $<$ 0.25 contains 23 sources from this work and 4 sources from Chin et al. (1996) between 6.5 and 9.5 kpc and the average galactocentric distance is 7.8. As shown in Table 5, average $^{32}$S$/$$^{34}$S values derived from different sets of $^{12}$C/$^{13}$C values were calculated. Furthermore, in order to compare the average $^{32}$S$/$$^{34}$S with the ratio of the solar system, we also adjust those average values to a galactocentric distance of 8.122 kpc by using those gradients obtained from the different samples mentioned above. By using the carbon isotope ratios of \citet{2019ApJ...877..154Y}, we obtain average $^{32}$S$/$$^{34}$S values in the small range between 19.1 $\pm$ 3.1 and 19.4 $\pm$ 3.3, which is consistent with the result obtained directly from the gradient within the error limits. Instead of the subsolar values, the results derived from the carbon isotope ratios of  \citet{2005ApJ...634.1126M} with different molecules show larger values, which are in the range between 22.0 $\pm$ 4.9 and 29.2 $\pm$ 5.0.

\subsubsection{Sulfur Nucleosynthesis and Galactic Chemical Evolution}
$^{32}$S is a primary isotope, i.e. the stellar production does not strongly depend on the initial metallicity of the
stellar model \citep{1995ApJS..101..181W}. The synthesis of $^{32}$S is related to oxygen-burning
\begin{equation}
^{16}\rm O + ^{16}\rm O  \longrightarrow  ^{28}\rm Si + ^{4}\rm He,    ^{28}\rm Si + ^{4}\rm He \longrightarrow ^{32}\rm S.
\end{equation}
\citep{1995ApJS..101..181W, 2008MNRAS.390.1710H}, where two $^{16}\rm O$ atoms collide to form $^{28}\rm Si$ and $^{4}\rm He$ and these products subsequently create $^{32}$S by nuclear fusion. The oxygen-burning process includes hydrostatic oxygen-burning proceeding a type \uppercase\expandafter{\romannumeral2} supernova and explosive oxygen-burning in type \uppercase\expandafter{\romannumeral1} supernova events. Most of the $^{32}$S is created from the former sources which eject approximately 10 times the amount synthesized in the latter case and occur about five times more often \citep{2008MNRAS.390.1710H}. $^{34}$S is also related to type \uppercase\expandafter{\romannumeral2} and type \uppercase\expandafter{\romannumeral1} supernovae \citep{1995ApJS..101..181W}. $^{34}$S is a by-product of oxygen burning, which is produced by newly synthesized $^{32}$S and $^{33}$S followed by neutron capture. In this case, $^{34}$S is partly a secondary isotope since its yield increases with increasing metallicity \citep{2008MNRAS.390.1710H}.

In view of the ``primary'' and partially ``secondary'' origin of the nuclei, it is reasonable to predict a weak $^{32}$S$/$$^{34}$S gradient as a function of galactocentric distance. With the chemical evolution model, \citet{2001ApJ...554.1044C} proposed an `inside-out' formation scenario of our Galaxy. Thus the timescale for Galactic disk formation and chemical abundance ratios would form a linear correlation with respect to galactocentric distance. \citet{2008MNRAS.390.1710H} indicated that $^{32}$S$/$$^{34}$S abundance ratios tend to decrease with time and predict a positive radial gradient in $^{32}$S$/$$^{34}$S, reflecting the `inside-out' Galaxy formation frame. The $^{32}$S$/$$^{34}$S abundance gradient we determined here (Figs. 7 -- 11) is powerful evidence to support this scenario.

Similar to $^{32}$S, $^{33}$S is a primary isotope, which is synthesized in explosive oxygen- and neon-burning, related to massive stars. As mentioned above, $^{34}$S is not a clean primary isotope. Therefore a $^{34}$S$/$$^{33}$S gradient can be expected, i.e., the isotope ratio should decrease with increasing galactocentric distance. However, such a $^{34}$S$/$$^{33}$S gradient is not seen, neither in our work nor in that of \citet{1996A&A...305..960C} (see Fig. 7(b)).

To summarize, the results we have displayed here provide an important insight into sulfur nucleosynthesis and Galactic chemical evolution. However, more data are still needed to support the $^{32}$S$/$$^{34}$S gradient. Our sample only includes one source in the Galactic center and is devoid of sources in the far outer Galactic disk beyond the Perseus arm. It would be an extremely interesting issue to find out whether the sulfur isotopic abundance in the Galactic center regions and the outermost regions of the disk follow the $^{32}$S$/$$^{34}$S gradient reported here or provide similar $^{34}$S$/$$^{33}$S values as those deduced for the galactocentric distances surveyed. Other sulfur isotopic abundance ratios, such as $^{32}$S$/$$^{36}$S and $^{34}$S$/$$^{36}$S, are also interesting and needed. Such measurements will also be carried out toward our sample.

\section{Conclusions} \label{sec:conclusions}
Using the ARO 12-m and IRAM 30-m telescope, we have performed systematic measurements on CS, C$^{34}$S, $^{13}$CS and C$^{33}$S J=2$-$1 lines toward 95 sources, detecting all four lines in 61 of them. The main conclusions in this work can be summarized as follows:

1. Taking new $^{12}$C/$^{13}$C results and accurate distances obtained from parallax measurements, $^{32}$S$/$$^{34}$S isotope abundance ratios have been determined from integrated $^{13}$CS/C$^{34}$S line intensity ratios. Beam size effects do not play a role as is indicated by a comparison of ARO 12-m with IRAM 30-m data.

2. Our RADEX non-LTE model calculations and analysis of line intensity ratios and line widths show that our results on isotope ratios are not affected significantly by optical depth effects.

3. The sources of our sample have been selected to possess accurate galactocentric distances from the trigonometric parallax method measuring H$_2$O and CH$_3$OH masers with very long baseline interferometry. A $^{32}$S$/$$^{34}$S gradient, first suggested by \citet{1996A&A...305..960C}, is confirmed and determined with higher accuracy, covering sources extending from the Galactic center out to the Perseus arm. A least squares fit to our data yields $^{32}\rm S / ^{34}\rm S$ = (1.56 $\pm$ 0.17)$\rm D_{\rm GC}$ + (6.75 $\pm$ 1.22) with a correlation coefficient of 0.77. Combining our data from the northern hemisphere with those from the southern hemisphere (for the latter, see \citet{1996A&A...305..960C} and analyzing them in the same way, we obtain $^{32}\rm S / ^{34}\rm S$ = (1.50 $\pm$ 0.17)$D_{\rm GC}$ + (7.90 $\pm$ 1.17), with a correlation coefficient of 0.71. More data in the Galactic center and outermost regions of the Galactic disk are expected to even better confine this gradient or to constrain its extent.

4. The presence of a positive $^{32}$S$/$$^{34}$S gradient as a function of increasing galactocentric radius requires local interstellar ratios that are smaller than than in the solar system. This can only be reproduced through equation (2) if the carbon isotope ratios from \citet{2019ApJ...877..154Y} are taken. Using instead the older carbon isotope ratios derived from CN, C$^18$O and H$_2$CO as summarized by \citet{2005ApJ...634.1126M} would lead to $^{32}$S$/$$^{34}$S ratios in the local ISM which are larger than those in the solar system. The most likely interpretation of these contradicting results is that the new carbon isotope ratios, based on a large number of sources and being supported by parallax measurements providing accurate distances are superior to the older data.

5. The average $^{34}$S$/$$^{33}$S ratio derived from our data is 5.9 $\pm$ 1.5, the error denoting the standard deviation of an individual source. No $^{34}$S$/$$^{33}$S gradient can be found in our measurements, which is unexpected considering current models. Thus, more observational and theoretical work is still needed. Additional observations of CS and its isotopomers toward a larger sample, especially toward sources in the Galactic center region and the far outer Galactic disk, will determine further the global Galactic distribution of $^{32}$S$/$$^{36}$S and $^{34}$S$/$$^{36}$S, which would provide additional constraints on oxygen burning nucleosynthesis models.

This work is supported by the Natural Science Foundation of China (No. 11590782). We would like to thank the staff of the ARO 12-m and IRAM 30-m for their kind assistance and advice during our observations. Y.T.Y. is a member of the International Max Planck Research School (IMPRS) for Astronomy and Astrophysics at the Universities of Bonn and Cologne. Y.T.Y.  would like to thank the China Scholarship Council (CSC) for the support.

\startlongtable
\begin{deluxetable*}{ccccccccccc}
\tabletypesize{\scriptsize}
\tablewidth{650pt}
\tablecaption{Observational parameters of CS and three of its rare isotopologues measured by the ARO 12-m telescope. \label{tab:table}}
\tablehead{
\colhead{Source} & \colhead{R.A.} & \colhead{Dec}  & \colhead{Molecule} & \colhead{time} & \colhead{r.m.s} &  \colhead{$\Delta{V}$}&  \colhead{$v_{\rm LSR}$}& \colhead{$\Delta$$v_{1/2}$}  & \colhead{$ \int{T_{\rm mb}{\rm{d}} v}$}  & \colhead{$T_{\rm mb}$}  \\ & (hh:mm:ss)& (dd:mm:ss )  & & (min) & (mK) & ($\rm km \, \rm s^{-1}$) & ($\rm km \, \rm s^{-1}$)& ($\rm km \, \rm s^{-1}$) &(${\rm K} \, {\rm km} \, {\rm s}^{-1}$) &  (K) }
\decimalcolnumbers
\startdata
G121.29+00.65	&	00:36:47.00	&	63:29:02.20	&	$^{12}$CS	&	32	&	33 	&	0.48 	&	-17.5 	$\pm$	0.01 	&	4.0 	$\pm$	0.03 	&	14.41 	$\pm$	0.09 	&	3.44 	\\
	&		&		&	C$^{34}$S	&	28	&	13 	&	0.48 	&	-17.4 	$\pm$	0.02 	&	2.5 	$\pm$	0.04 	&	1.39 	$\pm$	0.03 	&	0.52 	\\
	&		&		&	$^{13}$CS	&	52	&	12 	&	0.51 	&	-17.4 	$\pm$	0.01 	&	2.6 	$\pm$	0.10 	&	0.57 	$\pm$	0.02 	&	0.21 	\\
	&		&		&	C$^{33}$S	&	32	&	10 	&	0.96 	&	-16.9 	$\pm$	0.14 	&	2.7 	$\pm$	0.33 	&	0.25 	$\pm$	0.03 	&	0.08 	\\
G123.06-06.30	&	00:52:24.70	&	+56:33:50.50	&	$^{12}$CS	&	32	&	12 	&	0.48 	&	-30.6 	$\pm$	0.01 	&	4.0 	$\pm$	0.02 	&	11.95 	$\pm$	0.05 	&	2.79 	\\
	&		&		&	C$^{34}$S	&	12	&	17 	&	0.48 	&	-30.4 	$\pm$	0.05 	&	3.5 	$\pm$	0.13 	&	1.25 	$\pm$	0.03 	&	0.33 	\\
	&		&		&	$^{13}$CS	&	24	&	22 	&	0.51 	&	-30.3 	$\pm$	0.11 	&	3.0 	$\pm$	0.28 	&	0.42 	$\pm$	0.03 	&	1.12 	\\
	&		&		&	C$^{33}$S	&	32	&	4 	&	0.96 	&	-29.9 	$\pm$	0.15 	&	2.9 	$\pm$	0.39 	&	0.19 	$\pm$	0.02 	&	0.06 	\\
G133.94+01.06	&	02:27:03.82	&	+61:52:25.20	&	$^{12}$CS	&	8	&	21 	&	0.48 	&	-47.1 	$\pm$	0.01 	&	5.1 	$\pm$	0.02 	&	31.41 	$\pm$	0.08 	&	5.82 	\\
	&		&		&	C$^{34}$S	&	12	&	17 	&	0.48 	&	-46.7 	$\pm$	0.02 	&	4.0 	$\pm$	0.04 	&	4.80 	$\pm$	0.04 	&	1.13 	\\
	&		&		&	$^{13}$CS	&	20	&	12 	&	0.51 	&	-46.8 	$\pm$	0.03 	&	4.0 	$\pm$	0.08 	&	1.86 	$\pm$	0.03 	&	0.44 	\\
	&		&		&	C$^{33}$S	&	8	&	10 	&	0.96 	&	-46.4 	$\pm$	0.24 	&	5.4 	$\pm$	0.75 	&	0.97 	$\pm$	0.07 	&	0.19 	\\
G209.00-19.38	&	05:35:15.80	&	-05:23:14.10	&	$^{12}$CS	&	24	&	15 	&	0.48 	&	8.3 	$\pm$	0.01 	&	3.7 	$\pm$	0.04 	&	31.80 	$\pm$	0.25 	&	8.06 	\\
	&		&		&	C$^{34}$S	&	12	&	22 	&	0.48 	&	8.2 	$\pm$	0.02 	&	3.4 	$\pm$	0.07 	&	3.44 	$\pm$	0.06 	&	0.94 	\\
	&		&		&	$^{13}$CS	&	20	&	13 	&	0.51 	&	8.1 	$\pm$	0.05 	&	3.6 	$\pm$	0.12 	&	1.26 	$\pm$	0.03 	&	0.33 	\\
	&		&		&	C$^{33}$S	&	24	&	8 	&	0.96 	&	8.7 	$\pm$	0.17 	&	3.8 	$\pm$	0.42 	&	0.59 	$\pm$	0.04 	&	0.17 	\\
G183.72-03.66	&	05:40:24.23	&	+23:50:54.70	&	$^{12}$CS	&	52	&	8 	&	0.48 	&	2.4 	$\pm$	0.01 	&	3.5 	$\pm$	0.02 	&	7.85 	$\pm$	0.04 	&	2.10 	\\
	&		&		&	C$^{34}$S	&	12	&	23 	&	0.48 	&	2.5 	$\pm$	0.04 	&	2.0 	$\pm$	0.13 	&	0.60 	$\pm$	0.03 	&	0.29 	\\
	&		&		&	$^{13}$CS	&	24	&	16 	&	0.51 	&	2.0 	$\pm$	0.20 	&	4.1 	$\pm$	0.64 	&	0.28 	$\pm$	0.03 	&	0.07 	\\
	&		&		&	C$^{33}$S	&	52	&	3 	&	0.96 	&	2.7 	$\pm$	0.36 	&	3.7 	$\pm$	0.86 	&	0.08 	$\pm$	0.02 	&	0.03 	\\
G188.94+00.88	&	06:08:53.35	&	21:38:28.70	&	$^{12}$CS	&	16	&	21 	&	0.48 	&	3.4 	$\pm$	0.01 	&	3.6 	$\pm$	0.01 	&	23.14 	$\pm$	0.06 	&	5.98 	\\
	&		&		&	C$^{34}$S	&	24	&	12 	&	0.48 	&	3.2 	$\pm$	0.01 	&	2.9 	$\pm$	0.03 	&	2.33 	$\pm$	0.02 	&	0.76 	\\
	&		&		&	$^{13}$CS	&	32	&	12 	&	0.51 	&	3.2 	$\pm$	0.03 	&	2.9 	$\pm$	0.09 	&	0.76 	$\pm$	0.02 	&	0.24 	\\
	&		&		&	C$^{33}$S	&	16	&	14 	&	0.96 	&	3.9 	$\pm$	0.10 	&	2.8 	$\pm$	0.23 	&	0.41 	$\pm$	0.03 	&	0.14 	\\
G188.79+01.03	&	06:09:06.97	&	+21:50:41.40	&	$^{12}$CS	&	24	&	16 	&	0.48 	&	-0.5 	$\pm$	0.01 	&	2.9 	$\pm$	0.02 	&	9.48 	$\pm$	0.04 	&	3.11 	\\
	&		&		&	C$^{34}$S	&	12	&	15 	&	0.48 	&	-0.3 	$\pm$	0.02 	&	2.4 	$\pm$	0.05 	&	1.43 	$\pm$	0.03 	&	0.57 	\\
	&		&		&	$^{13}$CS	&	20	&	11 	&	0.51 	&	-0.2 	$\pm$	0.06 	&	2.5 	$\pm$	0.16 	&	0.46 	$\pm$	0.02 	&	0.18 	\\
	&		&		&	C$^{33}$S	&	24	&	6 	&	0.96 	&	0.5 	$\pm$	0.16 	&	2.2 	$\pm$	0.40 	&	0.22 	$\pm$	0.03 	&	0.09 	\\
G192.60-00.04	&	06:12:54.02	&	+17:59:23.30	&	$^{12}$CS	&	36	&	15 	&	0.48 	&	7.4 	$\pm$	0.01 	&	2.7 	$\pm$	0.02 	&	14.58 	$\pm$	0.05 	&	5.15 	\\
	&		&		&	C$^{34}$S	&	12	&	16 	&	0.48 	&	7.4 	$\pm$	0.01 	&	2.3 	$\pm$	0.03 	&	3.46 	$\pm$	0.03 	&	1.31 	\\
	&		&		&	$^{13}$CS	&	12	&	11 	&	0.51 	&	7.3 	$\pm$	0.02 	&	2.4 	$\pm$	0.06 	&	1.22 	$\pm$	0.03 	&	0.48 	\\
	&		&		&	C$^{33}$S	&	36	&	9 	&	0.96 	&	8.0 	$\pm$	0.08 	&	2.2 	$\pm$	0.18 	&	0.33 	$\pm$	0.03 	&	0.16 	\\
G236.81+01.98	&	07:44:28.24	&	-20:08:30.20	&	$^{12}$CS	&	28	&	12 	&	0.48 	&	52.5 	$\pm$	0.01 	&	3.4 	$\pm$	0.03 	&	3.57 	$\pm$	0.04 	&	1.00 	\\
	&		&		&	C$^{34}$S	&	16	&	18 	&	0.48 	&	52.8 	$\pm$	0.05 	&	2.1 	$\pm$	0.14 	&	0.44 	$\pm$	0.03 	&	0.20 	\\
	&		&		&	$^{13}$CS	&	32	&	17 	&	0.51 	&	52.8 	$\pm$	0.16 	&	2.3 	$\pm$	0.41 	&	0.16 	$\pm$	0.02 	&	0.07 	\\
	&		&		&	C$^{33}$S	&	28	&	9 	&	0.96 	&	53.5 	$\pm$	0.26 	&	2.9 	$\pm$	0.75 	&	0.13 	$\pm$	0.03 	&	0.04 	\\
G005.88-00.39	&	18:00:30.31	&	-24:04:04.50	&	$^{12}$CS	&	52	&	13 	&	0.48 	&	9.4 	$\pm$	0.02 	&	5.3 	$\pm$	0.06 	&	48.68 	$\pm$	0.43 	&	8.60 	\\
	&		&		&	C$^{34}$S	&	8	&	23 	&	0.48 	&	9.4 	$\pm$	0.03 	&	4.6 	$\pm$	0.08 	&	7.54 	$\pm$	0.10 	&	1.56 	\\
	&		&		&	$^{13}$CS	&	8	&	43 	&	0.51 	&	9.5 	$\pm$	0.05 	&	4.3 	$\pm$	0.13 	&	3.14 	$\pm$	0.08 	&	0.69 	\\
	&		&		&	C$^{33}$S	&	52	&	7 	&	0.96 	&	10.2 	$\pm$	0.11 	&	5.0 	$\pm$	0.36 	&	1.26 	$\pm$	0.07 	&	0.23 	\\
G009.62+00.19	&	18:06:14.66	&	-20:31:31.70	&	$^{12}$CS	&	52	&	11 	&	0.48 	&	4.7 	$\pm$	0.02 	&	7.4 	$\pm$	0.06 	&	26.87 	$\pm$	0.09 	&	3.44 	\\
	&		&		&	C$^{34}$S	&	8	&	22 	&	0.48 	&	4.3 	$\pm$	0.03 	&	5.6 	$\pm$	0.08 	&	5.07 	$\pm$	0.06 	&	0.85 	\\
	&		&		&	$^{13}$CS	&	12	&	27 	&	0.51 	&	4.2 	$\pm$	0.07 	&	5.8 	$\pm$	0.17 	&	2.34 	$\pm$	0.05 	&	0.39 	\\
	&		&		&	C$^{33}$S	&	52	&	8 	&	0.96 	&	4.8 	$\pm$	0.16 	&	6.0 	$\pm$	0.56 	&	1.05 	$\pm$	0.08 	&	0.15 	\\
G010.62-00.38	&	18:10:28.55	&	-19:55:48.60	&	$^{12}$CS	&	28	&	32 	&	0.48 	&	-3.1 	$\pm$	0.01 	&	7.3 	$\pm$	0.01 	&	65.75 	$\pm$	0.05 	&	8.43 	\\
	&		&		&	C$^{34}$S	&	8	&	21 	&	0.48 	&	-3.0 	$\pm$	0.02 	&	6.0 	$\pm$	0.04 	&	14.52 	$\pm$	0.09 	&	2.29 	\\
	&		&		&	$^{13}$CS	&	8	&	33 	&	0.51 	&	-2.9 	$\pm$	0.03 	&	5.5 	$\pm$	0.08 	&	6.11 	$\pm$	0.07 	&	1.04 	\\
	&		&		&	C$^{33}$S	&	28	&	15 	&	0.96 	&	-2.1 	$\pm$	0.09 	&	6.4 	$\pm$	0.28 	&	2.80 	$\pm$	0.10 	&	0.41 	\\
G012.88+00.48	&	18:11:51.00	&	-17:31:29.00	&	$^{12}$CS	&	36	&	12 	&	0.48 	&	33.9 	$\pm$	0.03 	&	4.8 	$\pm$	0.06 	&	8.38 	$\pm$	0.08 	&	1.63 	\\
	&		&		&	C$^{34}$S	&	8	&	20 	&	0.48 	&	33.3 	$\pm$	0.03 	&	3.6 	$\pm$	0.07 	&	2.92 	$\pm$	0.05 	&	0.76 	\\
	&		&		&	$^{13}$CS	&	8	&	42 	&	0.51 	&	33.2 	$\pm$	0.06 	&	3.3 	$\pm$	0.13 	&	1.38 	$\pm$	0.05 	&	0.39 	\\
	&		&		&	C$^{33}$S	&	36	&	7 	&	0.96 	&	33.9 	$\pm$	0.09 	&	3.5 	$\pm$	0.21 	&	0.61 	$\pm$	0.03 	&	0.16 	\\
G012.68-00.18	&	18:13:54.75	&	-18:01:46.60	&	$^{12}$CS	&	32	&	14 	&	0.48 	&	57.8 	$\pm$	0.02 	&	3.4 	$\pm$	0.04 	&	4.96 	$\pm$	0.05 	&	1.38 	\\
	&		&		&	C$^{34}$S	&	8	&	21 	&	0.48 	&	56.0 	$\pm$	0.06 	&	4.3 	$\pm$	0.17 	&	1.82 	$\pm$	0.05 	&	0.40 	\\
	&		&		&	$^{13}$CS	&	20	&	18 	&	0.51 	&	55.9 	$\pm$	0.08 	&	3.6 	$\pm$	0.18 	&	0.82 	$\pm$	0.04 	&	0.21 	\\
	&		&		&	C$^{33}$S	&	32	&	7 	&	0.96 	&	56.4 	$\pm$	0.14 	&	3.0 	$\pm$	0.29 	&	0.33 	$\pm$	0.03 	&	0.10 	\\
G011.91-00.61	&	18:13:58.12	&	-18:54:20.30	&	$^{12}$CS	&	32	&	15 	&	0.48 	&	35.9 	$\pm$	0.07 	&	10.5 	$\pm$	0.15 	&	12.56 	$\pm$	0.15 	&	1.43 	\\
	&		&		&	C$^{34}$S	&	8	&	24 	&	0.48 	&	36.5 	$\pm$	0.12 	&	5.6 	$\pm$	0.28 	&	1.64 	$\pm$	0.07 	&	2.68 	\\
	&		&		&	$^{13}$CS	&	20	&	22 	&	0.51 	&	36.9 	$\pm$	0.15 	&	4.6 	$\pm$	0.43 	&	0.64 	$\pm$	0.04 	&	0.13 	\\
	&		&		&	C$^{33}$S	&	32	&	5 	&	0.96 	&	36.5 	$\pm$	0.50 	&	7.6 	$\pm$	1.39 	&	0.45 	$\pm$	0.05 	&	0.04 	\\
G012.80-00.20	&	18:14:14.23	&	-17:55:40.50	&	$^{12}$CS	&	32	&	33 	&	0.48 	&	35.7 	$\pm$	0.01 	&	7.1 	$\pm$	0.01 	&	80.72 	$\pm$	0.14 	&	10.64 	\\
	&		&		&	C$^{34}$S	&	8	&	24 	&	0.48 	&	35.7 	$\pm$	0.01 	&	5.5 	$\pm$	0.03 	&	16.25 	$\pm$	0.06 	&	2.75 	\\
	&		&		&	$^{13}$CS	&	12	&	28 	&	0.51 	&	35.6 	$\pm$	0.02 	&	4.9 	$\pm$	0.05 	&	6.77 	$\pm$	0.06 	&	1.30 	\\
	&		&		&	C$^{33}$S	&	32	&	14 	&	0.96 	&	36.3 	$\pm$	0.04 	&	5.5 	$\pm$	0.12 	&	2.96 	$\pm$	0.08 	&	0.50 	\\
G012.90-00.26	&	18:14:39.57	&	-17:52:00.04	&	$^{12}$CS	&	64	&	10 	&	0.48 	&	34.7 	$\pm$	0.02 	&	8.4 	$\pm$	0.06 	&	13.18 	$\pm$	0.08 	&	1.47 	\\
	&		&		&	C$^{34}$S	&	16	&	16 	&	0.48 	&	34.5 	$\pm$	0.07 	&	6.7 	$\pm$	0.20 	&	2.20 	$\pm$	0.05 	&	0.31 	\\
	&		&		&	$^{13}$CS	&	16	&	10 	&	0.51 	&	34.8 	$\pm$	0.19 	&	5.8 	$\pm$	0.43 	&	0.65 	$\pm$	0.04 	&	0.11 	\\
	&		&		&	C$^{33}$S	&	64	&	6 	&	0.96 	&	35.2 	$\pm$	0.35 	&	6.3 	$\pm$	0.89 	&	0.26 	$\pm$	0.03 	&	0.04 	\\
G013.87+00.28	&	18:14:35.83	&	-16:45:35.90:	&	$^{12}$CS	&	32	&	13 	&	0.48 	&	49.2 	$\pm$	0.01 	&	4.8 	$\pm$	0.02 	&	12.23 	$\pm$	0.04 	&	2.40 	\\
	&		&		&	C$^{34}$S	&	8	&	23 	&	0.48 	&	48.8 	$\pm$	0.03 	&	3.0 	$\pm$	0.08 	&	2.60 	$\pm$	0.05 	&	0.77 	\\
	&		&		&	$^{13}$CS	&	20	&	26 	&	0.51 	&	48.9 	  $\pm$	0.05 	&	2.6 	  $\pm$	0.11 	&	1.09 	  $\pm$	0.04 	&	0.37 	\\
	&		&		&	C$^{33}$S	&	32	&	11 	&	0.96 	&	49.4 	$\pm$	0.06 	&	2.8 	$\pm$	0.15 	&	0.56 	$\pm$	0.03 	&	0.18 	\\
G011.49-01.48	&	18:16:22.00	&	-19:41:27.20	&	$^{12}$CS	&	32	&	14 	&	0.48 	&	10.7 	$\pm$	0.02 	&	3.1 	$\pm$	0.04 	&	7.24 	$\pm$	0.08 	&	2.22 	\\
	&		&		&	C$^{34}$S	&	8	&	24 	&	0.48 	&	10.7 	$\pm$	0.09 	&	2.5 	$\pm$	0.24 	&	0.64 	$\pm$	0.04 	&	0.22 	\\
	&		&		&	$^{13}$CS	&	20	&	22 	&	0.51 	&	11.0 	$\pm$	0.13 	&	1.6 	$\pm$	0.37 	&	0.17 	$\pm$	0.03 	&	0.09 	\\
	&		&		&	C$^{33}$S	&	32	&	8 	&	0.96 	&	11.8 	$\pm$	0.78 	&	2.5 	$\pm$	0.68 	&	0.38 	$\pm$	0.02 	&	0.04 	\\
G014.33-00.64	&	18:18:54.67	&	-16:47:50.30	&	$^{12}$CS	&	24	&	11 	&	0.48 	&	22.7 	$\pm$	0.03 	&	5.7 	$\pm$	0.07 	&	15.59 	$\pm$	0.14 	&	2.55 	\\
	&		&		&	C$^{34}$S	&	12	&	15 	&	0.48 	&	22.2 	$\pm$	0.02 	&	2.9 	$\pm$	0.04 	&	3.19 	$\pm$	0.04 	&	0.98 	\\
	&		&		&	$^{13}$CS	&	20	&	36 	&	0.51 	&	22.1 	$\pm$	0.06 	&	2.6 	$\pm$	0.15 	&	1.17 	$\pm$	0.05 	&	0.38 	\\
	&		&		&	C$^{33}$S	&	24	&	9 	&	0.96 	&	22.8 	$\pm$	0.07 	&	2.8 	$\pm$	0.18 	&	0.57 	$\pm$	0.03 	&	1.90 	\\
G014.63-00.57	&	18:19:15.54	&	-16:29:45.80	&	$^{12}$CS	&	24	&	10 	&	0.48 	&	18.6 	$\pm$	0.01 	&	4.4 	$\pm$	0.02 	&	11.45 	$\pm$	0.06 	&	2.45 	\\
	&		&		&	C$^{34}$S	&	8	&	21 	&	0.48 	&	18.6 	$\pm$	0.04 	&	2.8 	$\pm$	0.09 	&	1.48 	$\pm$	0.04 	&	0.49 	\\
	&		&		&	$^{13}$CS	&	12	&	43 	&	0.51 	&	18.5 	$\pm$	0.15 	&	2.8 	$\pm$	0.36 	&	0.58 	$\pm$	0.06 	&	0.20 	\\
	&		&		&	C$^{33}$S	&	24	&	10 	&	0.96 	&	18.7 	$\pm$	0.22 	&	3.9 	$\pm$	0.59 	&	0.33 	$\pm$	0.04 	&	0.08 	\\
G016.58-00.05	&	18:21:09.08	&	-14:31:48.80	&	$^{12}$CS	&	24	&	16 	&	0.48 	&	59.4 	$\pm$	0.02 	&	3.4 	$\pm$	0.05 	&	10.52 	$\pm$	0.13 	&	2.93 	\\
	&		&		&	C$^{34}$S	&	8	&	10 	&	0.48 	&	59.5 	$\pm$	0.04 	&	3.1 	$\pm$	0.09 	&	2.04 	$\pm$	0.05 	&	0.61 	\\
	&		&		&	$^{13}$CS	&	24	&	22 	&	0.51 	&	59.4 	$\pm$	0.09 	&	3.3 	$\pm$	0.22 	&	0.71 	$\pm$	0.04 	&	0.20 	\\
	&		&		&	C$^{33}$S	&	24	&	35 	&	0.96 	&	60.1 	$\pm$	0.19 	&	3.9 	$\pm$	0.52 	&	0.34 	$\pm$	0.04 	&	0.08 	\\
G023.43-00.18	&	18:34:39.29	&	-08:31:25.40	&	$^{12}$CS	&	32	&	14 	&	0.48 	&	102.6 	$\pm$	0.02 	&	6.7 	$\pm$	0.52 	&	8.37 	$\pm$	0.12 	&	1.18 	\\
	&		&		&	C$^{34}$S	&	20	&	23 	&	0.48 	&	101.6 	$\pm$	0.06 	&	6.7 	$\pm$	0.17 	&	2.48 	$\pm$	0.05 	&	0.35 	\\
	&		&		&	$^{13}$CS	&	24	&	15 	&	0.51 	&	101.7 	$\pm$	0.16 	&	6.6 	$\pm$	0.47 	&	0.93 	$\pm$	0.05 	&	0.15 	\\
	&		&		&	C$^{33}$S	&	32	&	6 	&	0.96 	&	102.1 	$\pm$	0.33 	&	7.6 	$\pm$	1.15 	&	0.45 	$\pm$	0.05 	&	0.06 	\\
G027.36-00.16	&	18:41:51.06	&	-05:01:43.40	&	$^{12}$CS	&	32	&	12 	&	0.48 	&	91.9 	$\pm$	0.02 	&	4.8 	$\pm$	0.06 	&	13.91 	$\pm$	0.13 	&	2.71 	\\
	&		&		&	C$^{34}$S	&	20	&	8 	&	0.48 	&	92.3 	$\pm$	0.04 	&	4.3 	$\pm$	0.10 	&	1.90 	$\pm$	0.04 	&	0.42 	\\
	&		&		&	$^{13}$CS	&	24	&	17 	&	0.51 	&	92.3 	$\pm$	0.11 	&	4.5 	$\pm$	0.30 	&	0.67 	$\pm$	0.04 	&	0.14 	\\
	&		&		&	C$^{33}$S	&	32	&	21 	&	0.96 	&	93.3 	$\pm$	0.31 	&	6.2 	$\pm$	0.86 	&	0.41 	$\pm$	0.04 	&	0.06 	\\
G028.86+00.06	&	18:43:46.22	&	-03:35:29.60	&	$^{12}$CS	&	32	&	23 	&	0.48 	&	103.7 	$\pm$	0.02 	&	4.4 	$\pm$	0.06 	&	9.01 	$\pm$	0.10 	&	1.94 	\\
	&		&		&	C$^{34}$S	&	20	&	17 	&	0.48 	&	103.4 	$\pm$	0.03 	&	3.5 	$\pm$	0.08 	&	1.97 	$\pm$	0.04 	&	0.53 	\\
	&		&		&	$^{13}$CS	&	24	&	18 	&	0.51 	&	103.4 	$\pm$	0.06 	&	3.5 	$\pm$	0.17 	&	0.84 	$\pm$	0.03 	&	0.23 	\\
	&		&		&	C$^{33}$S	&	32	&	7 	&	0.96 	&	104.0 	$\pm$	0.18 	&	3.9 	$\pm$	0.45 	&	0.32 	$\pm$	0.03 	&	0.08 	\\
G029.86-00.04	&	18:45:59.57	&	-02:45:06.70	&	$^{12}$CS	&	32	&	15 	&	0.48 	&	101.2 	$\pm$	0.01 	&	4.2 	$\pm$	0.03 	&	9.89 	$\pm$	0.06 	&	2.23 	\\
	&		&		&	C$^{34}$S	&	20	&	20 	&	0.48 	&	101.0 	$\pm$	0.05 	&	3.8 	$\pm$	0.11 	&	1.28 	$\pm$	0.03 	&	0.31 	\\
	&		&		&	$^{13}$CS	&	24	&	13 	&	0.51 	&	100.8 	$\pm$	0.12 	&	3.6 	$\pm$	0.29 	&	0.39 	$\pm$	0.03 	&	0.10 	\\
	&		&		&	C$^{33}$S	&	32	&	8 	&	0.96 	&	102.2 	$\pm$	0.63 	&	4.9 	$\pm$	0.89 	&	0.16 	$\pm$	0.04 	&	0.03 	\\
G029.95-00.01	&	18:46:03.74	&	-02:39:22.30:	&	$^{12}$CS	&	56	&	8 	&	0.48 	&	97.6 	$\pm$	0.01 	&	4.5 	$\pm$	0.02 	&	22.90 	$\pm$	0.08 	&	4.81 	\\
	&		&		&	C$^{34}$S	&	8	&	18 	&	0.48 	&	97.8 	$\pm$	0.02 	&	3.9 	$\pm$	0.05 	&	5.08 	$\pm$	0.06 	&	1.21 	\\
	&		&		&	$^{13}$CS	&	8	&	24 	&	0.51 	&	97.8 	$\pm$	0.03 	&	3.9 	$\pm$	0.08 	&	2.17 	$\pm$	0.04 	&	0.52 	\\
	&		&		&	C$^{33}$S	&	56	&	7 	&	0.96 	&	98.5 	$\pm$	0.06 	&	4.0 	$\pm$	0.15 	&	0.89 	$\pm$	0.03 	&	0.21 	\\
G031.28+00.06	&	18:48:12.39	&	-01:26:30.70	&	$^{12}$CS	&	56	&	18 	&	0.48 	&	108.8 	$\pm$	0.02 	&	7.2 	$\pm$	0.04 	&	12.62 	$\pm$	0.06 	&	1.64 	\\
	&		&		&	C$^{34}$S	&	8	&	23 	&	0.48 	&	108.8 	$\pm$	0.04 	&	4.7 	$\pm$	0.10 	&	2.86 	$\pm$	0.05 	&	0.57 	\\
	&		&		&	$^{13}$CS	&	32	&	10 	&	0.51 	&	108.7 	$\pm$	0.04 	&	4.3 	$\pm$	0.09 	&	1.12 	$\pm$	0.02 	&	0.25 	\\
	&		&		&	C$^{33}$S	&	56	&	8 	&	0.96 	&	109.6 	$\pm$	0.10 	&	3.9 	$\pm$	0.26 	&	0.46 	$\pm$	0.03 	&	0.11 	\\
G031.58+00.07	&	18:48:41.68	&	-01:09:59.00	&	$^{12}$CS	&	32	&	15 	&	0.48 	&	96.1 	$\pm$	0.02 	&	4.3 	$\pm$	0.04 	&	7.50 	$\pm$	0.06 	&	1.62 	\\
	&		&		&	C$^{34}$S	&	20	&	19 	&	0.48 	&	96.0 	$\pm$	0.03 	&	3.2 	$\pm$	0.08 	&	1.44 	$\pm$	0.03 	&	0.42 	\\
	&		&		&	$^{13}$CS	&	24	&	18 	&	0.51 	&	96.2 	$\pm$	0.10 	&	3.2 	$\pm$	0.23 	&	0.54 	$\pm$	0.03 	&	0.16 	\\
	&		&		&	C$^{33}$S	&	32	&	9 	&	0.96 	&	96.5 	$\pm$	0.21 	&	3.5 	$\pm$	0.61 	&	0.25 	$\pm$	0.03 	&	0.07 	\\
G032.04+00.05	&	18:49:36.58	&	-00:45:47	&	$^{12}$CS	&	32	&	13 	&	0.48 	&	94.6 	$\pm$	0.02 	&	6.0 	$\pm$	0.05 	&	12.91 	$\pm$	0.08 	&	2.01 	\\
	&		&		&	C$^{34}$S	&	12	&	18 	&	0.48 	&	95.3 	$\pm$	0.07 	&	5.1 	$\pm$	0.19 	&	1.56 	$\pm$	0.05 	&	0.29 	\\
	&		&		&	$^{13}$CS	&	32	&	15 	&	0.51 	&	94.9 	$\pm$	0.09 	&	5.5 	$\pm$	0.24 	&	0.67 	$\pm$	0.02 	&	0.11 	\\
	&		&		&	C$^{33}$S	&	32	&	7 	&	0.96 	&	95.4 	$\pm$	0.28 	&	3.5 	$\pm$	0.73 	&	0.18 	$\pm$	0.03 	&	0.05 	\\
G034.39+00.22	&	18:53:19.00	&	01:24:08.80	&	$^{12}$CS	&	56	&	13 	&	0.48 	&	57.3 	$\pm$	0.02 	&	5.2 	$\pm$	0.00 	&	11.41 	$\pm$	0.07 	&	2.05 	\\
	&		&		&	C$^{34}$S	&	12	&	19 	&	0.48 	&	57.3 	$\pm$	0.05 	&	4.2 	$\pm$	0.13 	&	1.64 	$\pm$	0.04 	&	0.37 	\\
	&		&		&	$^{13}$CS	&	32	&	14 	&	0.51 	&	57.4 	$\pm$	0.06 	&	3.9 	$\pm$	0.16 	&	0.68 	$\pm$	0.02 	&	0.16 	\\
	&		&		&	C$^{33}$S	&	56	&	7 	&	0.96 	&	58.1 	$\pm$	0.26 	&	3.7 	$\pm$	0.63 	&	0.18 	$\pm$	0.03 	&	0.05 	\\
G035.02+00.34	&	18:54:00.67	&	+02:01:19.20	&	$^{12}$CS	&	8	&	21 	&	0.48 	&	53.3 	$\pm$	0.01 	&	4.2 	$\pm$	0.04 	&	10.39 	$\pm$	0.07 	&	2.32 	\\
	&		&		&	C$^{34}$S	&	20	&	20 	&	0.48 	&	52.9 	$\pm$	0.03 	&	3.5 	$\pm$	0.06 	&	1.85 	$\pm$	0.03 	&	0.50 	\\
	&		&		&	$^{13}$CS	&	24	&	21 	&	0.51 	&	52.9 	$\pm$	0.12 	&	3.5 	$\pm$	0.34 	&	0.64 	$\pm$	0.05 	&	0.17 	\\
	&		&		&	C$^{33}$S	&	8	&	9 	&	0.96 	&	53.5 	$\pm$	0.25 	&	3.8 	$\pm$	0.55 	&	0.34 	$\pm$	0.05 	&	0.08 	\\
G037.43+01.51	&	18:54:14.35	&	+04:41:41.70	&	$^{12}$CS	&	32	&	13 	&	0.48 	&	44.1 	$\pm$	0.01 	&	3.3 	$\pm$	0.01 	&	13.18 	$\pm$	0.04 	&	3.73 	\\
	&		&		&	C$^{34}$S	&	20	&	18 	&	0.48 	&	44.1 	$\pm$	0.03 	&	2.8 	$\pm$	0.07 	&	1.19 	$\pm$	0.03 	&	0.39 	\\
	&		&		&	$^{13}$CS	&	24	&	14 	&	0.51 	&	44.0 	$\pm$	0.08 	&	2.8 	$\pm$	0.19 	&	0.41 	$\pm$	0.02 	&	0.14 	\\
	&		&		&	C$^{33}$S	&	32	&	6 	&	0.96 	&	44.8 	$\pm$	0.29 	&	3.3 	$\pm$	0.66 	&	0.15 	$\pm$	0.03 	&	0.04 	\\
G035.19-00.74	&	18:58:13.05	&	+01:40:35.70	&	$^{12}$CS	&	8	&	24 	&	0.48 	&	34.7 	$\pm$	0.01 	&	7.7 	$\pm$	0.03 	&	22.81 	$\pm$	0.08 	&	5.42 	\\
	&		&		&	C$^{34}$S	&	20	&	19 	&	0.48 	&	34.2 	$\pm$	0.03 	&	4.9 	$\pm$	0.07 	&	2.91 	$\pm$	0.04 	&	0.56 	\\
	&		&		&	$^{13}$CS	&	24	&	15 	&	0.51 	&	34.2 	$\pm$	0.06 	&	4.8 	$\pm$	0.15 	&	1.22 	$\pm$	0.03 	&	0.24 	\\
	&		&		&	C$^{33}$S	&	8	&	13 	&	0.96 	&	34.7 	$\pm$	0.33 	&	5.4 	$\pm$	0.97 	&	0.55 	$\pm$	0.07 	&	0.10 	\\
G035.20-01.73	&	19:01:45.54	&	+01:13:32.50	&	$^{12}$CS	&	8	&	23 	&	0.48 	&	43.2 	$\pm$	0.01 	&	5.3 	$\pm$	0.02 	&	19.83 	$\pm$	0.08 	&	3.51 	\\
	&		&		&	C$^{34}$S	&	20	&	17 	&	0.48 	&	43.5 	$\pm$	0.02 	&	4.3 	$\pm$	0.05 	&	2.72 	$\pm$	0.03 	&	0.59 	\\
	&		&		&	$^{13}$CS	&	24	&	16 	&	0.51 	&	43.6 	$\pm$	0.05 	&	4.3 	$\pm$	0.13 	&	1.11 	$\pm$	0.03 	&	0.24 	\\
	&		&		&	C$^{33}$S	&	8	&	14 	&	0.96 	&	44.4 	$\pm$	0.33 	&	4.3 	$\pm$	0.95 	&	0.43 	$\pm$	0.07 	&	0.09 	\\
G043.16+00.01	&	19:10:13.00	&	+09:06:12.80	&	$^{12}$CS	&	44	&	14 	&	0.48 	&	11.9 	$\pm$	0.04 	&	7.5 	$\pm$	0.07 	&	36.12 	$\pm$	0.65 	&	4.53 	\\
	&		&		&		&		&		&		&	3.6 	$\pm$	0.05 	&	8.5 	$\pm$	0.08 	&	27.25 	$\pm$	0.34 	&	3.01 	\\
	&		&		&	C$^{34}$S	&	8	&	18 	&	0.48 	&	12.2 	$\pm$	0.07 	&	6.8 	$\pm$	0.15 	&	5.39 	$\pm$	0.13 	&	0.75 	\\
	&		&		&		&		&		&		&	3.6 	$\pm$	0.14 	&	7.2 	$\pm$	0.32 	&	2.52 	$\pm$	0.11 	&	0.33 	\\
	&		&		&	$^{13}$CS	&	32	&	9 	&	0.51 	&	12.3 	$\pm$	0.05 	&	6.2 	$\pm$	1.01 	&	1.86 	$\pm$	0.07 	&	0.28 	\\
	&		&		&		&		&		&		&	3.5 	$\pm$	0.10 	&	6.8 	$\pm$	0.26 	&	0.95 	$\pm$	0.03 	&	0.13 	\\
	&		&		&	C$^{33}$S	&	44	&	6 	&	0.96 	&	12.7 	$\pm$	0.15 	&	7.9 	$\pm$	0.51 	&	1.20 	$\pm$	0.06 	&	0.14 	\\
	&		&		&		&		&		&		&	3.0 	$\pm$	0.17 	&	6.9 	$\pm$	0.44 	&	0.60 	$\pm$	0.04 	&	0.08 	\\
G043.79-00.12	&	19:11:53.99	&	+09:35:50.30:	&	$^{12}$CS	&	44	&	12 	&	0.48 	&	44.2 	$\pm$	0.01 	&	6.5 	$\pm$	0.02 	&	11.48 	$\pm$	0.04 	&	1.66 	\\
	&		&		&	C$^{34}$S	&	12	&	16 	&	0.48 	&	44.0 	$\pm$	0.08 	&	5.8 	$\pm$	0.19 	&	1.64 	$\pm$	0.05 	&	0.27 	\\
	&		&		&	$^{13}$CS	&	32	&	9 	&	0.51 	&	44.0 	$\pm$	0.09 	&	5.0 	$\pm$	0.19 	&	0.60 	$\pm$	0.02 	&	0.11 	\\
	&		&		&	C$^{33}$S	&	44	&	5 	&	0.96 	&	44.5 	$\pm$	0.29 	&	6.5 	$\pm$	0.75 	&	0.32 	$\pm$	0.03 	&	0.05 	\\
G045.07+00.13	&	19:13:22.04	&	+10:50:53.30	&	$^{12}$CS	&	32	&	13 	&	0.48 	&	59.1 	$\pm$	0.02 	&	5.5 	$\pm$	0.05 	&	13.35 	$\pm$	0.10 	&	2.29 	\\
	&		&		&	C$^{34}$S	&	20	&	18 	&	0.48 	&	59.3 	$\pm$	0.04 	&	5.1 	$\pm$	0.11 	&	2.02 	$\pm$	0.03 	&	0.37 	\\
	&		&		&	$^{13}$CS	&	24	&	20 	&	0.51 	&	59.1 	$\pm$	0.13 	&	5.8 	$\pm$	0.41 	&	0.98 	$\pm$	0.05 	&	0.16 	\\
	&		&		&	C$^{33}$S	&	32	&	7 	&	0.96 	&	59.9 	$\pm$	0.21 	&	4.2 	$\pm$	0.51 	&	0.29 	$\pm$	0.03 	&	0.07 	\\
G045.45+00.05	&	19:14:21.27	&	+11:09:15.90	&	$^{12}$CS	&	56	&	12 	&	0.48 	&	58.7 	$\pm$	0.01 	&	5.5 	$\pm$	0.03 	&	10.07 	$\pm$	0.04 	&	1.71 	\\
	&		&		&	C$^{34}$S	&	20	&	17 	&	0.48 	&	58.6 	$\pm$	0.05 	&	4.4 	$\pm$	0.13 	&	1.29 	$\pm$	0.03 	&	0.27 	\\
	&		&		&	$^{13}$CS	&	24	&	14 	&	0.51 	&	58.5 	$\pm$	0.13 	&	4.6 	$\pm$	0.42 	&	0.49 	$\pm$	0.03 	&	0.10 	\\
	&		&		&	C$^{33}$S	&	56	&	6 	&	0.96 	&	59.6 	$\pm$	0.33 	&	5.5 	$\pm$	0.78 	&	0.22 	$\pm$	0.03 	&	0.04 	\\
G043.89-00.78	&	19:14:26.39	&	+09:22:36.50	&	$^{12}$CS	&	44	&	10 	&	0.48 	&	54.2 	$\pm$	0.02 	&	6.5 	$\pm$	0.06 	&	7.74 	$\pm$	0.05 	&	1.12 	\\
	&		&		&	C$^{34}$S	&	16	&	13 	&	0.48 	&	54.6 	$\pm$	0.10 	&	4.6 	$\pm$	0.29 	&	0.78 	$\pm$	0.04 	&	0.16 	\\
	&		&		&	$^{13}$CS	&	32	&	15 	&	0.51 	&	54.7 	$\pm$	0.14 	&	3.4 	$\pm$	0.36 	&	0.23 	$\pm$	0.02 	&	0.06 	\\
	&		&		&	C$^{33}$S	&	44	&	10 	&	0.96 	&	54.8 	$\pm$	0.31 	&	3.6 	$\pm$	0.60 	&	0.14 	$\pm$	0.02 	&	0.04 	\\
G048.60+00.02	&	19:20:31.18	&	+13:55:25.20	&	$^{12}$CS	&	48	&	13 	&	0.48 	&	18.3 	$\pm$	0.01 	&	4.9 	$\pm$	0.03 	&	8.10 	$\pm$	0.04 	&	1.56 	\\
	&		&		&	C$^{34}$S	&	20	&	17 	&	0.48 	&	18.3 	$\pm$	0.07 	&	4.1 	$\pm$	0.17 	&	0.89 	$\pm$	0.03 	&	0.20 	\\
	&		&		&	$^{13}$CS	&	24	&	15 	&	0.51 	&	17.8 	$\pm$	0.20 	&	5.1 	$\pm$	0.58 	&	0.40 	$\pm$	0.02 	&	0.07 	\\
	&		&		&	C$^{33}$S	&	48	&	6 	&	0.96 	&	18.6 	$\pm$	0.30 	&	4.1 	$\pm$	0.69 	&	0.17 	$\pm$	0.03 	&	0.04 	\\
G049.48-00.36	&	19:23:39.82	&	14:31:05.00	&	$^{12}$CS	&	38	&	8 	&	0.48 	&	61.4 	$\pm$	0.01 	&	8.1 	$\pm$	0.01 	&	73.28 	$\pm$	0.07 	&	8.54 	\\
	&		&		&		&		&		&		&	50.4 	$\pm$	0.48 	&	3.9 	$\pm$	0.48 	&	8.21 	$\pm$	0.65 	&	1.95 	\\
	&		&		&	C$^{34}$S	&	16	&	13 	&	0.48 	&	61.6 	$\pm$	0.03 	&	7.1 	$\pm$	0.07 	&	9.63 	$\pm$	0.07 	&	1.26 	\\
	&		&		&		&		&		&		&	50.7 	$\pm$	0.10 	&	4.6 	$\pm$	0.24 	&	1.45 	$\pm$	0.23 	&	0.30 	\\
	&		&		&	$^{13}$CS	&	20	&	9 	&	0.51 	&	61.8 	$\pm$	0.03 	&	7.2 	$\pm$	0.07 	&	3.86 	$\pm$	0.03 	&	0.50 	\\
	&		&		&		&		&		&		&	50.5 	$\pm$	0.13 	&	4.8 	$\pm$	0.30 	&	0.48 	$\pm$	0.08 	&	0.10 	\\
	&		&		&	C$^{33}$S	&	38	&	6 	&	0.96 	&	62.4 	$\pm$	0.10 	&	9.9 	$\pm$	0.24 	&	2.15 	$\pm$	0.05 	&	0.20 	\\
	&		&		&		&		&		&		&	51.3 	$\pm$	0.13 	&	5.0 	$\pm$	0.31 	&	0.39 	$\pm$	0.02 	&	0.07 	\\
G049.48-00.38	&	19:23:43.87	&	+14:30:29.50	&	$^{12}$CS	&	38	&	14 	&	0.48 	&	57.1 	$\pm$	0.01 	&	10.6 	$\pm$	0.02 	&	107.25 	$\pm$	0.12 	&	9.44 	\\
	&		&		&	C$^{34}$S	&	16	&	14 	&	0.48 	&	57.3 	$\pm$	0.02 	&	9.6 	$\pm$	0.04 	&	20.01 	$\pm$	0.05 	&	1.96 	\\
	&		&		&	$^{13}$CS	&	20	&	14 	&	0.51 	&	57.1 	$\pm$	0.03 	&	9.8 	$\pm$	0.07 	&	8.96 	$\pm$	0.05 	&	0.86 	\\
	&		&		&	C$^{33}$S	&	38	&	10 	&	0.96 	&	57.5 	$\pm$	0.11 	&	11.9 	$\pm$	0.30 	&	4.89 	$\pm$	0.09 	&	0.34 	\\
G059.78+00.06	&	19:43:11.25	&	+23:44:03.30	&	$^{12}$CS	&	32	&	13 	&	0.48 	&	22.8 	$\pm$	0.01 	&	2.8 	$\pm$	0.02 	&	11.65 	$\pm$	0.05 	&	3.91 	\\
	&		&		&	C$^{34}$S	&	8	&	32 	&	0.48 	&	22.8 	$\pm$	0.02 	&	2.0 	$\pm$	0.06 	&	1.66 	$\pm$	0.03 	&	0.75 	\\
	&		&		&	$^{13}$CS	&	24	&	18 	&	0.51 	&	22.9 	$\pm$	0.04 	&	2.3 	$\pm$	0.10 	&	0.68 	$\pm$	0.02 	&	0.27 	\\
	&		&		&	C$^{33}$S	&	32	&	11 	&	0.96 	&	23.6 	$\pm$	0.09 	&	2.1 	$\pm$	0.22 	&	0.26 	$\pm$	0.02 	&	0.12 	\\
G069.54-00.97	&	20:10:09.07	&	31:31:36.00	&	$^{12}$CS	&	44	&	8 	&	0.48 	&	11.7 	$\pm$	0.01 	&	5.4 	$\pm$	0.03 	&	13.73 	$\pm$	0.05 	&	2.40 	\\
	&		&		&	C$^{34}$S	&	16	&	16 	&	0.48 	&	11.2 	$\pm$	0.06 	&	4.8 	$\pm$	0.16 	&	1.90 	$\pm$	0.04 	&	0.37 	\\
	&		&		&	$^{13}$CS	&	24	&	27 	&	0.51 	&	11.3 	$\pm$	0.12 	&	3.6 	$\pm$	0.27 	&	0.62 	$\pm$	0.04 	&	0.16 	\\
	&		&		&	C$^{33}$S	&	44	&	5 	&	0.96 	&	11.9 	$\pm$	0.16 	&	3.4 	$\pm$	0.31 	&	0.21 	$\pm$	0.02 	&	0.06 	\\
G078.12+03.63	&	20:14:26.07	&	+41:13:32.70	&	$^{12}$CS	&	88	&	7 	&	0.48 	&	-3.4 	$\pm$	0.01 	&	3.2 	$\pm$	0.03 	&	7.68 	$\pm$	0.05 	&	2.29 	\\
	&		&		&	C$^{34}$S	&	8	&	32 	&	0.48 	&	-3.6 	$\pm$	0.11 	&	2.8 	$\pm$	0.26 	&	0.57 	$\pm$	0.04 	&	0.19 	\\
	&		&		&	$^{13}$CS	&	24	&	22 	&	0.51 	&	-4.0 	$\pm$	0.36 	&	3.7 	$\pm$	0.42 	&	0.17 	$\pm$	0.04 	&	0.05 	\\
	&		&		&	C$^{33}$S	&	88	&	3 	&	0.96 	&	-3.0 	$\pm$	0.20 	&	2.9 	$\pm$	0.56 	&	0.08 	$\pm$	0.01 	&	0.03 	\\
G075.76+00.33	&	20:21:41.09	&	+37:25:29.30	&	$^{12}$CS	&	88	&	7 	&	0.48 	&	-1.1 	$\pm$	0.01 	&	5.3 	$\pm$	0.02 	&	22.59 	$\pm$	0.08 	&	4.01 	\\
	&		&		&	C$^{34}$S	&	8	&	32 	&	0.48 	&	-1.1 	$\pm$	0.03 	&	3.8 	$\pm$	0.08 	&	2.78 	$\pm$	0.05 	&	0.69 	\\
	&		&		&	$^{13}$CS	&	24	&	20 	&	0.51 	&	-1.0 	$\pm$	0.06 	&	3.4 	$\pm$	0.13 	&	0.92 	$\pm$	0.03 	&	0.25 	\\
	&		&		&	C$^{33}$S	&	88	&	4 	&	0.96 	&	-0.5 	$\pm$	0.10 	&	3.7 	$\pm$	0.27 	&	0.40 	$\pm$	0.02 	&	0.10 	\\
G075.78+00.34	&	20:21:44.01	&	37:26:37.50	&	$^{12}$CS	&	44	&	11 	&	0.48 	&	0.3 	$\pm$	0.02 	&	4.7 	$\pm$	0.05 	&	16.93 	$\pm$	0.15 	&	3.36 	\\
	&		&		&	C$^{34}$S	&	24	&	24 	&	0.48 	&	0.1 	$\pm$	0.02 	&	3.8 	$\pm$	0.08 	&	2.52 	$\pm$	0.03 	&	0.63 	\\
	&		&		&	$^{13}$CS	&	24	&	29 	&	0.51 	&	0.1 	$\pm$	0.11 	&	3.9 	$\pm$	0.28 	&	0.85 	$\pm$	0.04 	&	0.20 	\\
	&		&		&	C$^{33}$S	&	44	&	6 	&	0.96 	&	0.9 	$\pm$	0.14 	&	4.1 	$\pm$	0.49 	&	0.42 	$\pm$	0.03 	&	0.10 	\\
G074.03-01.71	&	20:25:07.11	&	+34:49:57.60	&	$^{12}$CS	&	8	&	31 	&	0.48 	&	5.9 	$\pm$	0.01 	&	3.8 	$\pm$	0.03 	&	10.83 	$\pm$	0.08 	&	2.65 	\\
	&		&		&	C$^{34}$S	&	8	&	16 	&	0.48 	&	6.2 	$\pm$	0.04 	&	2.5 	$\pm$	0.09 	&	1.21 	$\pm$	0.04 	&	0.46 	\\
	&		&		&	$^{13}$CS	&	24	&	15 	&	0.51 	&	6.1 	$\pm$	0.08 	&	3.1 	$\pm$	0.19 	&	0.55 	$\pm$	0.03 	&	0.16 	\\
	&		&		&	C$^{33}$S	&	8	&	31 	&	0.96 	&	6.8 	$\pm$	0.13 	&	1.8 	$\pm$	0.27 	&	0.24 	$\pm$	0.04 	&	0.13 	\\
G078.88+00.70	&	20:29:24.82	&	+40:11:19.60	&	$^{12}$CS	&	88	&	6 	&	0.48 	&	-5.9 	$\pm$	0.01 	&	3.9 	$\pm$	0.04 	&	15.21 	$\pm$	0.09 	&	3.68 	\\
	&		&		&	C$^{34}$S	&	20	&	19 	&	0.48 	&	-6.0 	$\pm$	0.02 	&	3.0 	$\pm$	0.06 	&	1.64 	$\pm$	0.03 	&	0.51 	\\
	&		&		&	$^{13}$CS	&	24	&	24 	&	0.51 	&	-6.0 	$\pm$	0.18 	&	3.2 	$\pm$	0.30 	&	0.48 	$\pm$	0.04 	&	0.14 	\\
	&		&		&	C$^{33}$S	&	88	&	5 	&	0.96 	&	-5.5 	$\pm$	0.12 	&	3.9 	$\pm$	0.35 	&	0.32 	$\pm$	0.02 	&	0.08 	\\
G079.87+01.17	&	20:30:29.14	&	+41:15:53.60	&	$^{12}$CS	&	40	&	10 	&	0.48 	&	-3.8 	$\pm$	0.00 	&	3.3 	$\pm$	0.01 	&	10.92 	$\pm$	0.02 	&	3.09 	\\
	&		&		&	C$^{34}$S	&	20	&	18 	&	0.48 	&	-3.8 	$\pm$	0.03 	&	2.9 	$\pm$	0.07 	&	1.33 	$\pm$	0.02 	&	0.43 	\\
	&		&		&	$^{13}$CS	&	24	&	16 	&	0.51 	&	-3.9 	$\pm$	0.07 	&	3.1 	$\pm$	0.19 	&	0.51 	$\pm$	0.03 	&	0.16 	\\
	&		&		&	C$^{33}$S	&	40	&	4 	&	0.96 	&	-3.3 	$\pm$	0.12 	&	3.2 	$\pm$	0.28 	&	0.22 	$\pm$	0.02 	&	0.07 	\\
G080.86+00.38	&	20:37:00.96	&	+41:34:55.70	&	$^{12}$CS	&	64	&	8 	&	0.48 	&	-1.3 	$\pm$	0.01 	&	3.1 	$\pm$	0.03 	&	7.04 	$\pm$	0.06 	&	2.16 	\\
	&		&		&	C$^{34}$S	&	20	&	16 	&	0.48 	&	-1.4 	$\pm$	0.05 	&	2.8 	$\pm$	0.11 	&	0.74 	$\pm$	0.02 	&	0.24 	\\
	&		&		&	$^{13}$CS	&	24	&	20 	&	0.51 	&	-1.5 	$\pm$	0.17 	&	3.0 	$\pm$	0.39 	&	0.25 	$\pm$	0.03 	&	0.08 	\\
	&		&		&	C$^{33}$S	&	64	&	4 	&	0.96 	&	-0.6 	$\pm$	0.16 	&	2.5 	$\pm$	0.33 	&	0.12 	$\pm$	0.01 	&	0.04 	\\
G081.87+00.78	&	20:38:36.43	&	+42:37:34.80	&	$^{12}$CS	&	8	&	28 	&	0.48 	&	9.8 	$\pm$	0.01 	&	4.7 	$\pm$	0.03 	&	34.71 	$\pm$	0.17 	&	7.02 	\\
	&		&		&	C$^{34}$S	&	20	&	22 	&	0.48 	&	9.7 	$\pm$	0.01 	&	3.4 	$\pm$	0.03 	&	4.37 	$\pm$	0.03 	&	1.21 	\\
	&		&		&	$^{13}$CS	&	24	&	16 	&	0.51 	&	9.6 	$\pm$	0.03 	&	3.2 	$\pm$	0.07 	&	1.51 	$\pm$	0.02 	&	0.45 	\\
	&		&		&	C$^{33}$S	&	8	&	19 	&	0.96 	&	10.4 	$\pm$	0.16 	&	3.7 	$\pm$	0.42 	&	0.75 	$\pm$	0.07 	&	0.19 	\\
G081.75+00.59	&	20:39:01.99	&	42:24:59.30	&	$^{12}$CS	&	56	&	8 	&	0.48 	&	-3.9 	$\pm$	0.01 	&	4.0 	$\pm$	0.02 	&	22.85 	$\pm$	0.10 	&	5.38 	\\
	&		&		&	C$^{34}$S	&	36	&	14 	&	0.48 	&	-3.8 	$\pm$	0.01 	&	2.7 	$\pm$	0.02 	&	3.16 	$\pm$	0.02 	&	1.09 	\\
	&		&		&	$^{13}$CS	&	44	&	19 	&	0.51 	&	-3.7 	$\pm$	0.03 	&	2.6 	$\pm$	0.08 	&	0.98 	$\pm$	0.02 	&	0.35 	\\
	&		&		&	C$^{33}$S	&	56	&	4 	&	0.96 	&	-3.2 	$\pm$	0.08 	&	2.9 	$\pm$	0.19 	&	0.50 	$\pm$	0.03 	&	0.16 	\\
G092.67+03.07	&	21:09:21.73	&	+52:22:37.10	&	$^{12}$CS	&	64	&	8 	&	0.48 	&	-6.2 	$\pm$	0.01 	&	2.8 	$\pm$	0.02 	&	12.07 	$\pm$	0.05 	&	4.11 	\\
	&		&		&	C$^{34}$S	&	20	&	20 	&	0.48 	&	-6.0 	$\pm$	0.02 	&	2.5 	$\pm$	0.05 	&	1.45 	$\pm$	0.03 	&	0.55 	\\
	&		&		&	$^{13}$CS	&	24	&	22 	&	0.51 	&	-6.1 	$\pm$	0.10 	&	2.6 	$\pm$	0.27 	&	0.42 	$\pm$	0.04 	&	0.15 	\\
	&		&		&	C$^{33}$S	&	64	&	5 	&	0.96 	&	-5.4 	$\pm$	0.10 	&	2.6 	$\pm$	0.23 	&	0.22 	$\pm$	0.02 	&	0.08 	\\
G097.53+03.18	&	21:32:12.43	&	55:53:49.70	&	$^{12}$CS	&	44	&	9 	&	0.48 	&	-71.0 	$\pm$	0.05 	&	5.9 	$\pm$	0.10 	&	5.11 	$\pm$	0.09 	&	0.82 	\\
	&		&		&	C$^{34}$S	&	28	&	5 	&	0.48 	&	-72.9 	$\pm$	0.12 	&	5.9 	$\pm$	0.27 	&	0.79 	$\pm$	0.03 	&	0.12 	\\
	&		&		&	$^{13}$CS	&	32	&	17 	&	0.51 	&	-70.7 	$\pm$	0.28 	&	4.8 	$\pm$	0.70 	&	0.25 	$\pm$	0.03 	&	0.05 	\\
	&		&		&	C$^{33}$S	&	44	&	19 	&	0.96 	&	-69.9 	$\pm$	0.35 	&	5.7 	$\pm$	1.09 	&	0.19 	$\pm$	0.03 	&	0.03 	\\
G108.59+00.49	&	22:52:38.30	&	60:00:52.00	&	$^{12}$CS	&	32	&	30 	&	0.48 	&	-51.2 	$\pm$	0.01 	&	3.4 	$\pm$	0.04 	&	8.26 	$\pm$	0.07 	&	2.28 	\\
	&		&		&	C$^{34}$S	&	28	&	20 	&	0.48 	&	-51.5 	$\pm$	0.06 	&	3.2 	$\pm$	0.15 	&	0.70 	$\pm$	0.03 	&	0.21 	\\
	&		&		&	$^{13}$CS	&	52	&	10 	&	0.51 	&	-51.1 	$\pm$	0.10 	&	2.9 	$\pm$	0.26 	&	0.19 	$\pm$	0.02 	&	0.06 	\\
	&		&		&	C$^{33}$S	&	32	&	9 	&	0.96 	&	-50.7 	$\pm$	0.58 	&	4.3 	$\pm$	1.03 	&	0.12 	$\pm$	0.03 	&	0.03 	\\
G109.87+02.11	&	22:56:18.00	&	+62:01:49.50	&	$^{12}$CS	&	8	&	23 	&	0.48 	&	-10.6 	$\pm$	0.01 	&	4.9 	$\pm$	0.03 	&	20.93 	$\pm$	0.07 	&	4.03 	\\
	&		&		&	C$^{34}$S	&	20	&	17 	&	0.48 	&	-10.3 	$\pm$	0.03 	&	3.4 	$\pm$	0.06 	&	2.11 	$\pm$	0.03 	&	0.57 	\\
	&		&		&	$^{13}$CS	&	24	&	15 	&	0.51 	&	-10.4 	$\pm$	0.07 	&	3.3 	$\pm$	0.16 	&	0.62 	$\pm$	0.03 	&	0.18 	\\
	&		&		&	C$^{33}$S	&	8	&	14 	&	0.96 	&	-9.7 	$\pm$	0.23 	&	3.8 	$\pm$	0.67 	&	0.40 	$\pm$	0.06 	&	0.10 	\\
G111.54+00.77	&	23:13:45.00	&	+61:28:10.60	&	$^{12}$CS	&	8	&	24 	&	0.48 	&	-57.1 	$\pm$	0.01 	&	4.6 	$\pm$	0.03 	&	36.33 	$\pm$	0.21 	&	7.34 	\\
	&		&		&	C$^{34}$S	&	20	&	22 	&	0.48 	&	-57.1 	$\pm$	0.02 	&	3.8 	$\pm$	0.04 	&	4.32 	$\pm$	0.04 	&	1.07 	\\
	&		&		&	$^{13}$CS	&	16	&	23 	&	0.51 	&	-57.2 	$\pm$	0.04 	&	3.7 	$\pm$	0.10 	&	1.63 	$\pm$	0.03 	&	0.42 	\\
	&		&		&	C$^{33}$S	&	8	&	15 	&	0.96 	&	-56.4 	$\pm$	0.15 	&	3.5 	$\pm$	0.38 	&	0.75 	$\pm$	0.07 	&	0.20 	\\
G111.25-00.76	&	23:16:10.00	&	59:55:28.50	&	$^{12}$CS	&	32	&	14 	&	0.48 	&	-44.4 	$\pm$	0.02 	&	3.6 	$\pm$	0.04 	&	6.20 	$\pm$	0.05 	&	1.59 	\\
	&		&		&	C$^{34}$S	&	28	&	5 	&	0.48 	&	-44.1 	$\pm$	0.06 	&	2.5 	$\pm$	0.13 	&	0.42 	$\pm$	0.02 	&	0.16 	\\
	&		&		&	$^{13}$CS	&	52	&	14 	&	0.51 	&	-44.6 	$\pm$	0.12 	&	3.4 	$\pm$	0.36 	&	0.23 	$\pm$	0.02 	&	0.06 	\\
	&		&		&	C$^{33}$S	&	32	&	14 	&	0.96 	&	-44.0 	$\pm$	0.61 	&	1.9 	$\pm$	0.31 	&	0.07 	$\pm$	0.02 	&	0.03 	\\
SGRB2	&	17:47:27.82	&	-28:23:10.4	&	$^{12}$CS	&	32	&	26 	&	0.48 	&	58.2 	$\pm$	0.15 	&	58.6 	$\pm$	0.39 	&	76.71 	$\pm$	0.32 	&	2.85 	\\
	&		&		&	C$^{34}$S	&	8	&	14 	&	0.48 	&	47.9 	$\pm$	0.21 	&	39.8 	$\pm$	0.57 	&	15.06 	$\pm$	0.17 	&	0.36 	\\
	&		&		&	$^{13}$CS	&	20	&	13 	&	0.51 	&	50.0 	$\pm$	0.26 	&	36.4 	$\pm$	0.64 	&	8.99 	$\pm$	0.13 	&	0.23 	\\
	&		&		&	C$^{33}$S	&	32	&	10 	&	0.96 	&	52.0 	$\pm$	0.51 	&	40.5 	$\pm$	1.35 	&	3.29 	$\pm$	0.08 	&	0.08 	\\
\enddata
\tablecomments{Column(1): Source name;  column(2): Right ascension (J2000);  column(3): Declination(J2000);  column(4): Molecular species;  column(5): Integration time;  column(6): The Root-Mean-Square (rms) noise value;  column(7): Corresponding channel width;  column(8): LSR velocity;  column(9): Linewidth (FWHM);  column(10): Integrated line intensity; column(11): \textbf{Peak $T_{\rm mb}$ value}.}
\end{deluxetable*}

\startlongtable
\begin{deluxetable*}{ccccccccccc}
\tabletypesize{\scriptsize}
\tablecaption{Observational parameters of CS and three of its rare isotopologues measured by the IRAM 30-m telescope. \label{tab:table}}
\tablehead{
\colhead{Source} & \colhead{R.A.} & \colhead{Dec}  & \colhead{Molecule} & \colhead{time} & \colhead{r.m.s} &  \colhead{$\Delta{V}$}&  \colhead{$v_{\rm LSR}$}& \colhead{$\Delta$$v_{1/2}$}  & \colhead{$ \int{T_{\rm mb}{\rm{d}} v}$}  & \colhead{$T_{\rm mb}$}  \\ & (hh:mm:ss)& (dd:mm:ss)  & & (min) & (mK) & ($\rm km \, \rm s^{-1}$) & ($\rm km \, \rm s^{-1}$)& ($\rm km \, \rm s^{-1}$) &(${\rm K} \, {\rm km} \, {\rm s}^{-1}$) &  (K) }
\decimalcolnumbers
\startdata
G035.19-00.74	&	18 58 12.62	&	+01 40 50.5	&	$^{12}$CS	&	18	&	17	&	0.60 	&	33.6 	$\pm$	0.12 	&	6.84 	$\pm$	0.09 	&	28.15 	$\pm$	0.33 	&	3.59 	\\
	&		&		&	C$^{34}$S	&	18	&	29	&	0.60 	&	33.7 	$\pm$	0.11 	&	4.06 	$\pm$	0.13 	&	4.33 	$\pm$	0.12 	&	1.00 	\\
	&		&		&	$^{13}$CS	&	18	&	16	&	0.60 	&	33.6 	$\pm$	0.10 	&	3.77 	$\pm$	0.13 	&	1.67 	$\pm$	0.15 	&	0.42 	\\
	&		&		&	C$^{33}$S	&	18	&	17	&	0.60 	&	34.3 	$\pm$	0.13 	&	3.67 	$\pm$	0.29 	&	0.77 	$\pm$	0.05 	&	0.20 	\\
G049.48-00.36	&	19 23 39.82	&	14 31 05	&	$^{12}$CS	&	18	&	32	&	0.60 	&	61.2 	$\pm$	0.14 	&	7.09 	$\pm$	0.01 	&	96.13 	$\pm$	0.14 	&	12.76 	\\
	&		&		&	C$^{34}$S	&	18	&	40	&	0.60 	&	61.2 	$\pm$	0.11 	&	6.86 	$\pm$	0.11 	&	13.39 	$\pm$	0.18 	&	1.84 	\\
	&		&		&	$^{13}$CS	&	18	&	14	&	0.60 	&	61.2 	$\pm$	0.04 	&	7.35 	$\pm$	0.09 	&	5.43 	$\pm$	0.06 	&	0.69 	\\
	&		&		&	C$^{33}$S	&	18	&	16	&	0.60 	&	61.8 	$\pm$	0.10 	&	8.44 	$\pm$	0.30 	&	2.92 	$\pm$	0.08 	&	0.32 	\\
G133.94+01.06	&	2 27 03.82	&	+61 52 25.20	&	$^{12}$CS	&	18	&	22	&	0.60 	&	-47.4 	$\pm$	0.00 	&	5.94 	$\pm$	0.00 	&	54.78 	$\pm$	0.08 	&	8.67 	\\
	&		&		&	C$^{34}$S	&	18	&	53	&	0.60 	&	-46.9 	$\pm$	0.03 	&	4.59 	$\pm$	0.11 	&	11.45 	$\pm$	0.22 	&	2.34 	\\
	&		&		&	$^{13}$CS	&	18	&	20	&	0.60 	&	-46.8 	$\pm$	0.03 	&	4.20 	$\pm$	0.08 	&	4.09 	$\pm$	0.06 	&	0.92 	\\
	&		&		&	C$^{33}$S	&	18	&	22	&	0.60 	&	-46.1 	$\pm$	0.10 	&	4.69 	$\pm$	0.27 	&	2.40 	$\pm$	0.11 	&	0.48 	\\
G031.28+00.06	&	18 48 12.39	&	-01 26 30.70	&	$^{12}$CS	&	38	&	25	&	0.60 	&	109.3 	$\pm$	0.03 	&	5.64 	$\pm$	0.04 	&	25.10 	$\pm$	0.14 	&	4.20 	\\
	&		&		&	C$^{34}$S	&	38	&	30	&	0.60 	&	109.1 	$\pm$	0.03 	&	4.38 	$\pm$	0.08 	&	8.65 	$\pm$	0.13 	&	1.86 	\\
	&		&		&	$^{13}$CS	&	38	&	13	&	0.60 	&	109.1 	$\pm$	0.02 	&	3.83 	$\pm$	0.05 	&	3.34 	$\pm$	0.04 	&	0.82 	\\
	&		&		&	C$^{33}$S	&	38	&	13	&	0.60 	&	109.7 	$\pm$	0.07 	&	3.90 	$\pm$	0.19 	&	1.73 	$\pm$	0.07 	&	0.42 	\\
G043.89-00.78	&	19 14 26.39	&	+09 22 36.50	&	$^{12}$CS	&	18	&	24	&	0.60 	&	53.3 	$\pm$	0.03 	&	10.63 	$\pm$	0.08 	&	17.73 	$\pm$	0.10 	&	1.52 	\\
	&		&		&	C$^{34}$S	&	9	&	20	&	0.60 	&	54.5 	$\pm$	0.08 	&	4.63 	$\pm$	0.23 	&	1.81 	$\pm$	0.07 	&	0.37 	\\
	&		&		&	$^{13}$CS	&	18	&	16	&	0.60 	&	54.9 	$\pm$	0.23 	&	5.42 	$\pm$	0.58 	&	0.61 	$\pm$	0.05 	&	0.11 	\\
	&		&		&	C$^{33}$S	&	9	&	72	&	0.60 	&	54.6 	$\pm$	0.84 	&	10.45 	$\pm$	2.21 	&	0.48 	$\pm$	0.08 	&	0.04 	\\
G059.78+00.06	&	19 43 11.00	&	+23 44 03.30	&	$^{12}$CS	&	18	&	23	&	0.60 	&	22.7 	$\pm$	0.00 	&	3.12 	$\pm$	0.01 	&	28.42 	$\pm$	0.05 	&	8.56 	\\
	&		&		&	C$^{34}$S	&	18	&	27	&	0.60 	&	22.9 	$\pm$	0.02 	&	2.46 	$\pm$	0.05 	&	4.90 	$\pm$	0.07 	&	1.87 	\\
	&		&		&	$^{13}$CS	&	18	&	15	&	0.60 	&	22.9 	$\pm$	0.02 	&	2.06 	$\pm$	0.04 	&	1.83 	$\pm$	0.03 	&	0.84 	\\
	&		&		&	C$^{33}$S	&	18	&	20	&	0.60 	&	23.6 	$\pm$	0.06 	&	2.25 	$\pm$	0.16 	&	0.85 	$\pm$	0.05 	&	0.35 	\\
\enddata
\tablecomments{Column(1): Source name;  column(2): Right ascension (J2000);  column(3): Declination(J2000);  column(4): Molecular species;  column(5): Integration time;  column(6): The Root-Mean-Square (rms) noise value;  column(7): Corresponding channel width;  column(8): LSR velocity;  column(9): Linewidth (FWHM);  column(10): Integrated line intensity; column(11): \textbf{Peak $T_{\rm mb}$ value}.}
\end{deluxetable*}

\startlongtable
\begin{deluxetable*}{cccccccccc}
\tabletypesize{}
\tablecaption{Sulfur isotope ratios and CS, C$^{34}$S and $^{13}$CS optical depths  \label{tab:table}}
\tablehead{
\colhead{Source} & \colhead{$D_{\rm Sun}$} & \colhead{$D_{\rm GC}$} & \colhead{$^{12}$C/$^{13}$C\tablenotemark{a}} & \colhead{$\tau_{\rm max}(^{12}\rm CS)$\tablenotemark{b}} & \colhead{$\tau_{\rm max}(^{13}\rm CS)$\tablenotemark{b}}& \colhead{$\tau_{\rm max}(\rm C^{34}S)$\tablenotemark{b}}  & \colhead{$^{32}\rm S/^{34}\rm S$\tablenotemark{c}} & \colhead{$^{34}\rm S/^{33}\rm S$\tablenotemark{d}} & \colhead{$T_{k}$} \\ &(kpc)& (kpc) &&&&&&& (K) }
\startdata
G121.29+00.65	&	0.93	&	8.6	&	56 	$\pm$	12 	&	3.4 	&	0.059 	&	0.155 	&	22.7 	$\pm$	6.0 	&	5.6 	$\pm$	0.7 	&	25.6 	\\
G123.06-06.30	&	2.82	&	10	&	63 	$\pm$	13 	&	2.5 	&	0.039 	&	0.115 	&	21.0 	$\pm$	6.3 	&	6.6 	$\pm$	1.0 	&	21.8 	\\
G133.94+01.06	&	1.95	&	9.6	&	61 	$\pm$	12 	&	4.7 	&	0.076 	&	0.211 	&	23.5 	$\pm$	5.5 	&	5.0 	$\pm$	0.4 	&	28.0 	\\
 &       &        &           &           &           &   &   21.6  $\pm$	5.2\tablenotemark{e}   &   4.8  $\pm$ 0.1\tablenotemark{e} \\
G209.00-19.38	&	0.42	&	8.5	&	55 	$\pm$	11 	&	2.0 	&	0.035 	&	0.104 	&	20.2 	$\pm$	5.1 	&	5.8 	$\pm$	0.5 	&	36.2 	\\
G183.72-03.66	&	1.59	&	9.7	&	61 	$\pm$	13 	&	1.7 	&	0.028 	&	0.118 	&	28.4 	$\pm$	10.6 	&	7.5 	$\pm$	1.7 	&	21.2 	\\
G188.94+00.88	&	2.1	&	10.2	&	64 	$\pm$	13 	&	2.5 	&	0.038 	&	0.124 	&	20.7 	$\pm$	5.0 	&	5.7 	$\pm$	0.5 	&	28.6 	\\
G188.79+01.03	&	2.02	&	10.1	&	63 	$\pm$	13 	&	3.6 	&	0.057 	&	0.196 	&	20.1 	$\pm$	5.5 	&	6.6 	$\pm$	1.1 	&	29.3 	\\
G192.60-00.04	&	1.59	&	9.7	&	61 	$\pm$	13 	&	6.0 	&	0.097 	&	0.292 	&	21.6 	$\pm$	5.1 	&	10.5 	$\pm$	0.9 	&	32.3 	\\
G236.81+01.98	&	3.36	&	10.4	&	65 	$\pm$	13 	&	4.4 	&	0.069 	&	0.225 	&	23.4 	$\pm$	9.6 	&	3.5 	$\pm$	0.9 	&	22.4 	\\
G005.88-00.39	&	2.99	&	5.2	&	38 	$\pm$	9 	&	3.1 	&	0.079 	&	0.188 	&	15.9 	$\pm$	4.3 	&	6.0 	$\pm$	0.4 	&	39.7 	\\
G009.62+00.19	&	5.15	&	3.2	&	28 	$\pm$	7 	&	3.4 	&	0.113 	&	0.267 	&	13.0 	$\pm$	3.9 	&	4.8 	$\pm$	0.4 	&	33.7 	\\
G010.62-00.38	&	4.95	&	3.4	&	29 	$\pm$	8 	&	4.0 	&	0.127 	&	0.306 	&	12.3 	$\pm$	3.5 	&	5.2 	$\pm$	0.2 	&	98.0 	\\
G012.88+00.48	&	2.5	&	5.7	&	41 	$\pm$	9 	&	11.5 	&	0.271 	&	0.627 	&	19.2 	$\pm$	5.4 	&	4.8 	$\pm$	0.3 	&	25.8 	\\
G012.68-00.18	&	2.4	&	5.8	&	41 	$\pm$	9 	&	7.1 	&	0.165 	&	0.338 	&	18.6 	$\pm$	5.6 	&	5.5 	$\pm$	0.6 	&	25.2 	\\
G011.91-00.61	&	3.37	&	4.9	&	37 	$\pm$	9 	&	4.2 	&	0.109 	&	0.374 	&	14.3 	$\pm$	4.9 	&	3.7 	$\pm$	0.5 	&	21.4 	\\
G012.80-00.20	&	2.92	&	5.3	&	39 	$\pm$	9 	&	5.2 	&	0.129 	&	0.297 	&	16.2 	$\pm$	3.9 	&	5.5 	$\pm$	0.2 	&	20.3 	\\
G012.90-00.26	&	2.45	&	5.8	&	41 	$\pm$	9 	&	4.5 	&	0.106 	&	0.349 	&	12.2 	$\pm$	3.8 	&	8.6 	$\pm$	1.2 	&	26.7 	\\
G013.87+00.28	&	3.94	&	4.4	&	34 	$\pm$	8 	&	6.1 	&	0.170 	&	0.392 	&	14.3 	$\pm$	4.3 	&	4.6 	$\pm$	0.3 	&	53.5 	\\
G011.49-01.48	&	1.25	&	6.9	&	47 	$\pm$	10 	&	1.4 	&	0.028 	&	0.071 	&	12.5 	$\pm$	6.0 	&	1.7 	$\pm$	0.2 	&	24.9 	\\
G014.33-00.64	&	1.12	&	7	&	47 	$\pm$	10 	&	8.7 	&	0.178 	&	0.537 	&	17.4 	$\pm$	4.7 	&	5.6 	$\pm$	0.4 	&	25.5 	\\
G014.63-00.57	&	1.83	&	6.4	&	44 	$\pm$	10 	&	4.0 	&	0.088 	&	0.237 	&	17.3 	$\pm$	6.1 	&	4.5 	$\pm$	0.6 	&	22.9 	\\
G016.58-00.05	&	3.58	&	4.8	&	36 	$\pm$	8 	&	2.5 	&	0.067 	&	0.218 	&	12.7 	$\pm$	4.1 	&	6.0 	$\pm$	0.8 	&	26.6 	\\
G023.43-00.18	&	5.88	&	3.6	&	30 	$\pm$	8 	&	3.7 	&	0.118 	&	0.308 	&	11.3 	$\pm$	3.7 	&	5.6 	$\pm$	0.7 	&		\\
G027.36-00.16	&	8	&	3.8	&	31 	$\pm$	8 	&	1.2 	&	0.035 	&	0.109 	&	11.0 	$\pm$	3.6 	&	4.7 	$\pm$	0.6 	&	28.4 	\\
G028.86+00.06	&	7.41	&	3.9	&	32 	$\pm$	8 	&	4.1 	&	0.124 	&	0.326 	&	13.5 	$\pm$	4.1 	&	6.1 	$\pm$	0.7 	&	35.8 	\\
G029.86-00.04	&	6.21	&	4.1	&	33 	$\pm$	8 	&	1.0 	&	0.029 	&	0.093 	&	10.0 	$\pm$	3.4 	&	8.1 	$\pm$	2.2 	&	23.1 	\\
G029.95-00.01	&	5.26	&	4.4	&	34 	$\pm$	8 	&	4.2 	&	0.117 	&	0.292 	&	14.6 	$\pm$	4.0 	&	5.7 	$\pm$	0.2 	&	35.8 	\\
G031.28+00.06	&	4.27	&	5	&	37 	$\pm$	9 	&	6.5 	&	0.167 	&	0.437 	&	14.6 	$\pm$	4.0 	&	6.3 	$\pm$	0.4 	&	33.3 	\\
&       &          &         &           &           &   &   14.4  $\pm$	3.7\tablenotemark{e}   &   5.0 $\pm$ 0.1\tablenotemark{e} \\
G031.58+00.07	&	4.9	&	4.7	&	36 	$\pm$	8 	&	3.9 	&	0.103 	&	0.304 	&	13.3 	$\pm$	4.3 	&	5.7 	$\pm$	0.9 	&	31.3 	\\
G032.04+00.05	&	5.18	&	4.6	&	35 	$\pm$	8 	&	2.5 	&	0.054 	&	0.140 	&	15.2 	$\pm$	5.4 	&	8.8 	$\pm$	1.7 	&	30.4 	\\
G034.39+00.22	&	1.56	&	6.9	&	47 	$\pm$	10 	&	4.1 	&	0.084 	&	0.202 	&	19.6 	$\pm$	5.4 	&	9.0 	$\pm$	1.5 	&	22.9 	\\
G035.02+00.34	&	2.33	&	6.4	&	44 	$\pm$	10 	&	3.6 	&	0.079 	&	0.245 	&	15.4 	$\pm$	4.8 	&	5.4 	$\pm$	0.8 	&	31.2 	\\
G037.43+01.51	&	1.88	&	6.7	&	46 	$\pm$	10 	&	1.3 	&	0.028 	&	0.083 	&	15.9 	$\pm$	4.8 	&	7.8 	$\pm$	1.5 	&	28.8 	\\
G035.19-00.74	&	2.19	&	6.5	&	45 	$\pm$	10 	&	3.9 	&	0.085 	&	0.213 	&	18.8 	$\pm$	4.8 	&	5.3 	$\pm$	0.8 	&	24.2 	\\
&       &          &         &           &           &   &   17.3  $\pm$	5.7\tablenotemark{e}   &   5.6 $\pm$ 0.2\tablenotemark{e} \\
G035.20-01.73	&	3.27	&	5.8	&	41 	$\pm$	9 	&	2.9 	&	0.069 	&	0.175 	&	16.9 	$\pm$	4.4 	&	6.4 	$\pm$	1.1 	&	20.0 	\\
G043.16+00.01	&	11.1	&	7.6	&	50 	$\pm$	11 	&	3.1 	&	0.061 	&	0.168 	&	17.5 	$\pm$	4.8 	&	4.5 	$\pm$	0.3 	&	35.0 	\\
G043.79-00.12	&	6.02	&	5.6	&	40 	$\pm$	9 	&	2.7 	&	0.066 	&	0.161 	&	14.8 	$\pm$	4.3 	&	5.1 	$\pm$	0.6 	&	38.2 	\\
G045.07+00.13	&	8	&	6.2	&	43 	$\pm$	9 	&	3.0 	&	0.068 	&	0.166 	&	21.1 	$\pm$	6.2 	&	6.9 	$\pm$	0.8 	&	51.6 	\\
G045.45+00.05	&	8.4	&	6.4	&	44 	$\pm$	10 	&	2.6 	&	0.057 	&	0.163 	&	17.0 	$\pm$	5.3 	&	5.8 	$\pm$	0.8 	&	34.3 	\\
G043.89-00.78	&	8.26	&	6.1	&	43 	$\pm$	9 	&	2.4 	&	0.055 	&	0.143 	&	12.5 	$\pm$	4.5 	&	5.7 	$\pm$	1.2 	&	20.7 	\\
&       &         &          &           &           &   &   14.5  $\pm$	5.0\tablenotemark{e}   &   3.8 $\pm$ 0.5\tablenotemark{e} \\
G048.60+00.02	&	10.75	&	8.1	&	53 	$\pm$	11 	&	3.0 	&	0.056 	&	0.159 	&	24.1 	$\pm$	7.0 	&	5.1 	$\pm$	0.9 	&	35.4 	\\
G049.48-00.36	&	5.13	&	6.2	&	43 	$\pm$	9 	&	2.4 	&	0.055 	&	0.145 	&	17.4 	$\pm$	4.1 	&	4.5 	$\pm$	0.1 	&	42.9 	\\
&       &         &          &           &           &   &   17.6  $\pm$	4.3\tablenotemark{e}   &   4.6 $\pm$ 0.1\tablenotemark{e} \\
G049.48-00.38	&	5.41	&	6.2	&	43 	$\pm$	9 	&	4.2 	&	0.094 	&	0.230 	&	19.4 	$\pm$	4.5 	&	4.1 	$\pm$	0.1 	&	30.0 	\\
G059.78+00.06	&	2.16	&	7.3	&	49 	$\pm$	10 	&	4.1 	&	0.081 	&	0.241 	&	19.9 	$\pm$	5.4 	&	6.4 	$\pm$	0.7 	&	22.1 	\\
&       &       &            &           &           &   &   18.3  $\pm$	4.5\tablenotemark{e}   &   5.8 $\pm$ 0.2\tablenotemark{e} \\
G069.54-00.97	&	2.46	&	7.6	&	50 	$\pm$	11 	&	3.6 	&	0.069 	&	0.169 	&	16.5 	$\pm$	4.9 	&	9.0 	$\pm$	1.0 	&	24.7 	\\
G078.12+03.63	&	1.64	&	8	&	53 	$\pm$	11 	&	0.3 	&	0.004 	&	0.017 	&	16.0 	$\pm$	7.9 	&	6.9 	$\pm$	1.5 	&	23.1 	\\
G075.76+00.33	&	3.51	&	8	&	53 	$\pm$	11 	&	3.3 	&	0.063 	&	0.182 	&	17.5 	$\pm$	4.6 	&	7.0 	$\pm$	0.5 	&		\\
G075.78+00.34	&	3.83	&	8.1	&	53 	$\pm$	11 	&	3.3 	&	0.062 	&	0.206 	&	17.9 	$\pm$	4.9 	&	6.0 	$\pm$	0.5 	&		\\
G074.03-01.71	&	1.59	&	7.8	&	51 	$\pm$	11 	&	3.3 	&	0.061 	&	0.182 	&	23.3 	$\pm$	7.0 	&	5.1 	$\pm$	0.9 	&	18.0 	\\
G078.88+00.70	&	3.33	&	8.2	&	54 	$\pm$	11 	&	1.5 	&	0.029 	&	0.109 	&	15.5 	$\pm$	4.8 	&	5.1 	$\pm$	0.4 	&	45.2 	\\
G079.87+01.17	&	1.61	&	8	&	53 	$\pm$	11 	&	2.6 	&	0.048 	&	0.136 	&	20.3 	$\pm$	5.7 	&	6.0 	$\pm$	0.6 	&	36.9 	\\
G080.86+00.38	&	1.46	&	8	&	53 	$\pm$	11 	&	1.5 	&	0.027 	&	0.087 	&	17.5 	$\pm$	6.3 	&	6.4 	$\pm$	1.0 	&	32.1 	\\
G081.87+00.78	&	1.3	&	8.1	&	53 	$\pm$	11 	&	3.3 	&	0.063 	&	0.180 	&	18.3 	$\pm$	4.2 	&	5.8 	$\pm$	0.6 	&	20.9 	\\
G081.75+00.59	&	1.5	&	8.1	&	53 	$\pm$	11 	&	3.5 	&	0.065 	&	0.219 	&	16.5 	$\pm$	4.0 	&	6.3 	$\pm$	0.4 	&		\\
G092.67+03.07	&	1.63	&	8.4	&	55 	$\pm$	11 	&	1.5 	&	0.028 	&	0.103 	&	15.8 	$\pm$	4.9 	&	6.7 	$\pm$	0.6 	&	28.9 	\\
G097.53+03.18	&	7.52	&	11.7	&	71 	$\pm$	14 	&	4.3 	&	0.061 	&	0.161 	&	23.1 	$\pm$	8.5 	&	4.2 	$\pm$	0.8 	&		\\
G108.59+00.49	&	2.51	&	9.2	&	59 	$\pm$	12 	&	1.2 	&	0.020 	&	0.069 	&	16.3 	$\pm$	5.4 	&	6.0 	$\pm$	1.8 	&	25.0 	\\
G109.87+02.11	&	0.7	&	8.4	&	55 	$\pm$	11 	&	2.2 	&	0.040 	&	0.134 	&	15.9 	$\pm$	4.3 	&	5.3 	$\pm$	0.8 	&	29.9 	\\
G111.54+00.77	&	2.65	&	9.4	&	60 	$\pm$	12 	&	3.4 	&	0.056 	&	0.150 	&	22.5 	$\pm$	5.3 	&	5.8 	$\pm$	0.6 	&	30.8 	\\
G111.25-00.76	&	3.4	&	9.9	&	62 	$\pm$	13 	&	2.3 	&	0.037 	&	0.098 	&	33.8 	$\pm$	11.3 	&	6.0 	$\pm$	2.3 	&	27.7 	\\
SGRB2	&	7.75	&	0	&	12 	$\pm$	7 	&	3.8 	&	0.006 	&	0.009 	&	7.1 	$\pm$	4.1 	&	4.6 	$\pm$	0.2 	&	51.0 	\\		
\enddata

\tablenotetext{a}{The $^{12}$C/$^{13}$C ratios were calculated using equations (1) and (3), based on the recent measurements by \citet{2019ApJ...877..154Y}); the applied propagated error is $\sigma$ = ($1.1^2$$\rm D_{GC}(kpc)^2$ + $6.6^2$)$^{1/2}$. }
\tablenotetext{b}{The estimated peak optical depths of $^{12}\rm CS$, C$^{34}$S and $^{13}\rm CS$ (see Eqs. 6, 7 and 8).}
\tablenotetext{c}{The $^{32}\rm S/^{34}\rm S$ isotope ratios from the integrated $^{13}$CS/C$^{34}$S line intensity ratios and the $^{12}$C/$^{13}$C isotope ratios (see Eqs. 2).}
\tablenotetext{d}{The $^{34}\rm S/^{33}\rm S$ ratios are obtained from the integrated line intensity ratios of C$^{34}$S and C$^{33}$S $J$$=$2$-$1 (see Eqs. 4).}
\tablenotetext{e}{Results from the IRAM 30-m telescope for comparison with the ARO 12-m data.}
\end{deluxetable*}

\begin{deluxetable*}{ccccccc}
\tablecaption{Estimated source sizes of those 6 sources with both ARO 12-m and IRAM 30-m observations\tablenotemark{a} \label{tab:table}}
\tablehead{ \colhead{Source} & \colhead{Beam size(ARO)} & \colhead{Beam size(IRAM)} & \colhead{Source size} & \colhead{Source size} & \colhead{Source size}& \colhead{Source size} \\ \colhead{} & \colhead{$\arcsec$} & \colhead{$\arcsec$} & \colhead{$\arcsec$} & \colhead{$\arcsec$} & \colhead{$\arcsec$} & \colhead{$\arcsec$}\\\colhead{} & \colhead{} & \colhead{} & \colhead{$\rm CS$ J$=$2$-$1} & \colhead{$\rm C^{34}S$ J$=$2$-$1} & \colhead{$\rm ^{13}CS$ J$=$2$-$1} & \colhead{$\rm C^{33}S$ J$=$2$-$1}}
\startdata
G035.19-00.74	&	63	&	25	&	107 	$\pm$	80 	&	60 	$\pm$	29 	&	62 	$\pm$	37 	&	50 	$\pm$	28 	\\
G049.48-00.36	&	63	&	25	&	78 	$\pm$	40 	&	82 	$\pm$	48 	&	88 	$\pm$	56 	&	71 	$\pm$	40 	\\
G133.94+01.06	&	63	&	25	&	79 	$\pm$	41 	&	50 	$\pm$	21 	&	49 	$\pm$	21 	&	39 	$\pm$	18 	\\
G031.28+00.06	&	63	&	25	&	39 	$\pm$	14 	&	29 	$\pm$	13 	&	28 	$\pm$	13 	&	24 	$\pm$	14 	\\
G043.89-00.78	&	63	&	25	&	94 	$\pm$	63 	&	44 	$\pm$	23 	&	66 	$\pm$	65 	&	126 	$\pm$	88 	\\
G059.78+00.06	&	63	&	25	&	47 	$\pm$	17 	&	40 	$\pm$	17 	&	31 	$\pm$	15 	&	32 	$\pm$	17 	\\
\enddata
\tablenotetext{a}{\textbf{Note that CS source sizes may be affected by optical depths beyond unity, while the C$^{33}$S source sizes are affected by lowest signal-to-noise ratios among the four isotopologues. Source size errors were derived from error propagation including the nominal uncertainties in $T$$\rm _{mb}$  and a 15$\%$ uncertainty in relative calibration.}}

\end{deluxetable*}

\begin{deluxetable*}{cccc}
\tablecaption{\textbf{Average $^{32}\rm S/^{34}\rm S$ values between galactocentric distances of 6.5\,kpc - 9.5\,kpc calculated by different sets of $^{12}$C/$^{13}$C values from different molecules.}}
\tablehead{ \colhead{Molecules} & \colhead{This work} & \colhead{This work and Chin et al. 1996} & \colhead{This work and Chin et al. 1996\tablenotemark{a}} \\ \colhead{} &  \colhead{7.6\,kpc\tablenotemark{b}}  & \colhead{7.9\,kpc\tablenotemark{b}}  & \colhead{7.8\,kpc\tablenotemark{b}} }
\startdata
CN\tablenotemark{c}	&	21.0 	$\pm$	4.5 	(22.0 $\pm$ 4.9)\tablenotemark{e}	&	21.9 	$\pm$	3.6 	(22.3 $\pm$ 3.7)	&	21.7 	$\pm$	3.7 	(22.3 $\pm$ 3.9)	\\
C$^{18}$O\tablenotemark{c}	&	21.7 	$\pm$	4.2 	(22.5 $\pm$ 4.9)	&	22.8 	$\pm$	3.7 	(23.1 $\pm$ 3.8)	&	22.5 	$\pm$	3.8 	(23.0 $\pm$ 4.0)	\\
H$_2$CO\tablenotemark{c}	&	27.4 	$\pm$	5.7 	(28.6 $\pm$ 6.1)	&	28.7 	$\pm$	4.7 	(29.2 $\pm$ 4.9)	&	28.4 	$\pm$	4.8 	(29.2 $\pm$ 5.0)	\\
H$_2$CO\tablenotemark{d}	&	18.3 	$\pm$	2.8 	(19.1 $\pm$ 3.1)	&	19.1 	$\pm$	3.1 	(19.4 $\pm$ 3.2)	&	18.9 	$\pm$	3.1 	(19.4$\pm$3.3)	\\
\enddata
\tablecomments{\textbf{The error represents the standard deviations of the mean.}}
\tablenotetext{a}{\textbf{A subsample with $\tau$(C$^{34}$S) $<$ 0.25.}}
\tablenotetext{b}{\textbf{Average galactocentric distance value.}}
\tablenotetext{c}{\textbf{From \citet{2005ApJ...634.1126M}}}
\tablenotetext{d}{\textbf{From \citet{2019ApJ...877..154Y}}}
\tablenotetext{e}{\textbf{Values in parenthesis were adjusted to a galactocentric distance of 8.122 kpc by using the gradients obtained from the respective entire sample of sources.} }
\end{deluxetable*}

\begin{figure*}[h]
\center
 \label{f_list_1}
  \includegraphics[width=85.5pt]{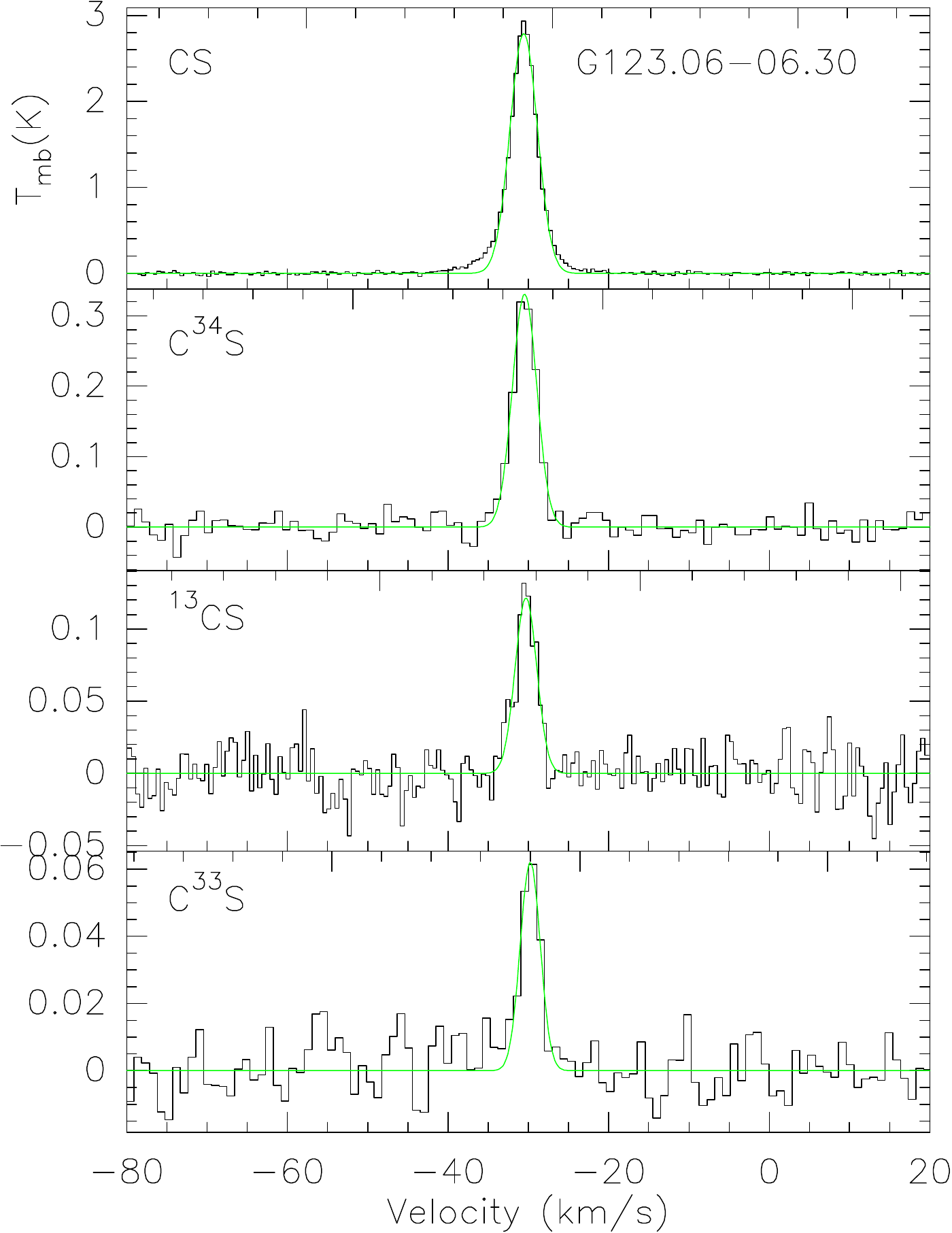}
  \includegraphics[width=81pt]{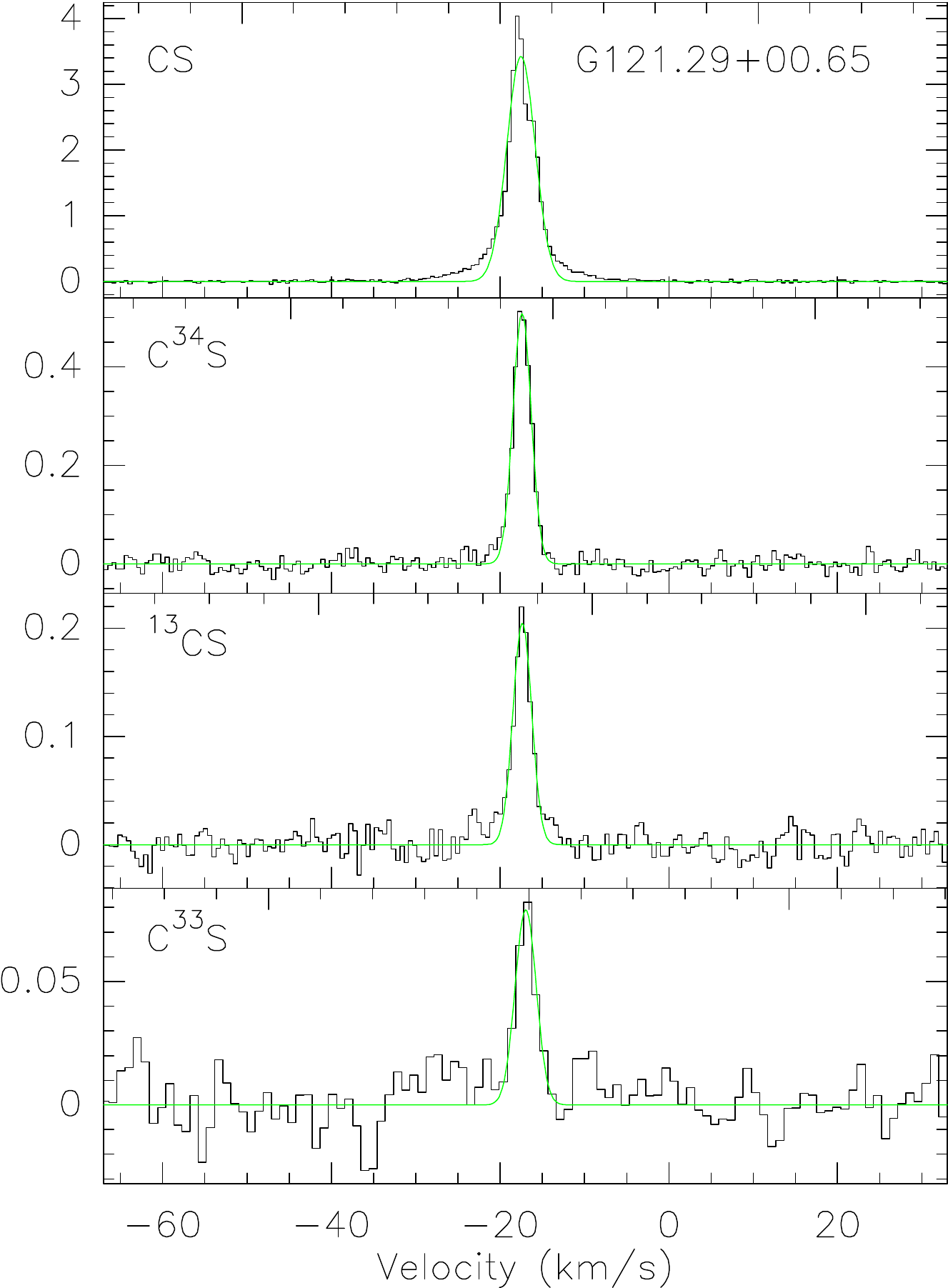}
  \includegraphics[width=79pt]{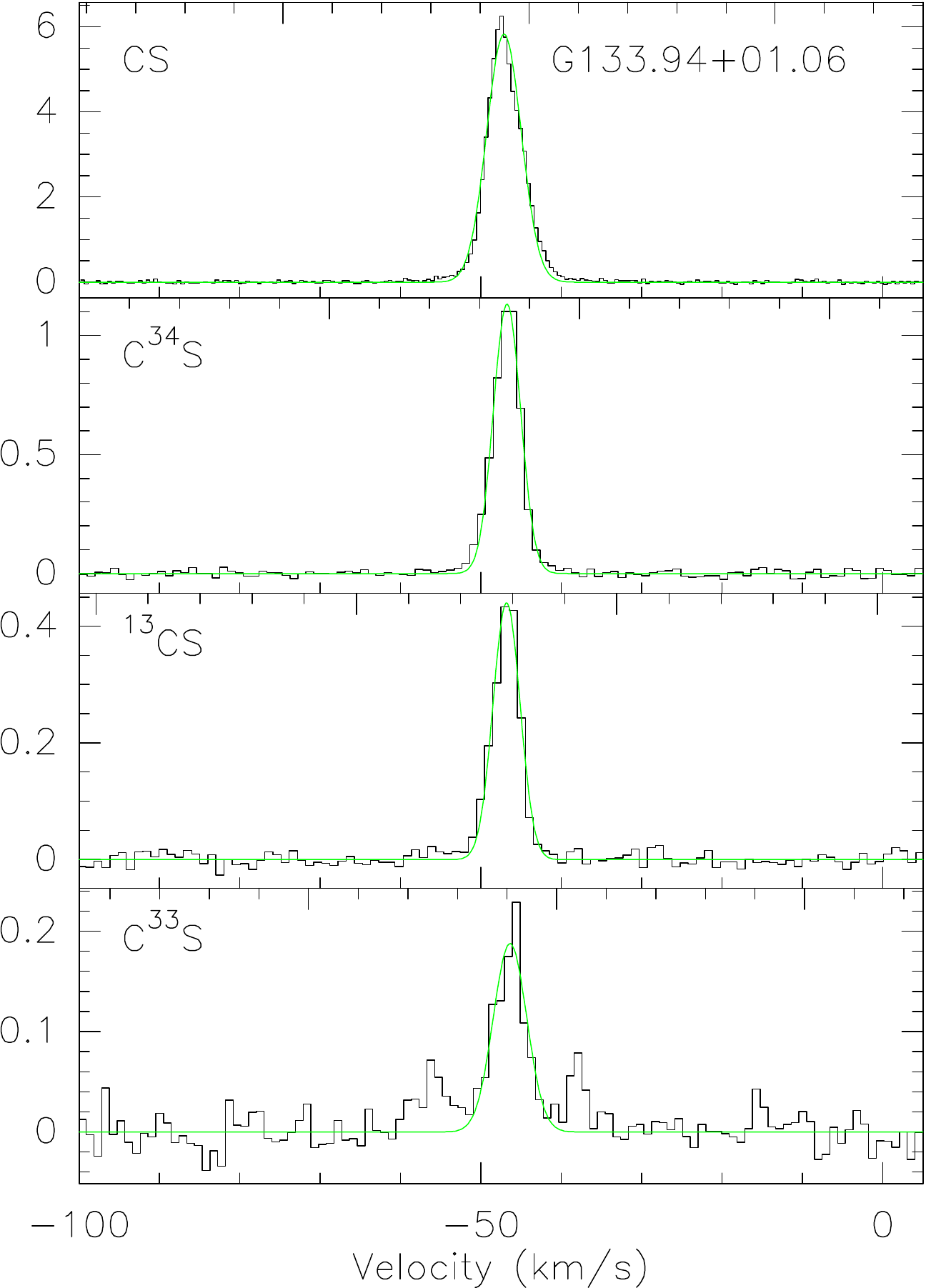}
  \includegraphics[width=83pt]{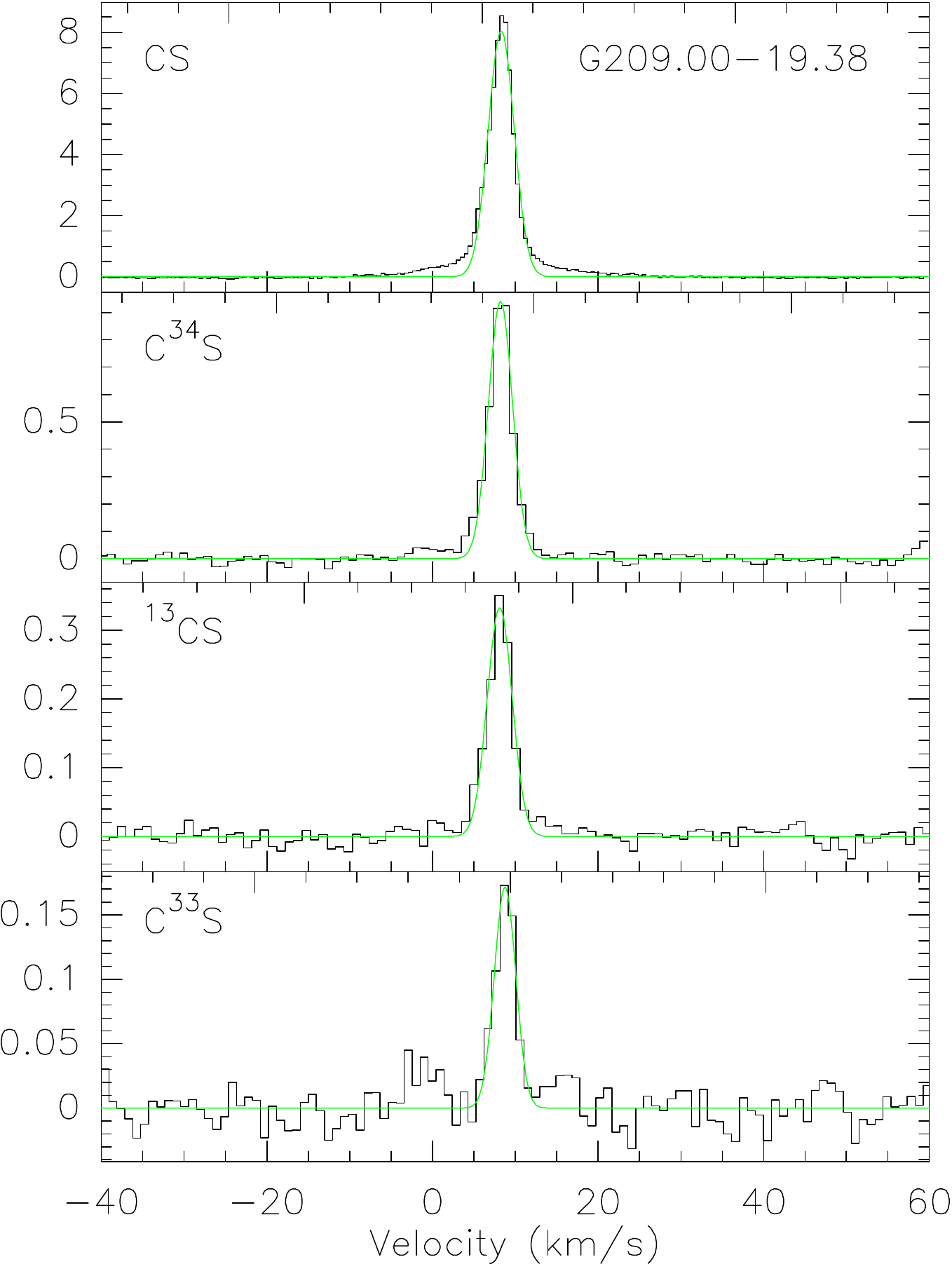}
  \includegraphics[width=81pt]{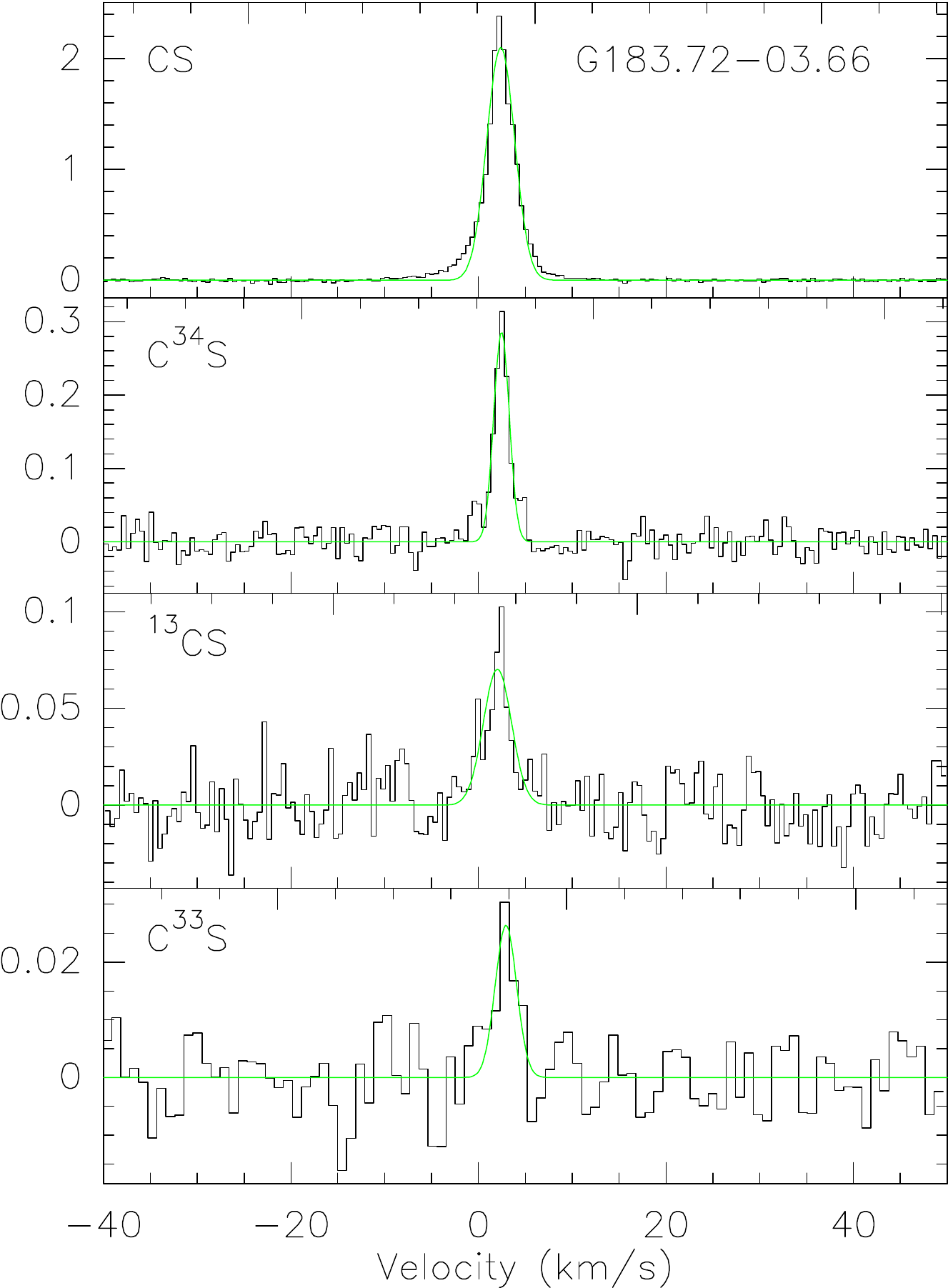}
  \includegraphics[width=80.5pt]{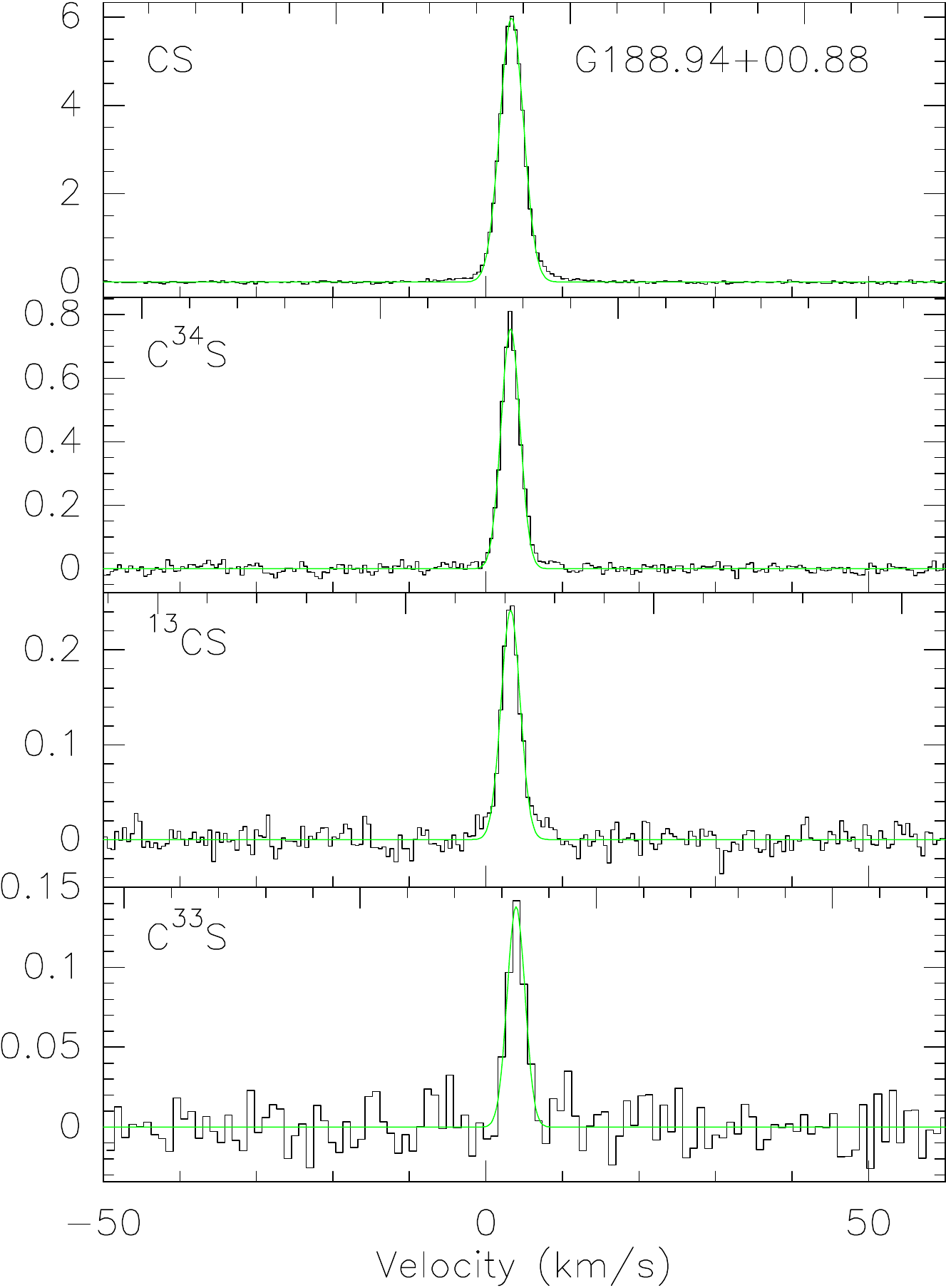}
  \includegraphics[width=82pt]{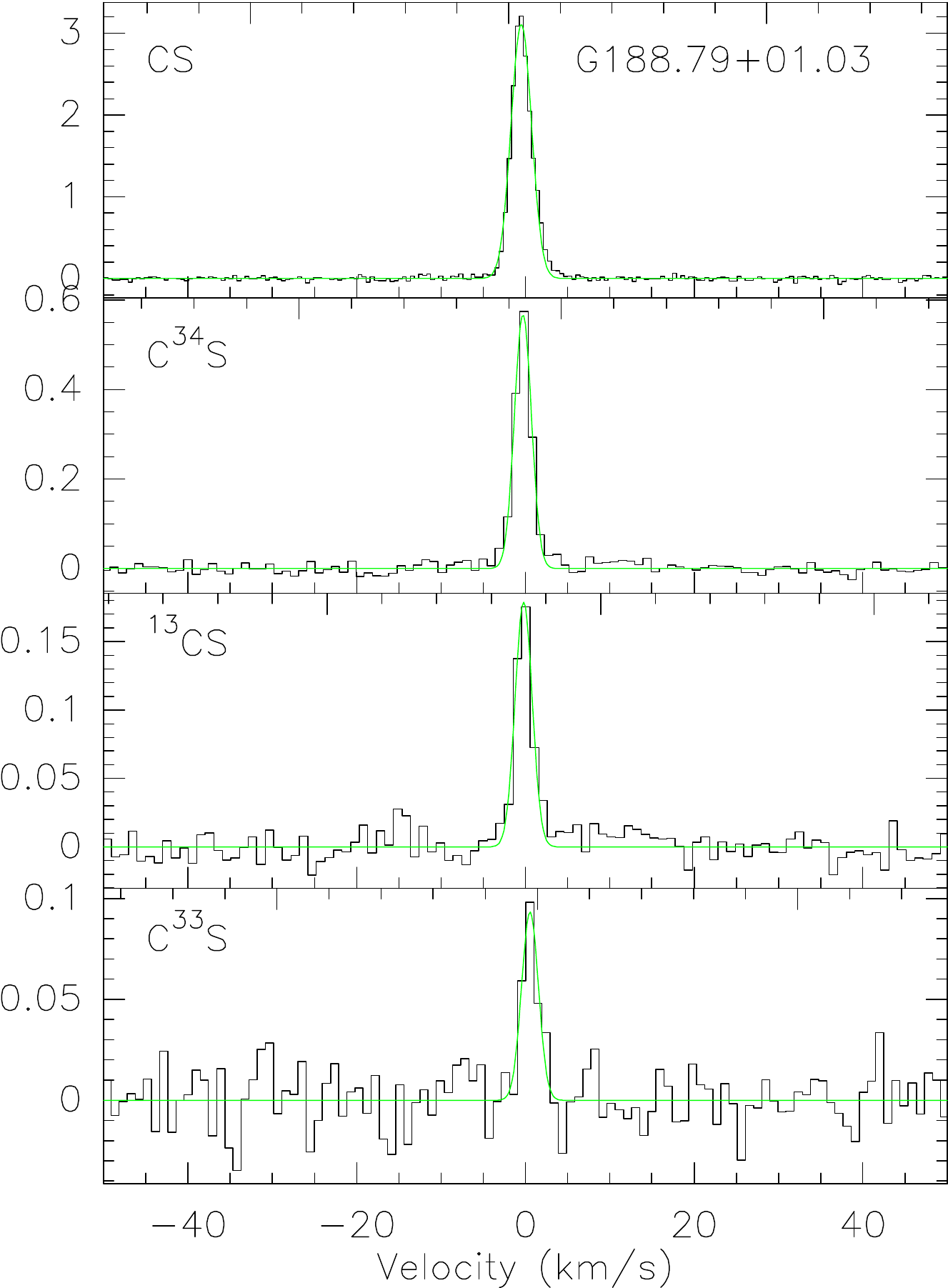}
  \includegraphics[width=82.5pt]{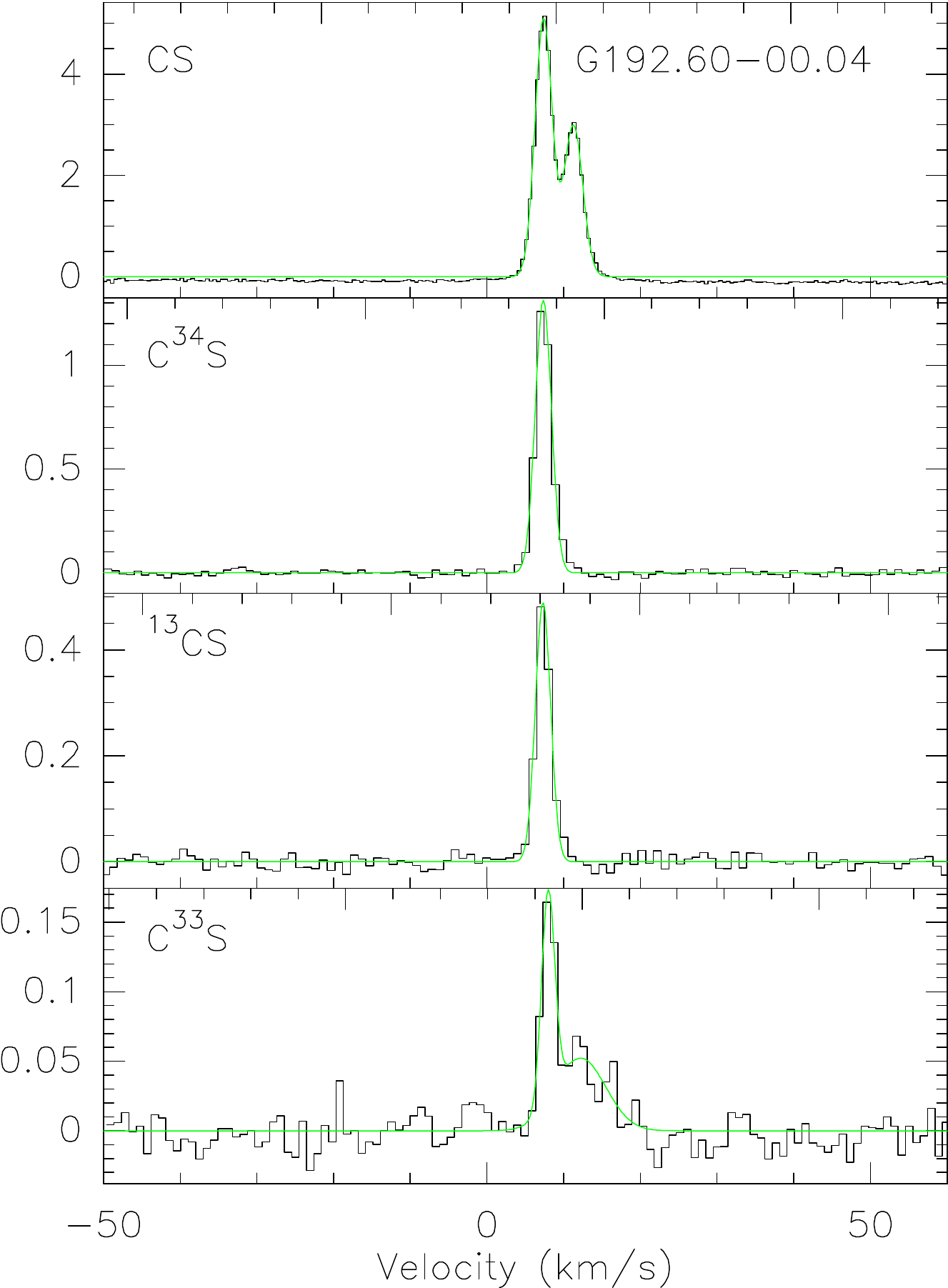}
  \includegraphics[width=86pt]{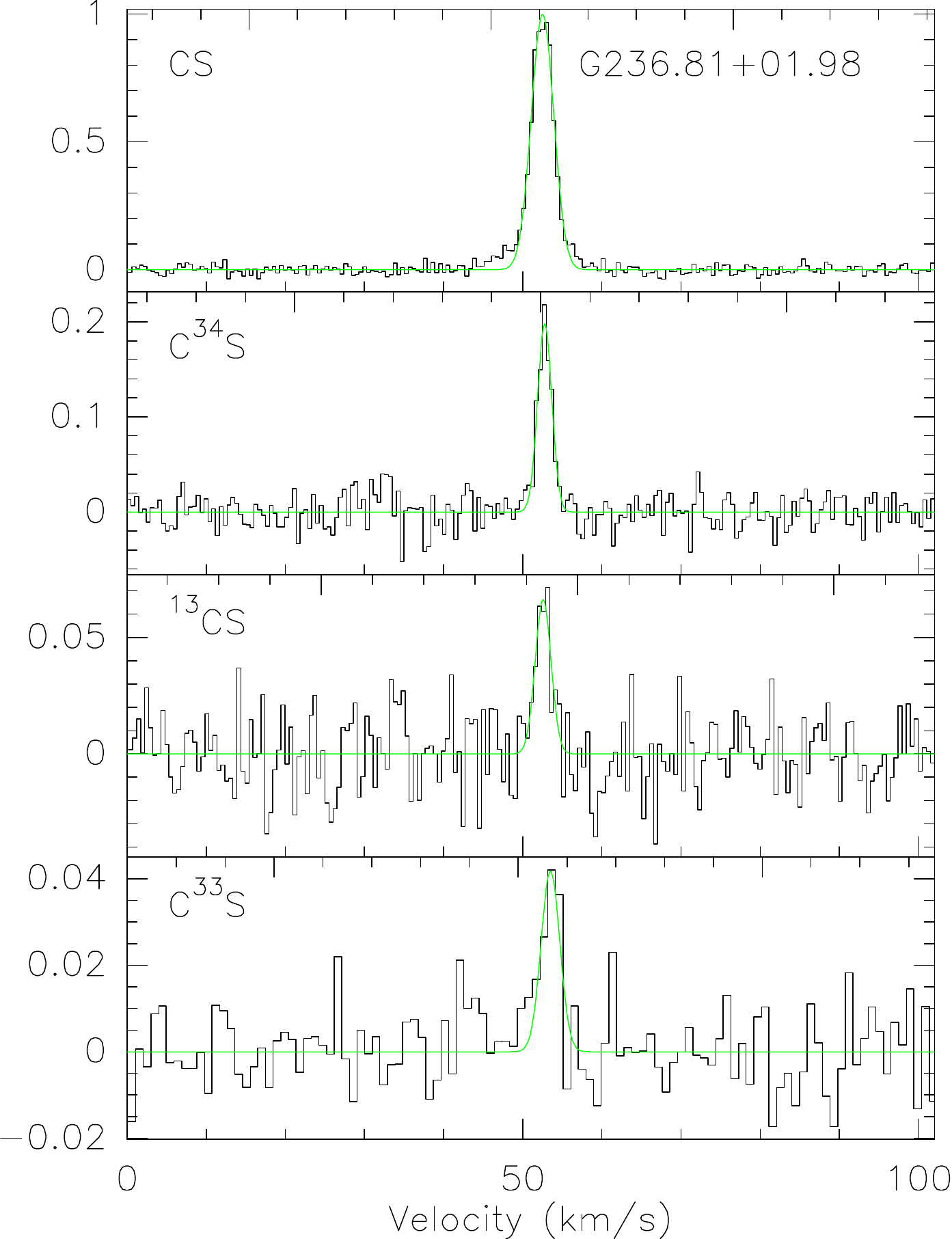}
  \includegraphics[width=81pt]{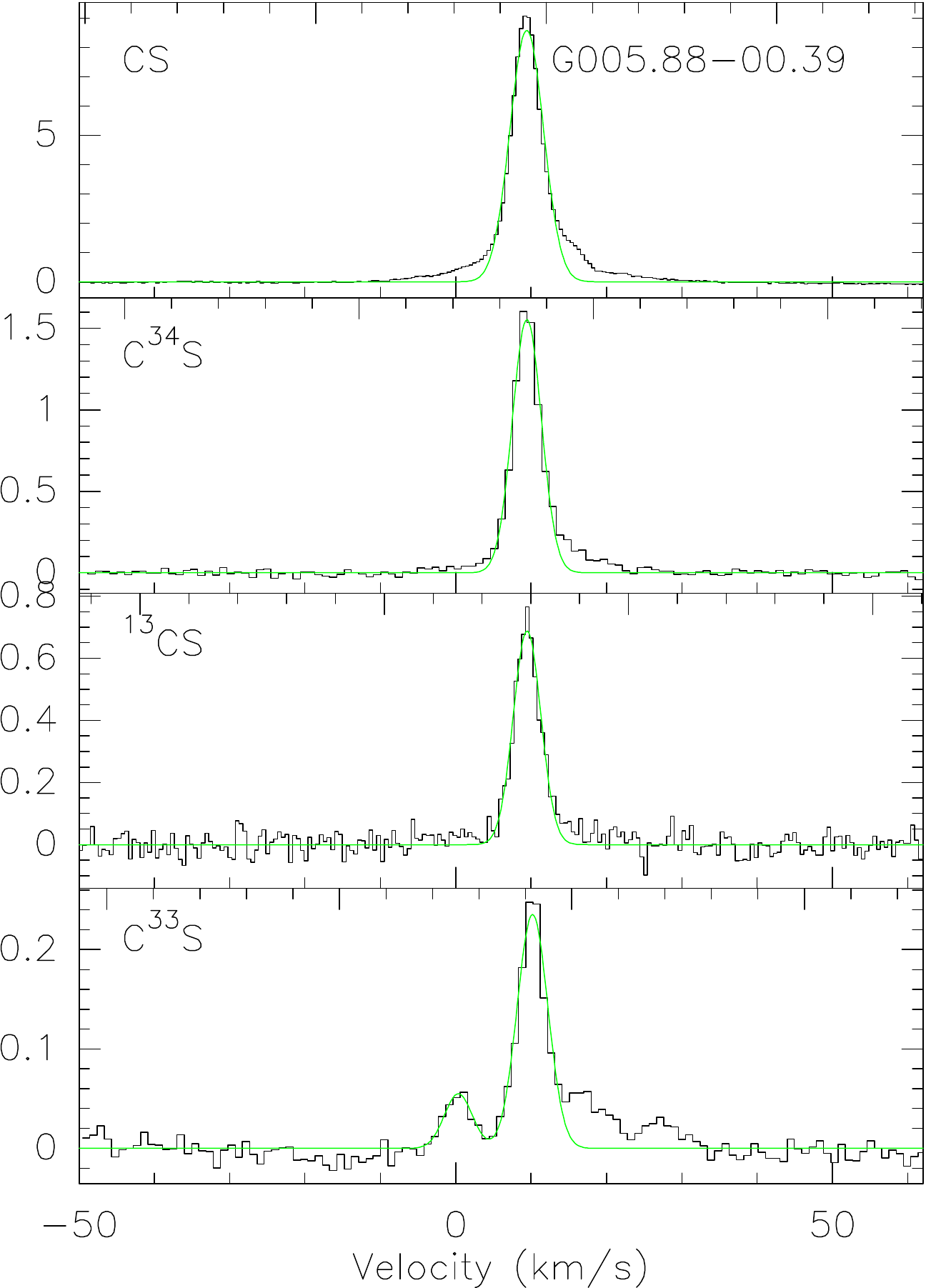}
  \includegraphics[width=83.5pt]{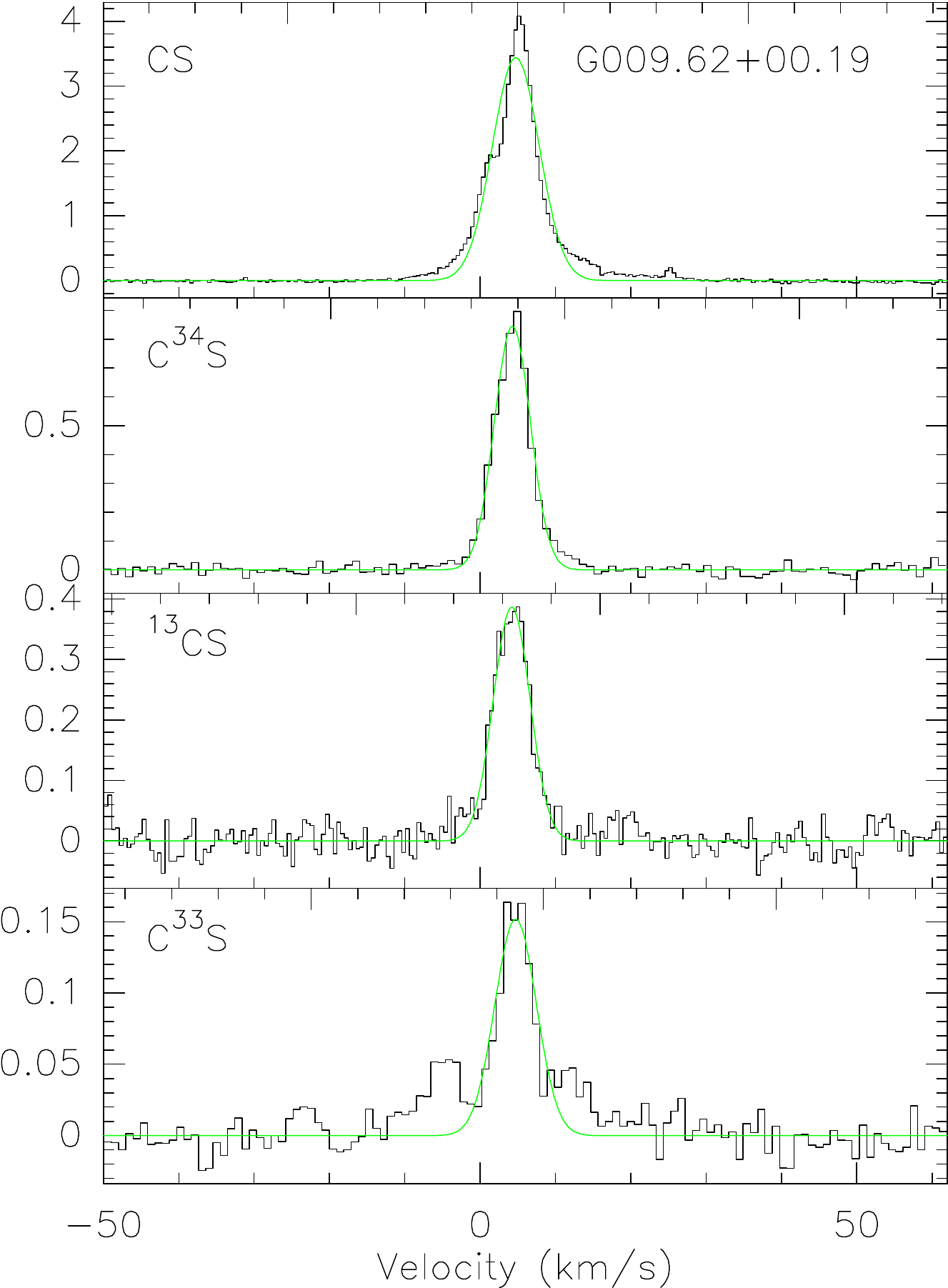}
  \includegraphics[width=83pt]{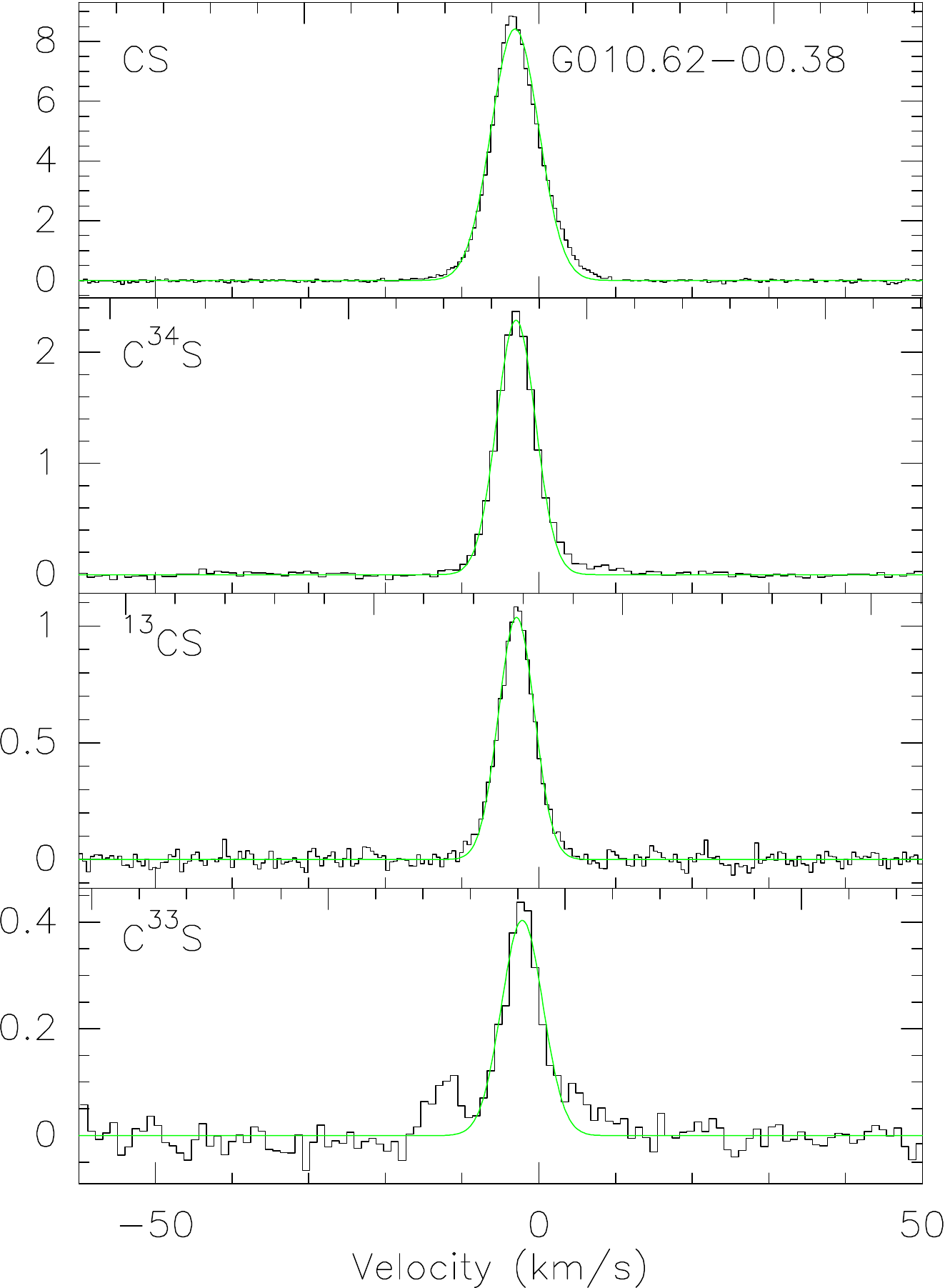}
  \includegraphics[width=83pt]{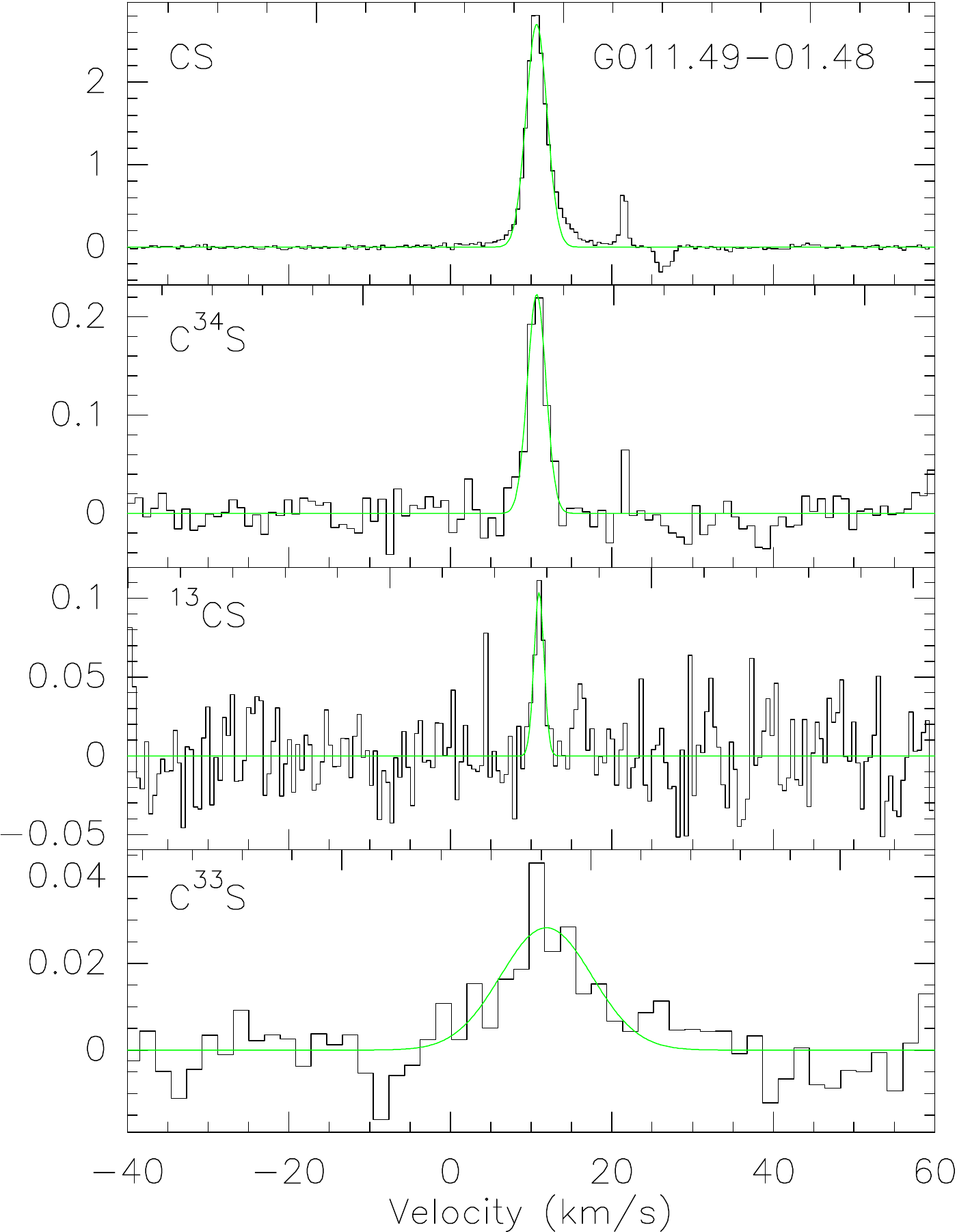}
  \includegraphics[width=83pt]{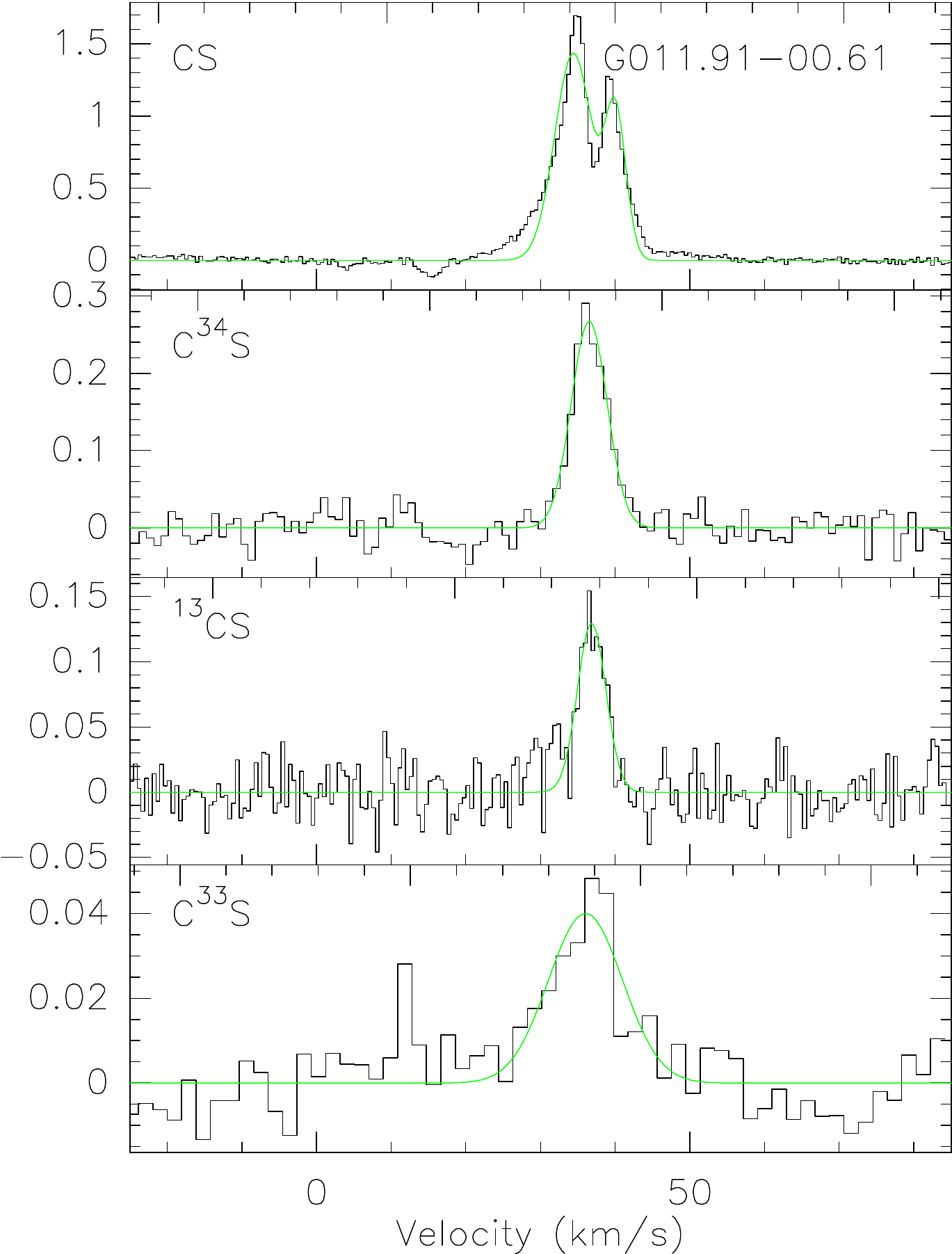}
  \includegraphics[width=83pt]{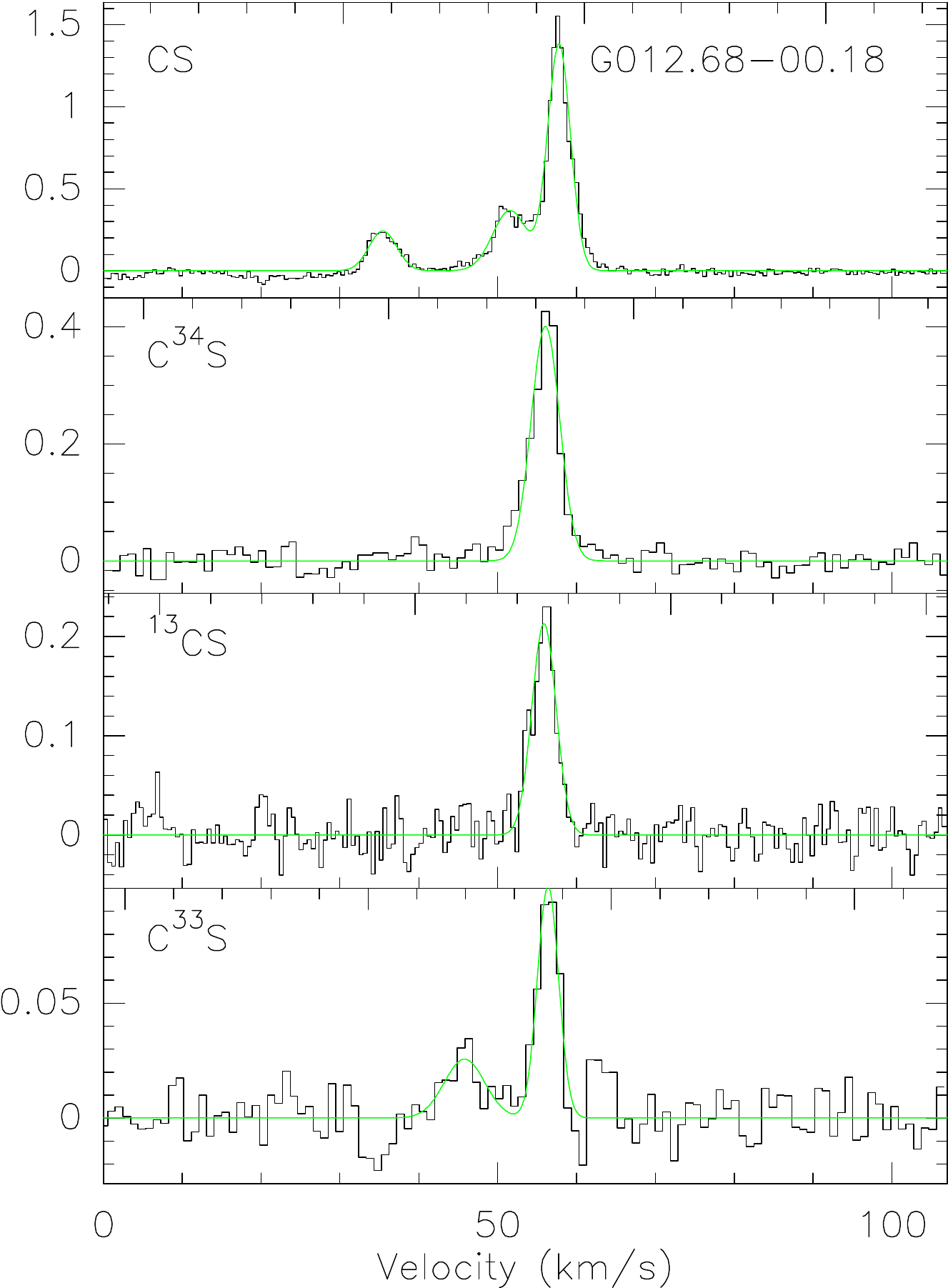}
  \includegraphics[width=83pt]{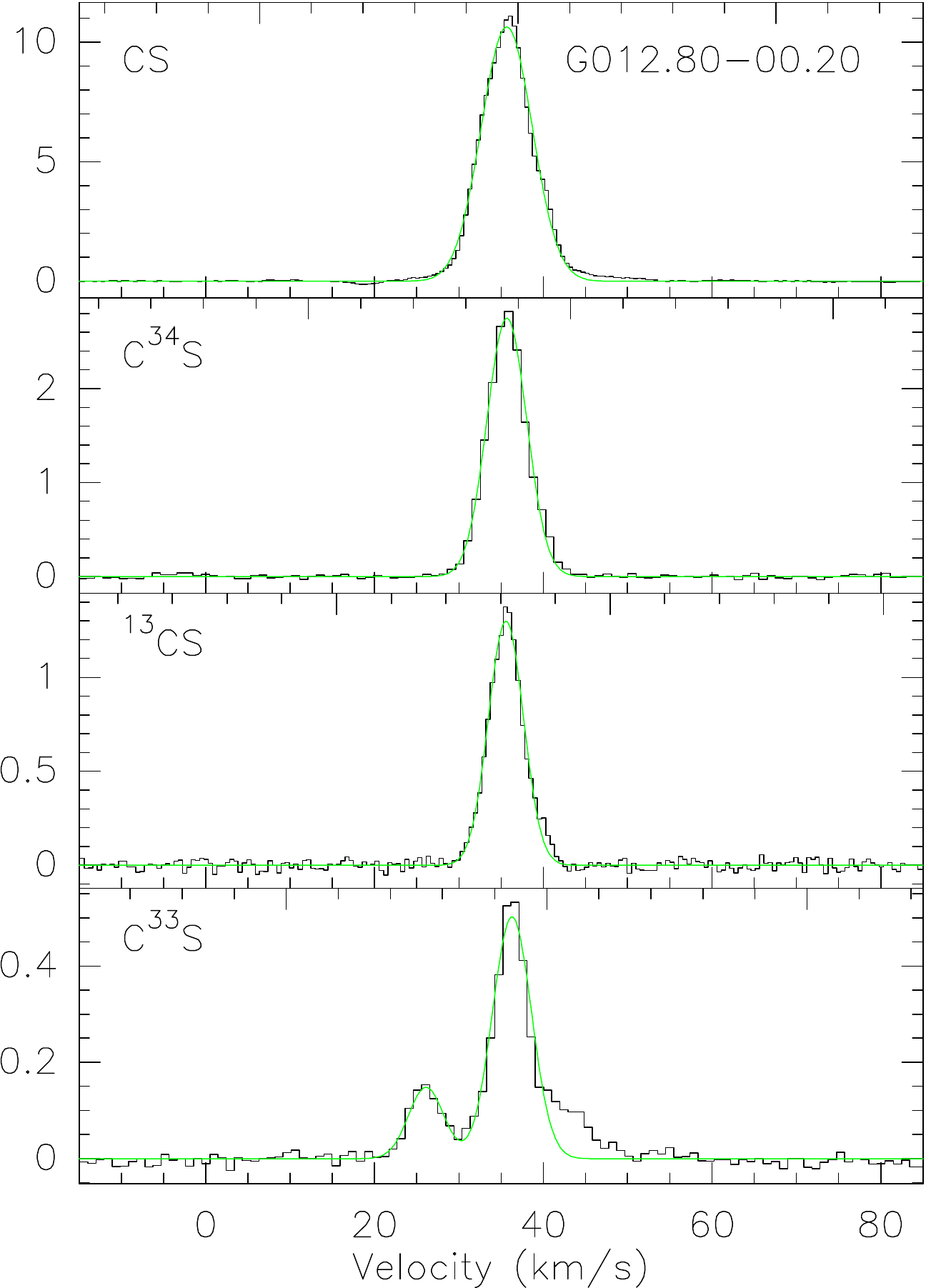}
  \includegraphics[width=83pt]{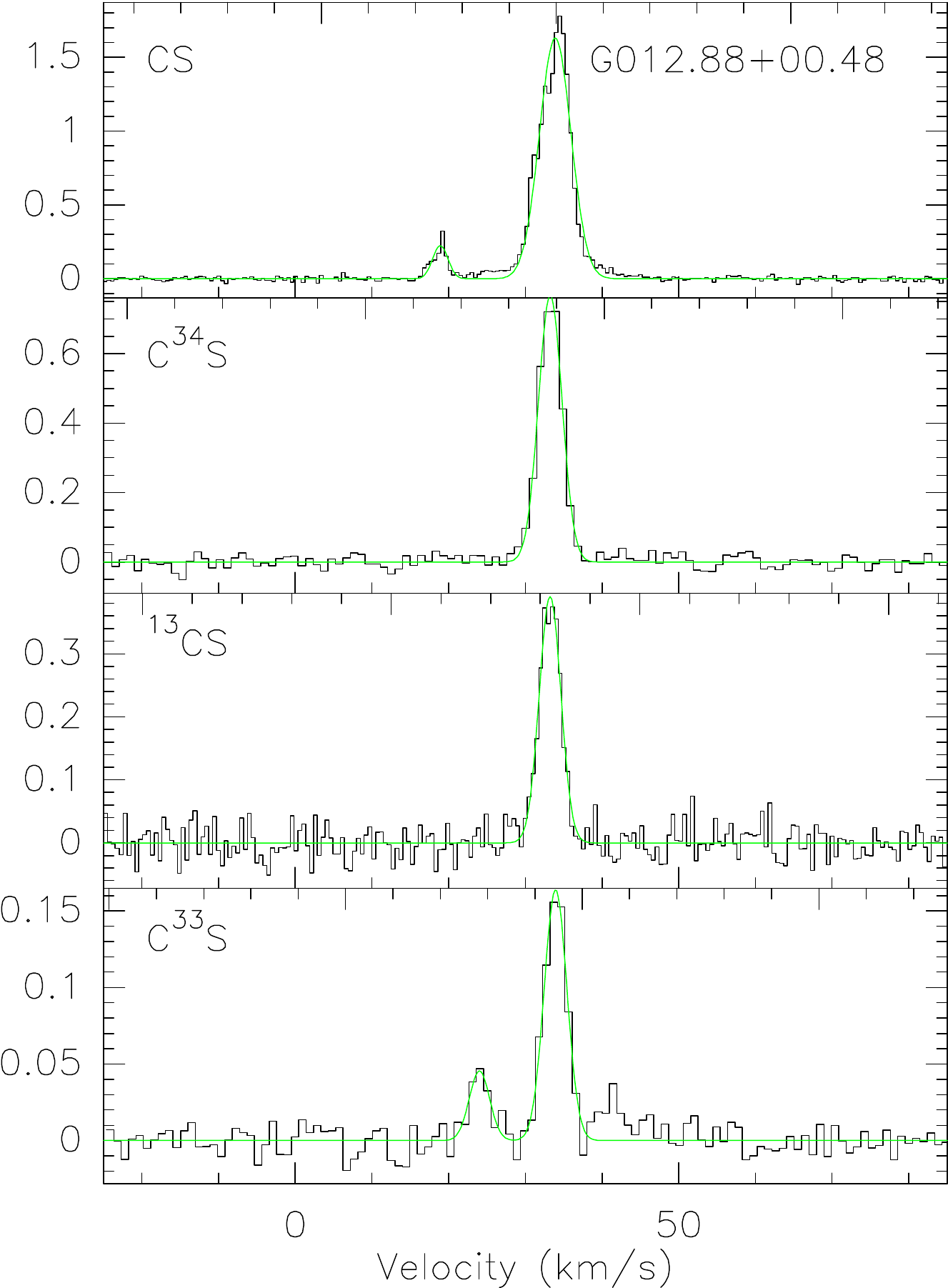}
  \includegraphics[width=83pt]{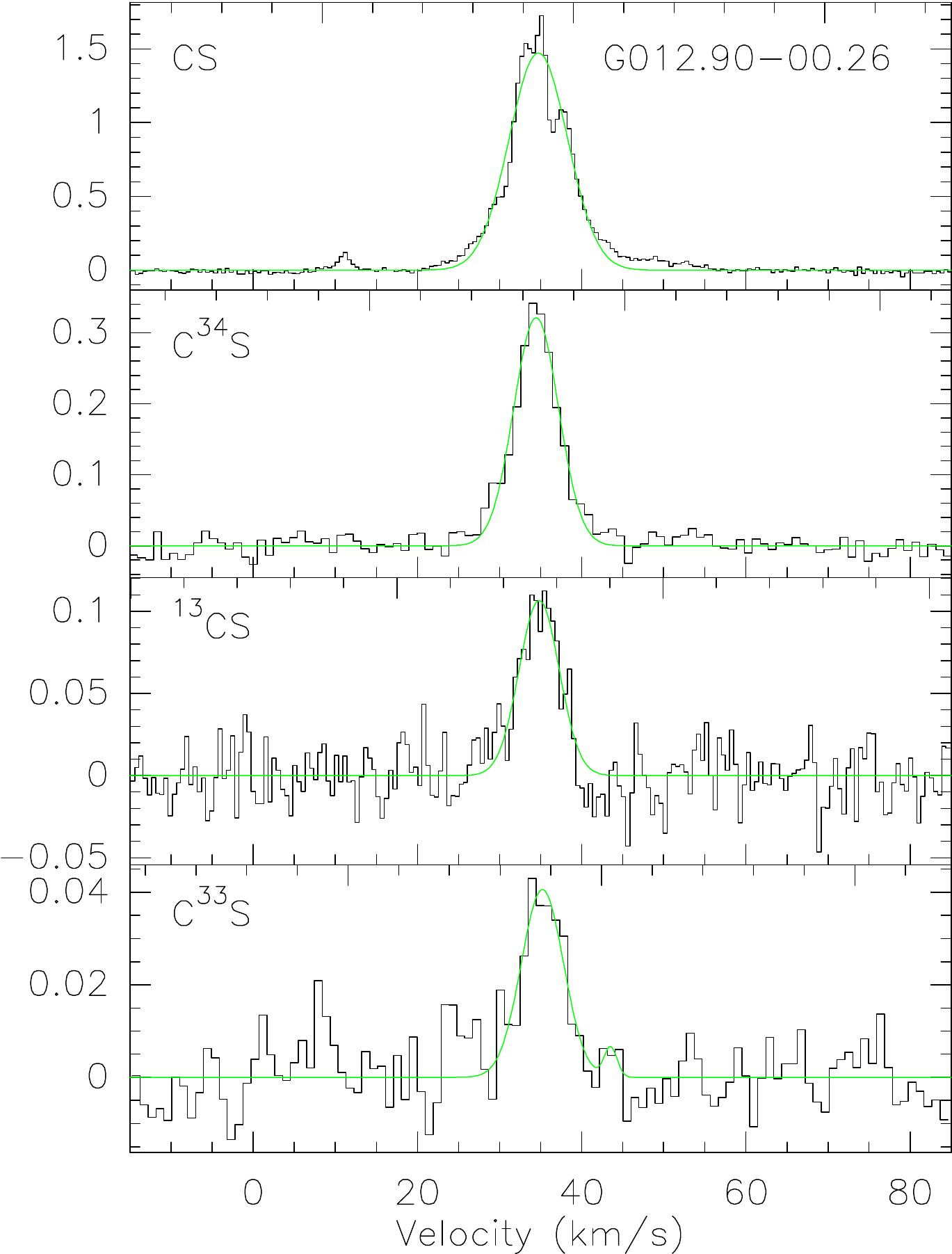}
  \includegraphics[width=83pt]{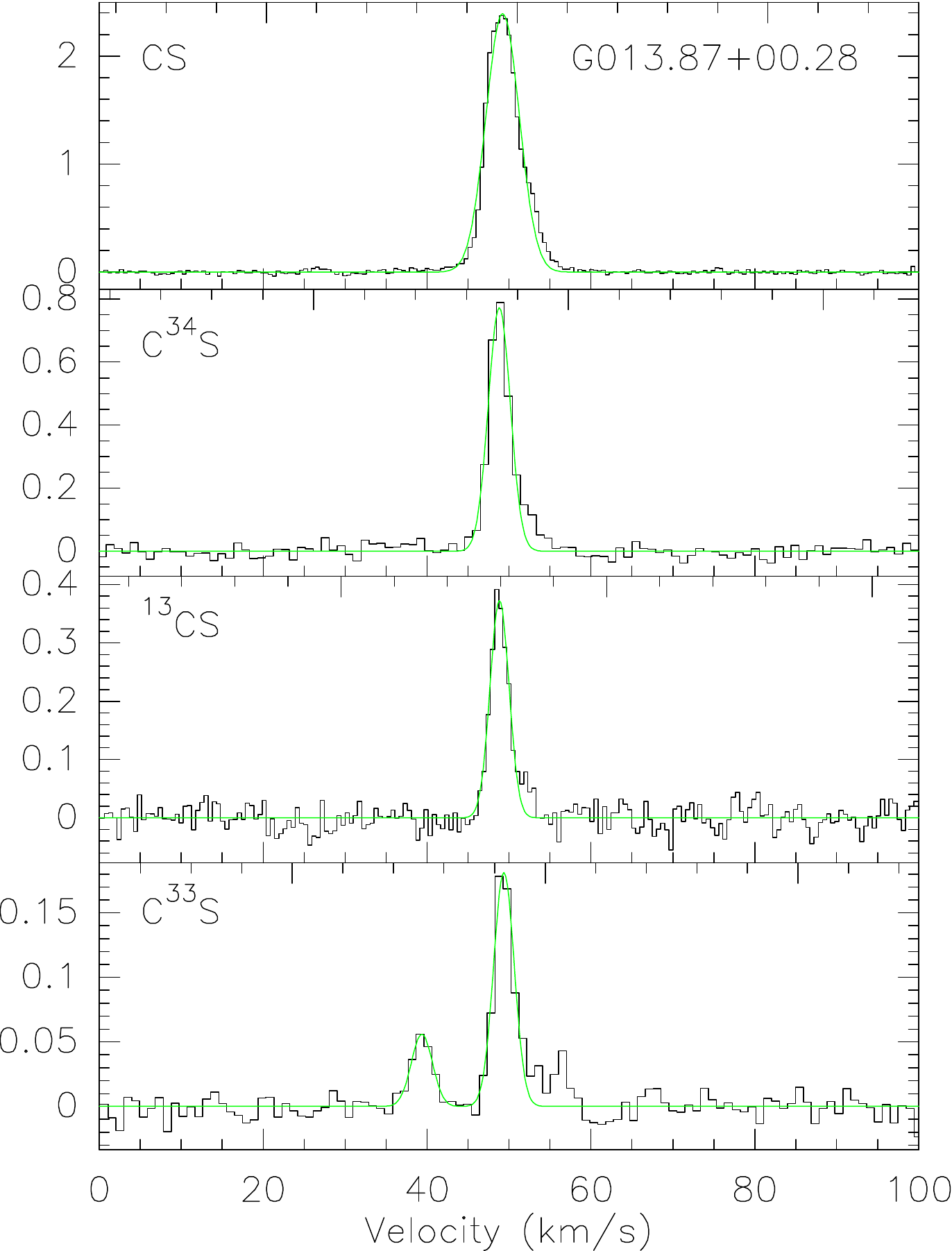}
  \includegraphics[width=83pt]{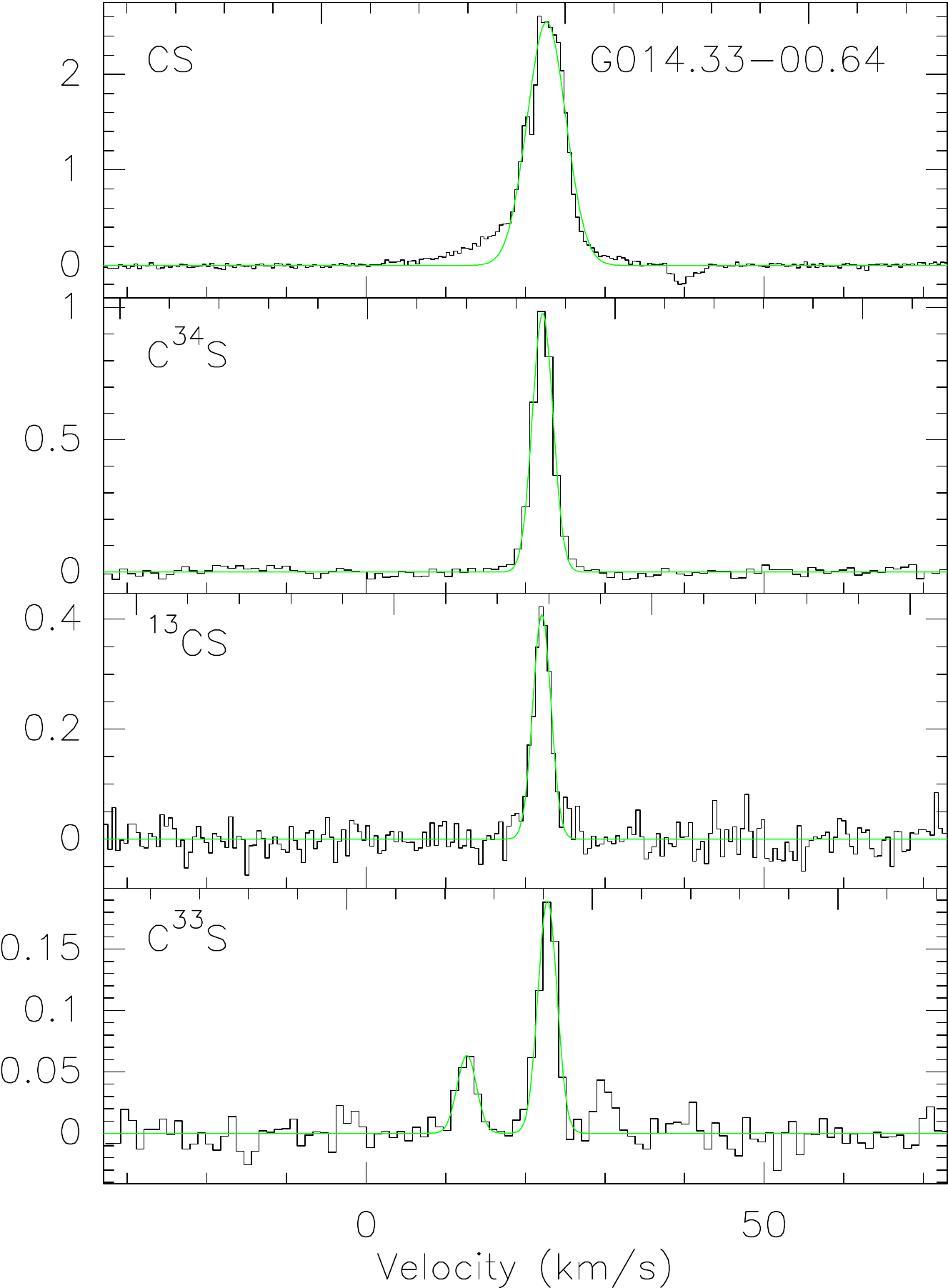}
  \includegraphics[width=83pt]{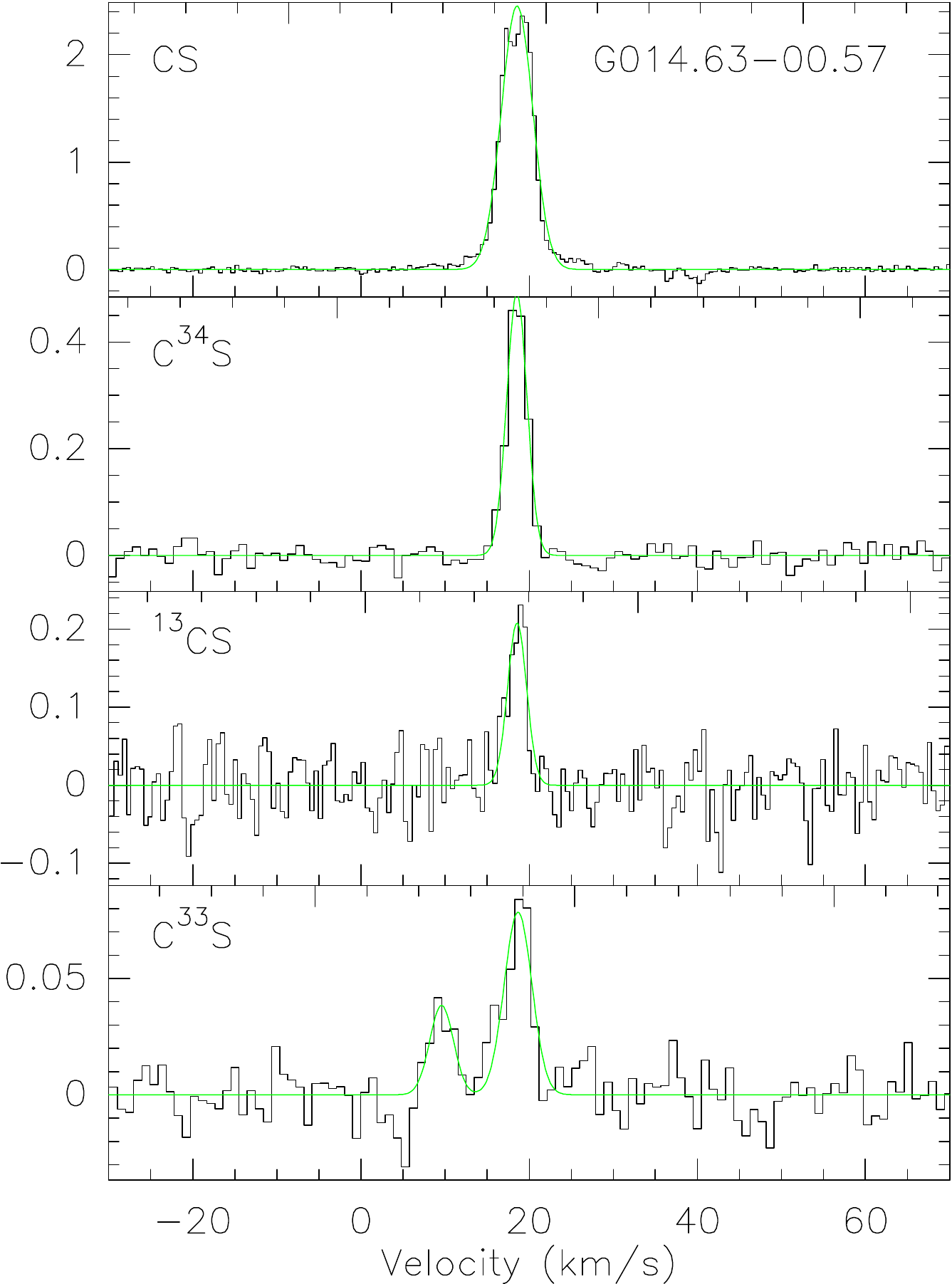}
  \includegraphics[width=83pt]{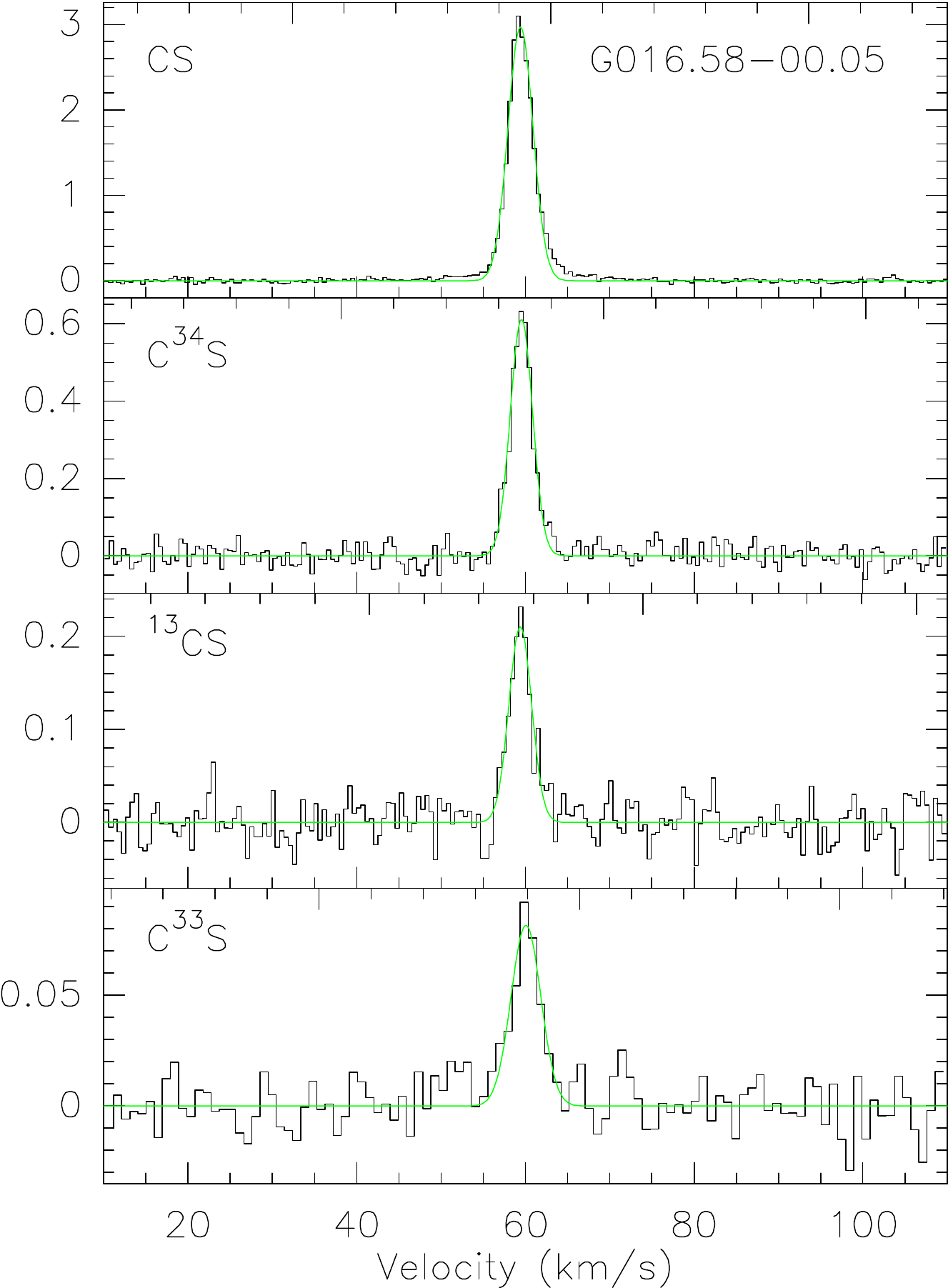}
  \includegraphics[width=83pt]{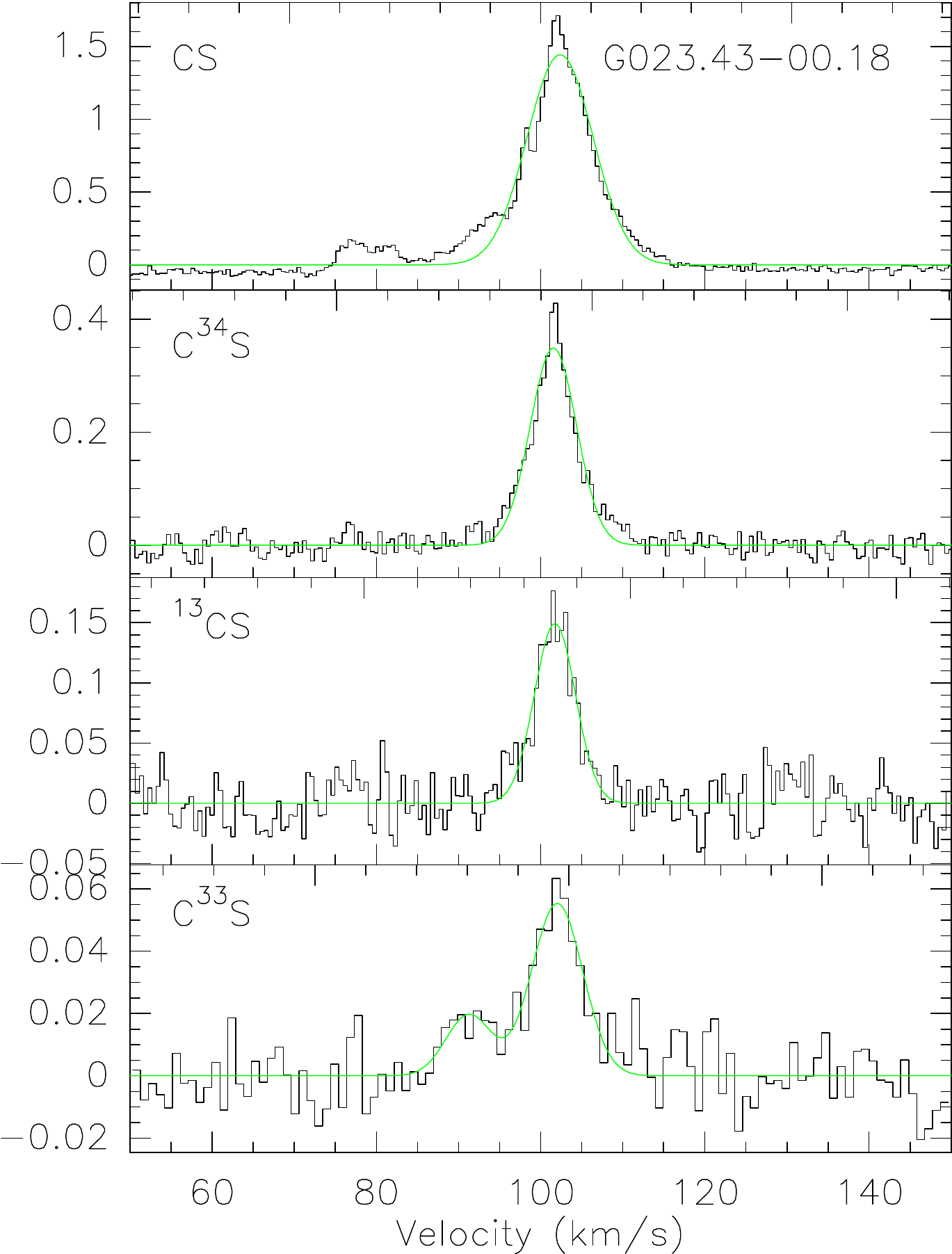}
  \includegraphics[width=83pt]{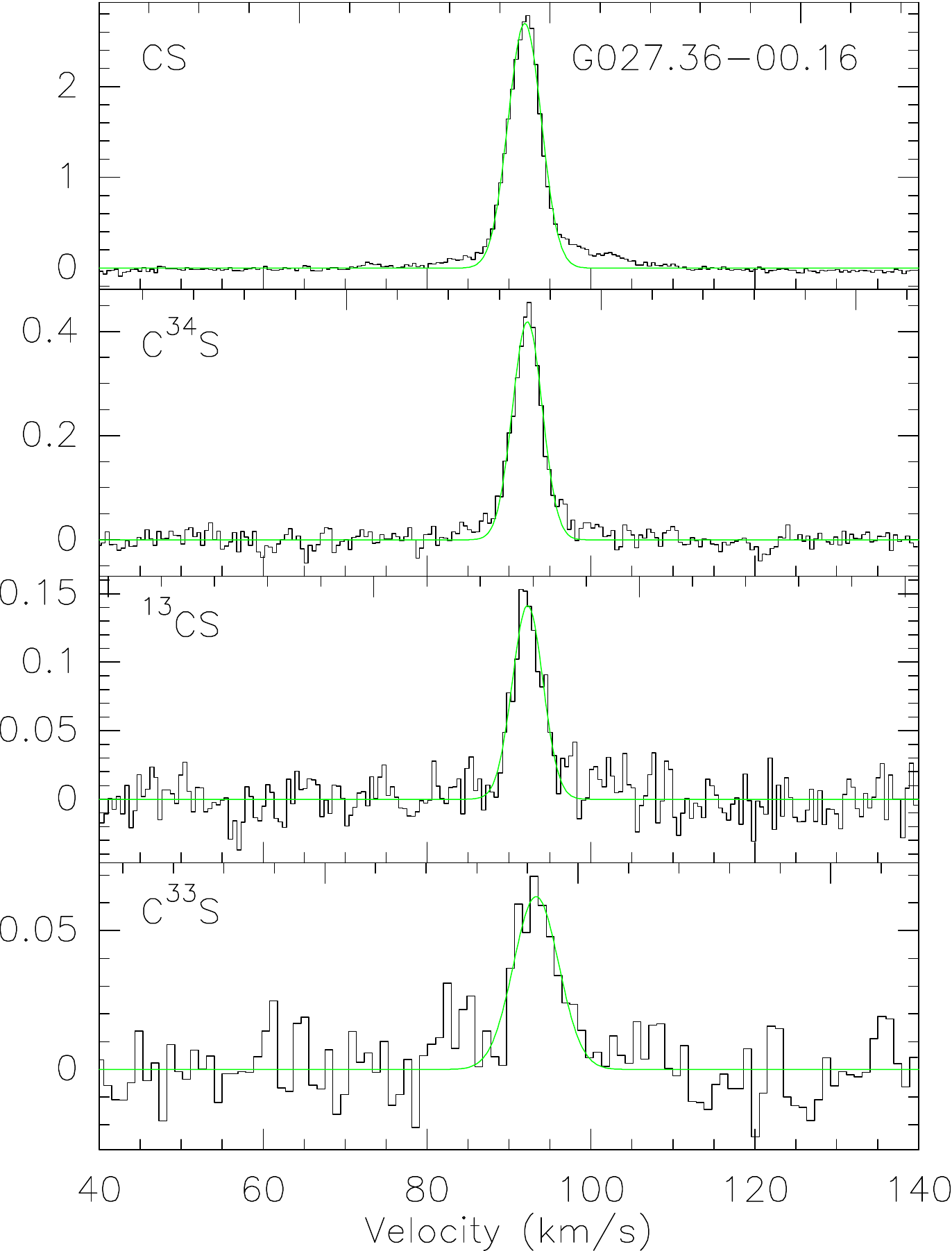}
  \includegraphics[width=83pt]{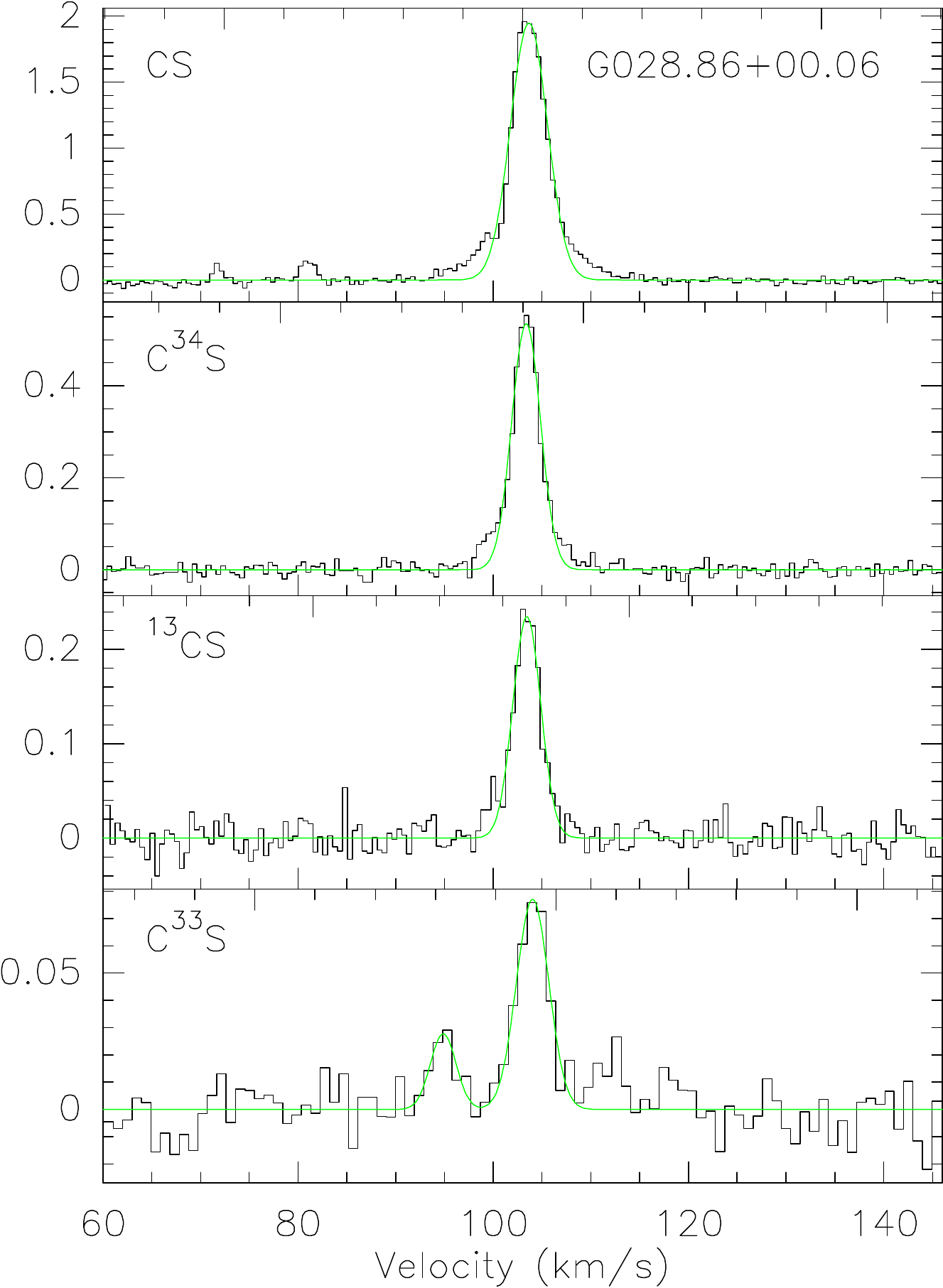}
  \includegraphics[width=83pt]{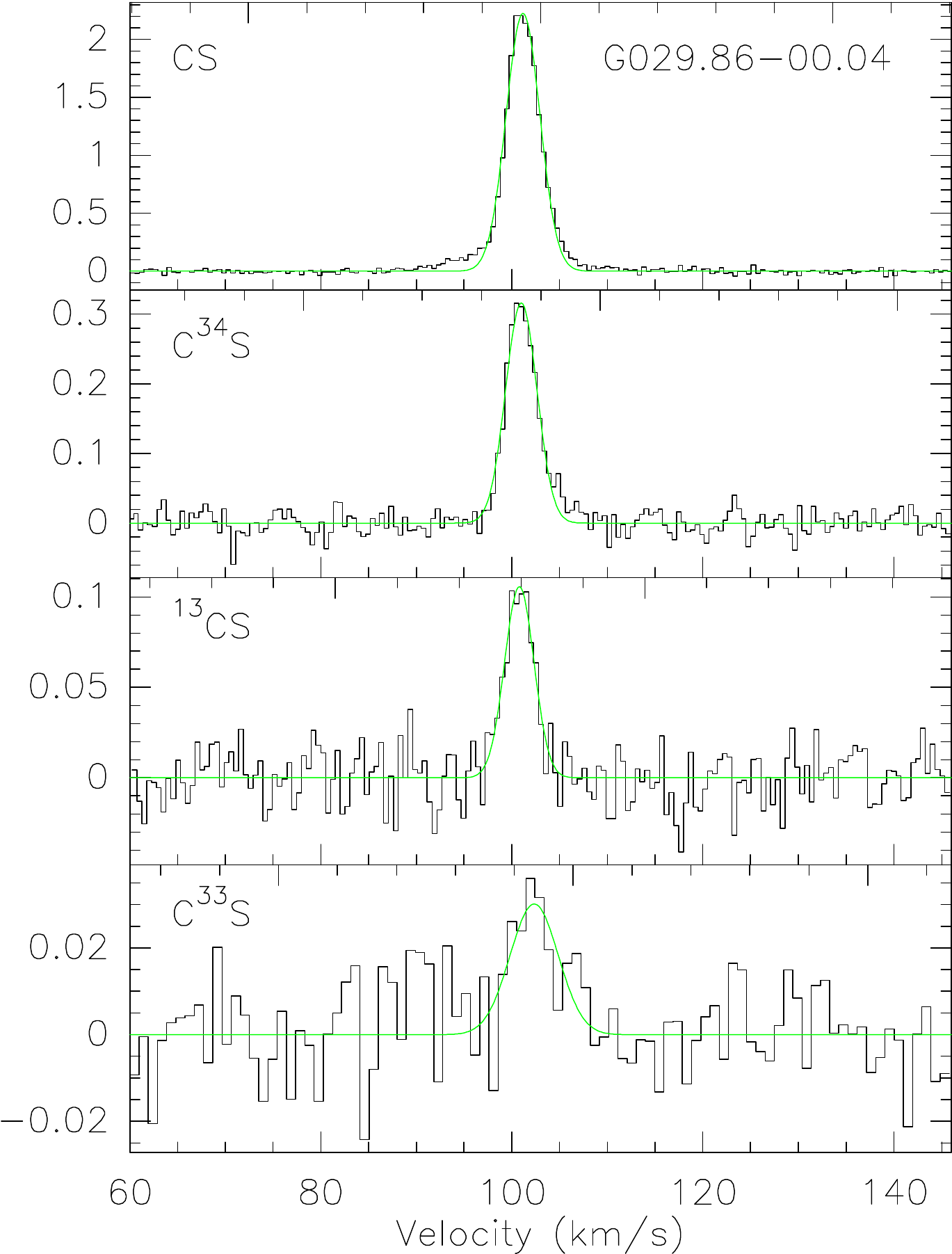}
  \includegraphics[width=83pt]{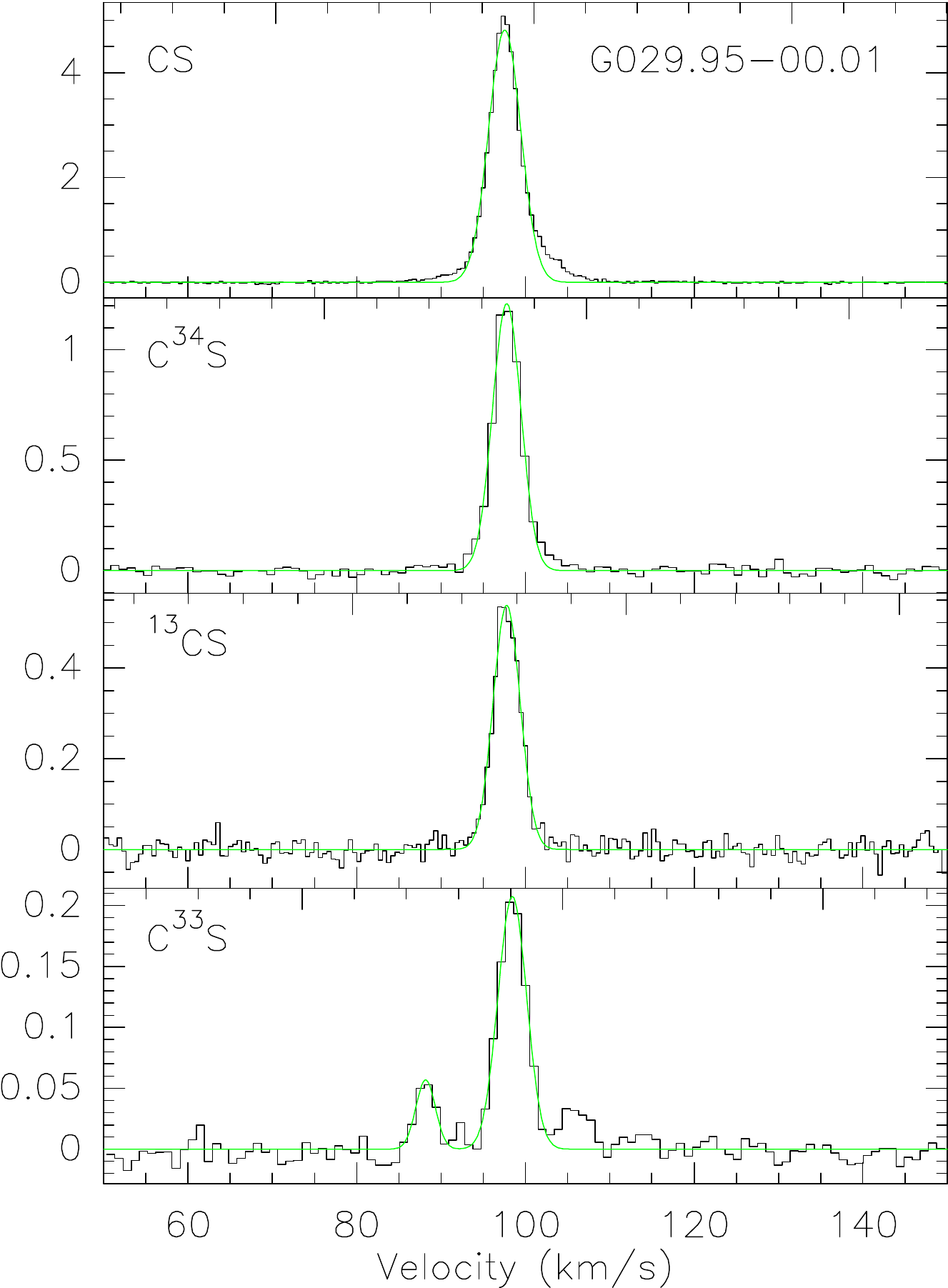}
  \includegraphics[width=83pt]{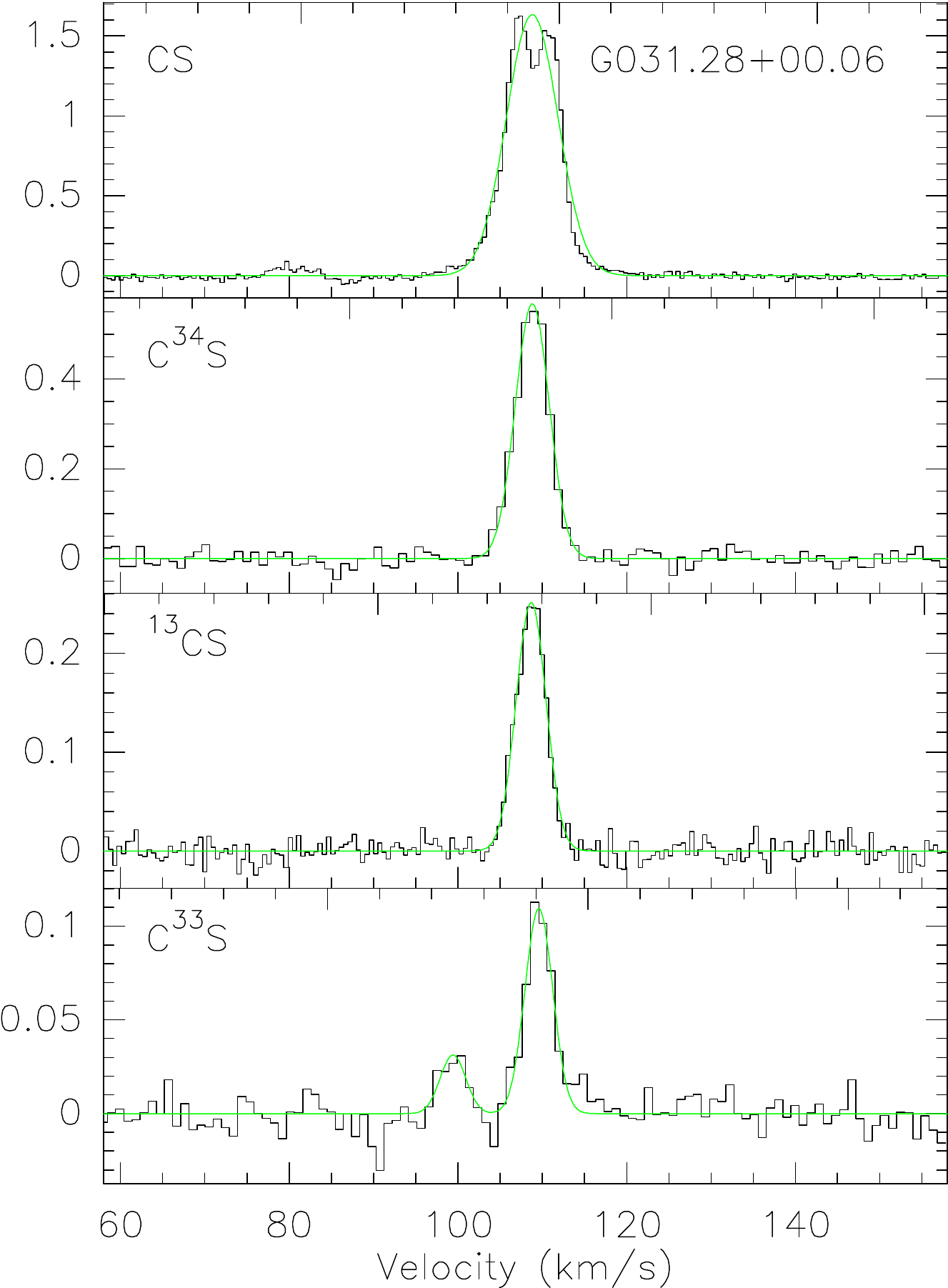}
  \includegraphics[width=83pt]{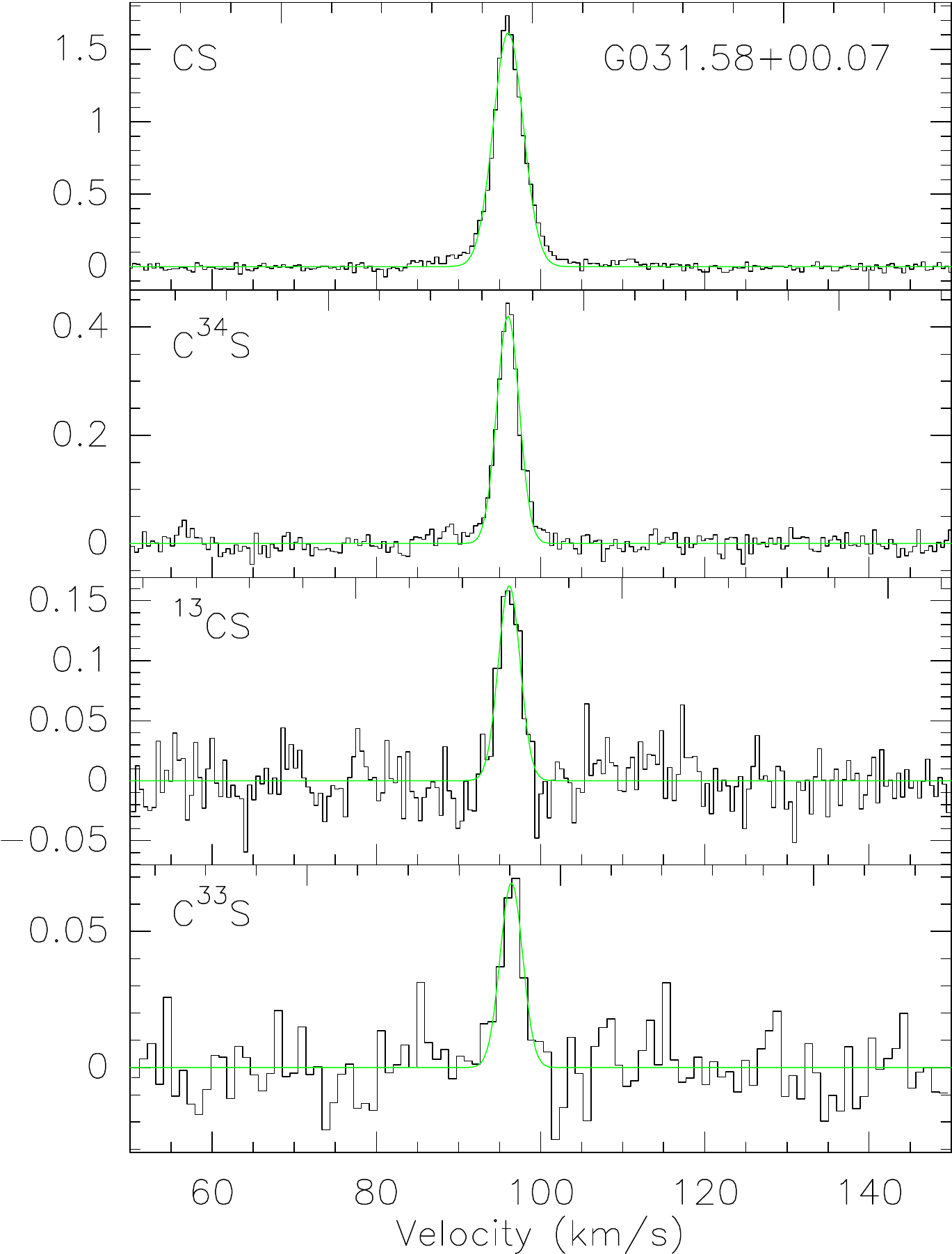}
  \includegraphics[width=83pt]{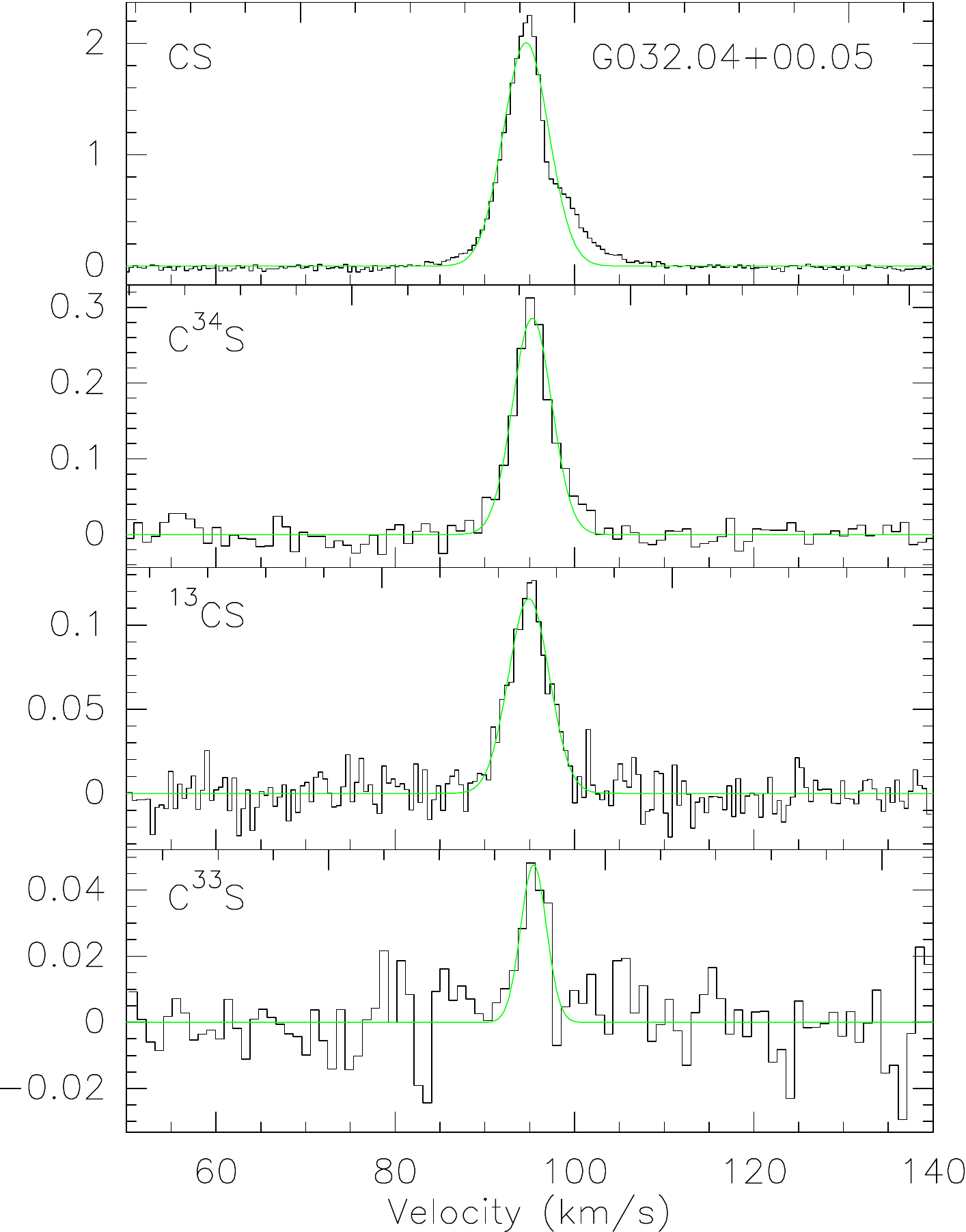}
  \includegraphics[width=85pt]{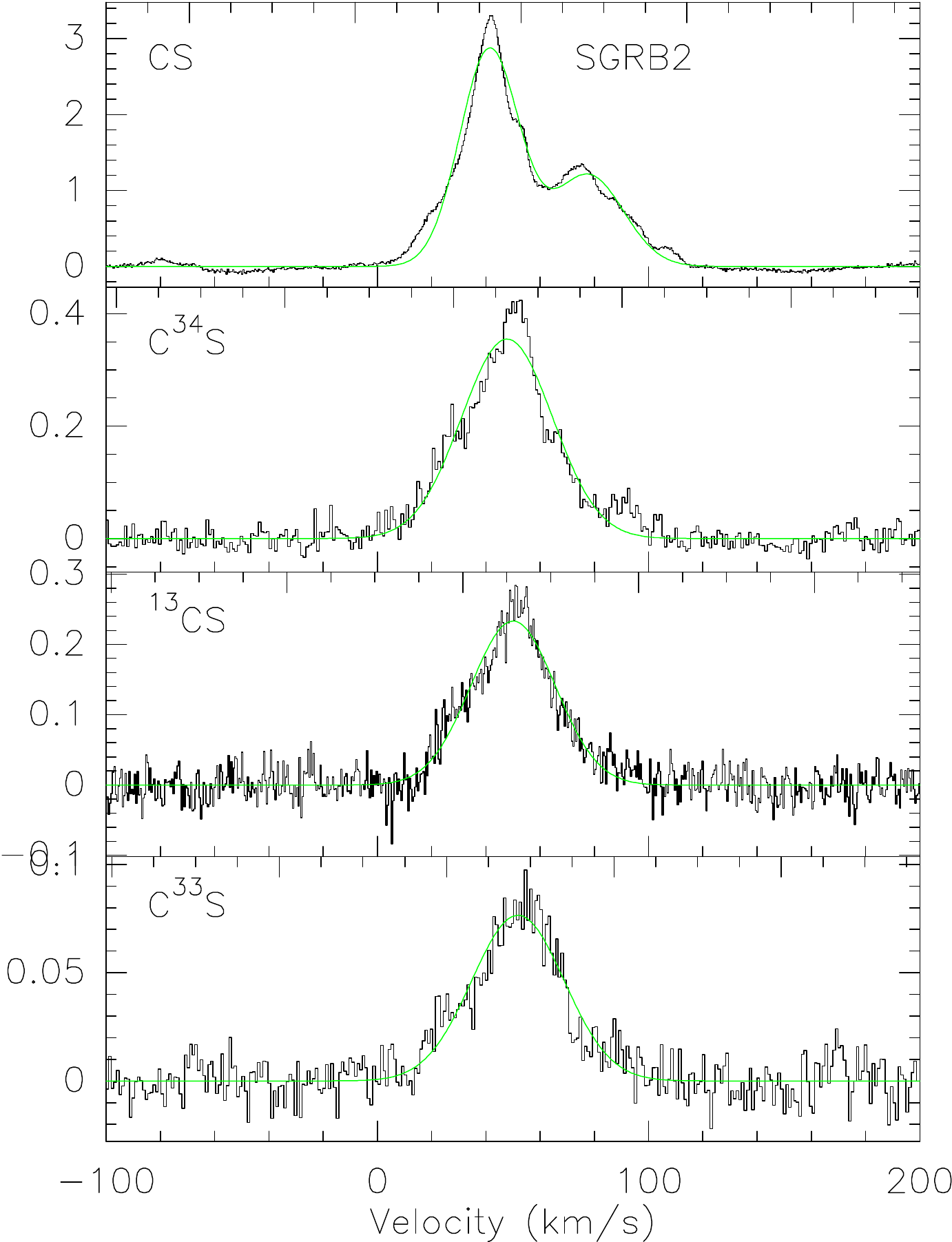}
\end{figure*}
\addtocounter{figure}{-1}
\begin{figure*}[h]
\center
\addtocounter{figure}{1}
\includegraphics[width=83pt]{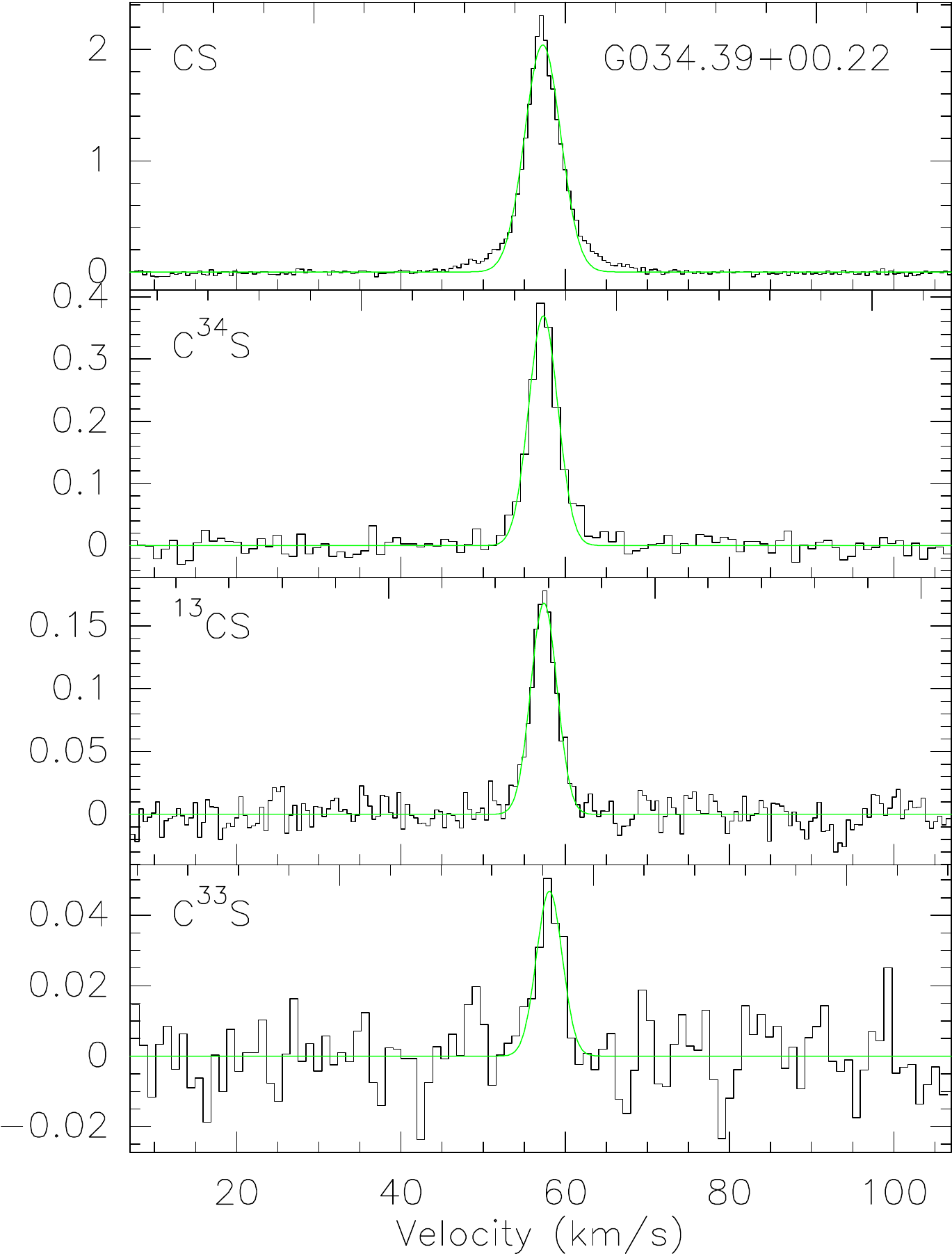}
  \includegraphics[width=80.6pt]{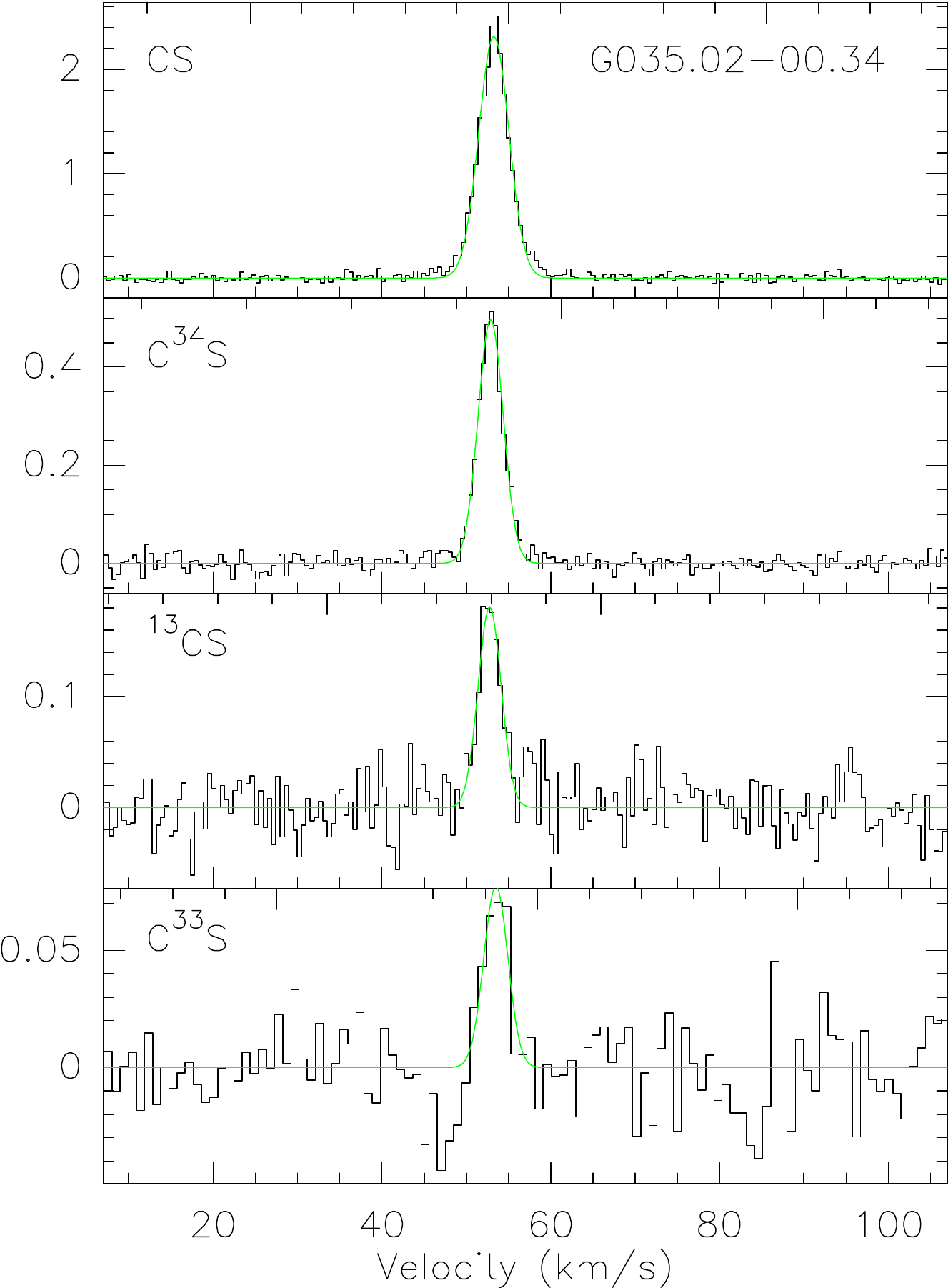}
  \includegraphics[width=83pt]{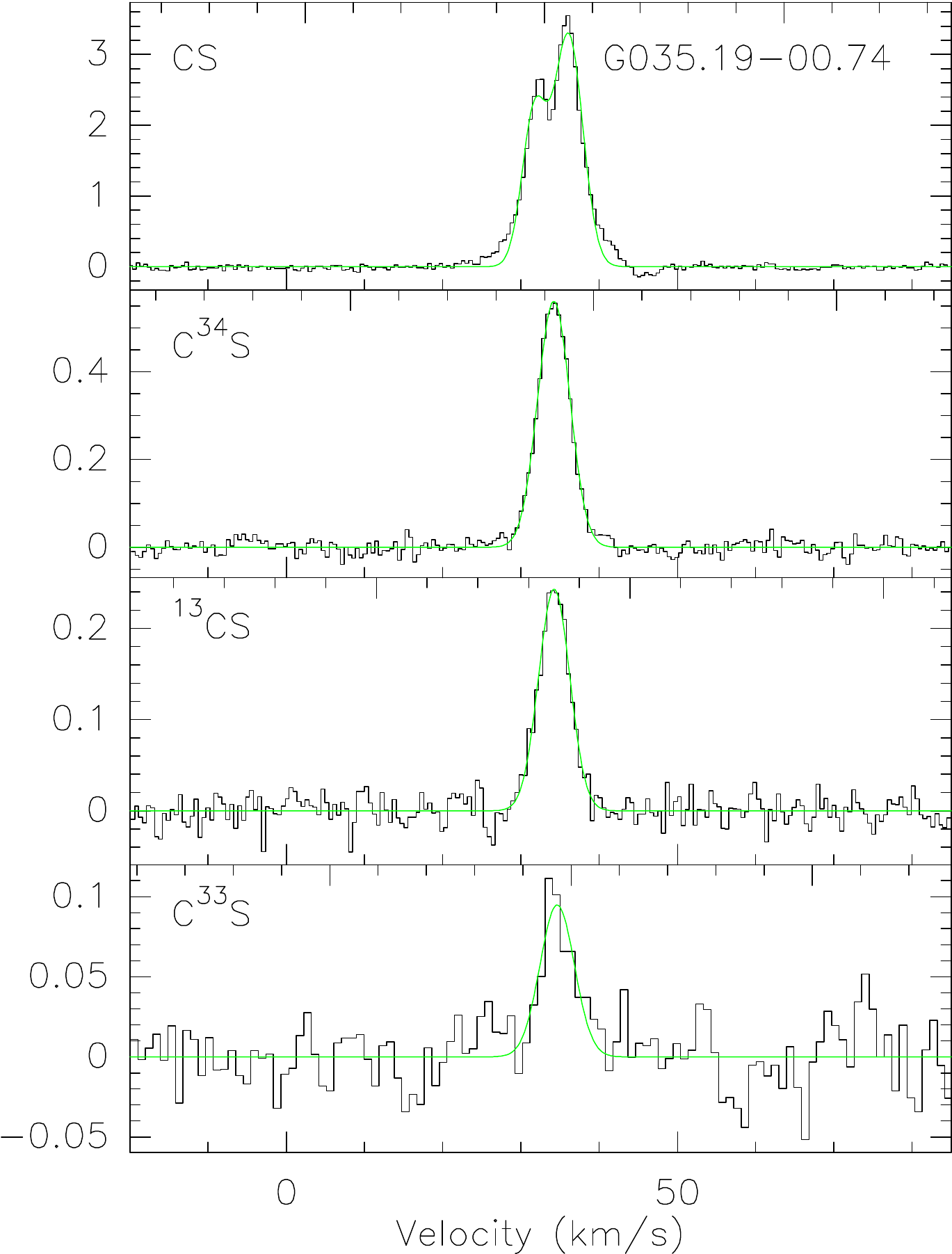}
  \includegraphics[width=83pt]{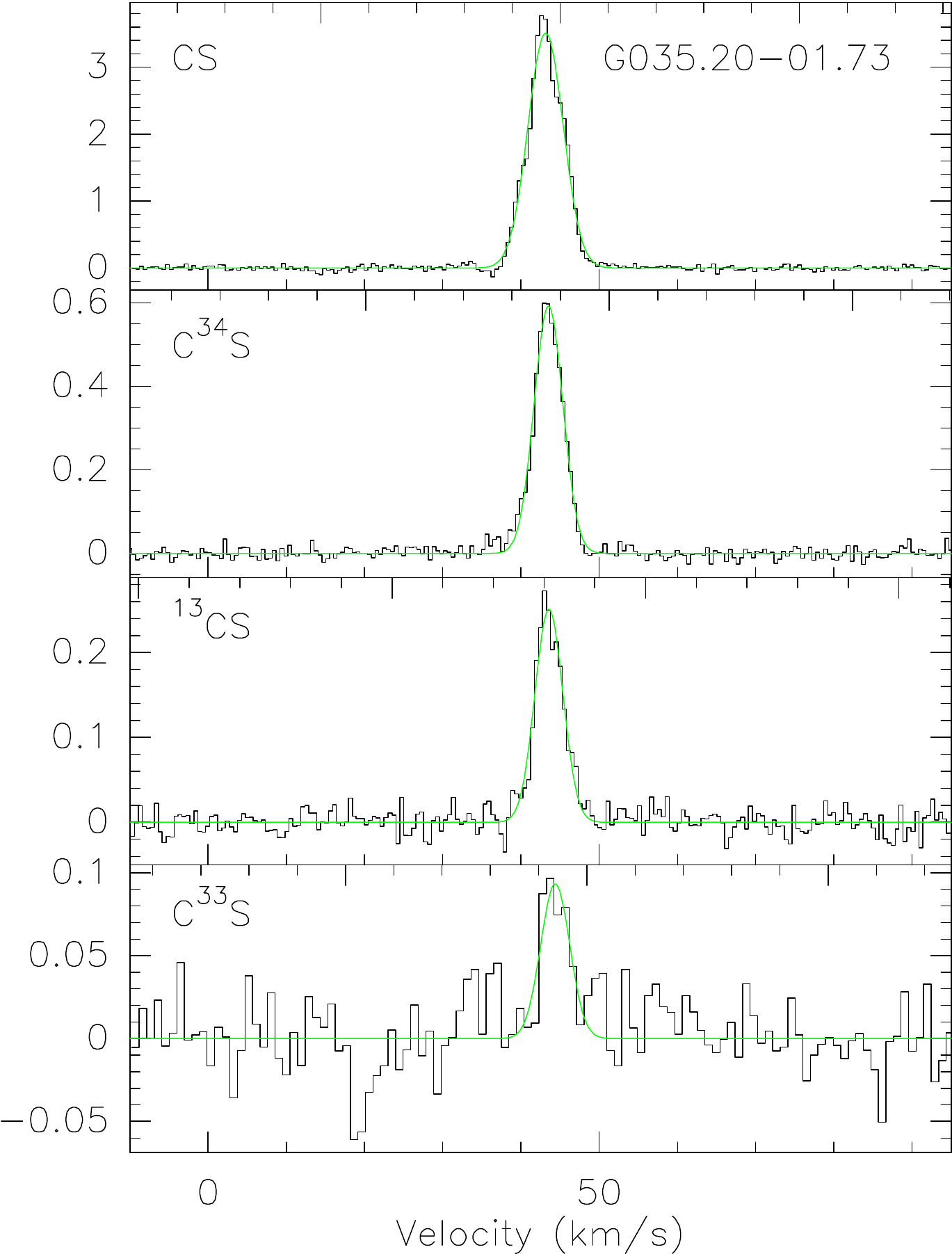}
  \includegraphics[width=83pt]{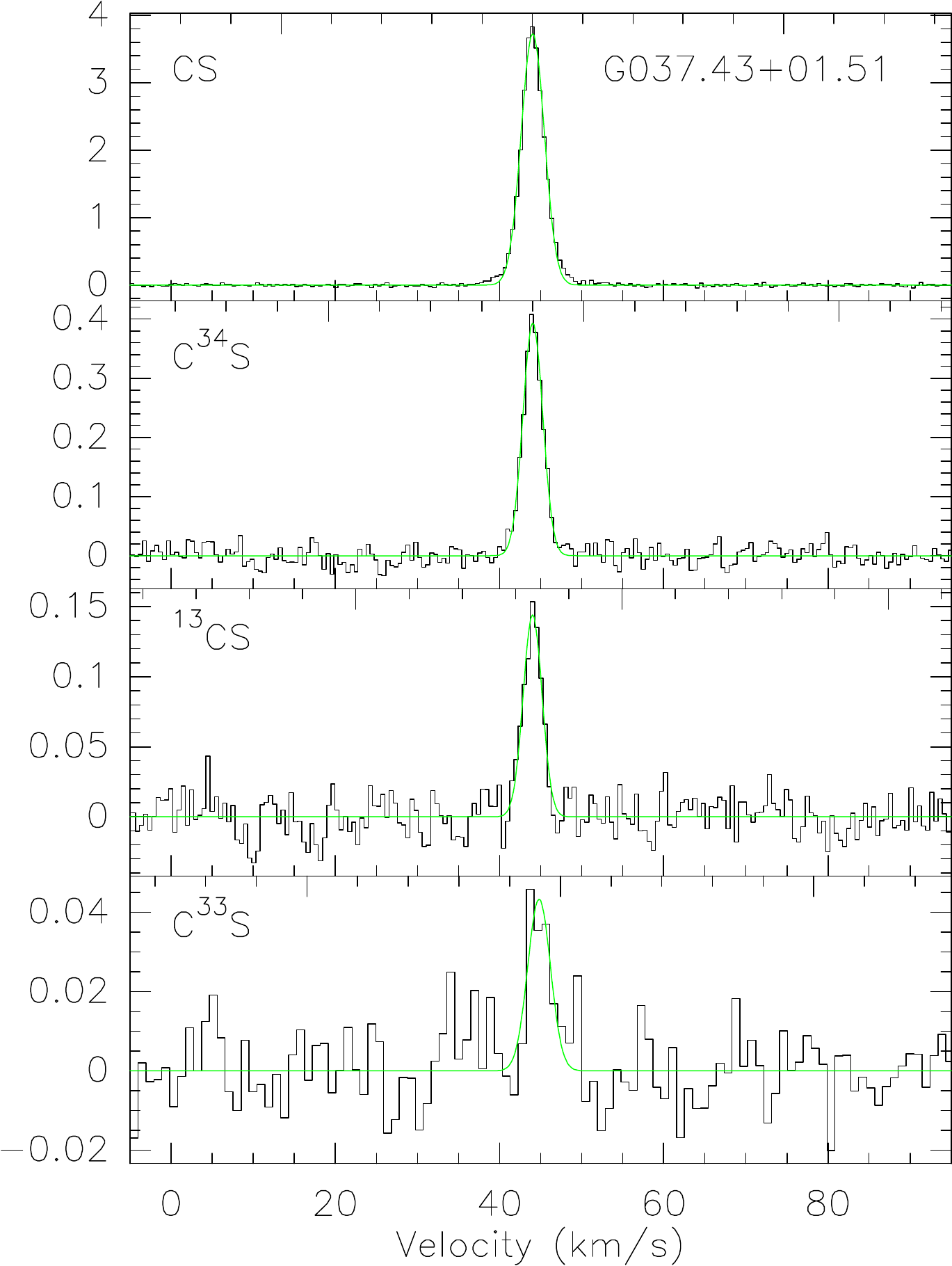}
  \includegraphics[width=83pt]{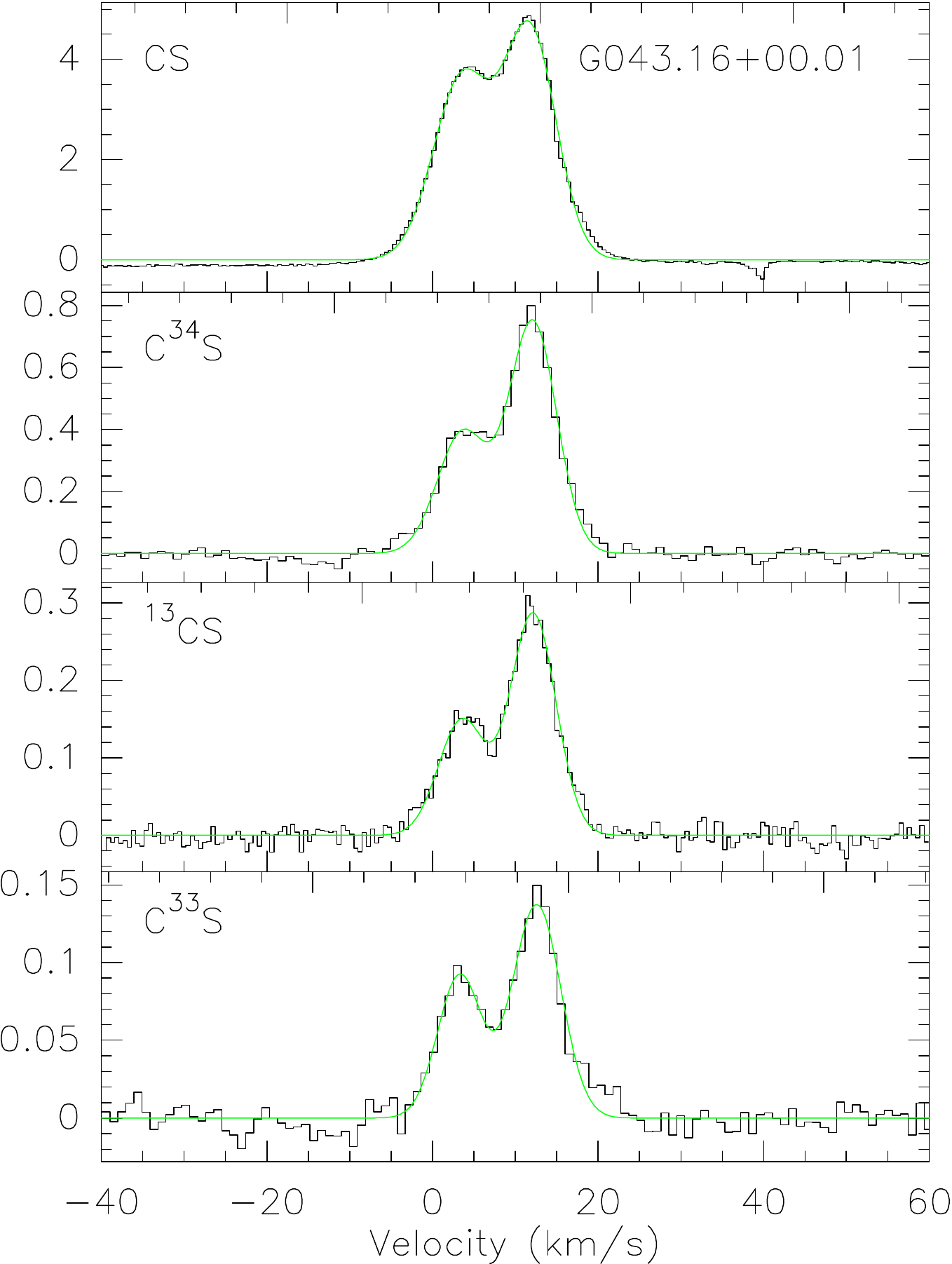}
  \includegraphics[width=83pt]{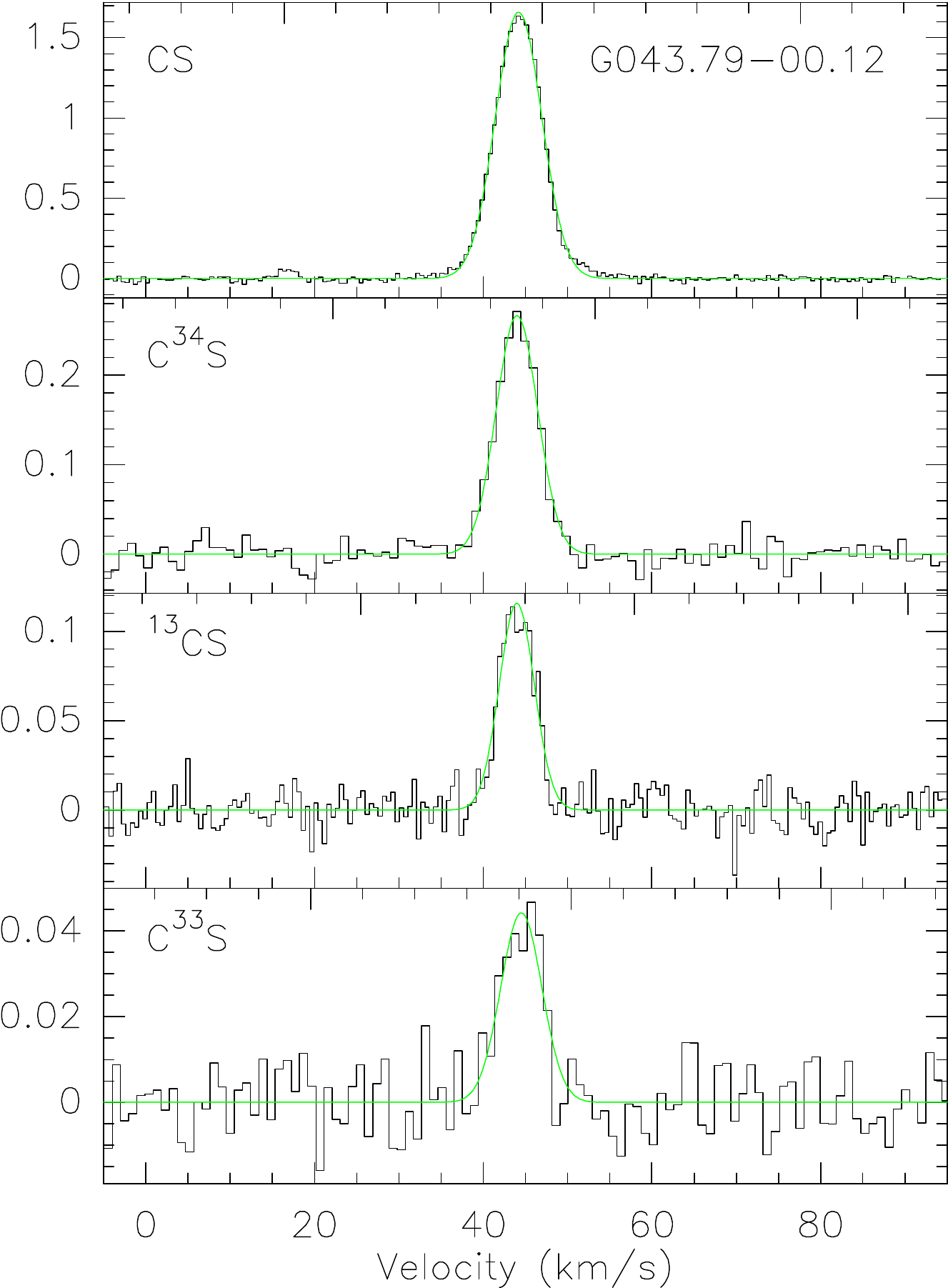}
  \includegraphics[width=83pt]{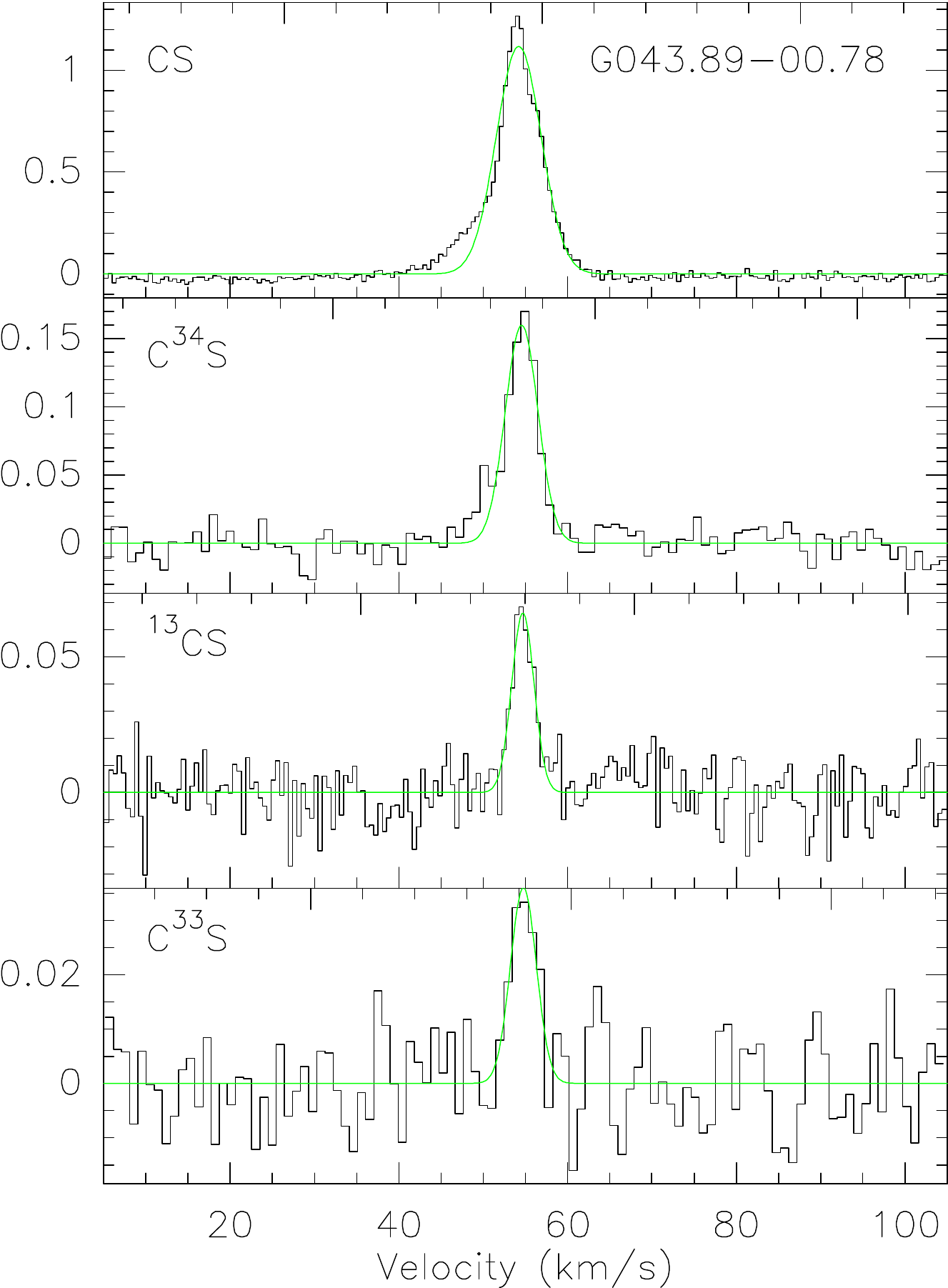}
  \includegraphics[width=83pt]{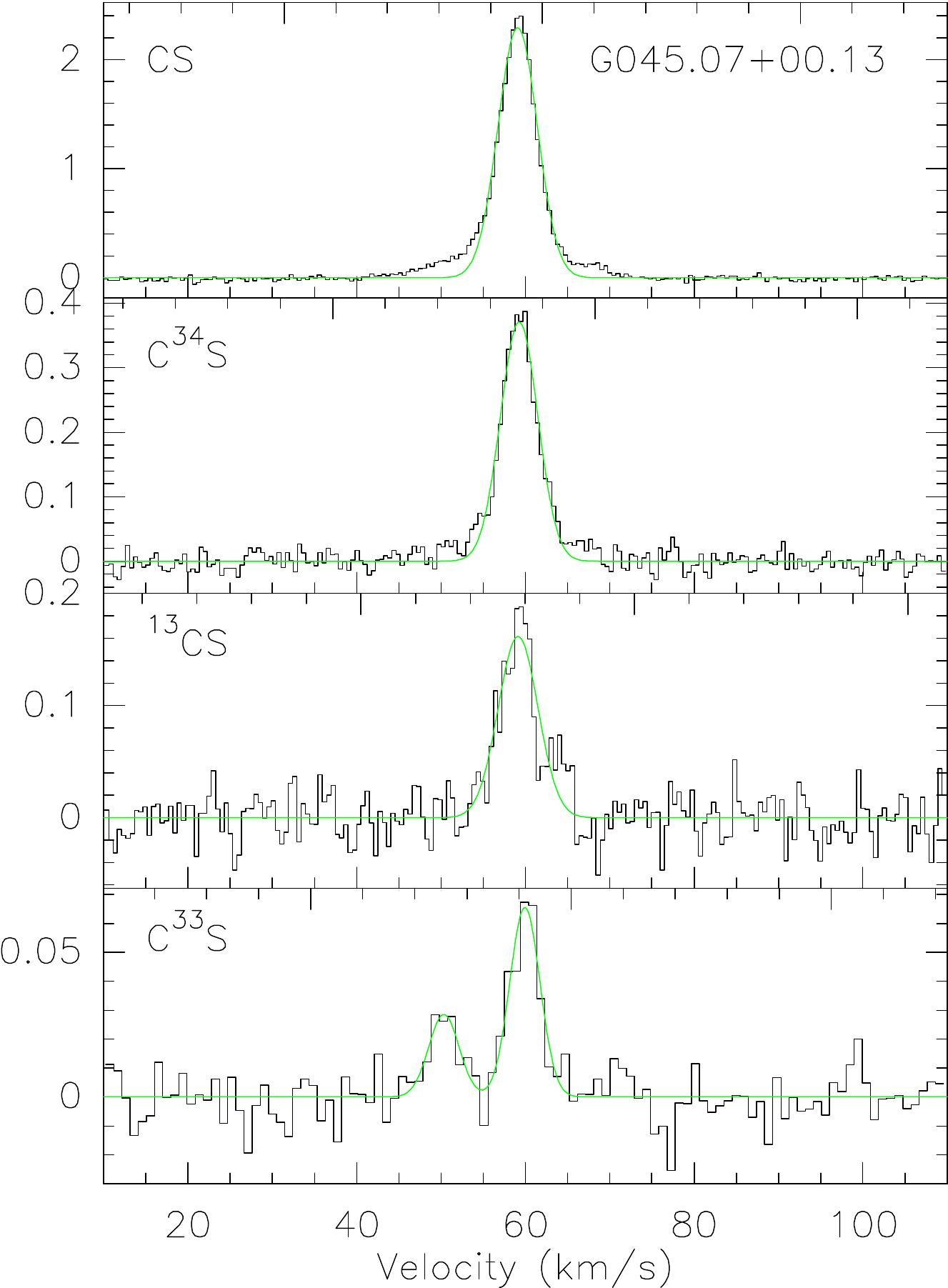}
  \includegraphics[width=83pt]{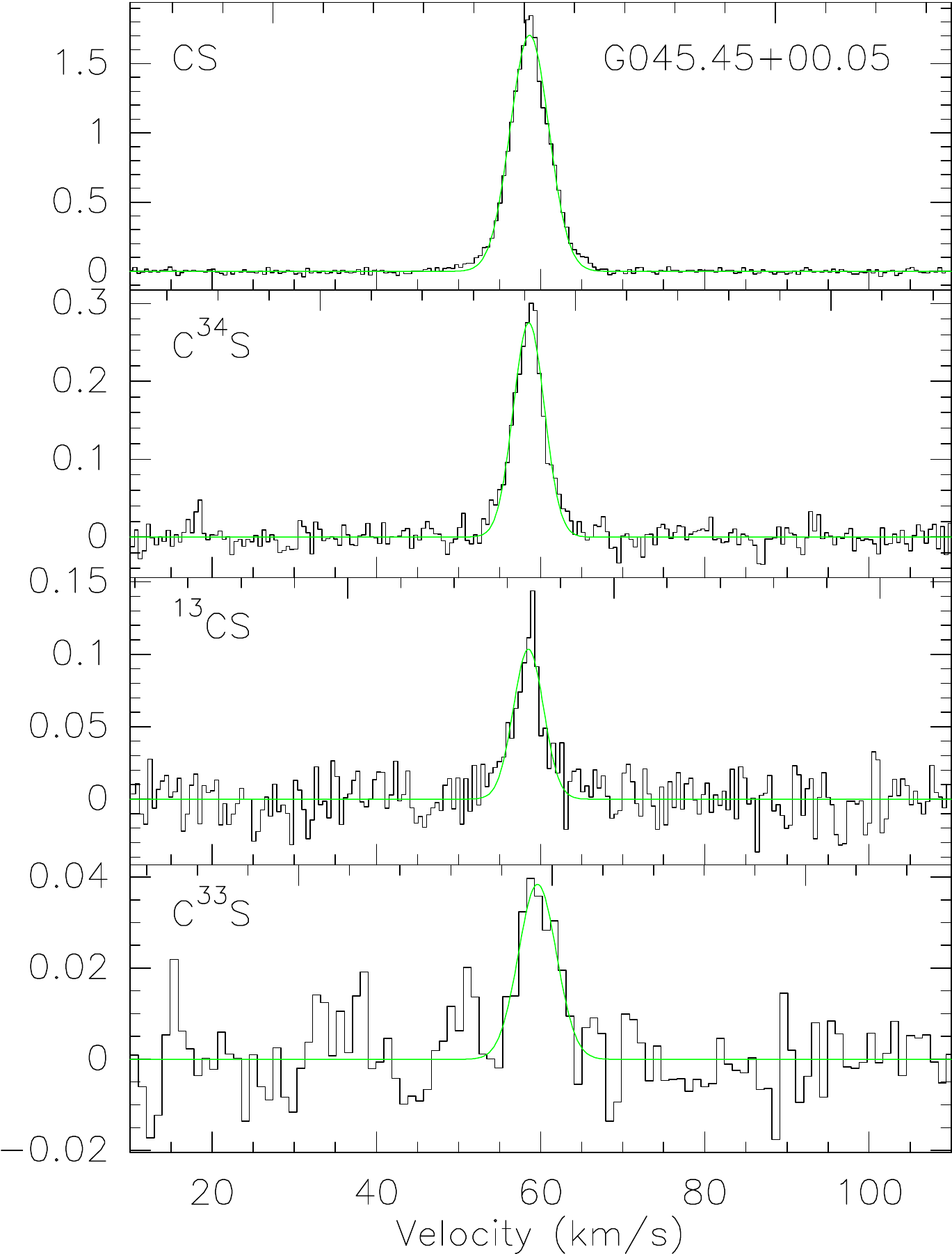}
  \includegraphics[width=83pt]{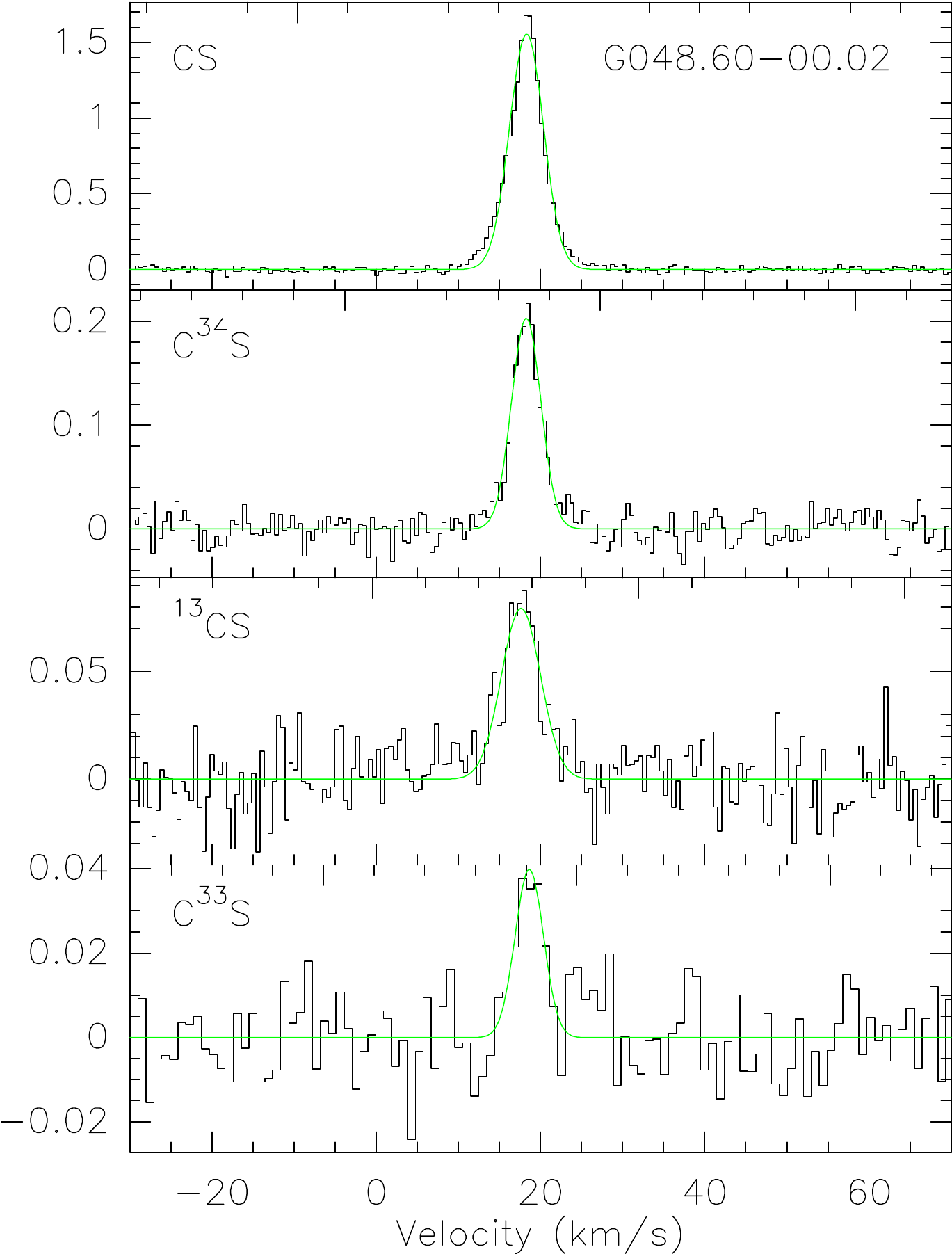}
  \includegraphics[width=83pt]{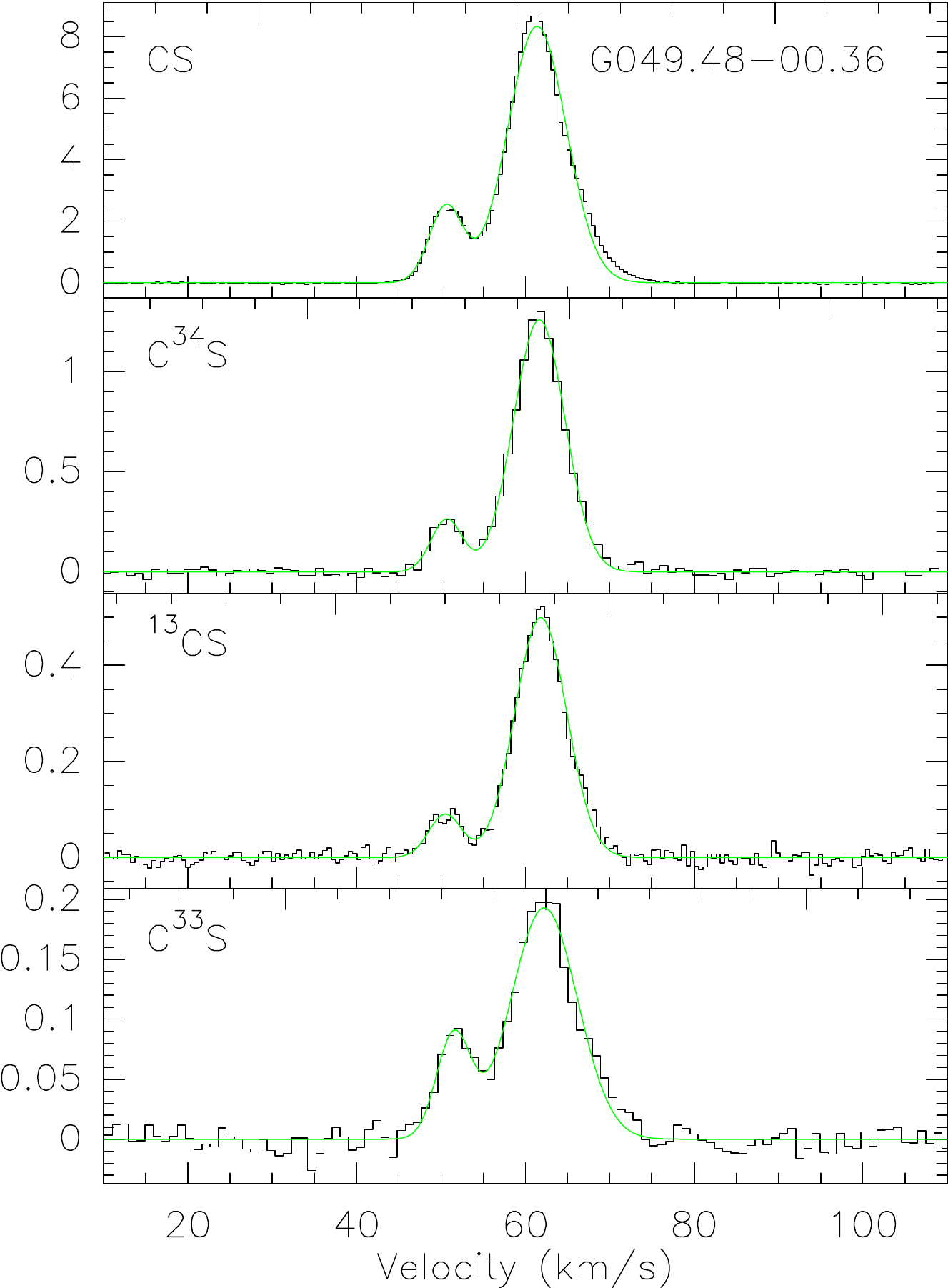}
  \includegraphics[width=83pt]{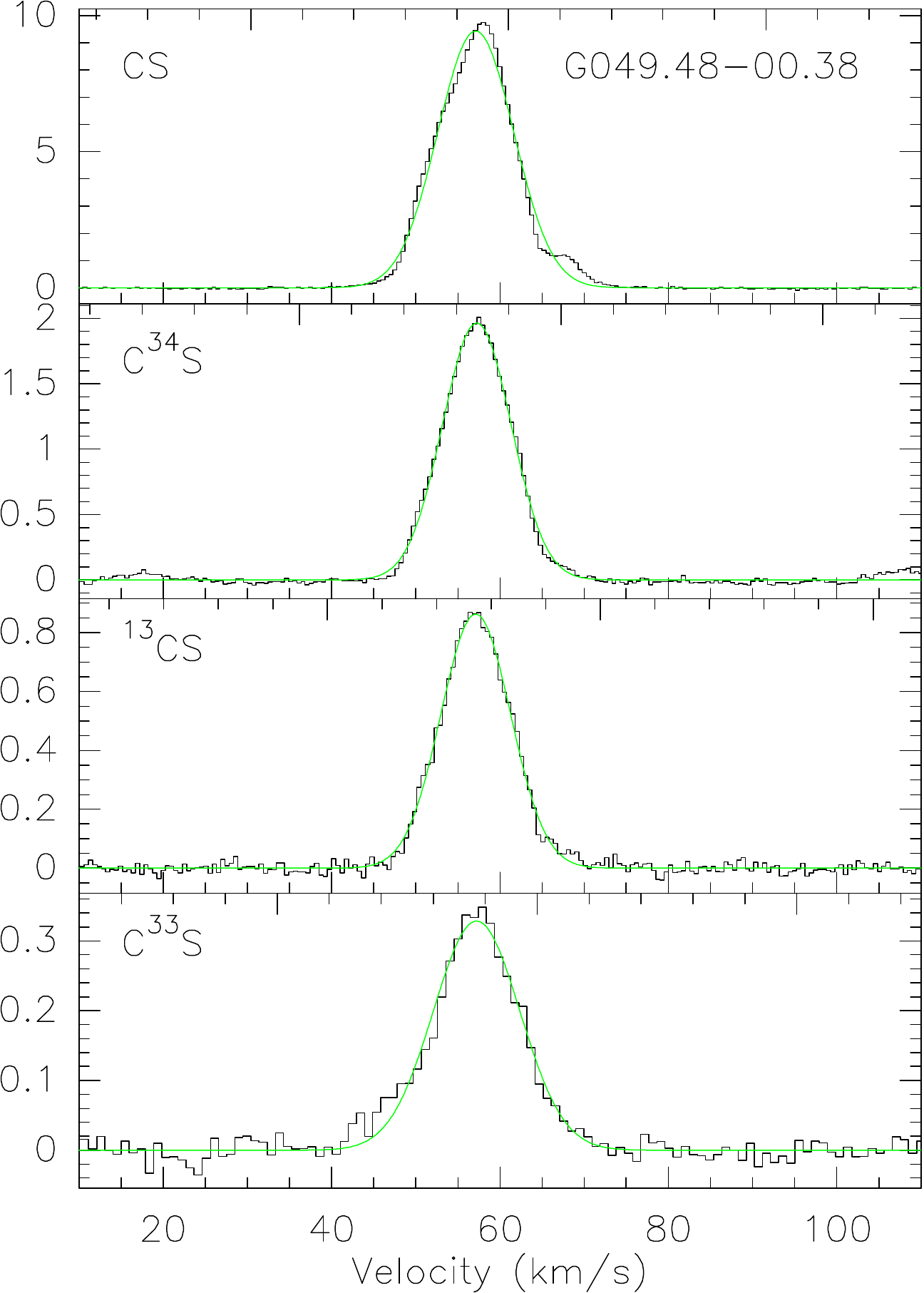}
  \includegraphics[width=83pt]{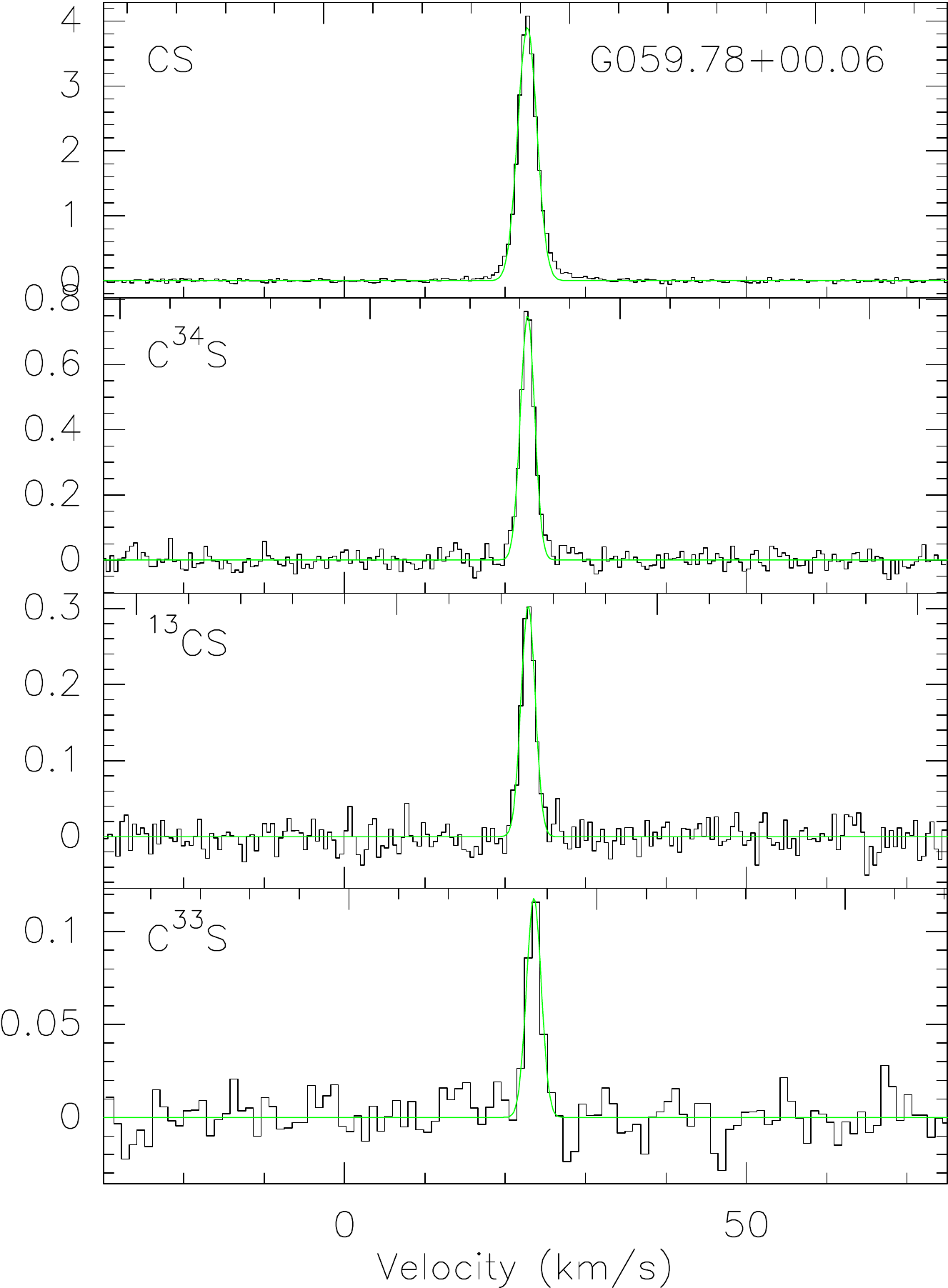}
  \includegraphics[width=83pt]{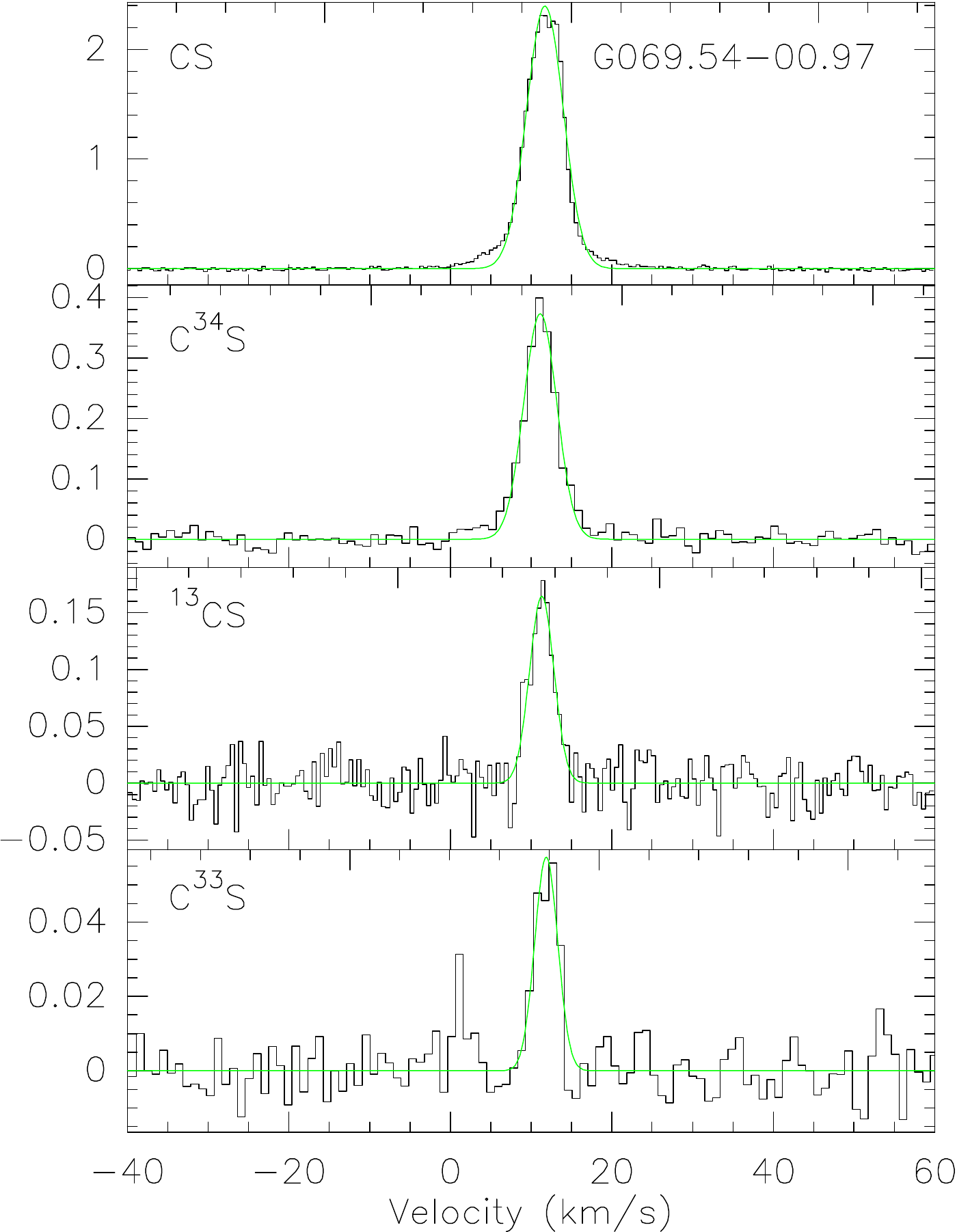}
  \includegraphics[width=83pt]{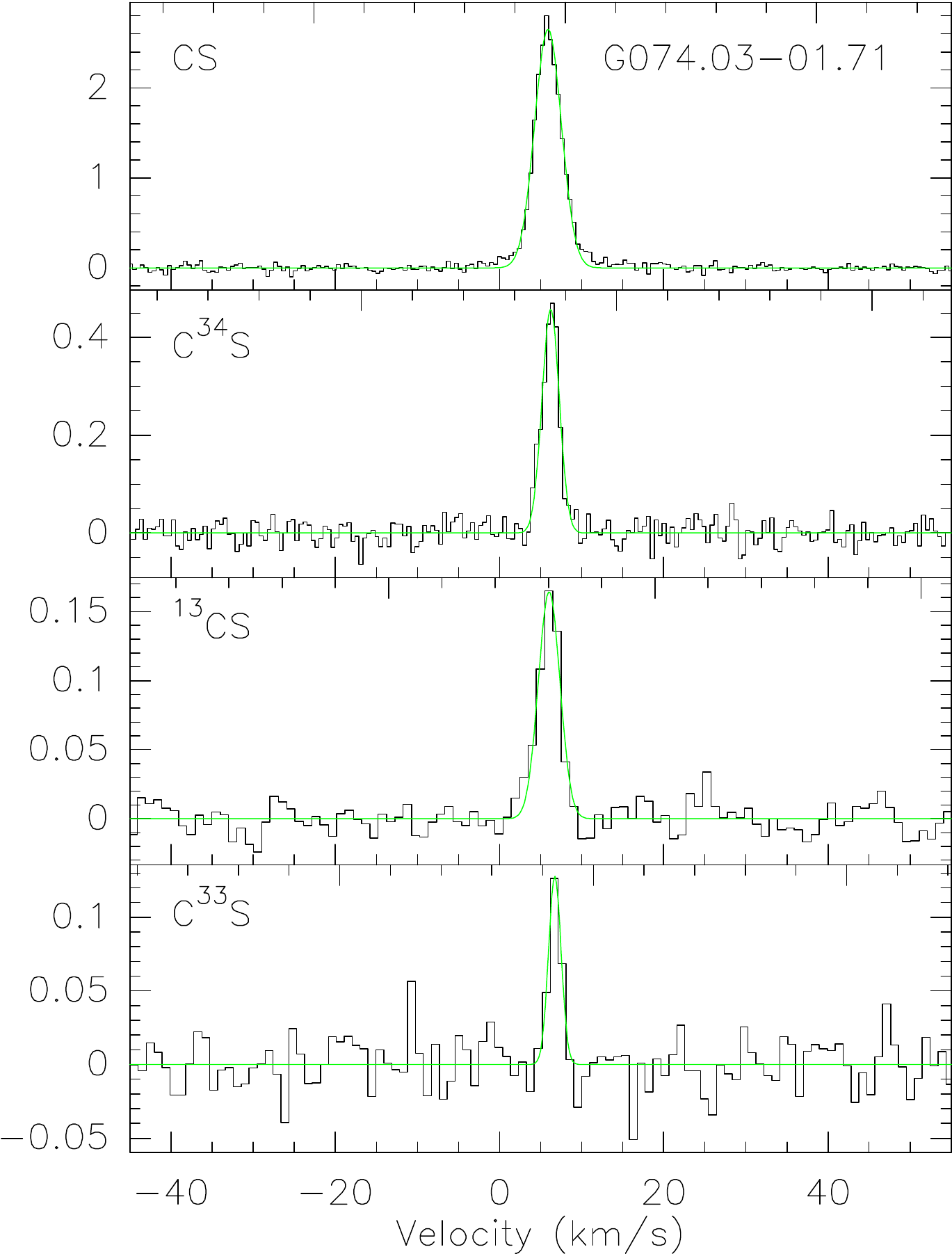}
  \includegraphics[width=83pt]{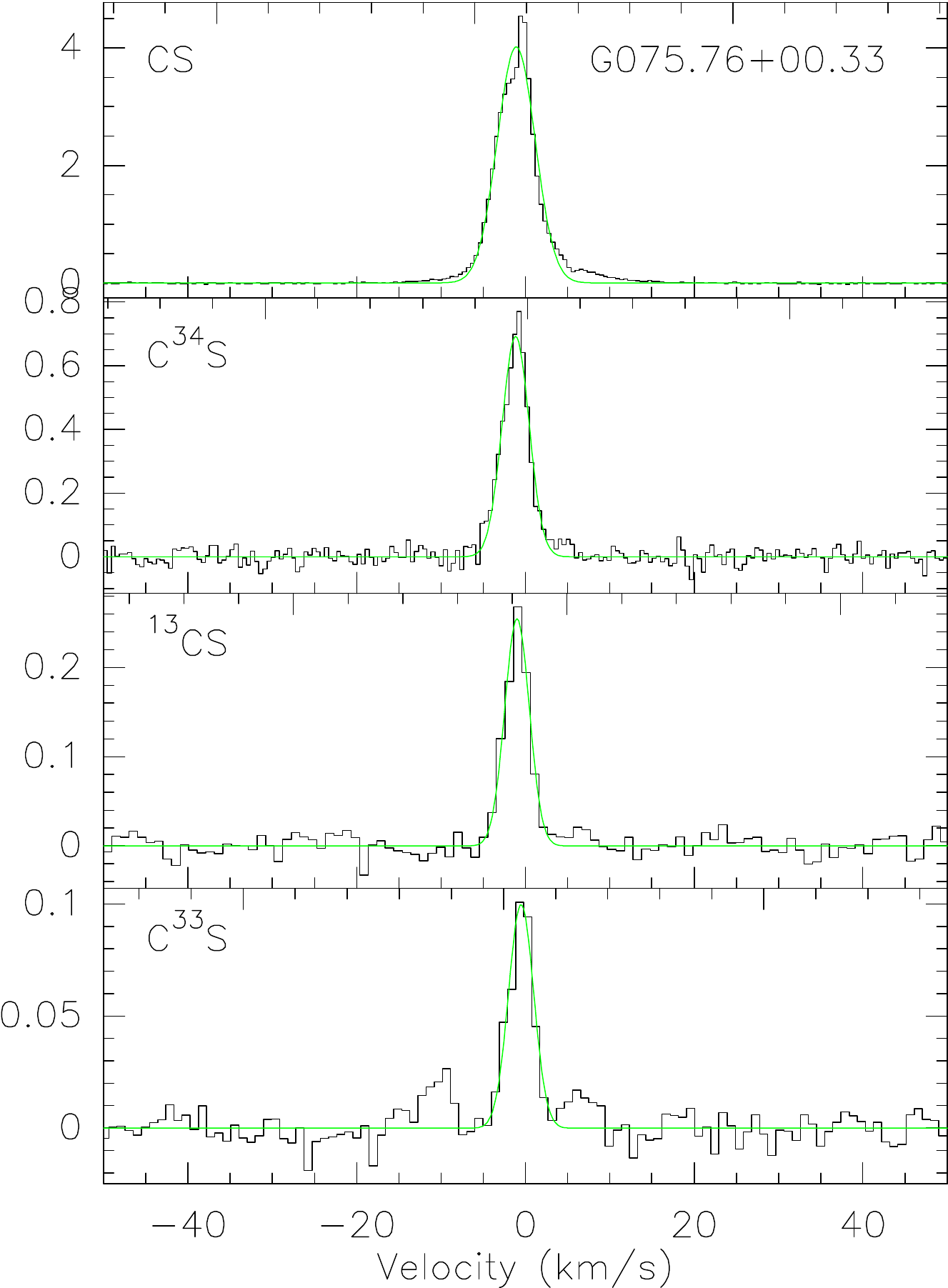}
  \includegraphics[width=83pt]{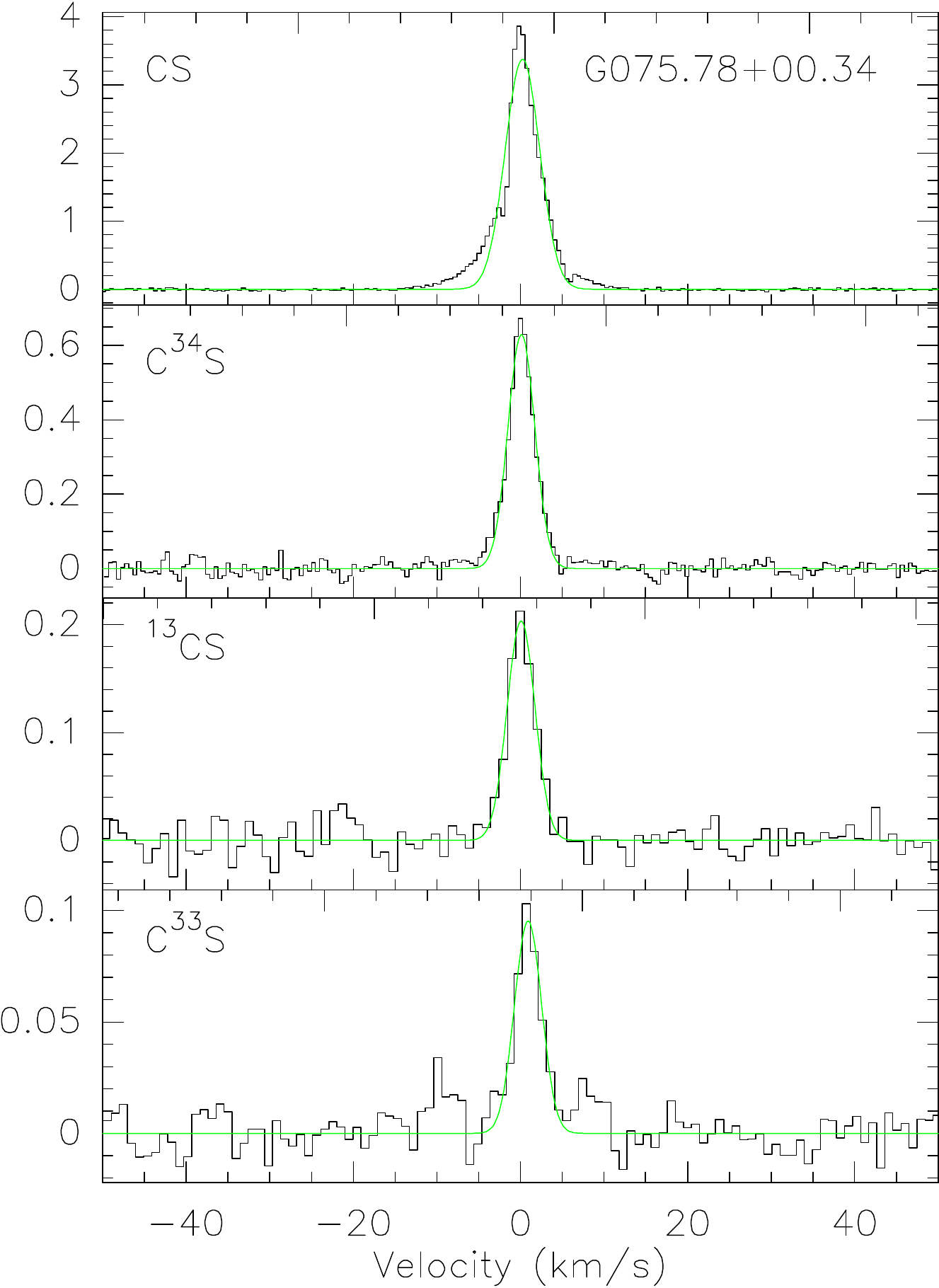}
  \includegraphics[width=83pt]{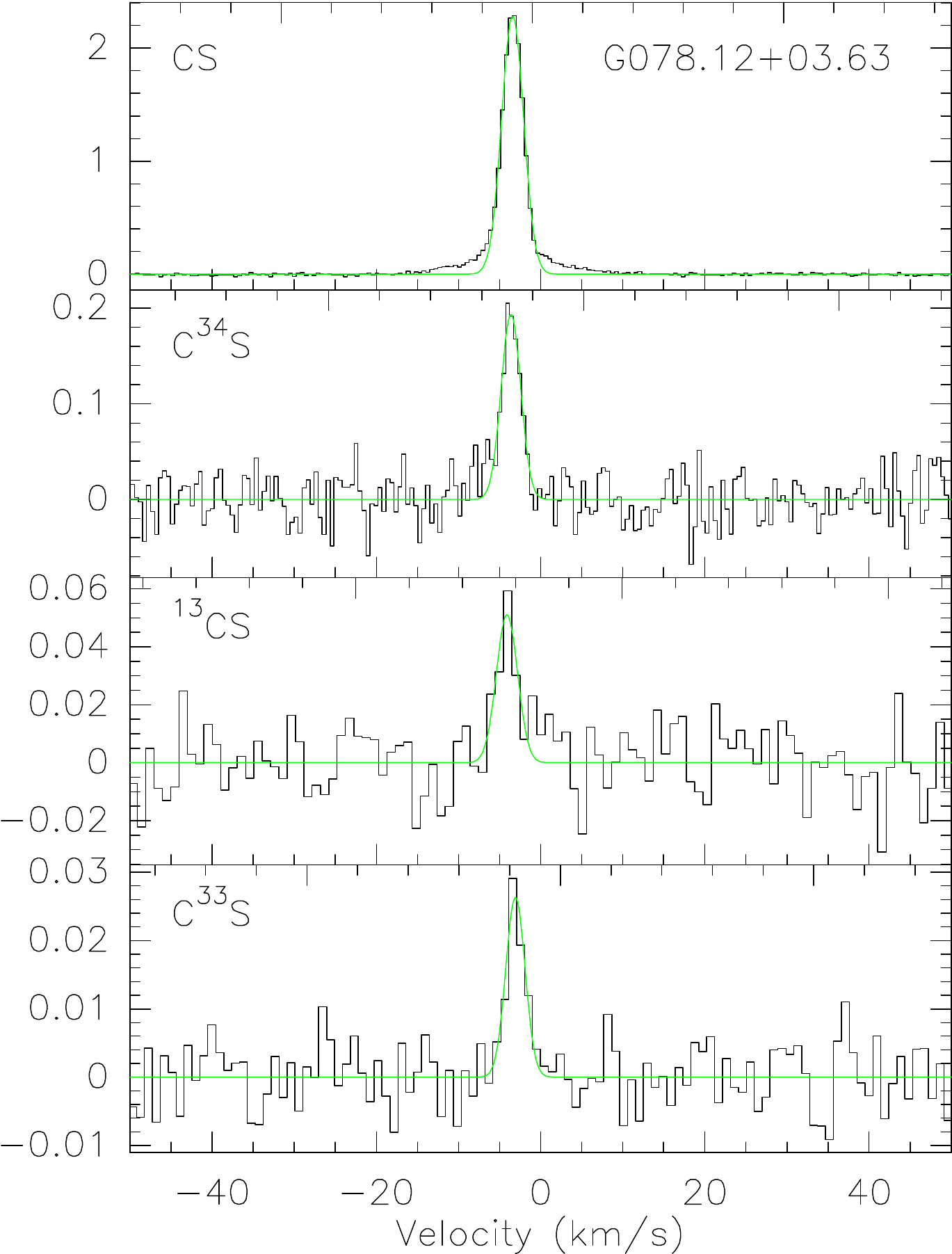}
  \includegraphics[width=83pt]{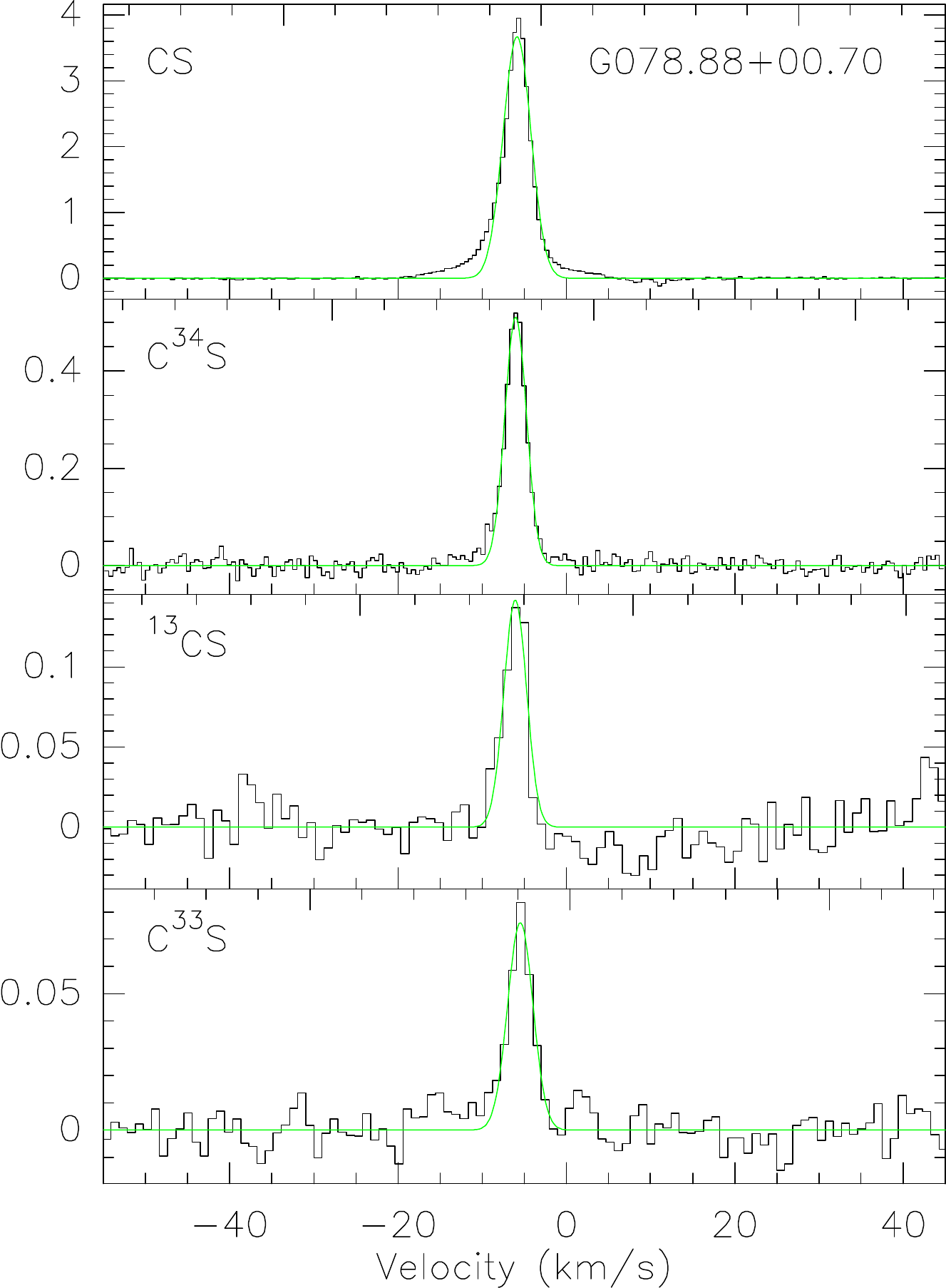}
  \includegraphics[width=83pt]{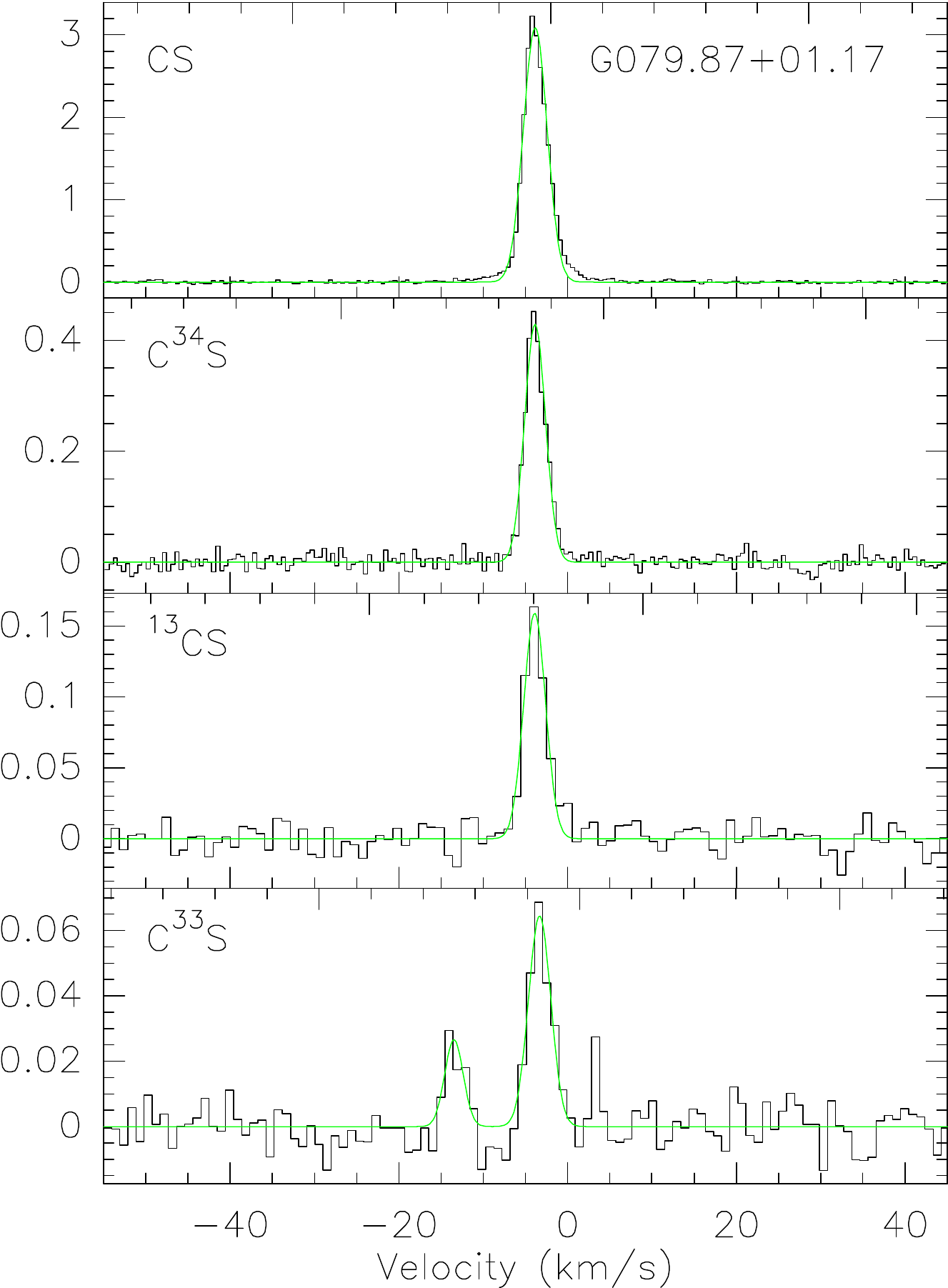}
  \includegraphics[width=83pt]{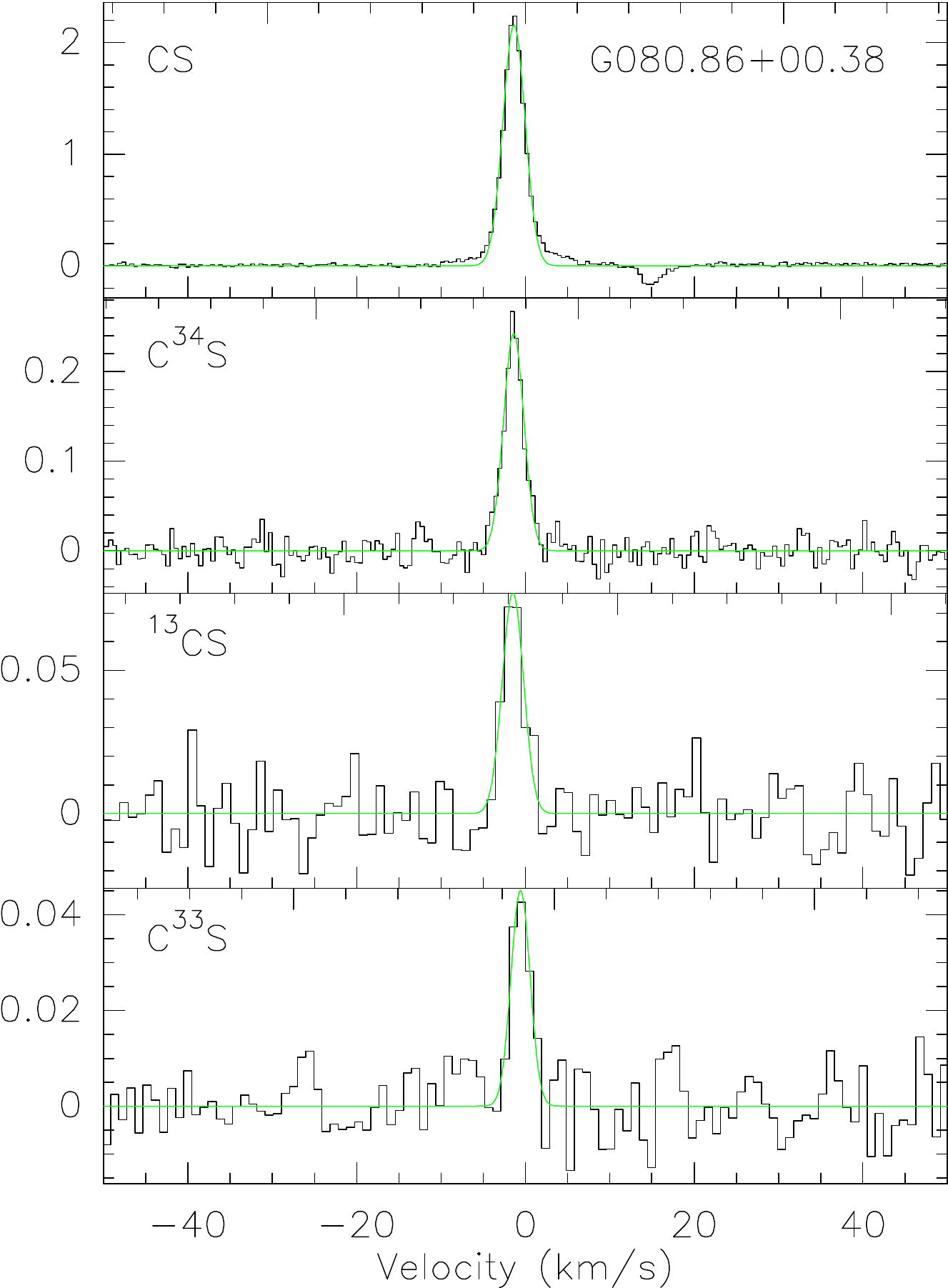}
  \includegraphics[width=83pt]{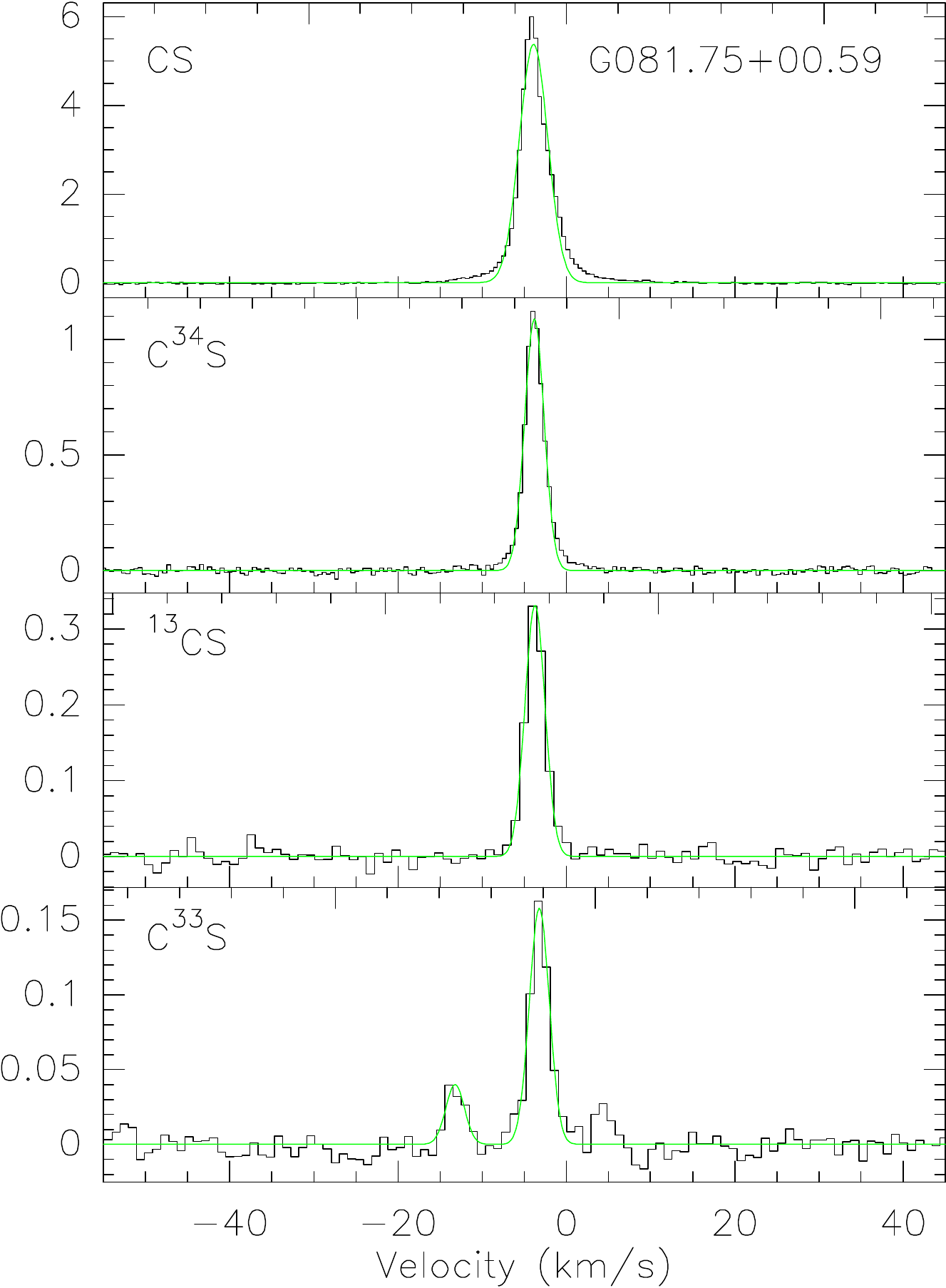}
  \includegraphics[width=83pt]{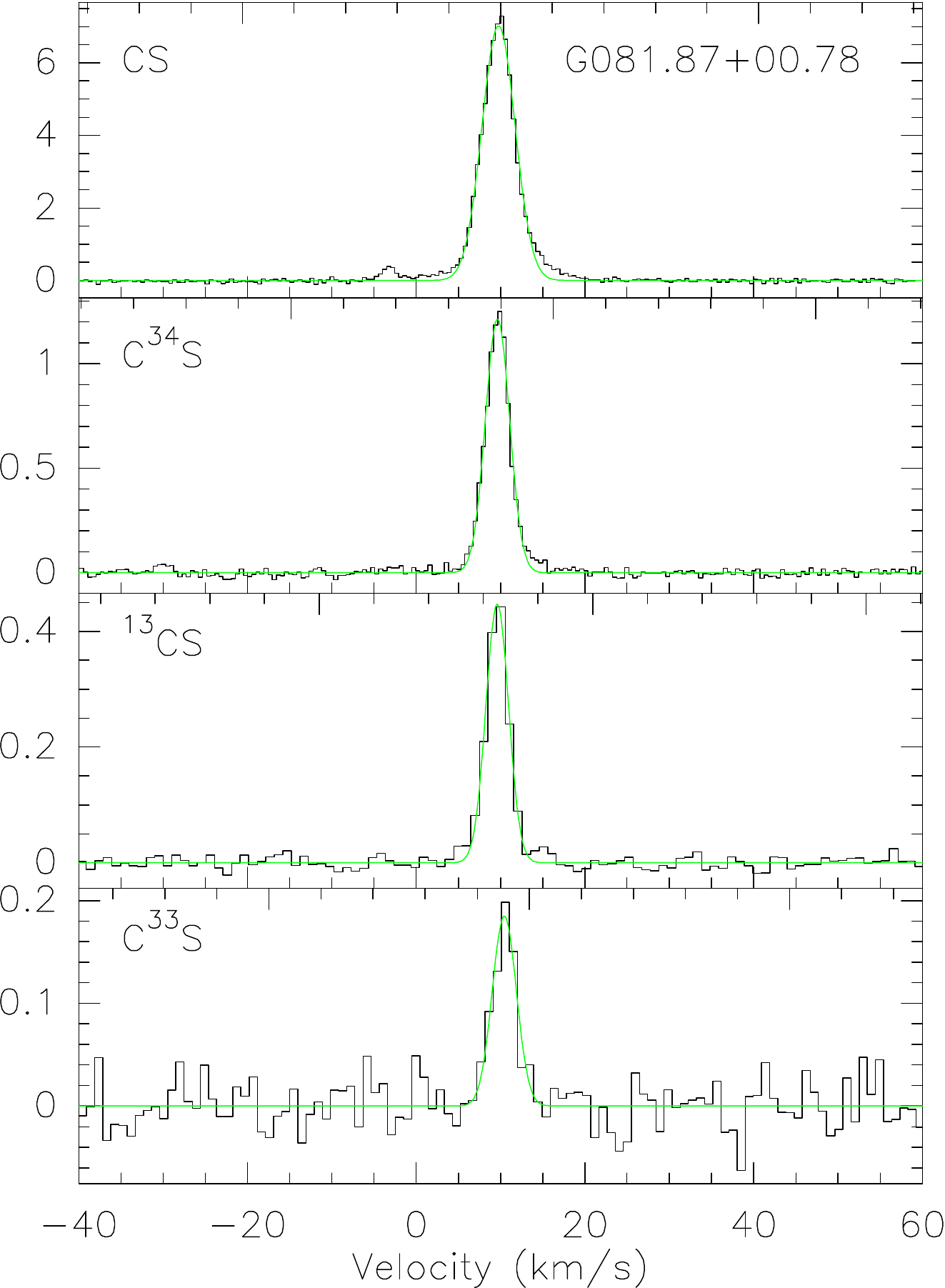}
  \includegraphics[width=83pt]{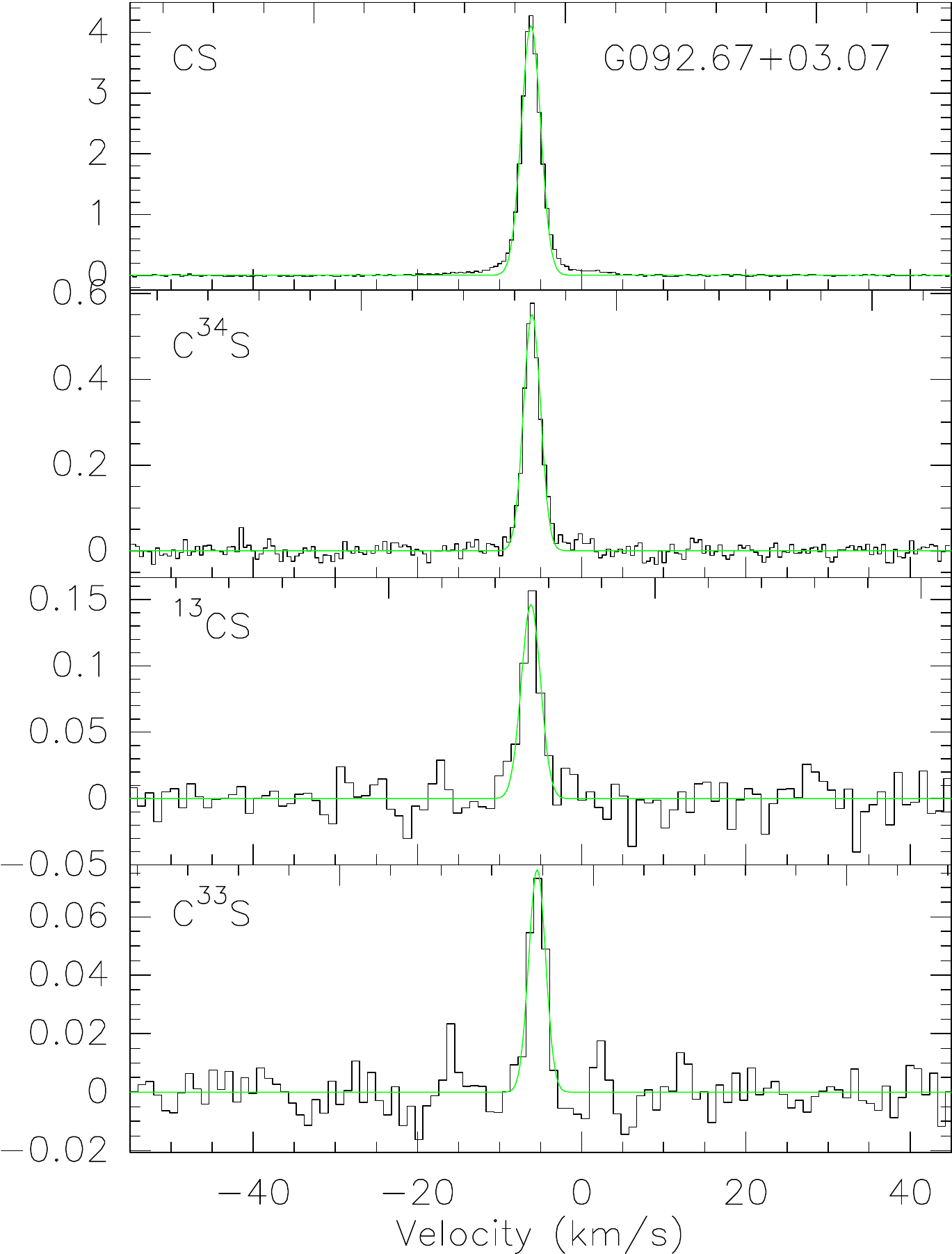}
  \includegraphics[width=83pt]{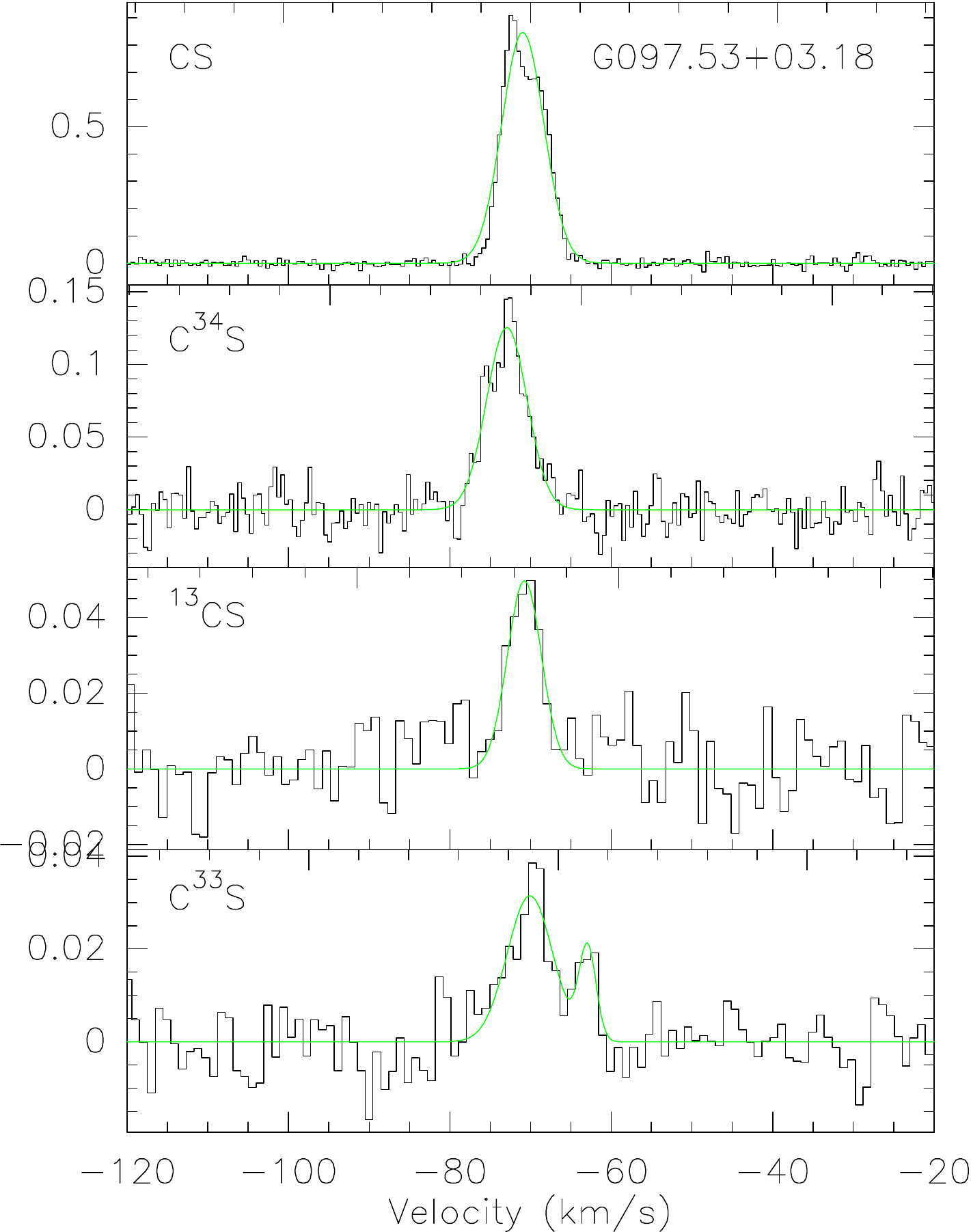}
  \includegraphics[width=83pt]{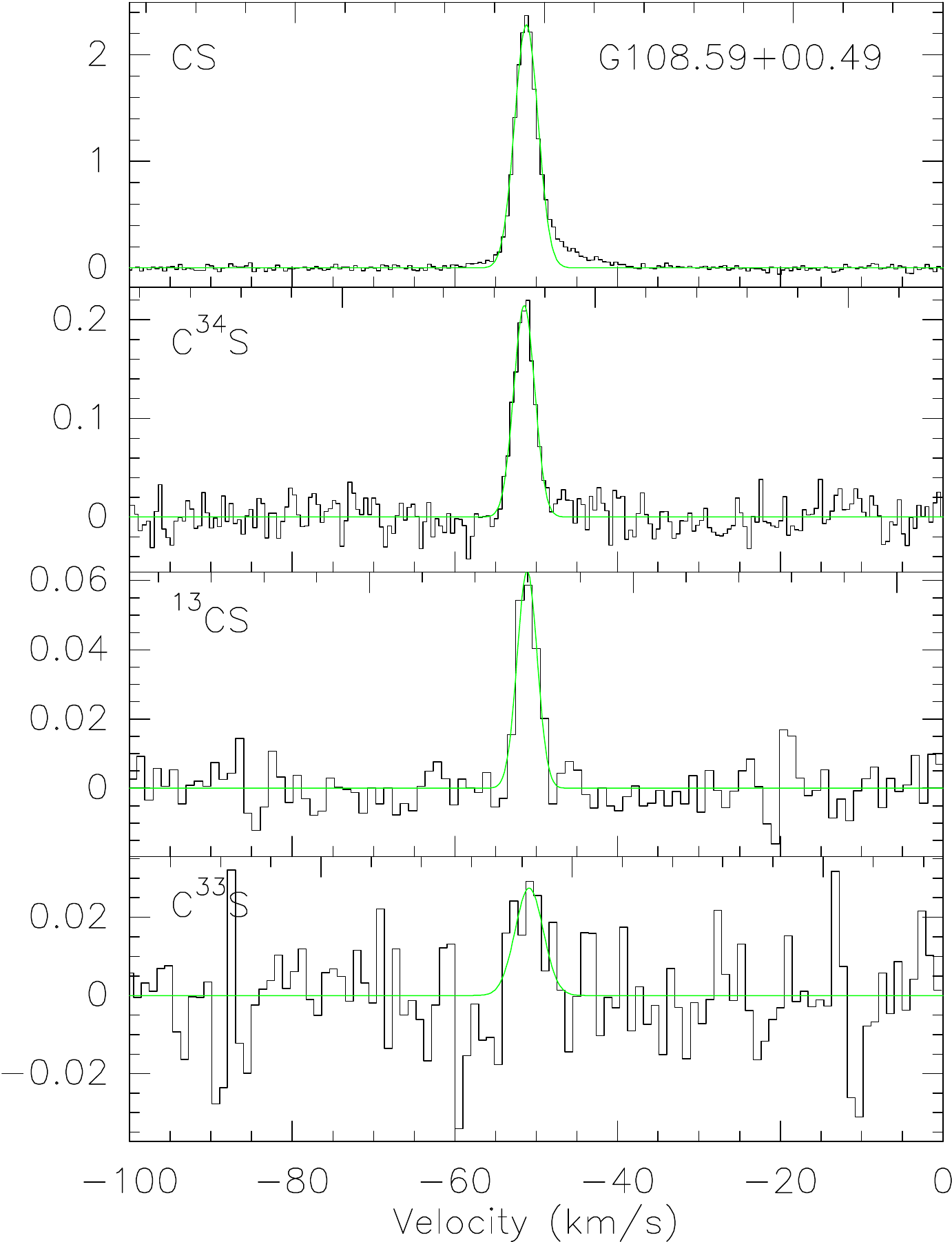}
  \includegraphics[width=83pt]{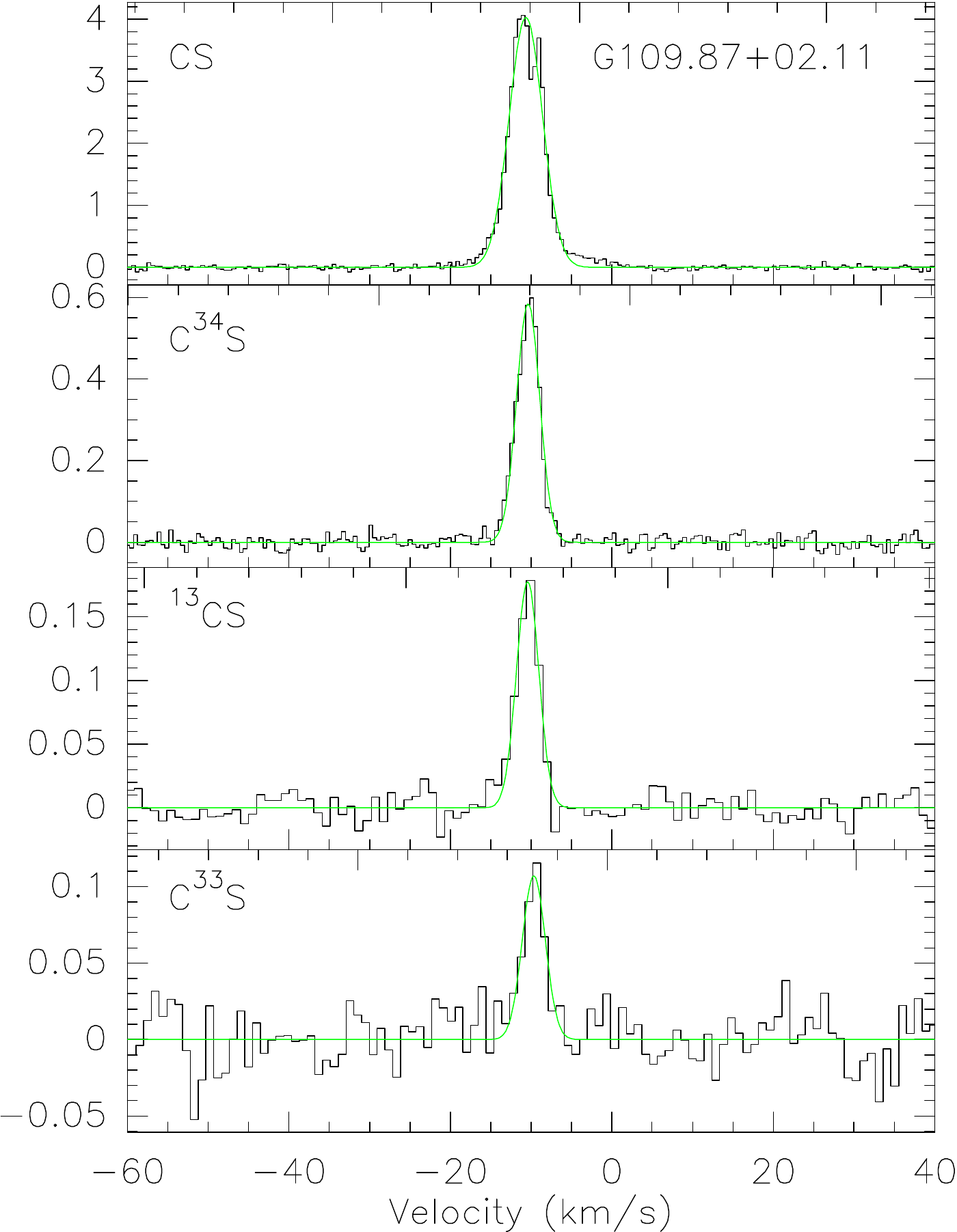}
  \includegraphics[width=83pt]{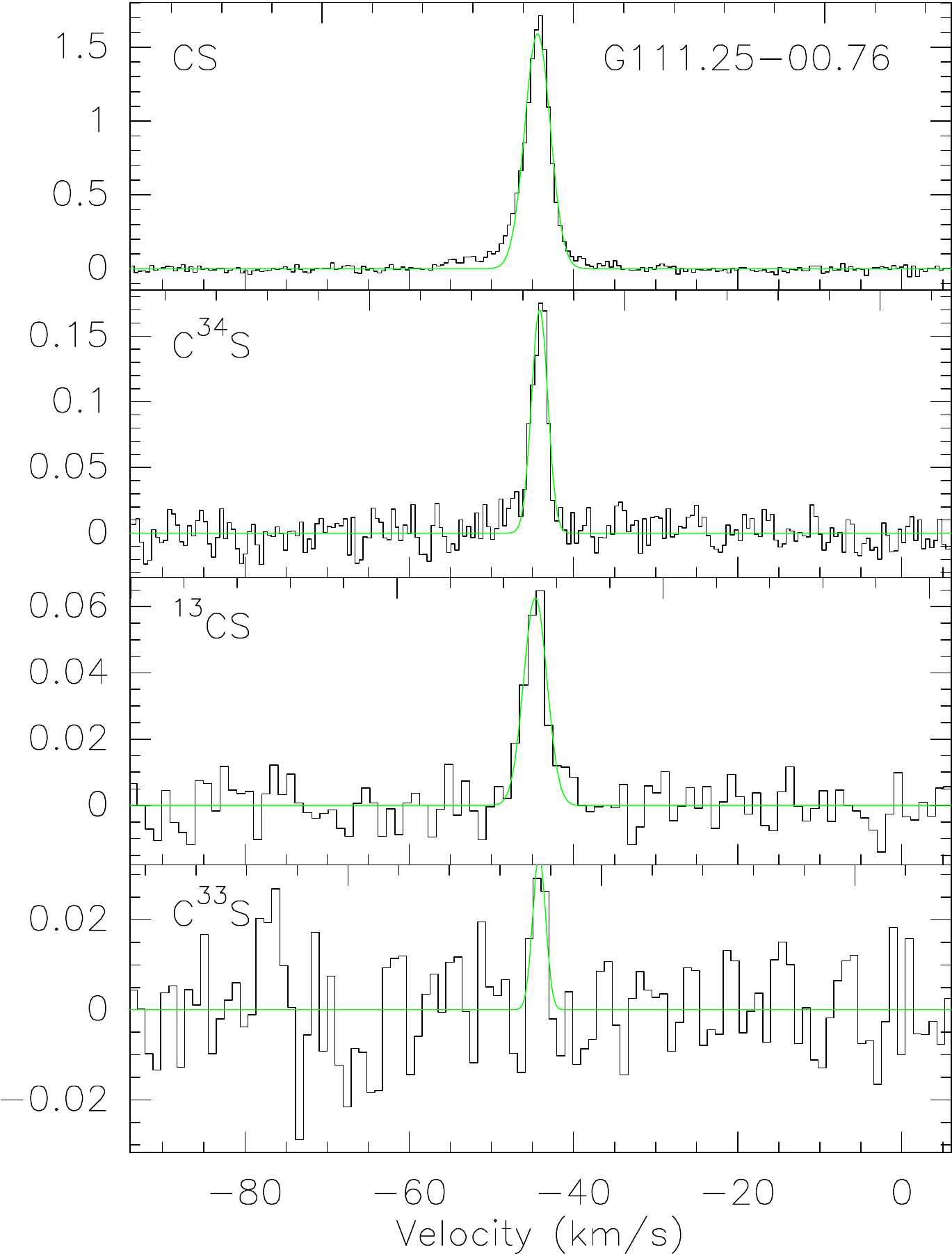}
  \includegraphics[width=83pt]{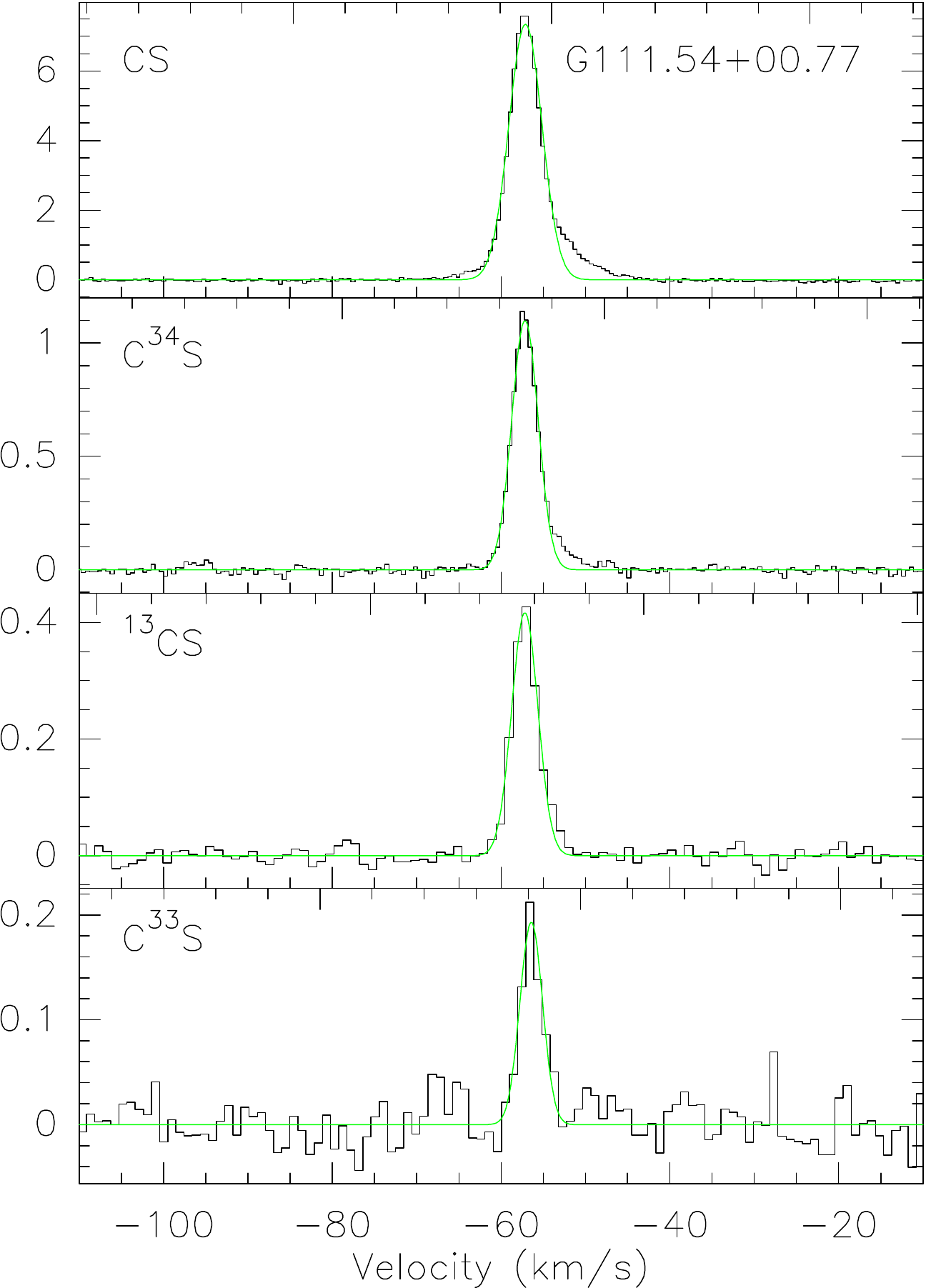}
\end{figure*}
 \begin{figure*}[h]
  \caption{ Spectra showing detected $J$=2$-$1 lines of CS isotopologues, i.e. CS, C$^{34}$S, $^{13}$CS and C$^{33}$S, toward 61 sources observed with the ARO 12-m telescope. }
\end{figure*}

\begin{figure*}[h]
\center
 \label{f_list_1}
  \includegraphics[width=82.3pt]{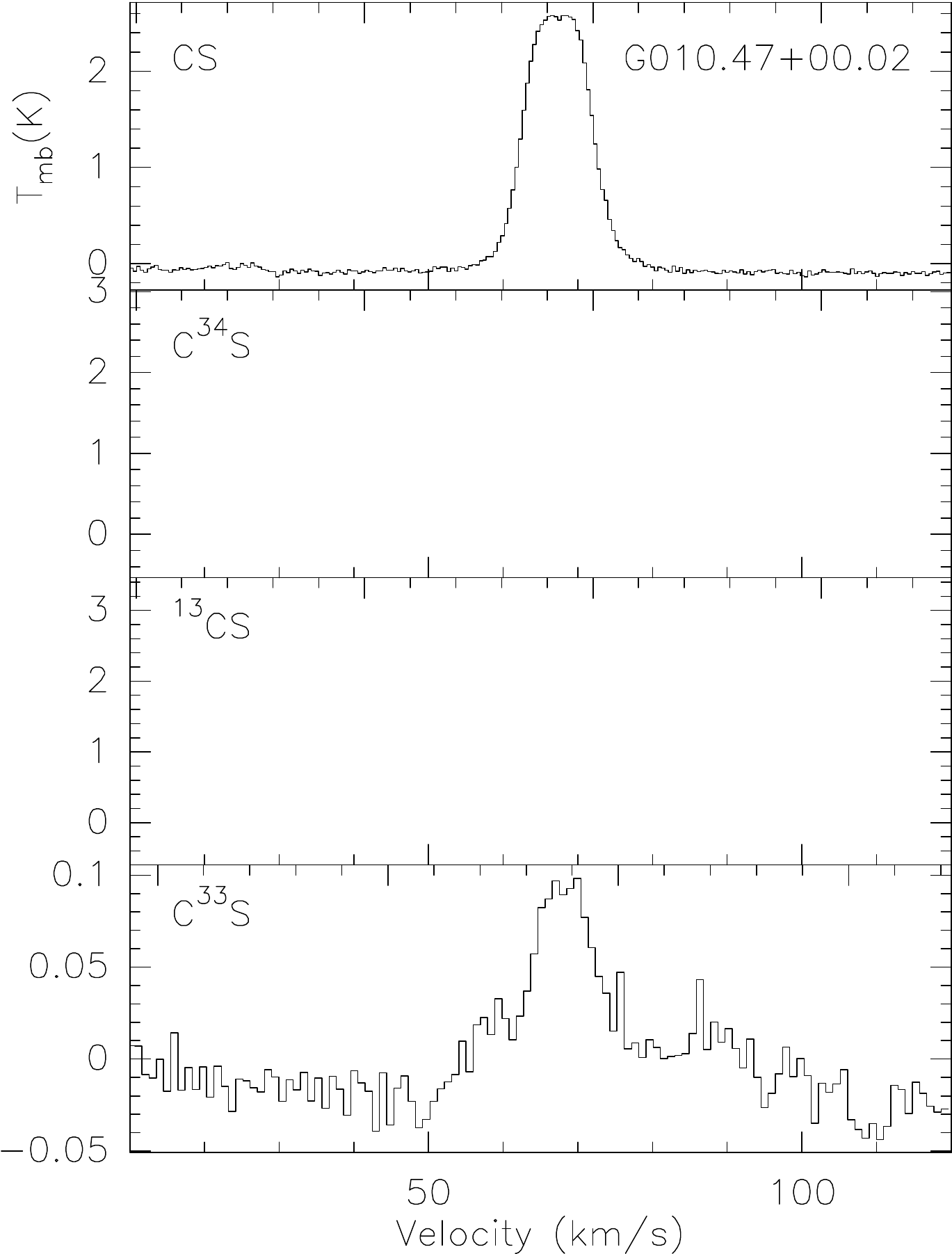}
  \includegraphics[width=85pt]{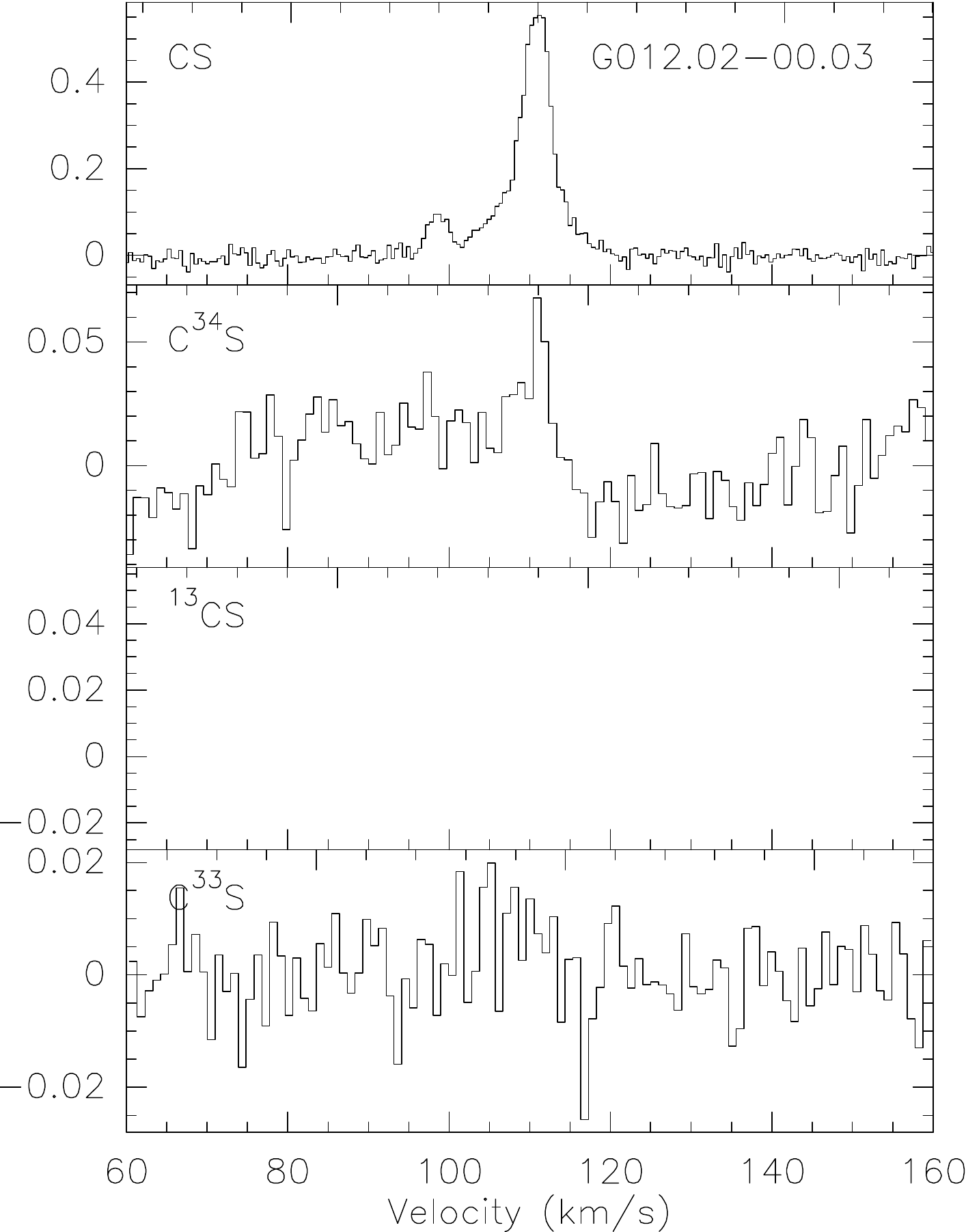}
  \includegraphics[width=80pt]{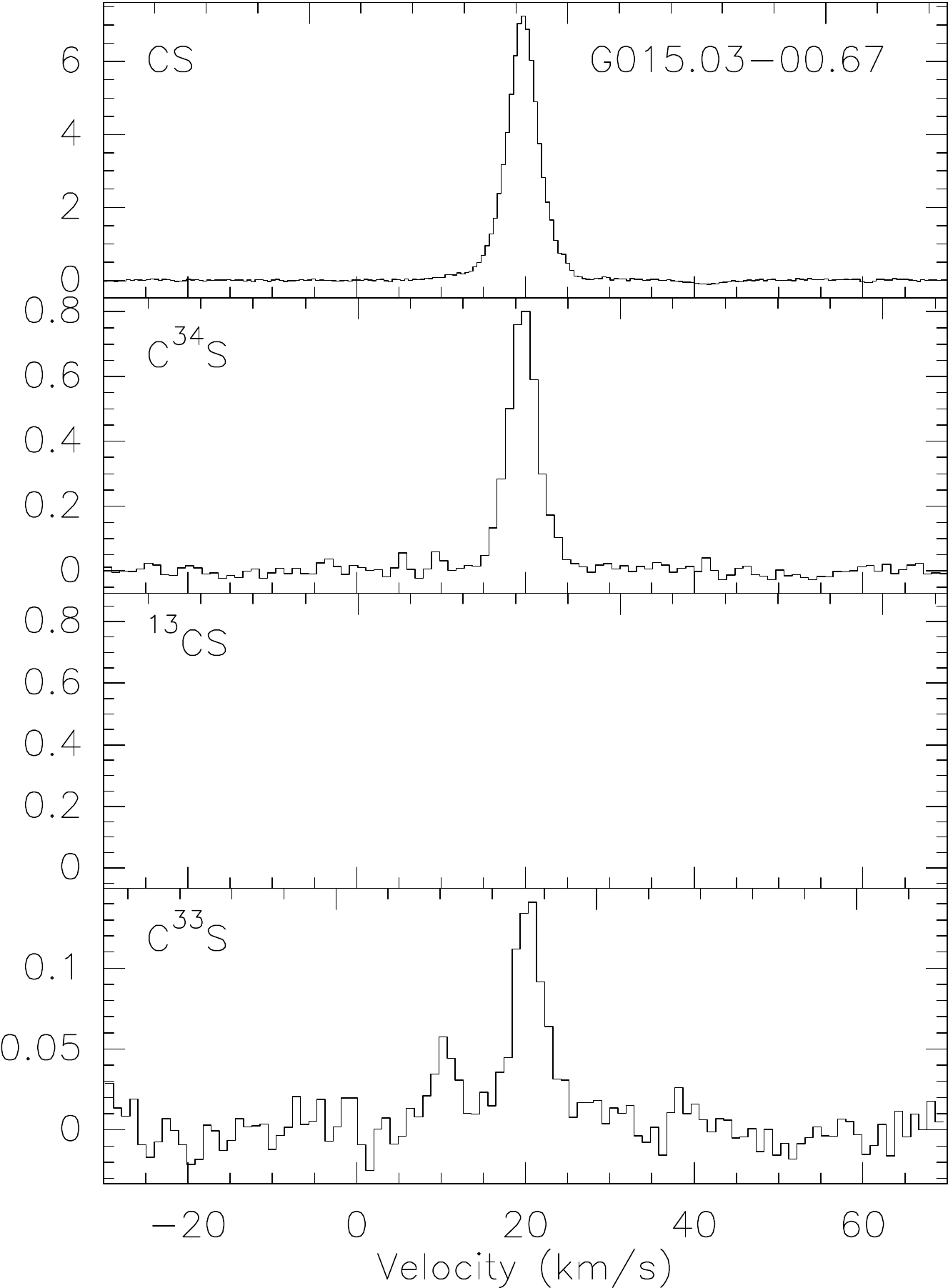}
  \includegraphics[width=82pt]{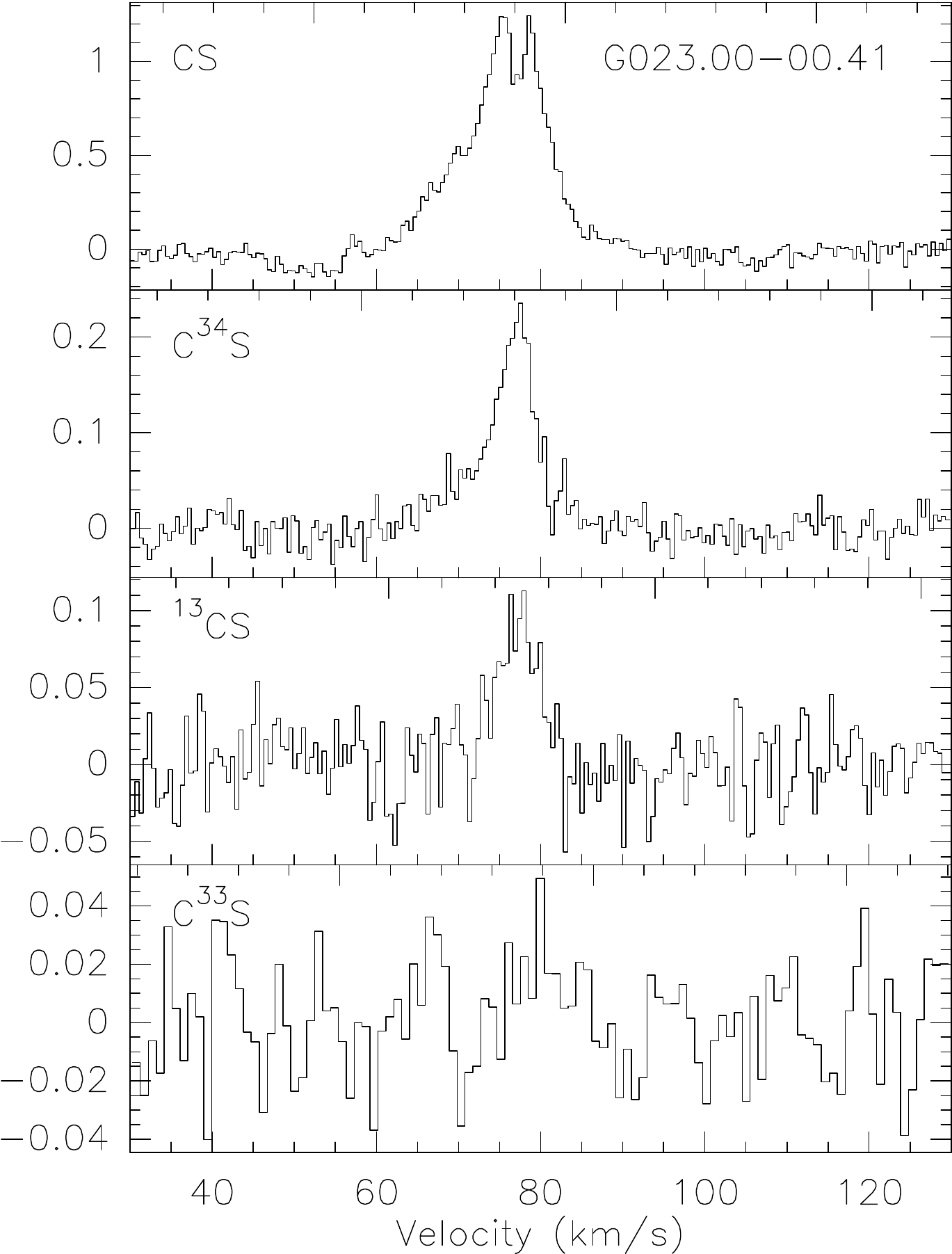}
  \includegraphics[width=82pt]{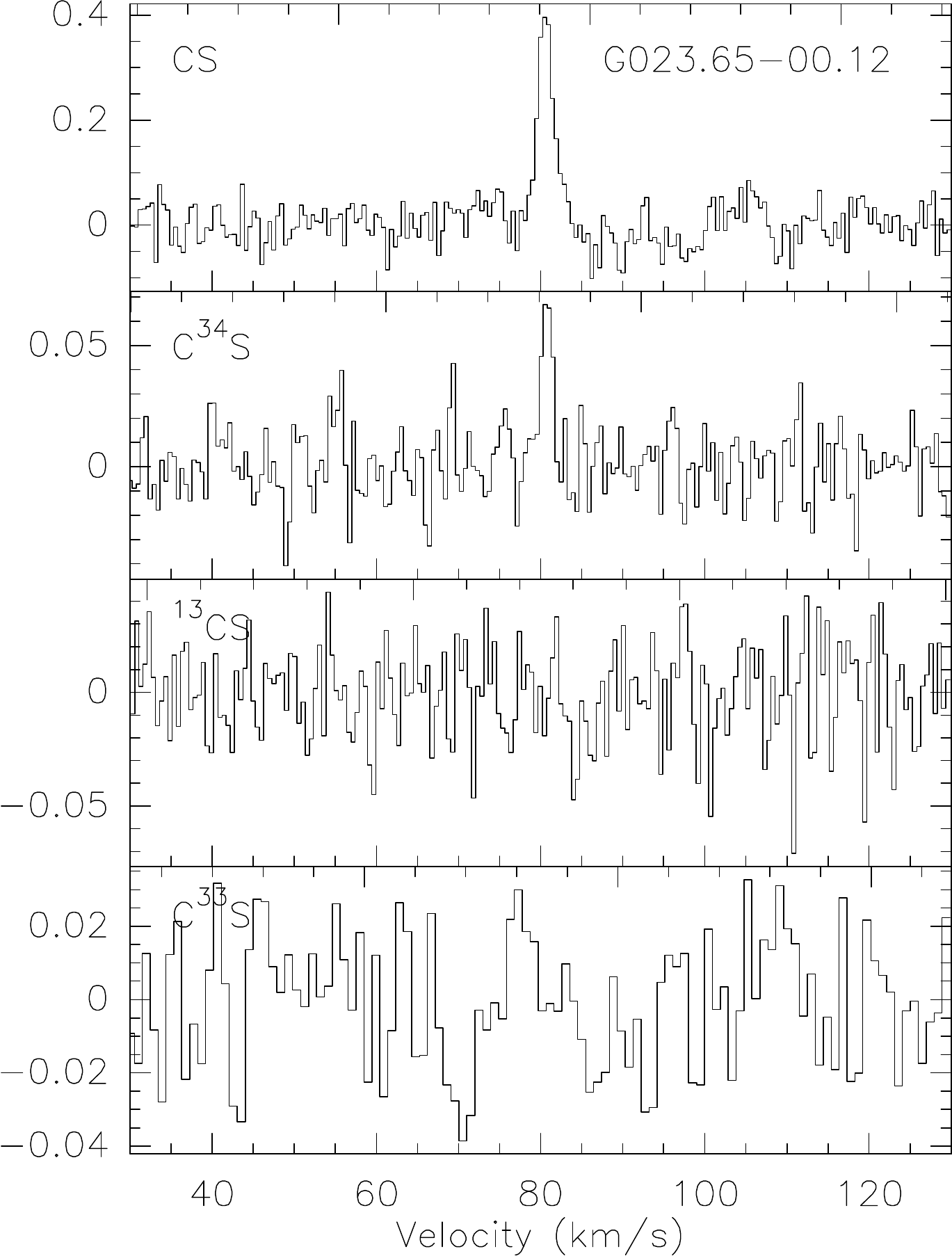}
  \includegraphics[width=84pt]{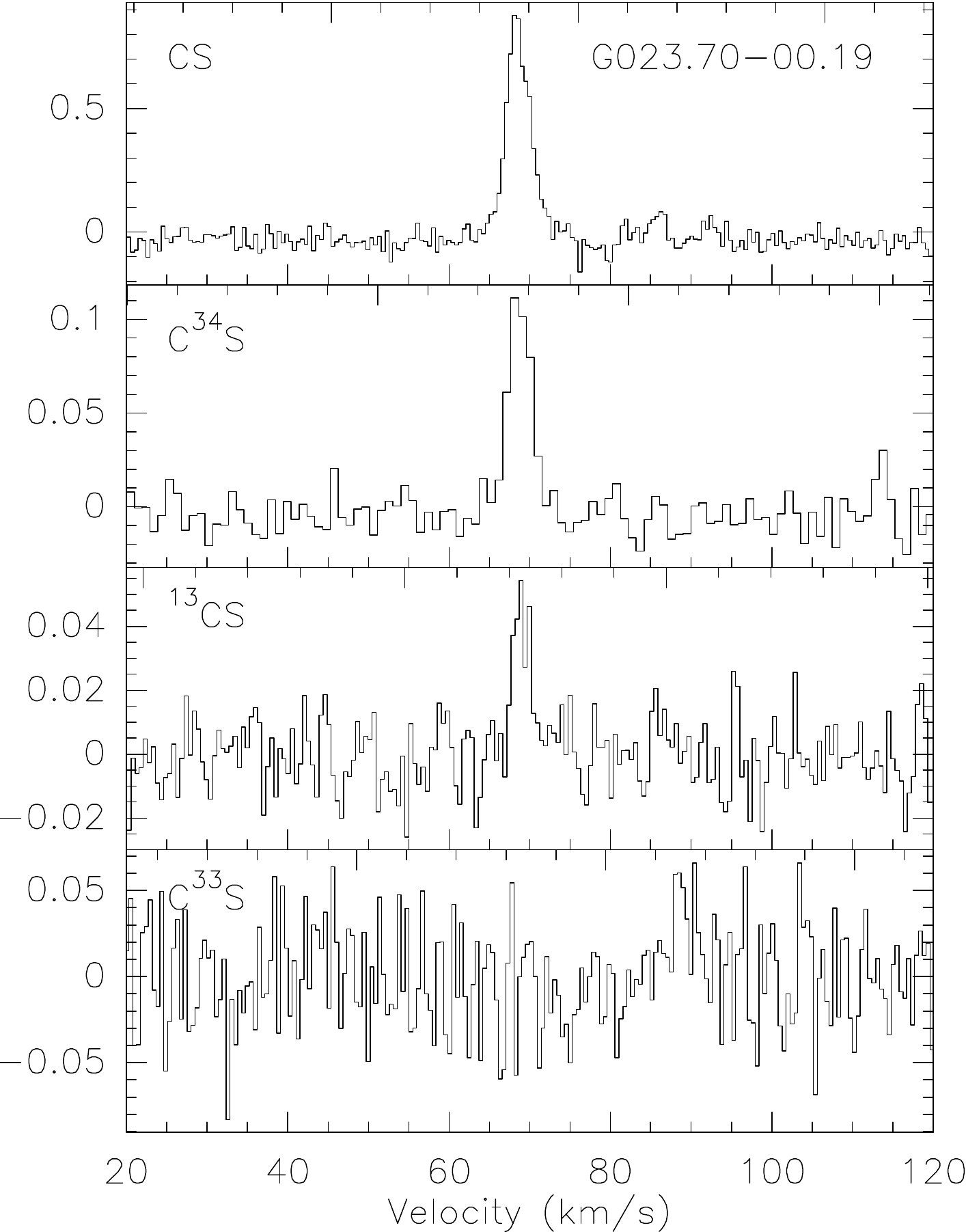}
  \includegraphics[width=83pt]{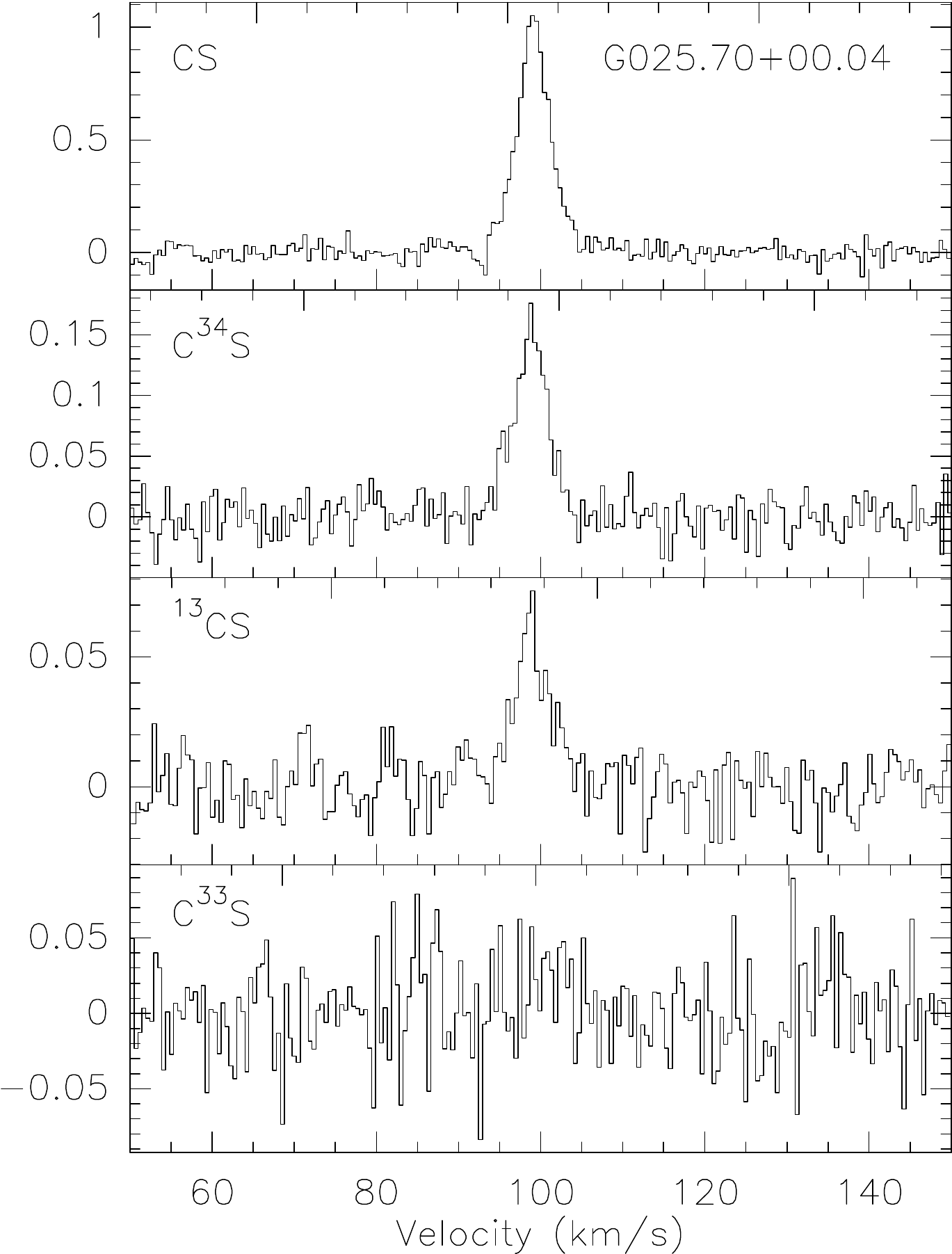}
  \includegraphics[width=83pt]{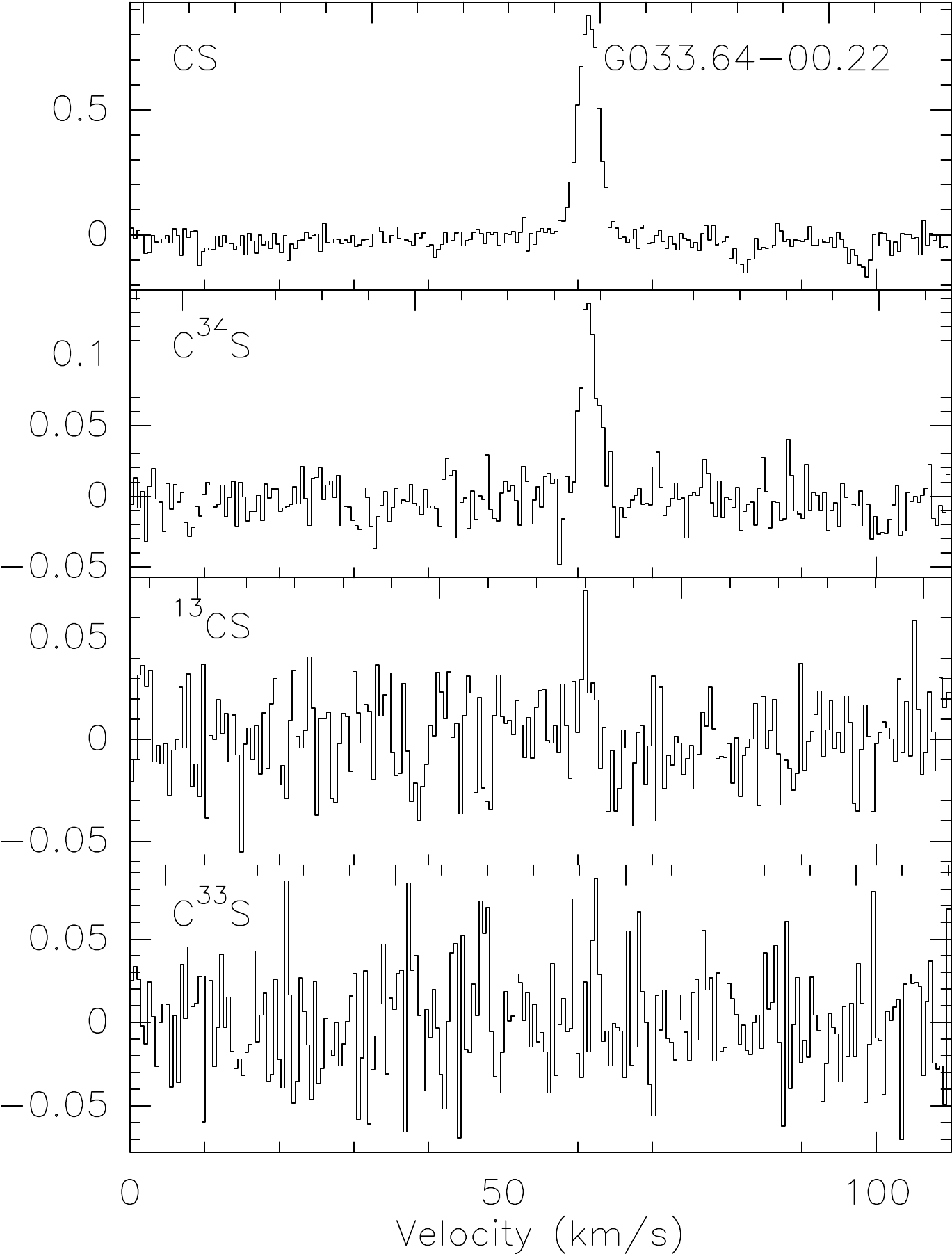}
  \includegraphics[width=83pt]{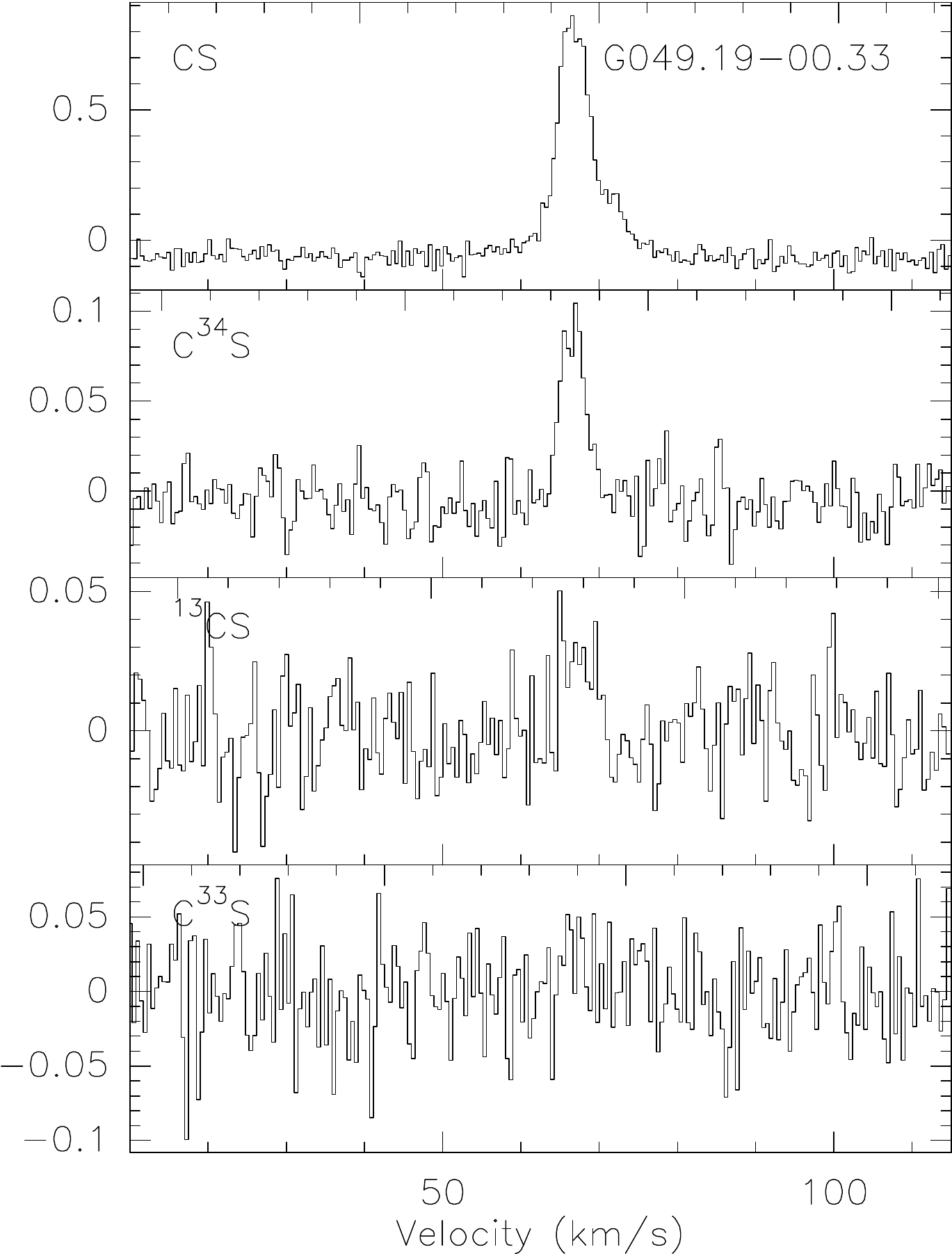}
  \includegraphics[width=83pt]{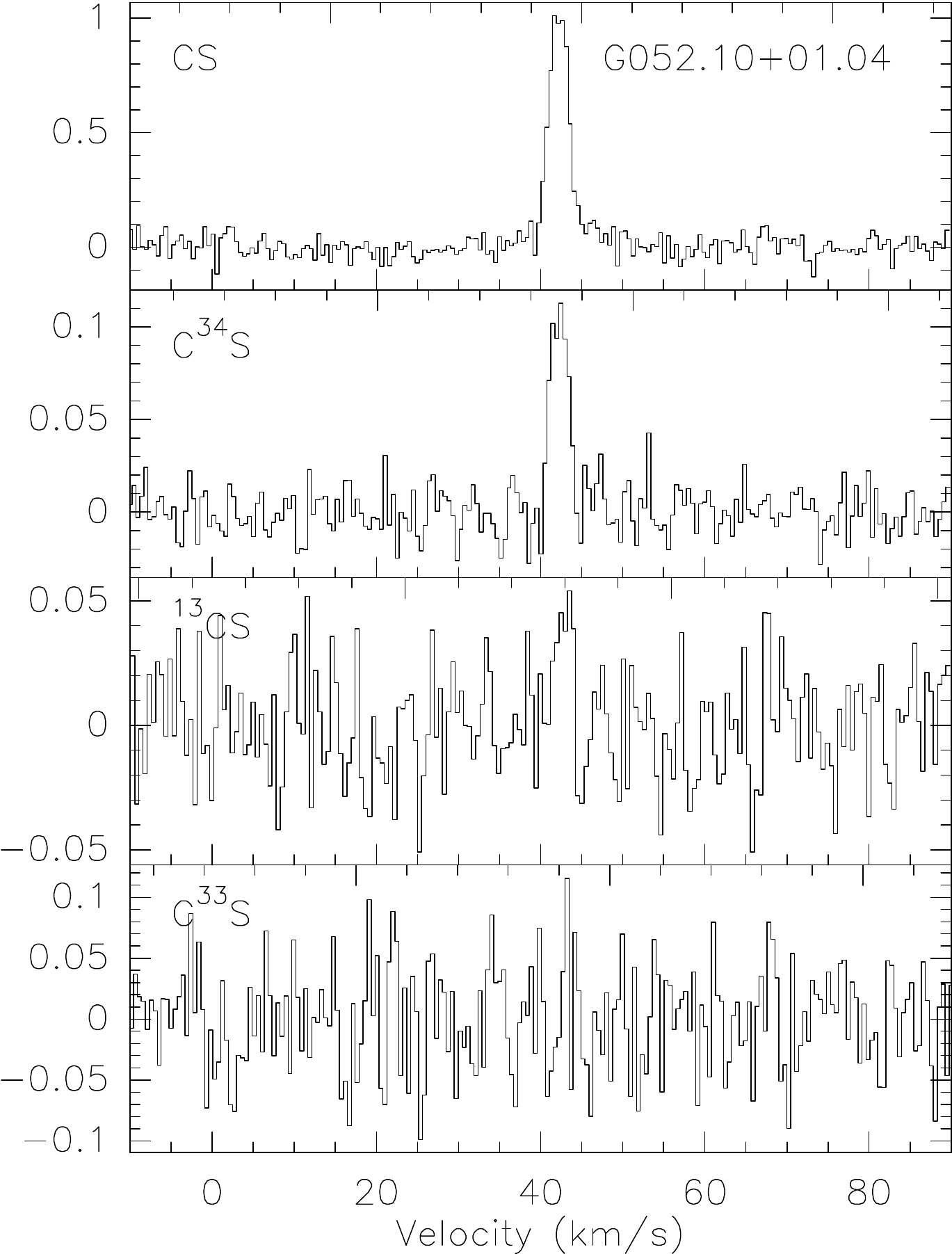}
  \includegraphics[width=83pt]{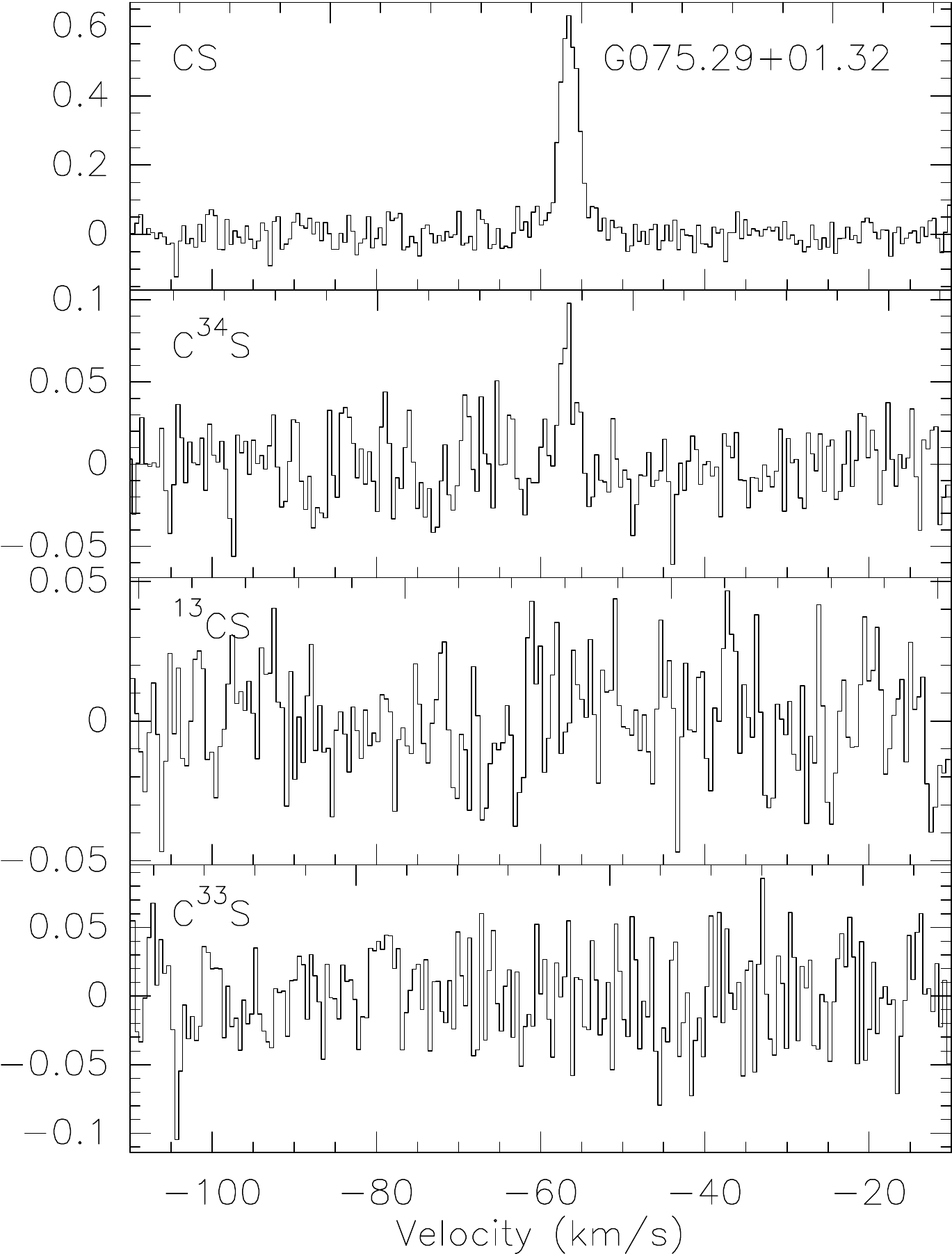}
  \includegraphics[width=83pt]{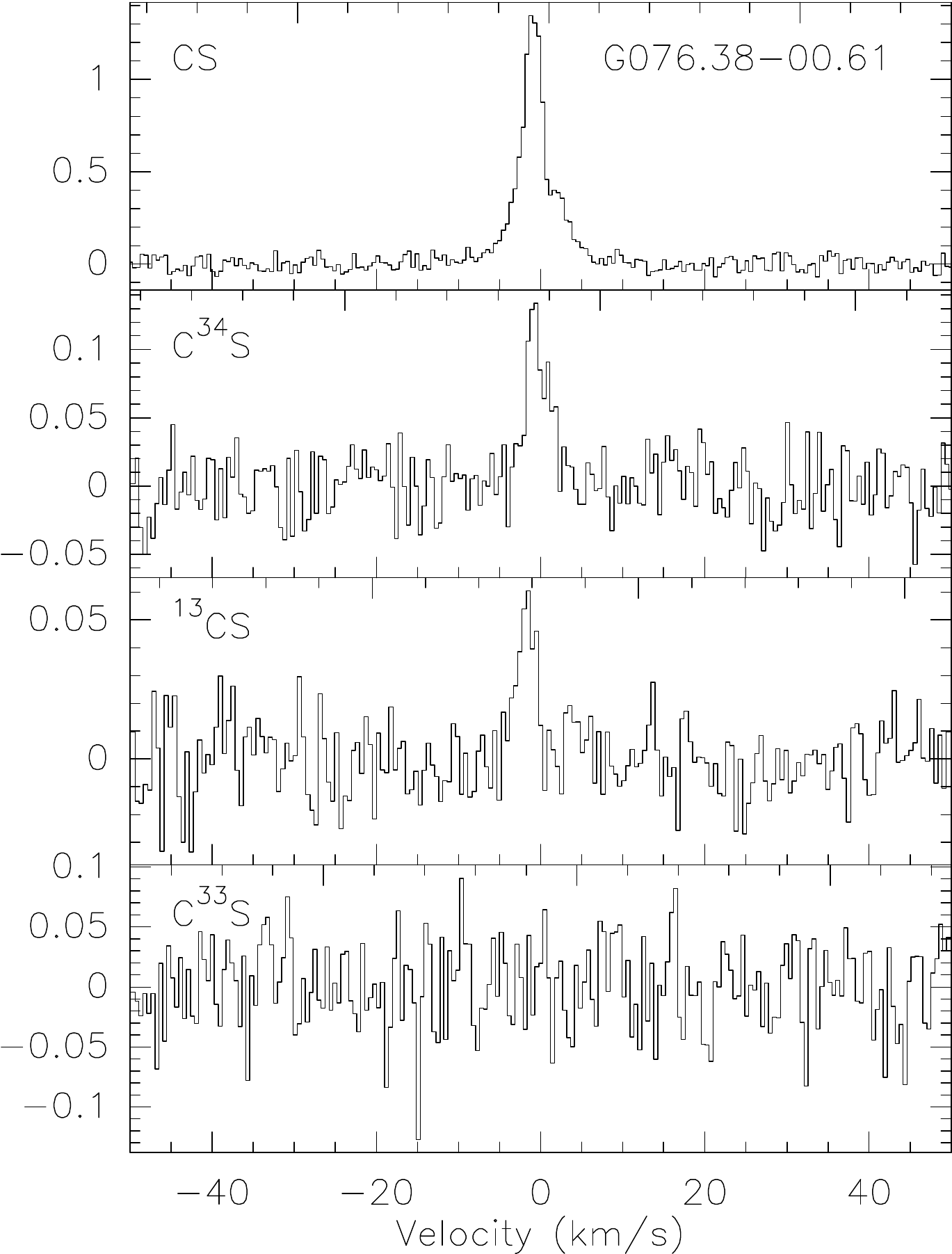}
  \includegraphics[width=83pt]{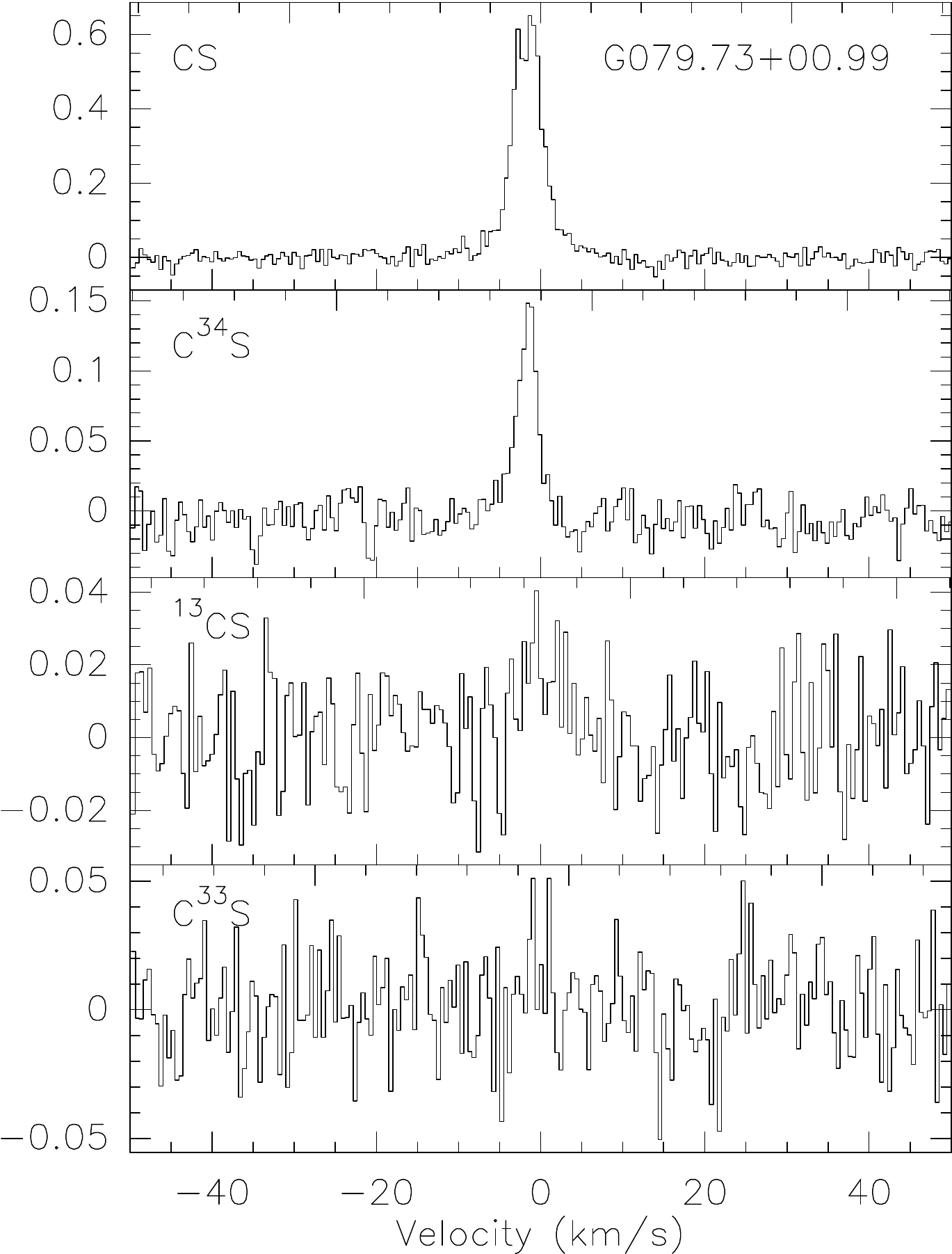}
  \includegraphics[width=83pt]{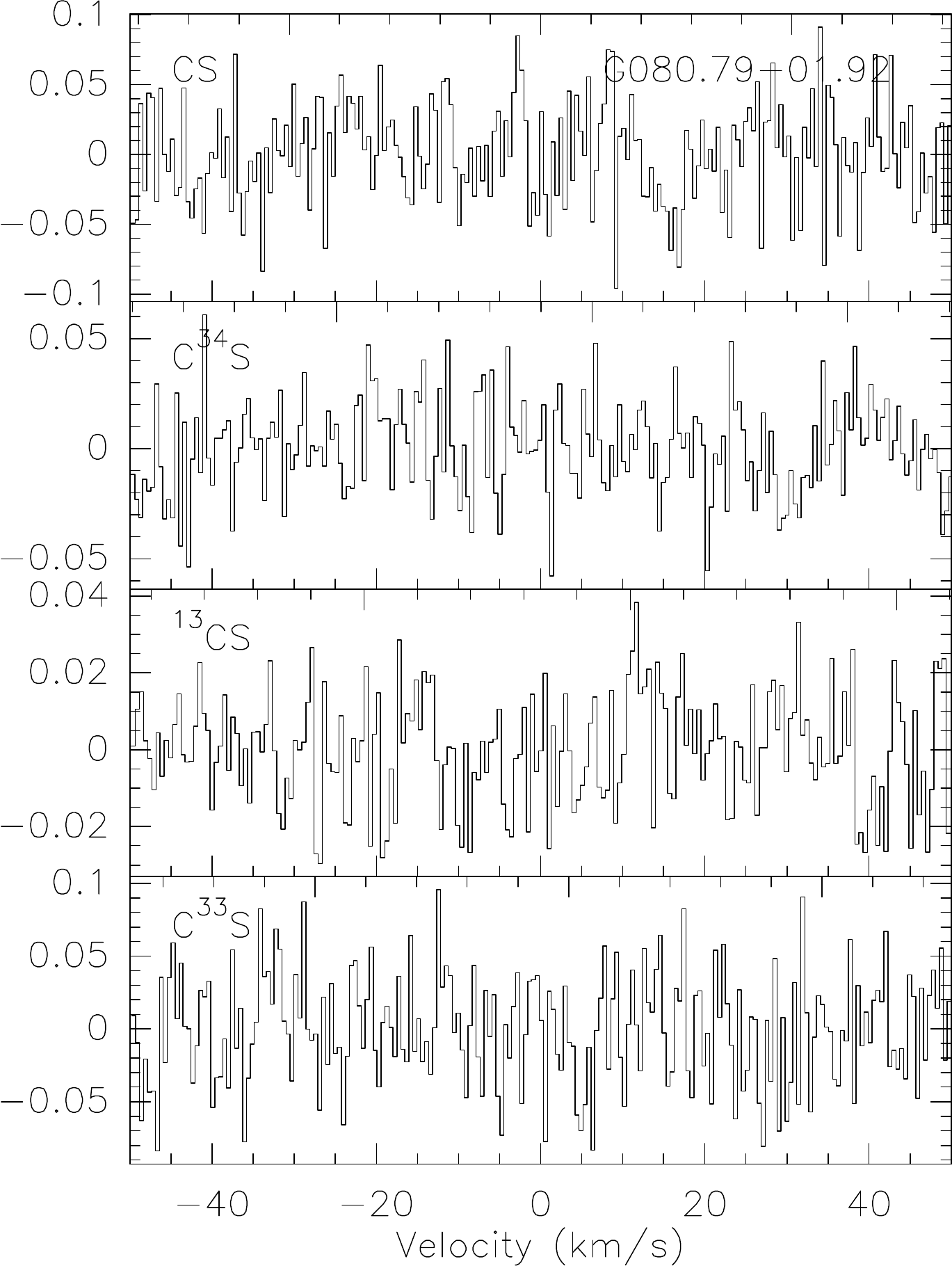}
  \includegraphics[width=83pt]{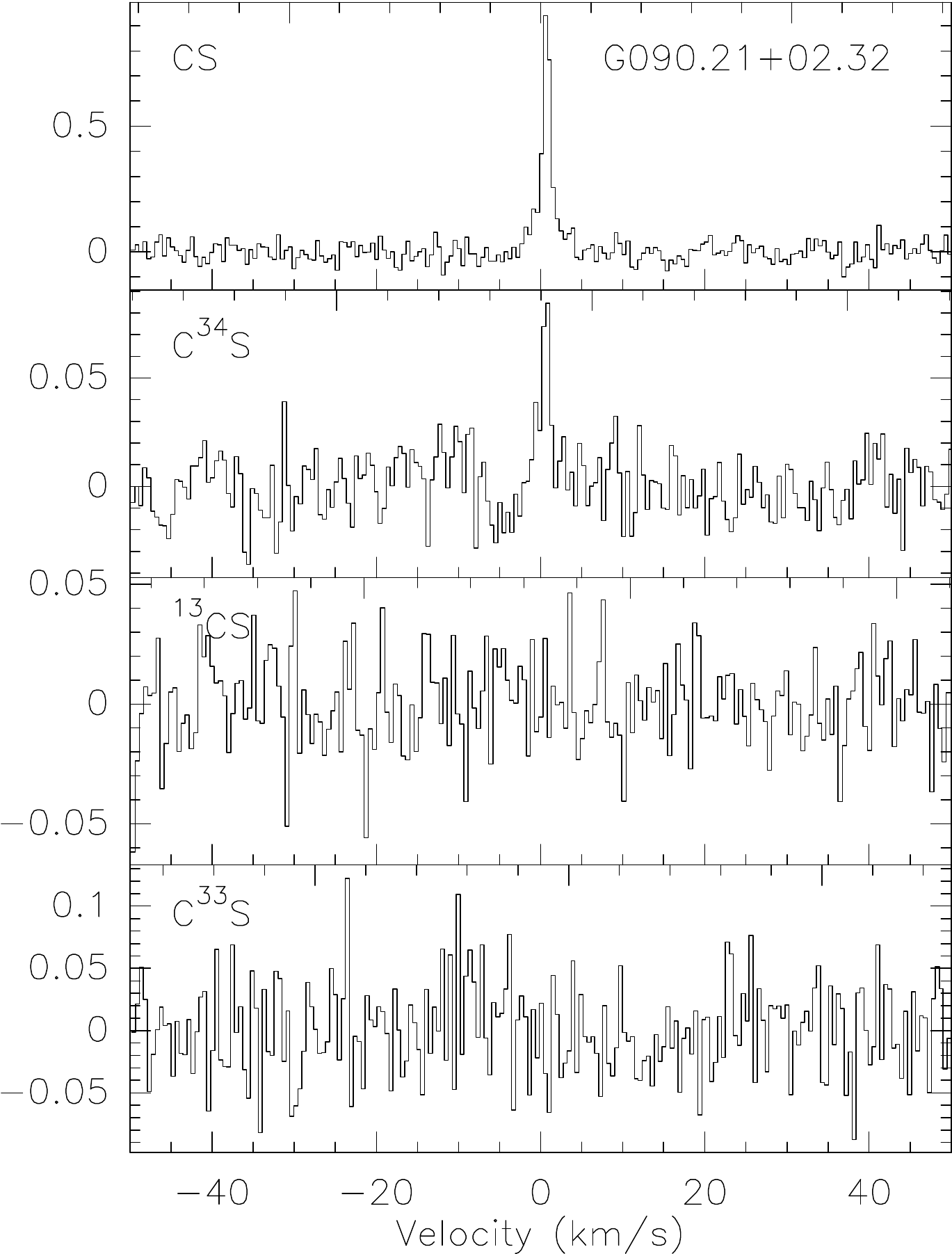}
  \includegraphics[width=83pt]{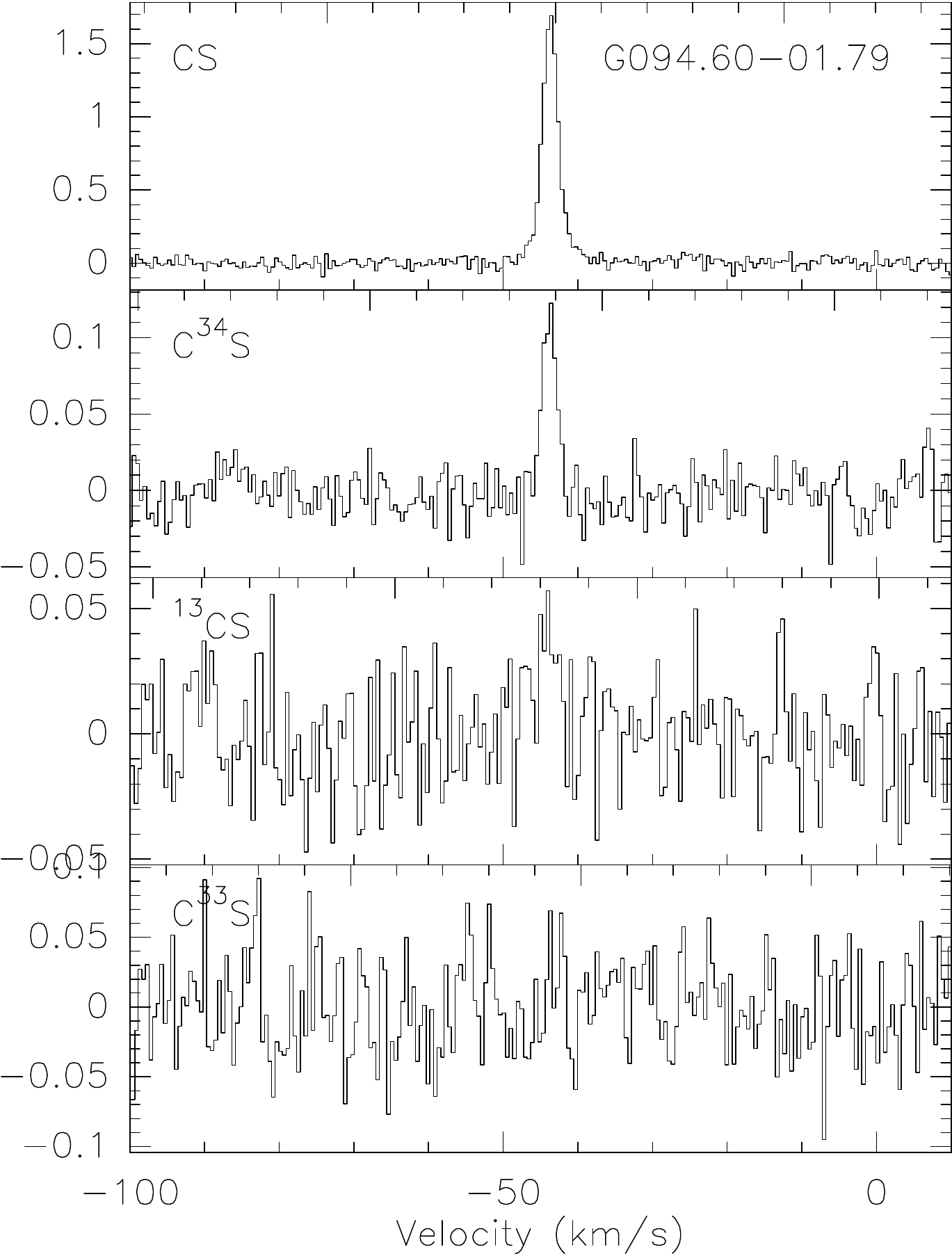}
  \includegraphics[width=83pt]{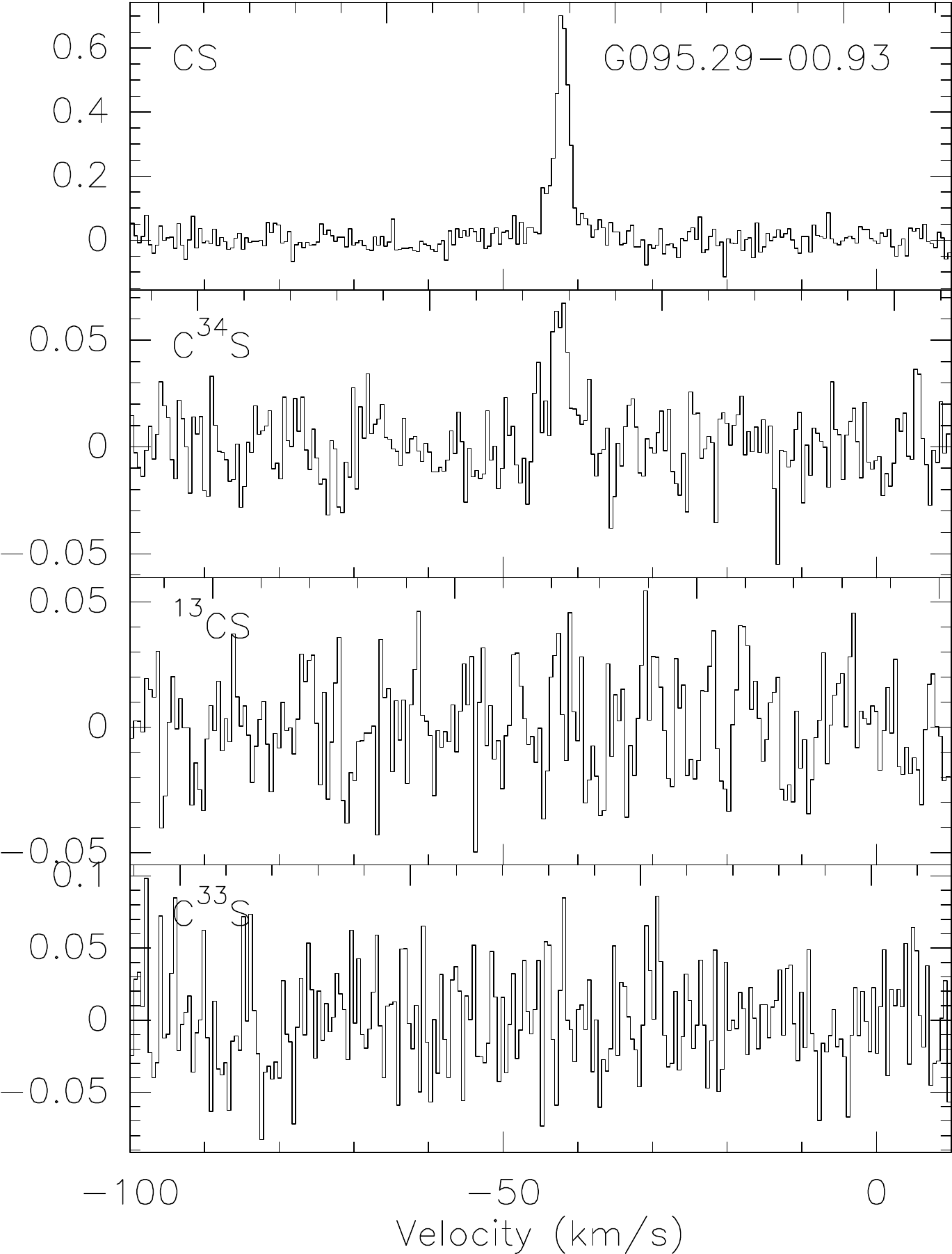}
  \includegraphics[width=83pt]{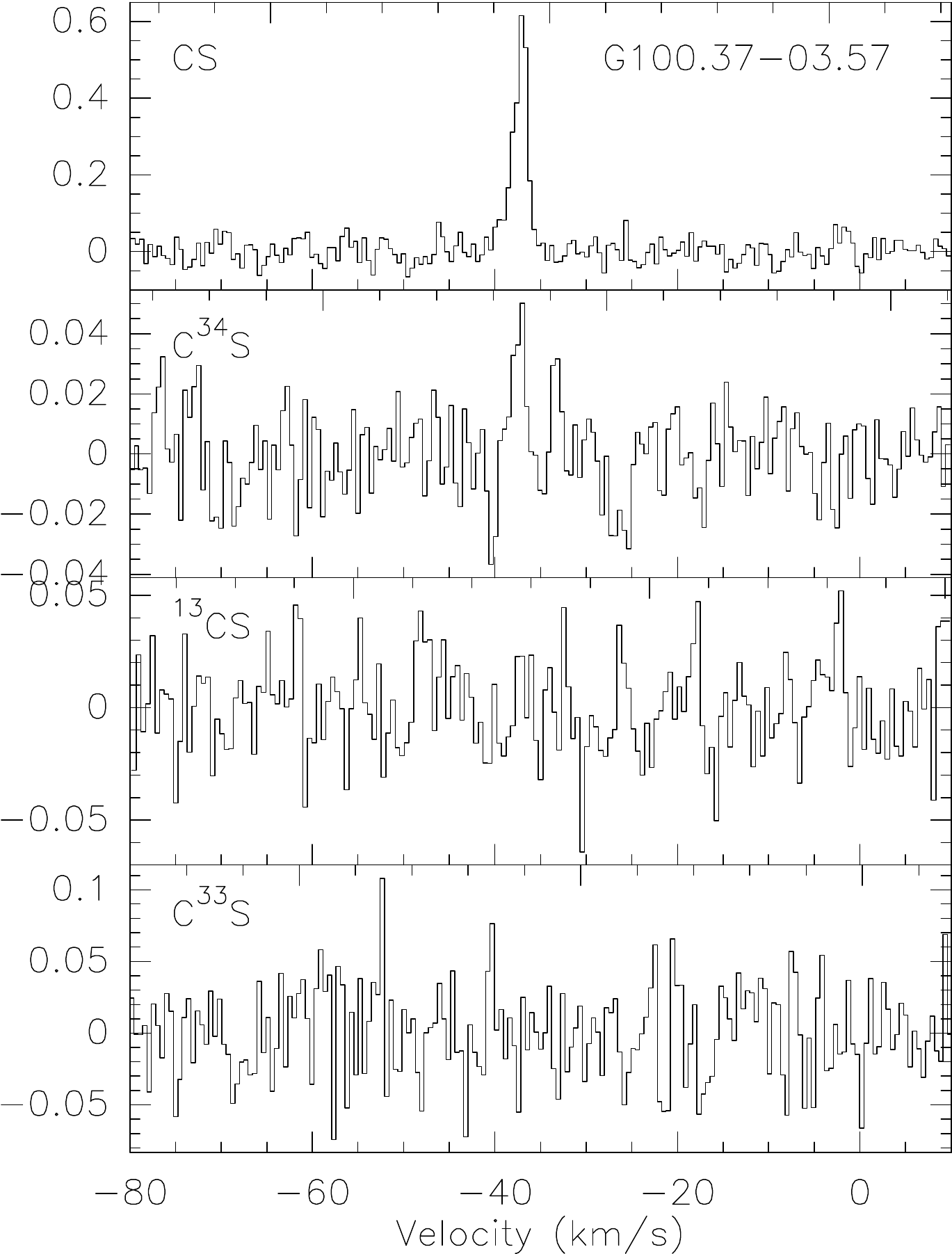}
  \includegraphics[width=83pt]{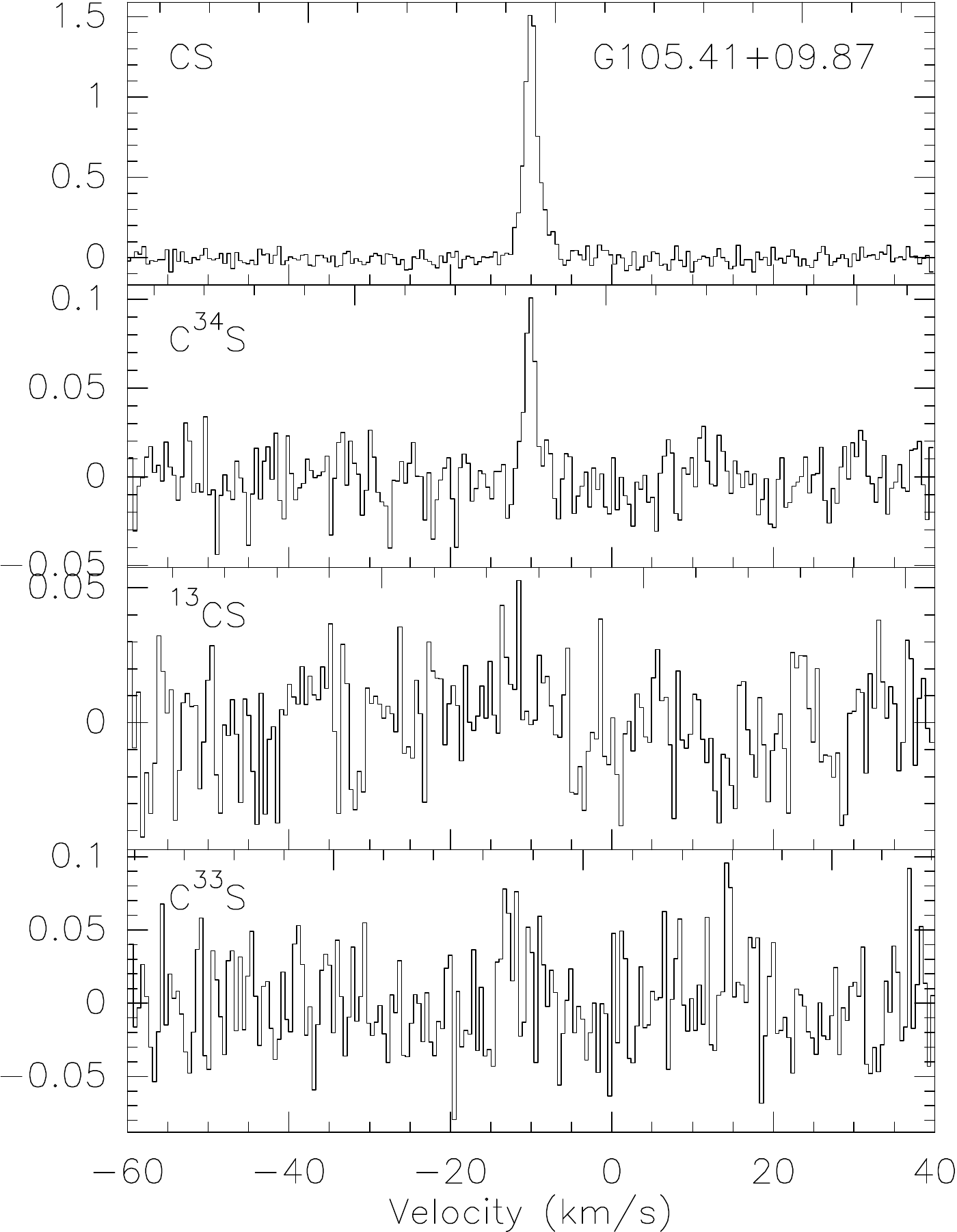}
  \includegraphics[width=83pt]{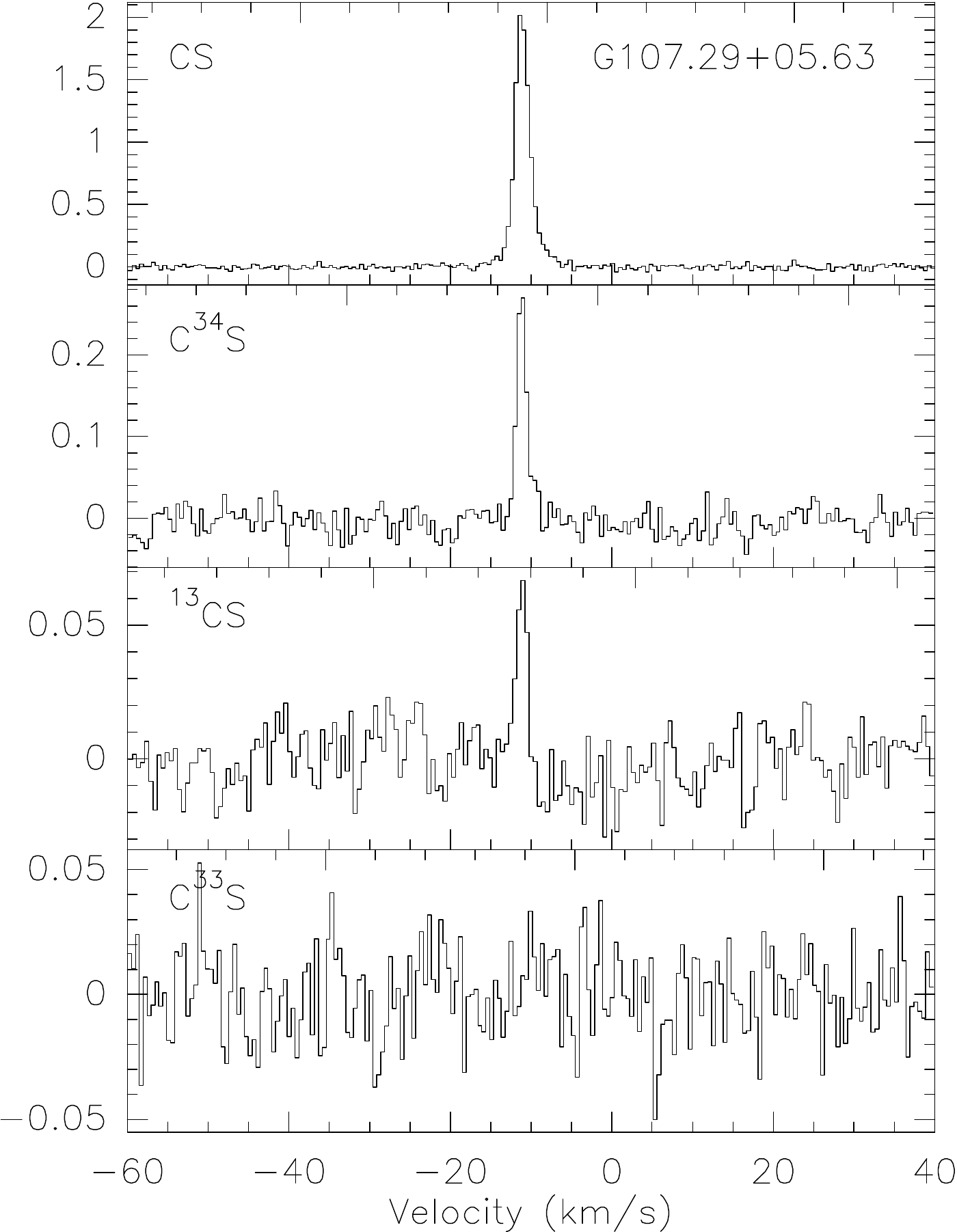}
  \includegraphics[width=83pt]{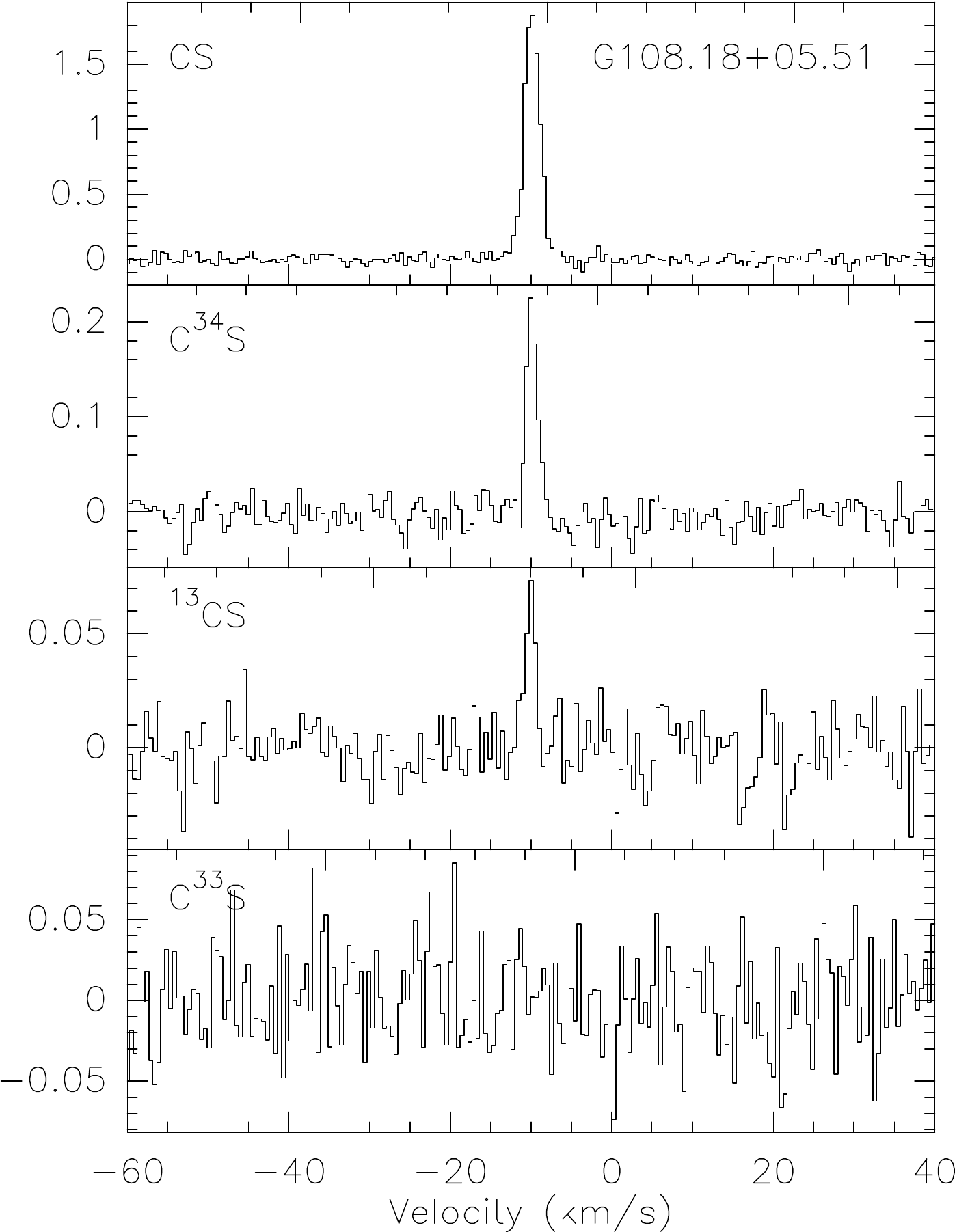}
  \includegraphics[width=83pt]{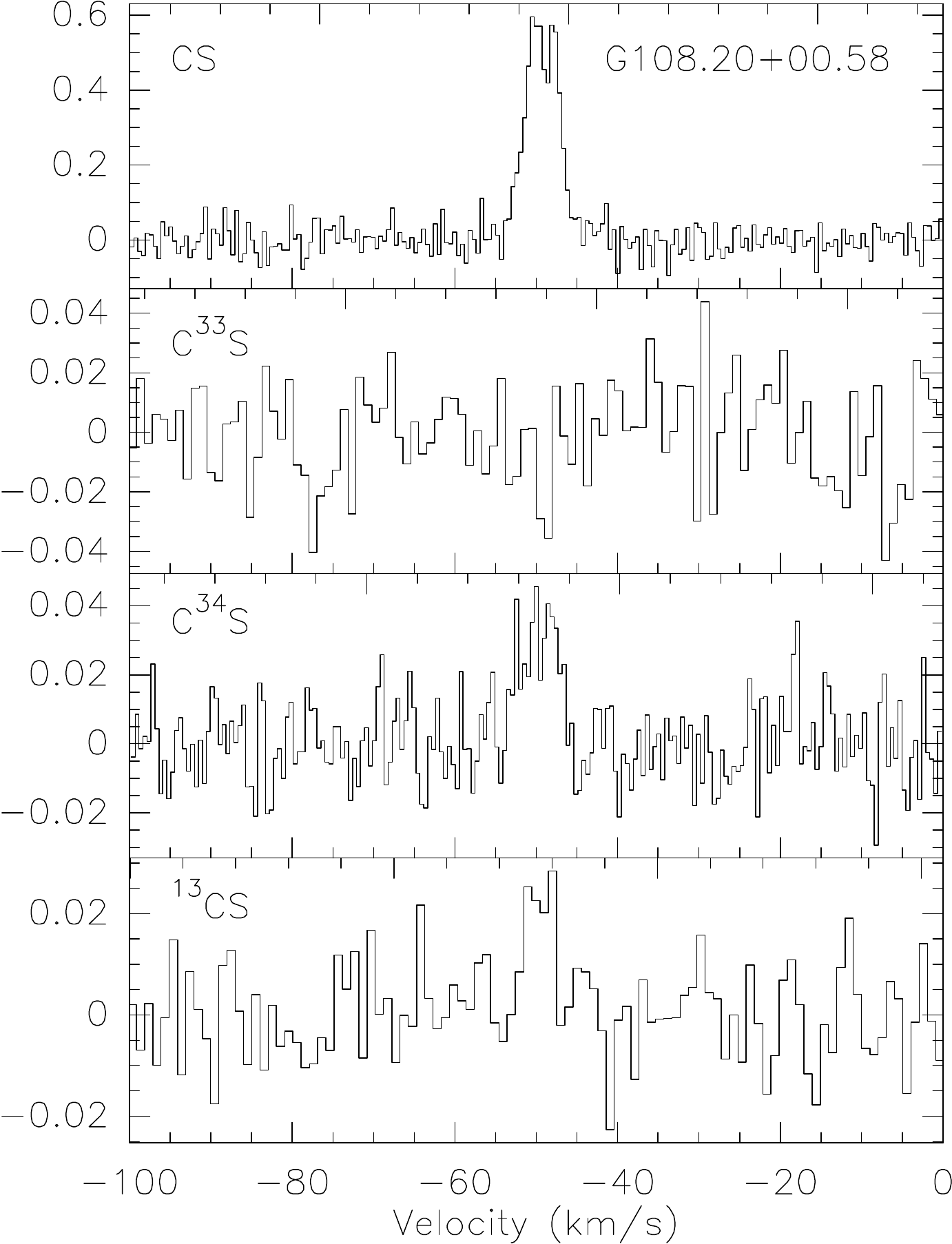}
  \includegraphics[width=83pt]{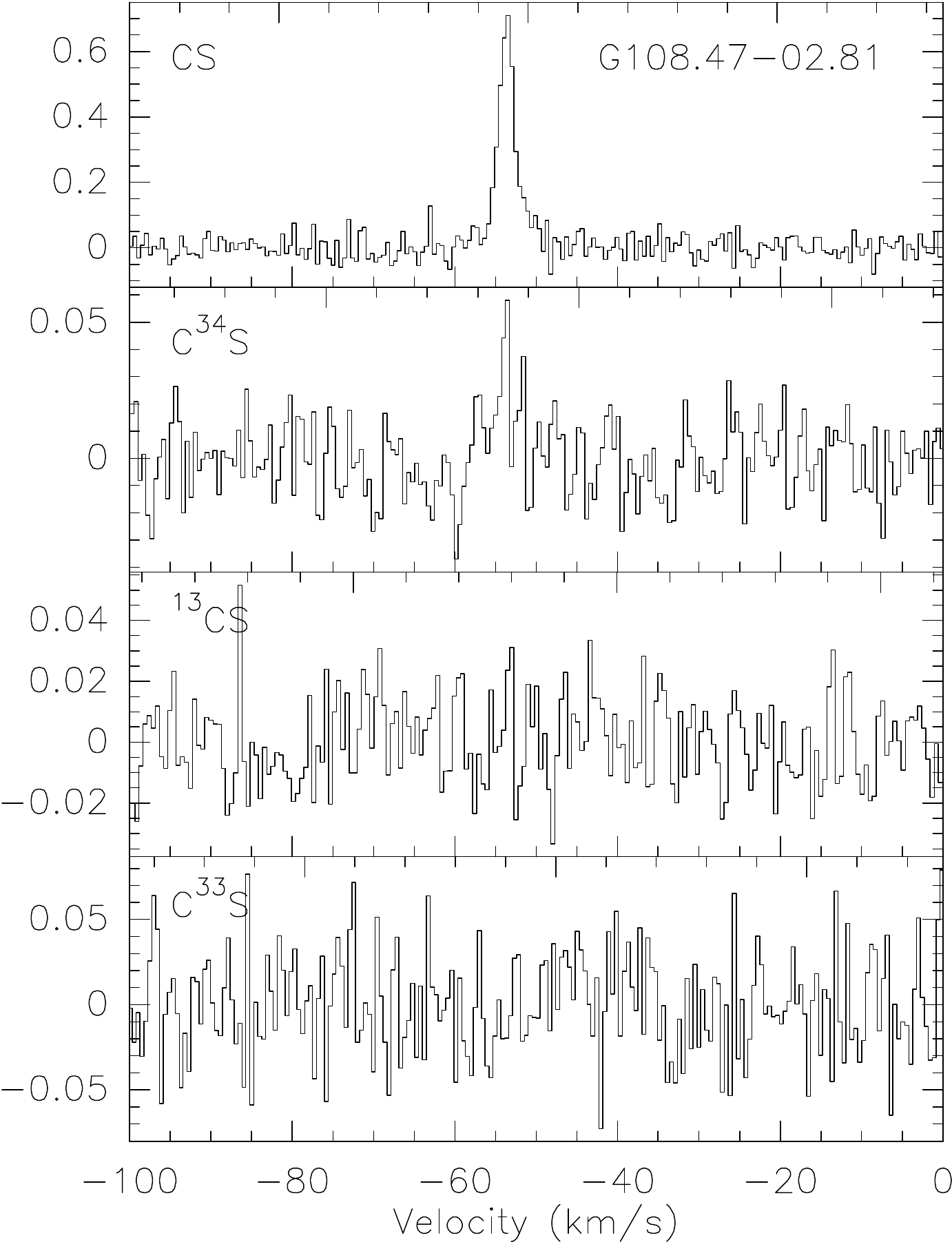}
  \includegraphics[width=83pt]{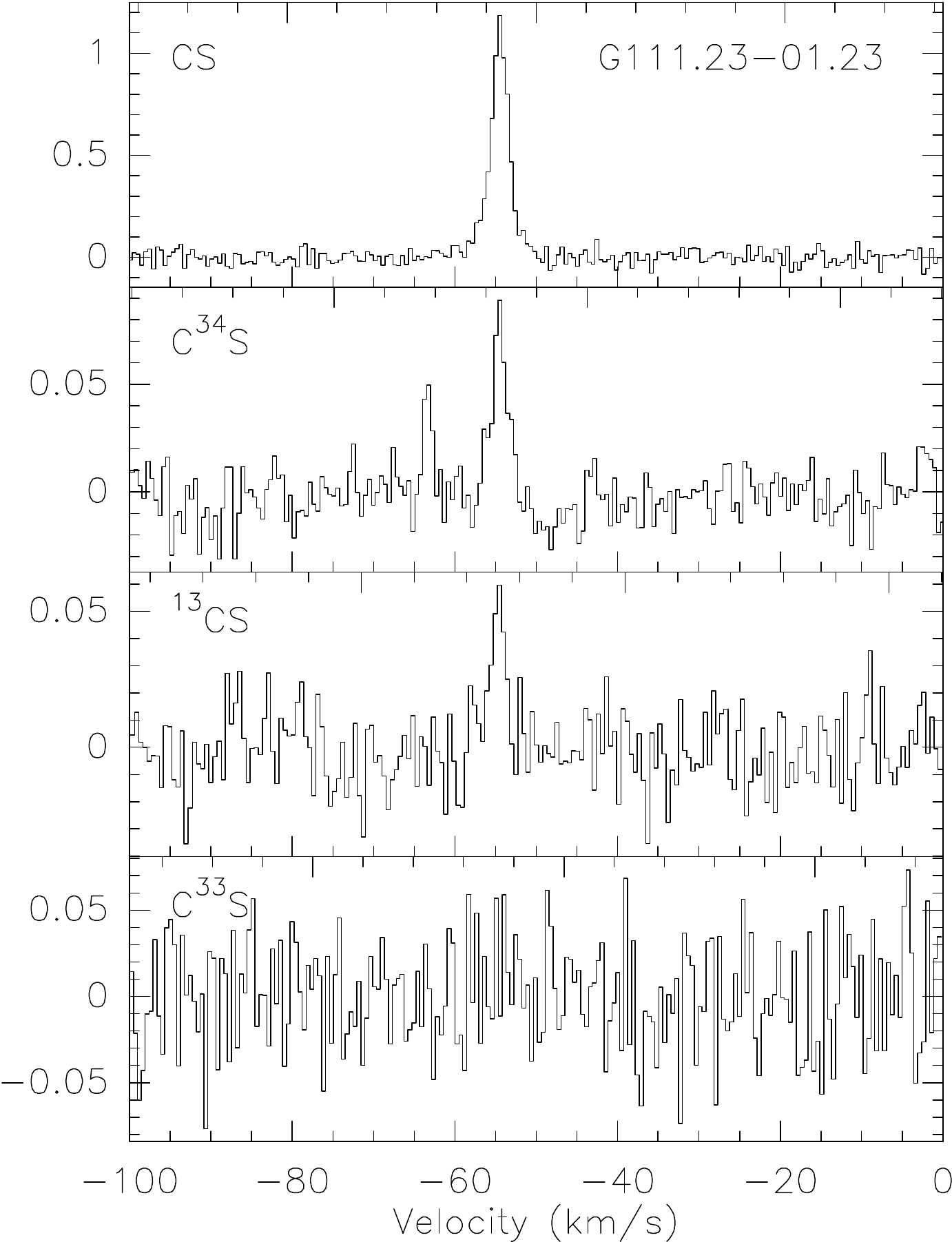}
  \includegraphics[width=83pt]{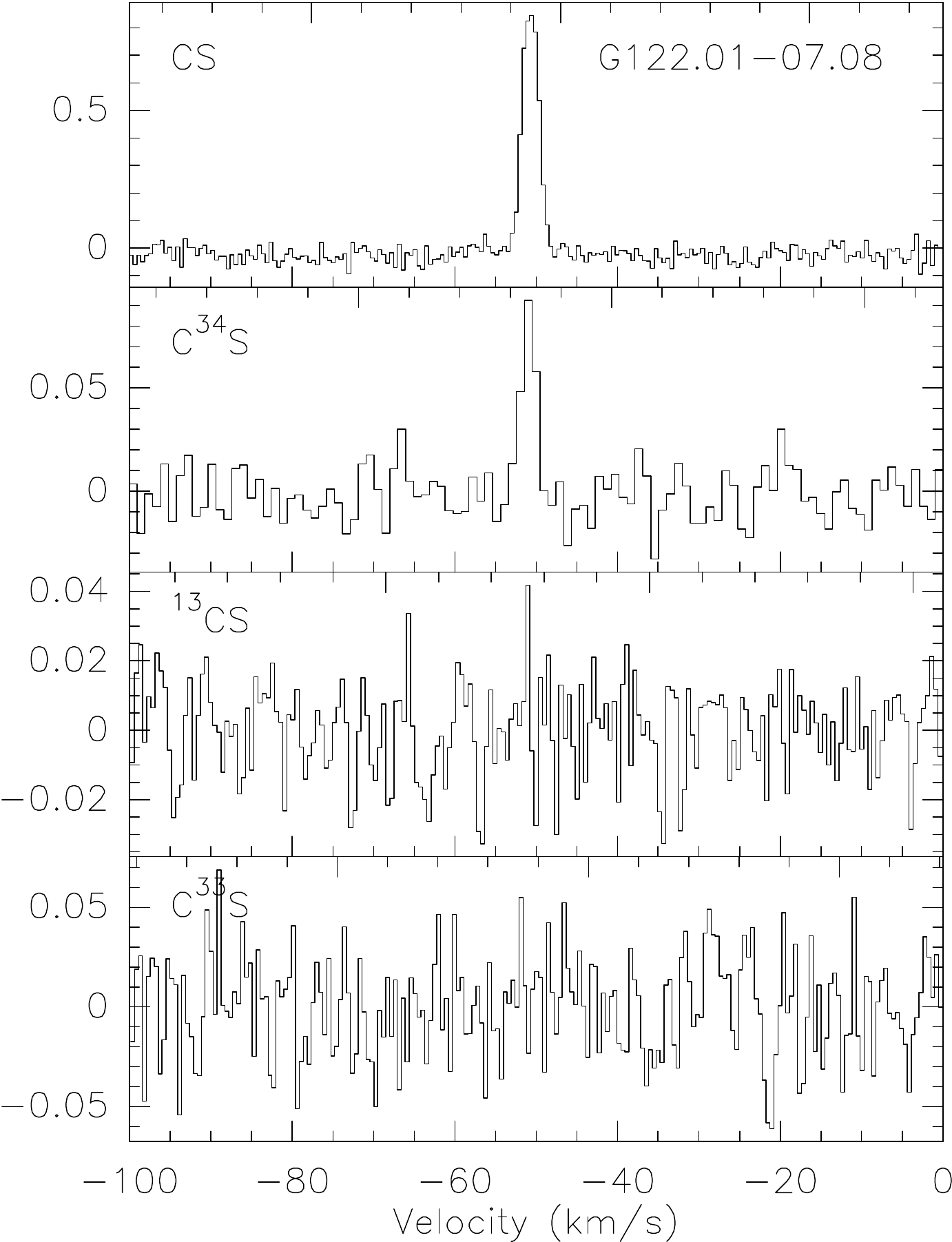}
  \includegraphics[width=83pt]{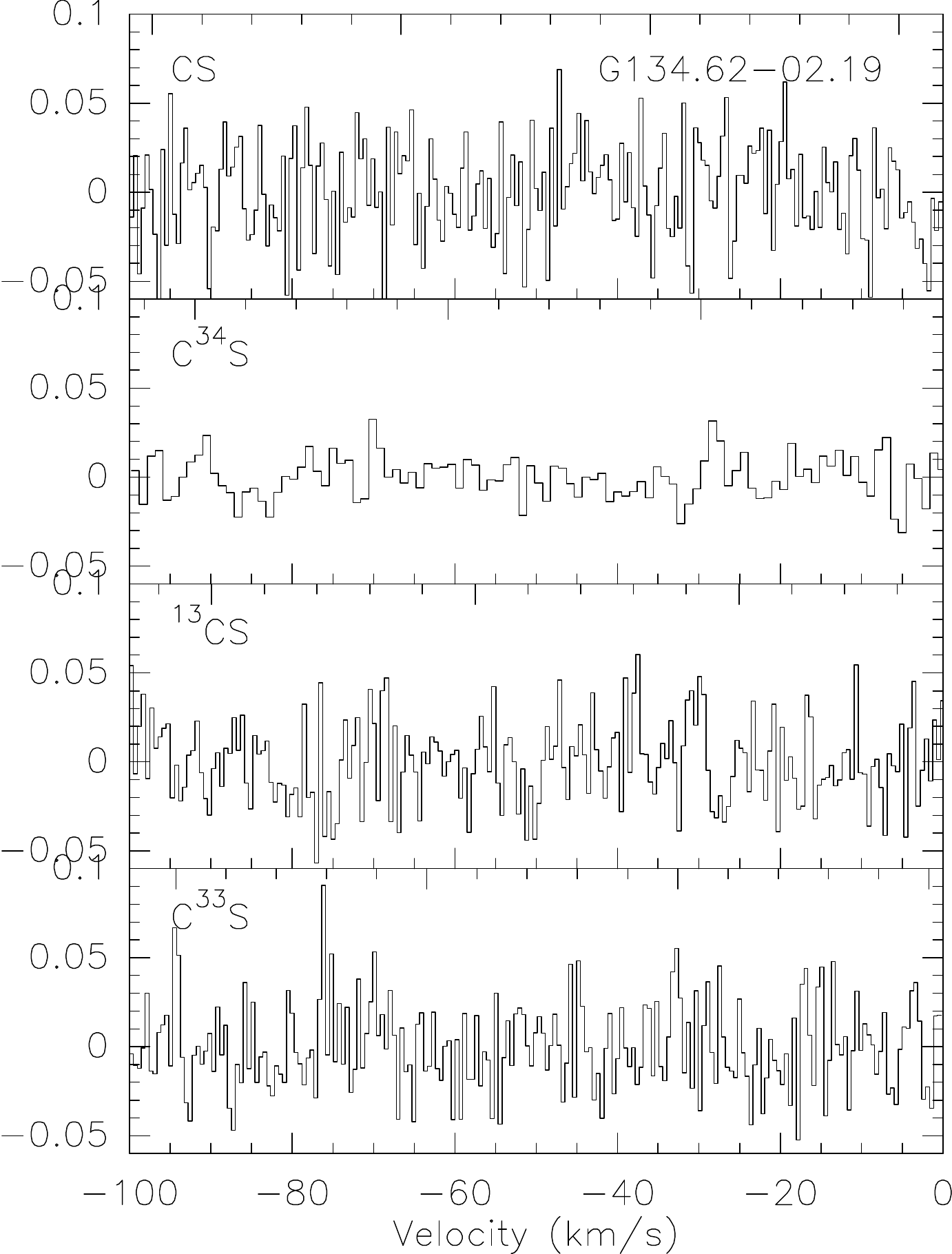}
  \includegraphics[width=83pt]{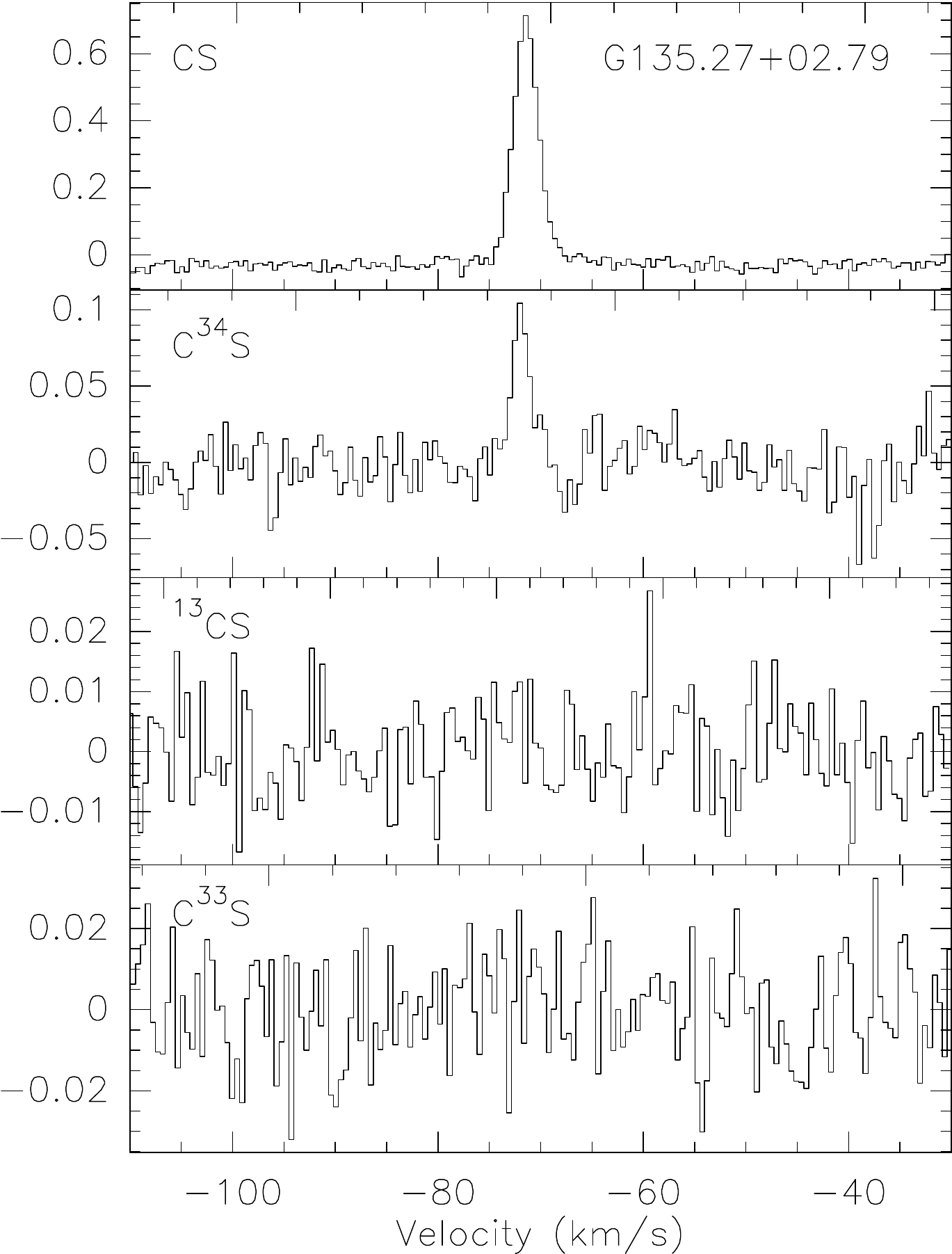}
  \includegraphics[width=83pt]{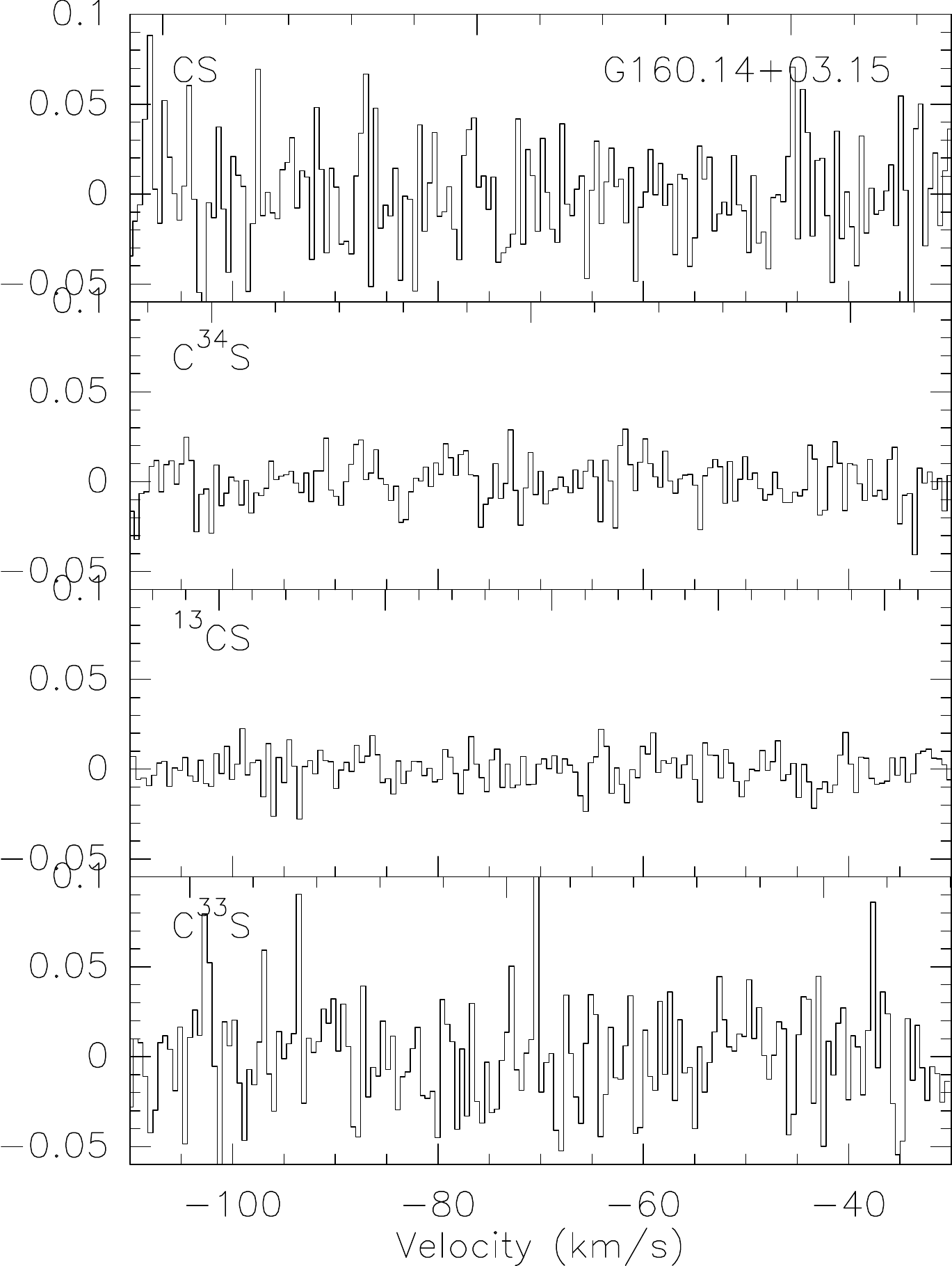}
  \includegraphics[width=83pt]{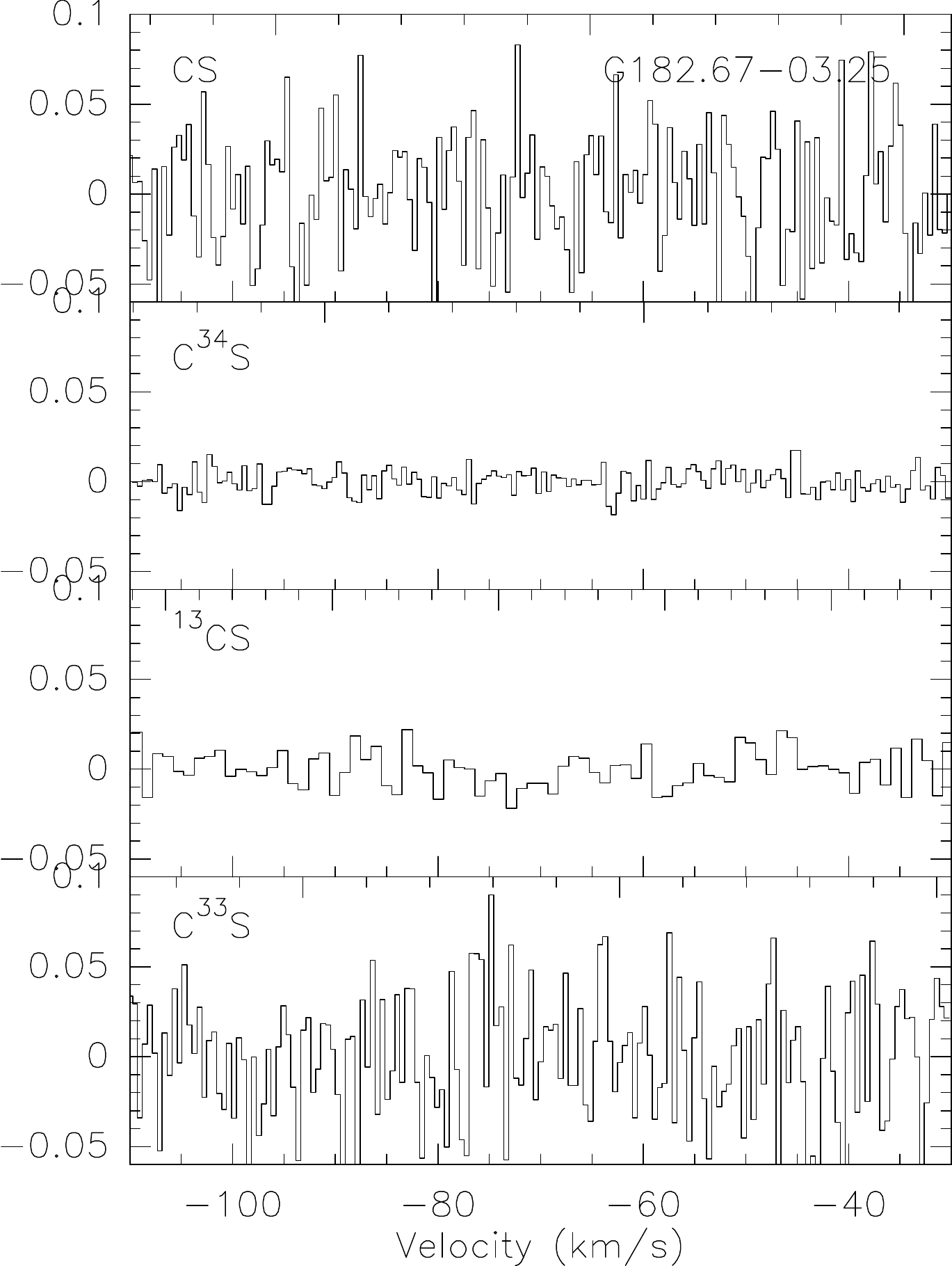}
  \includegraphics[width=83pt]{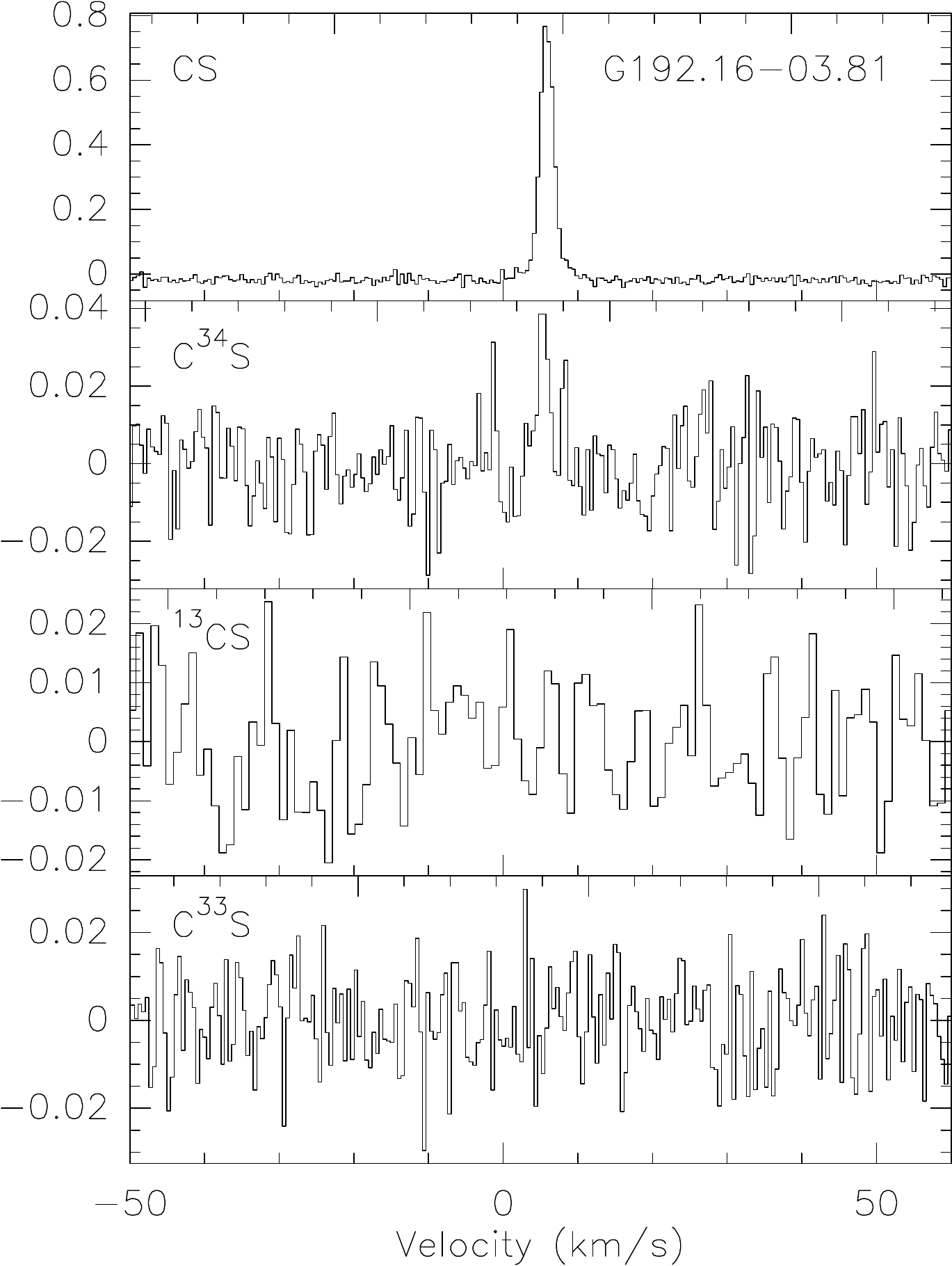}
  \includegraphics[width=81pt]{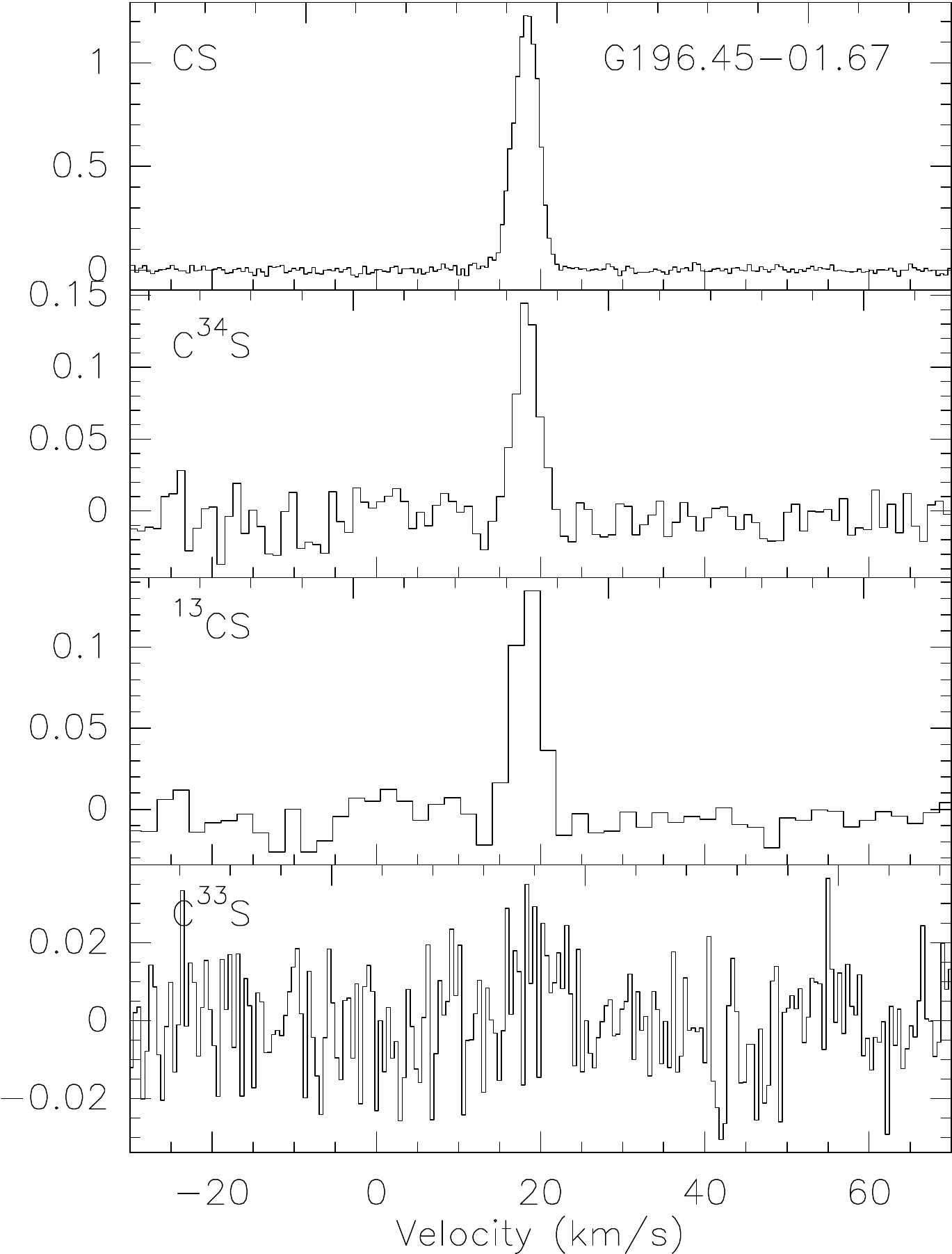}
  \includegraphics[width=83pt]{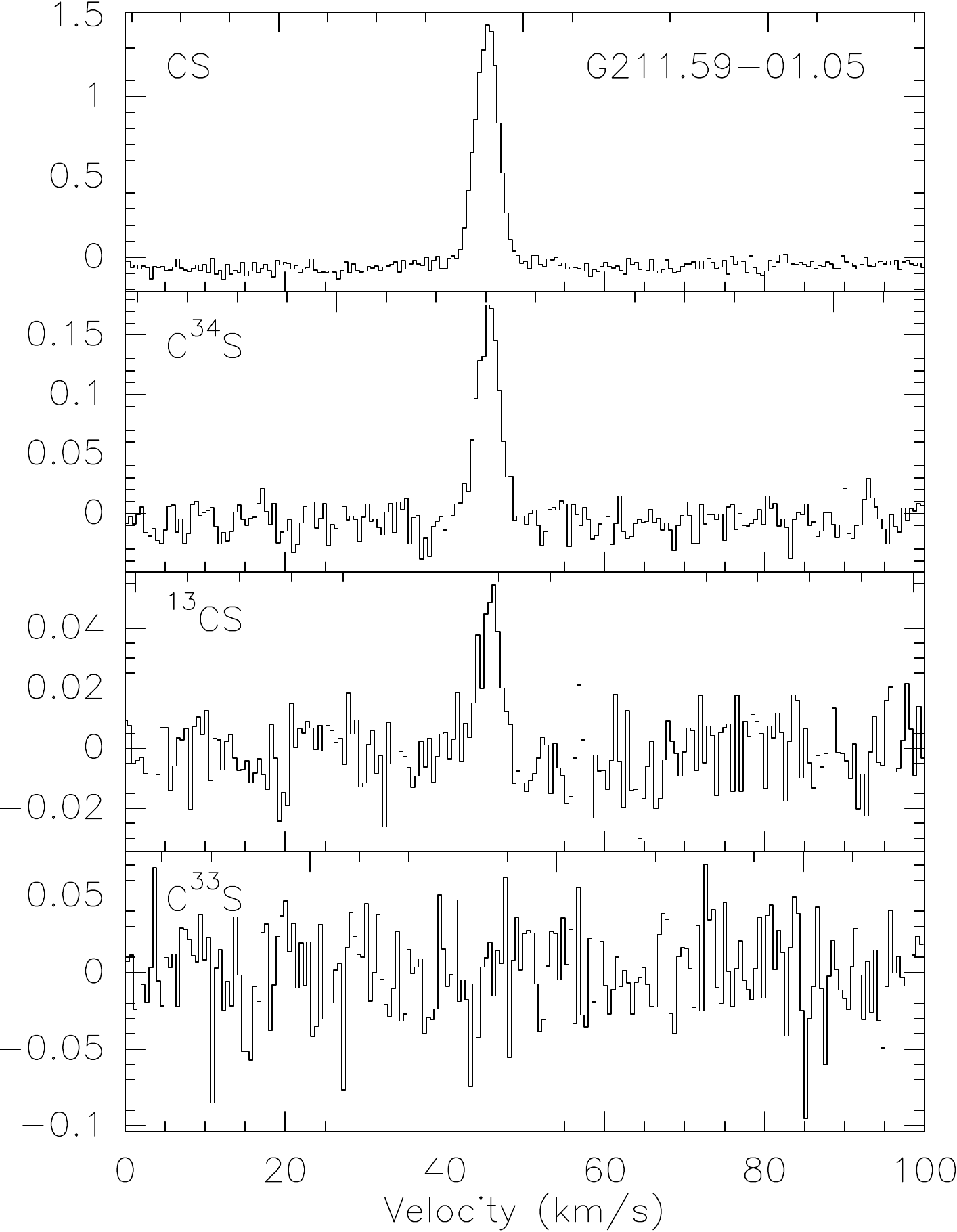}
  \includegraphics[width=83pt]{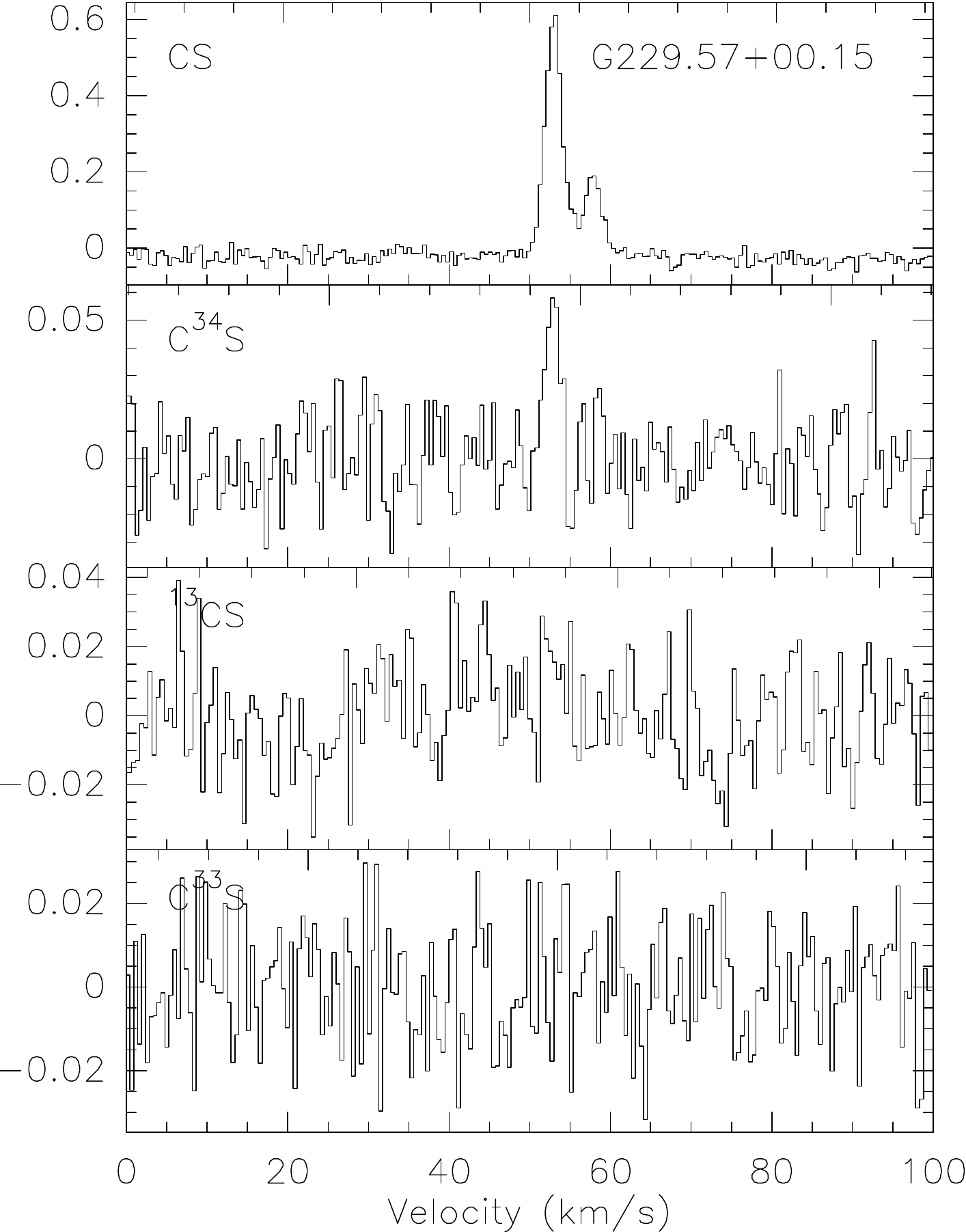}
  \includegraphics[width=83pt]{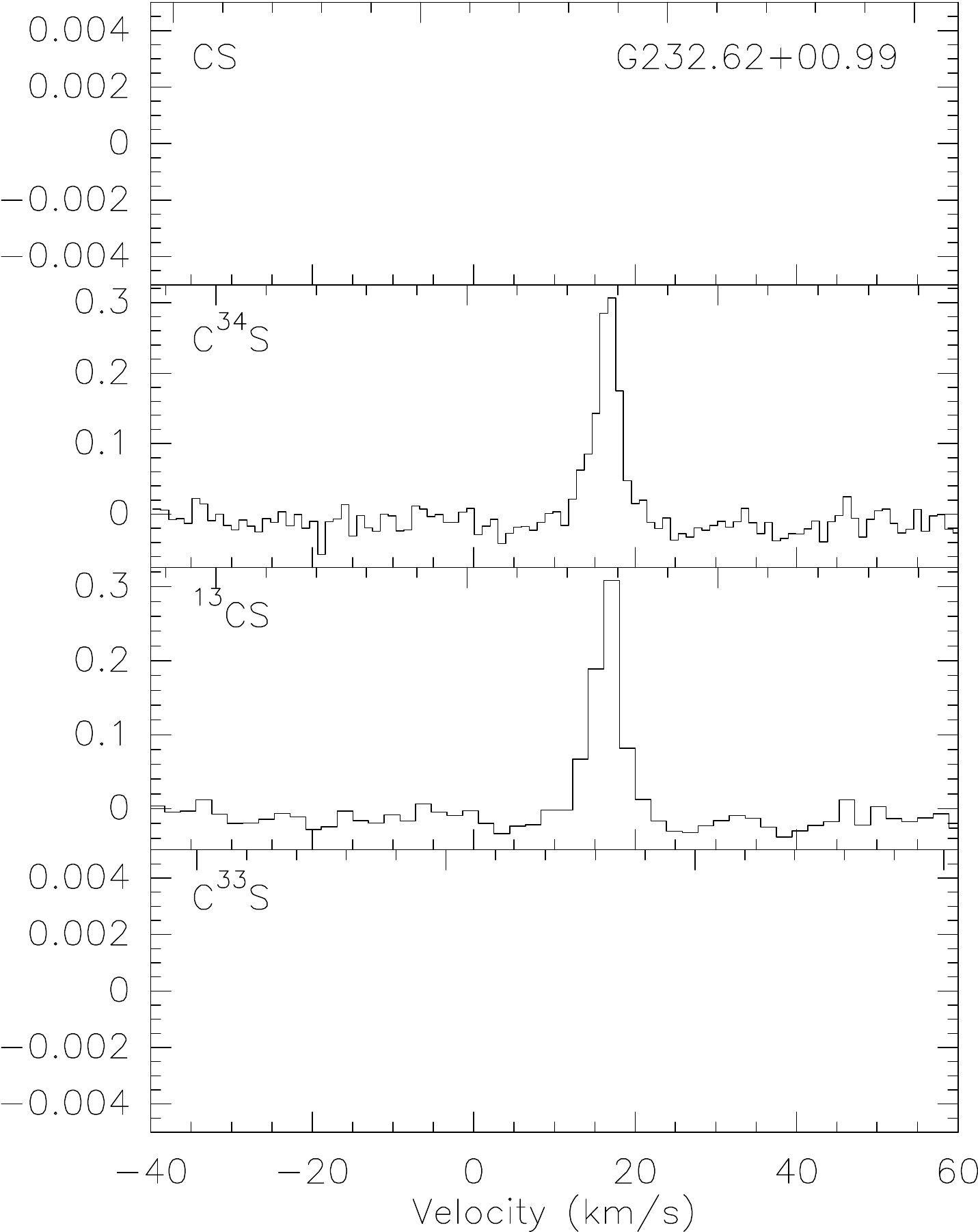}
\caption{ARO 12-m spectra towards those 34 sources, where not all four $J$=2$-$1 CS isotopologues were detected. }
\end{figure*}

\begin{figure*}[h]
\center
  \includegraphics[width=112.4pt]{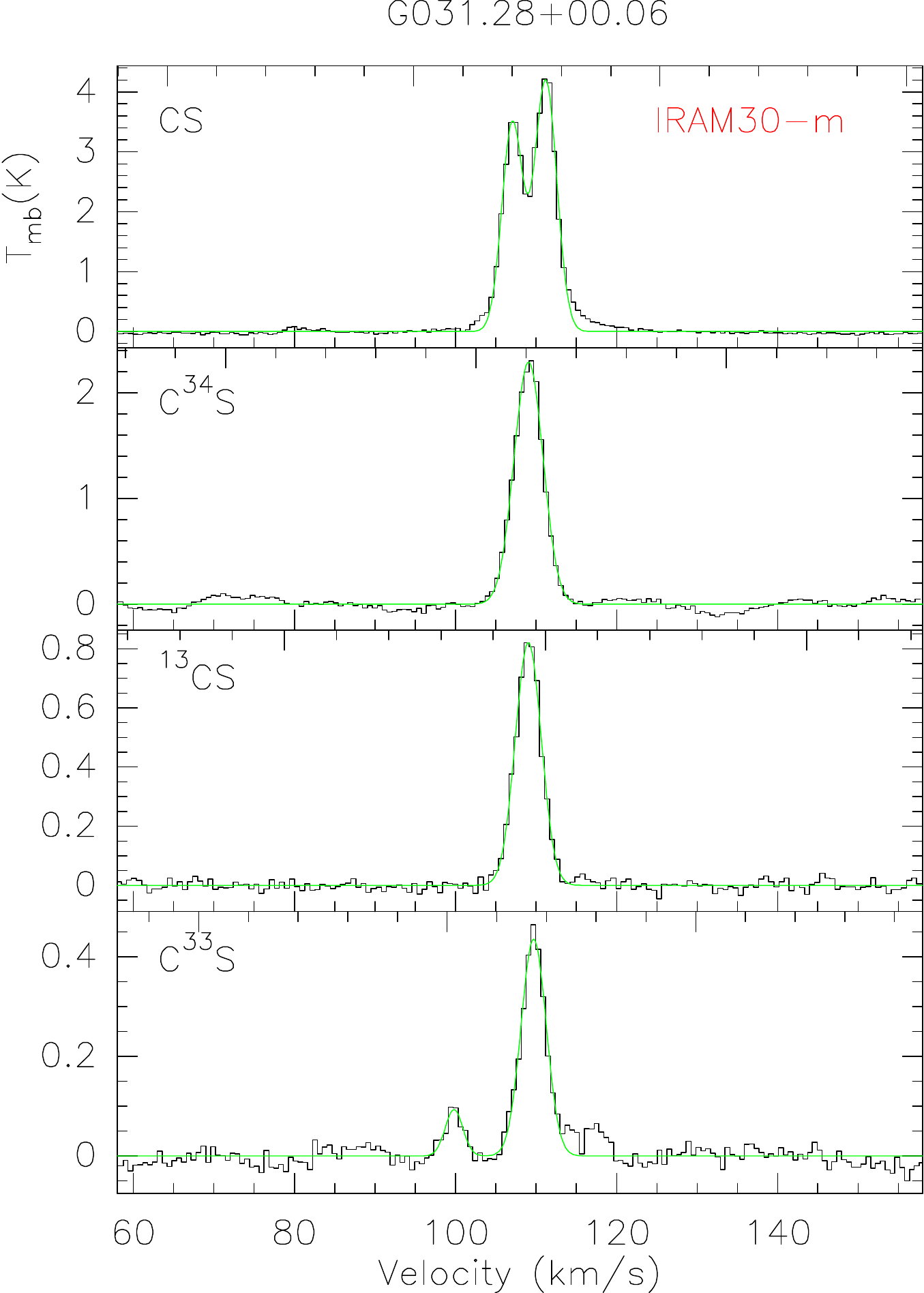}
  \includegraphics[width=110pt]{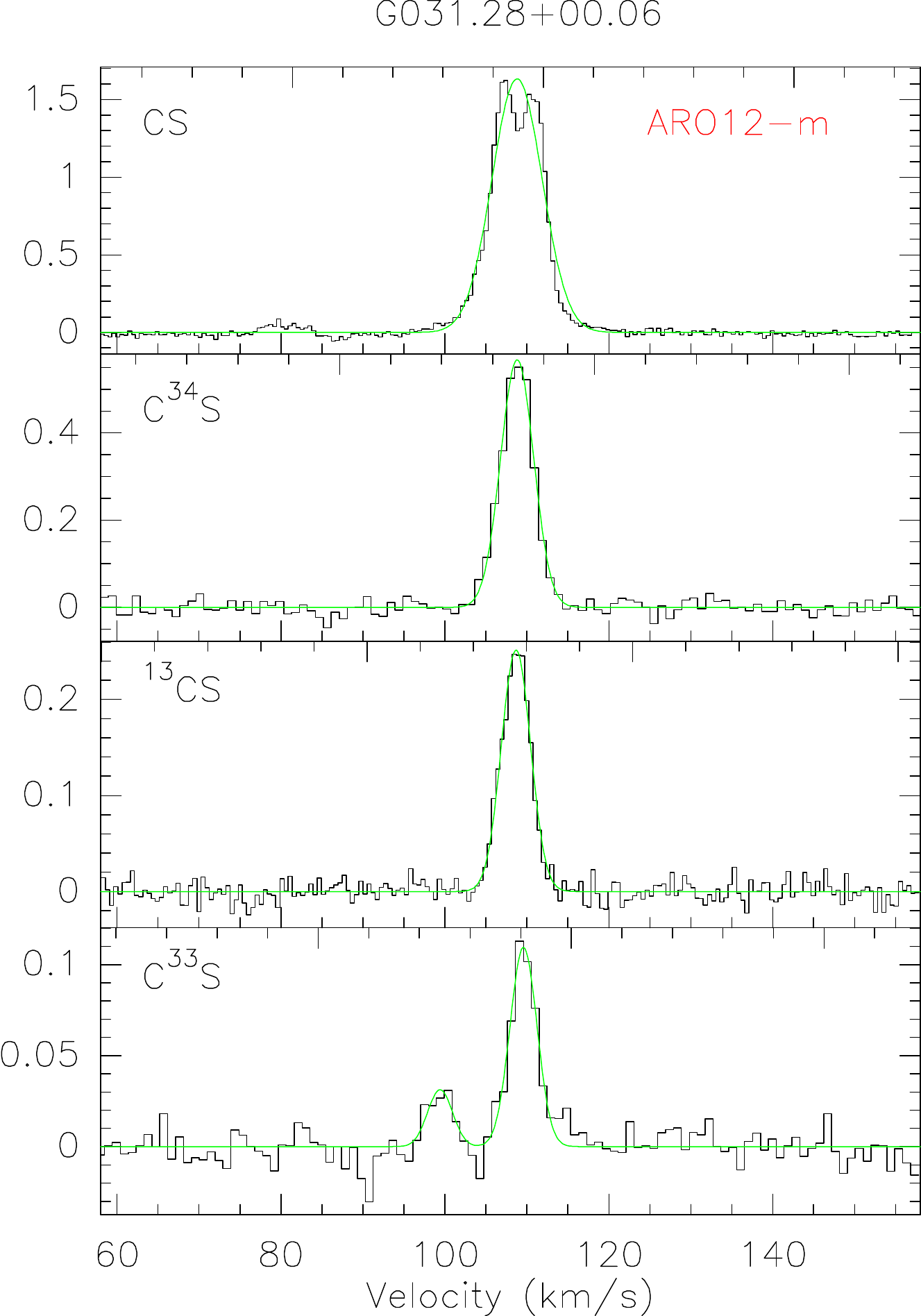}
  \includegraphics[width=107.5pt]{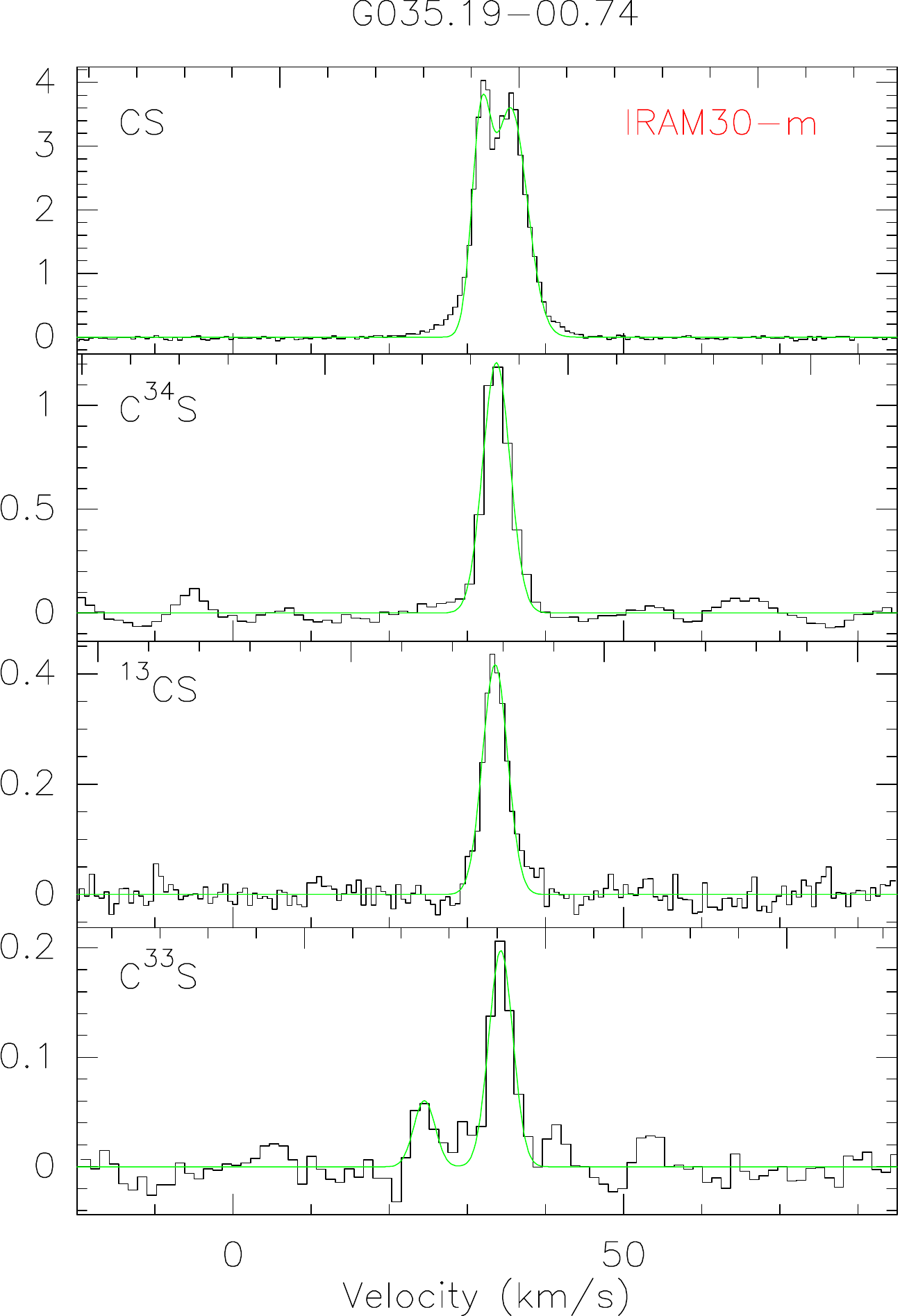}
  \includegraphics[width=114pt]{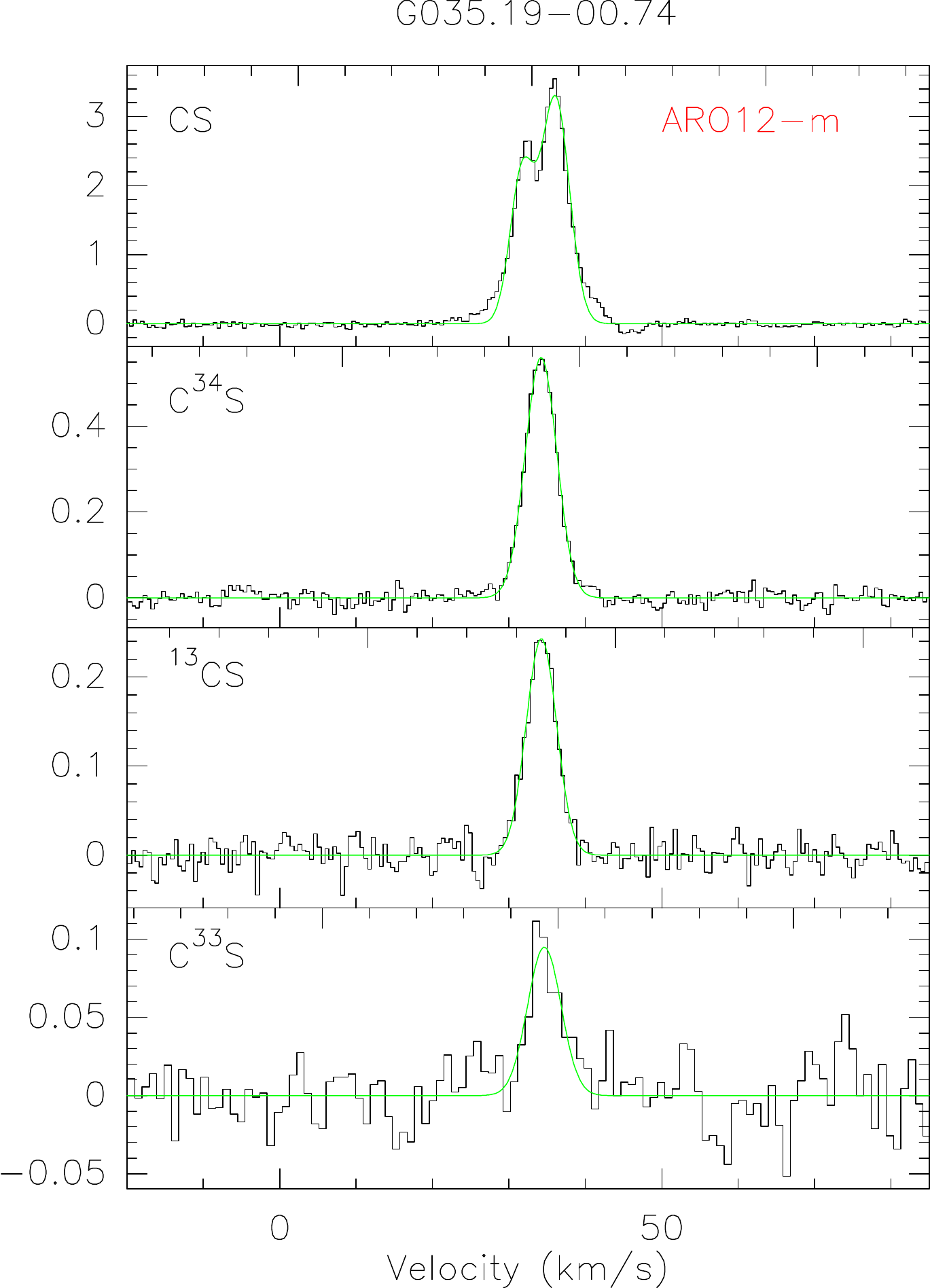}
  \includegraphics[width=112pt]{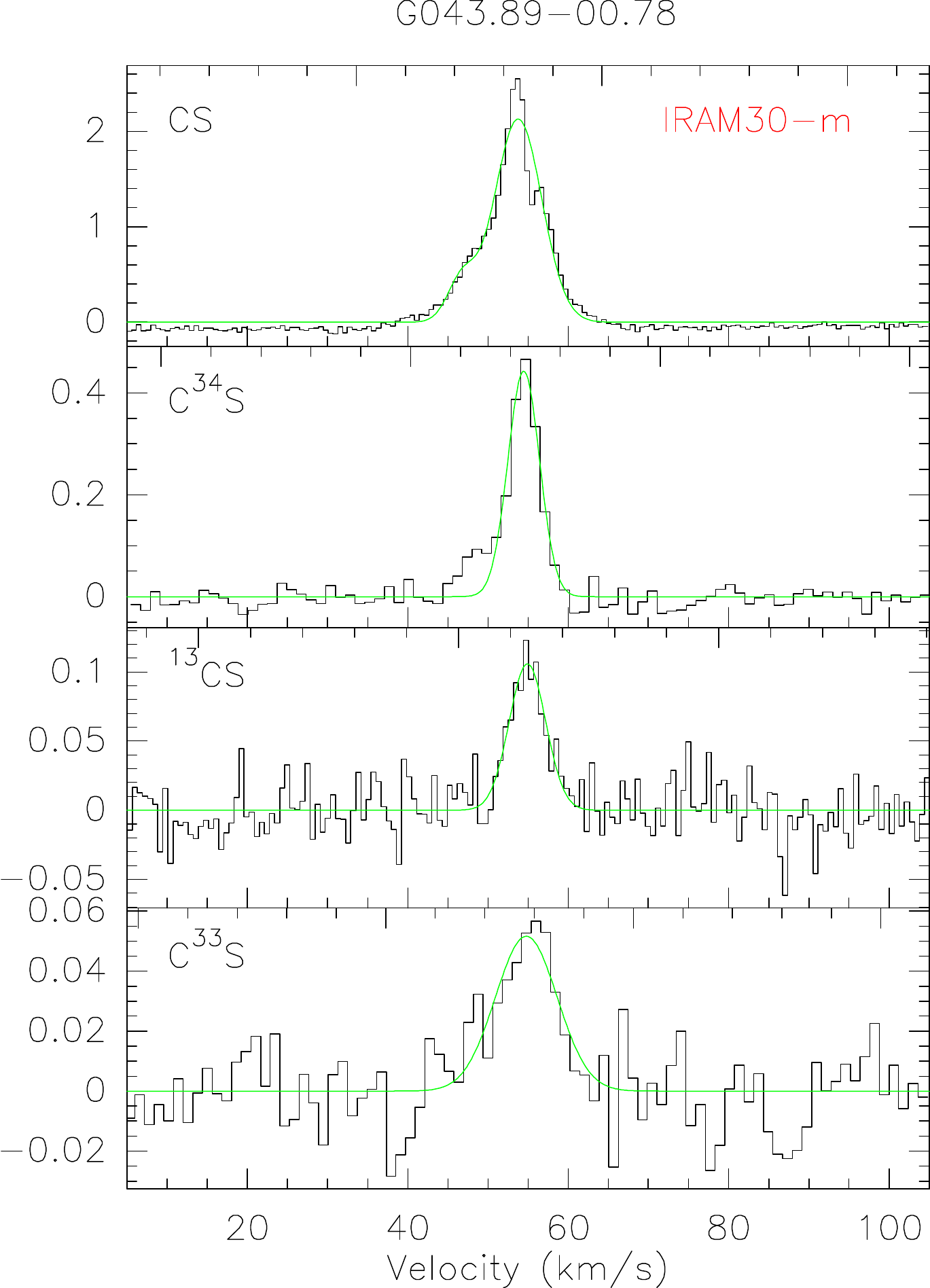}
  \includegraphics[width=109pt]{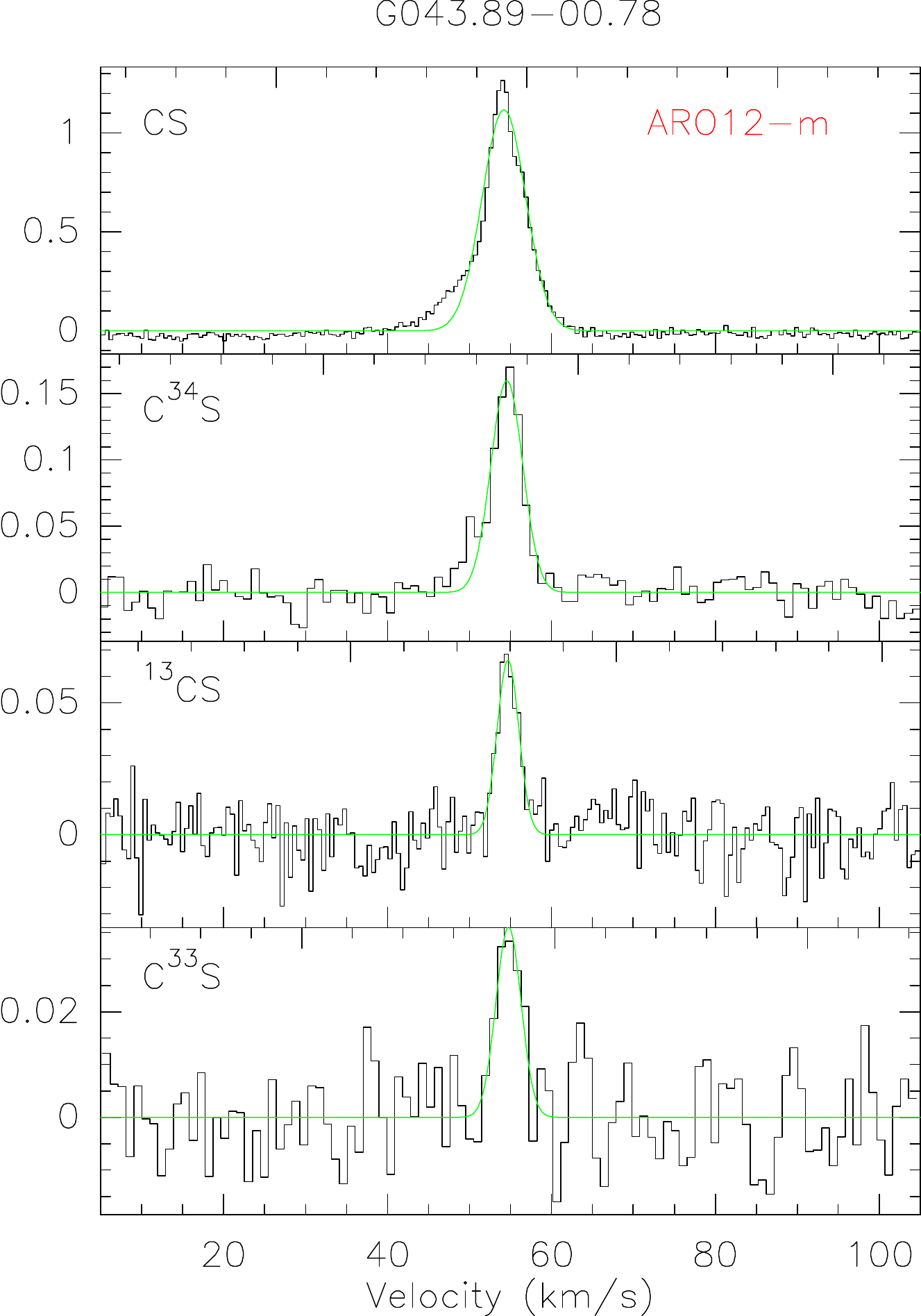}
  \includegraphics[width=106.5pt]{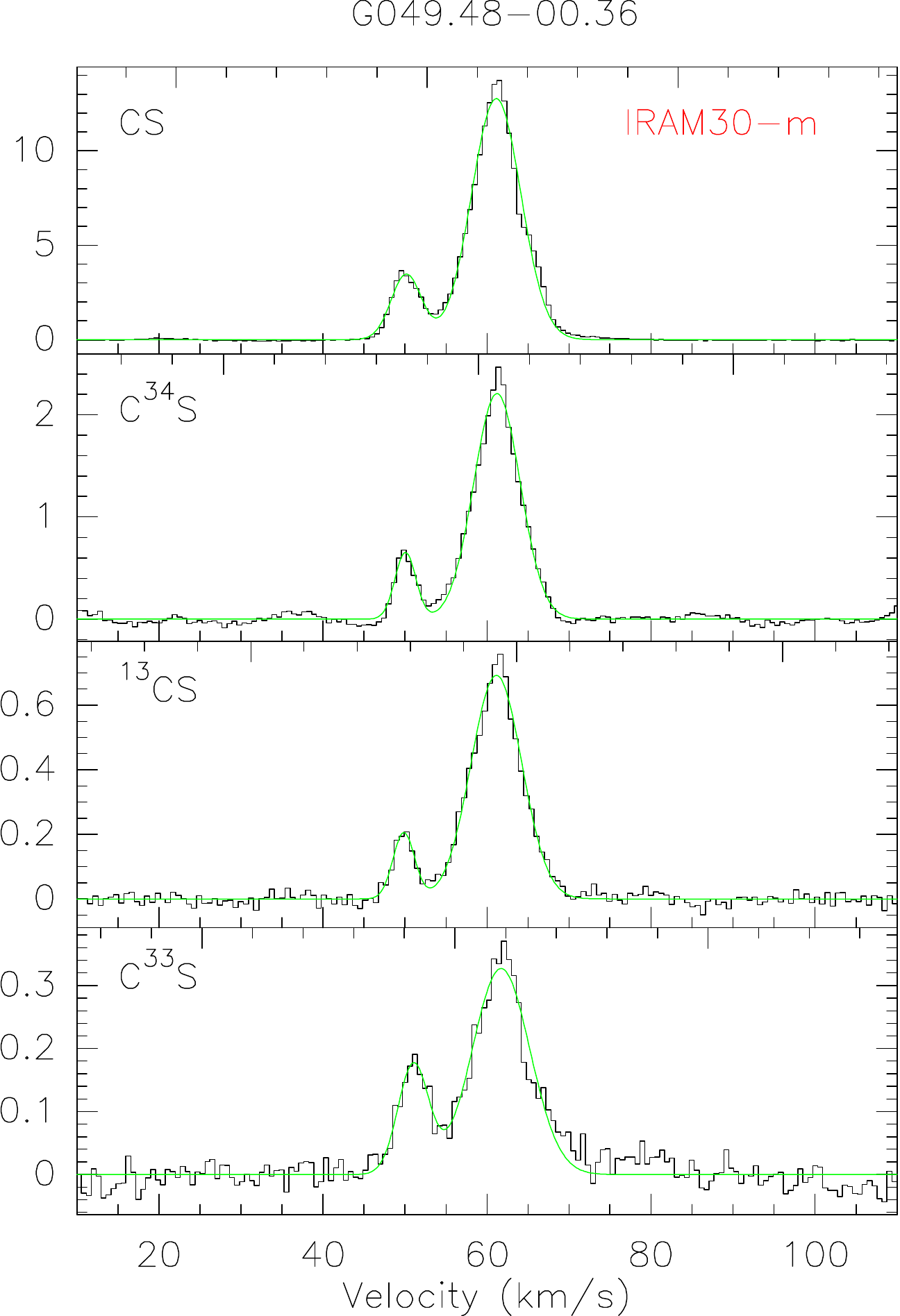}
  \includegraphics[width=109pt]{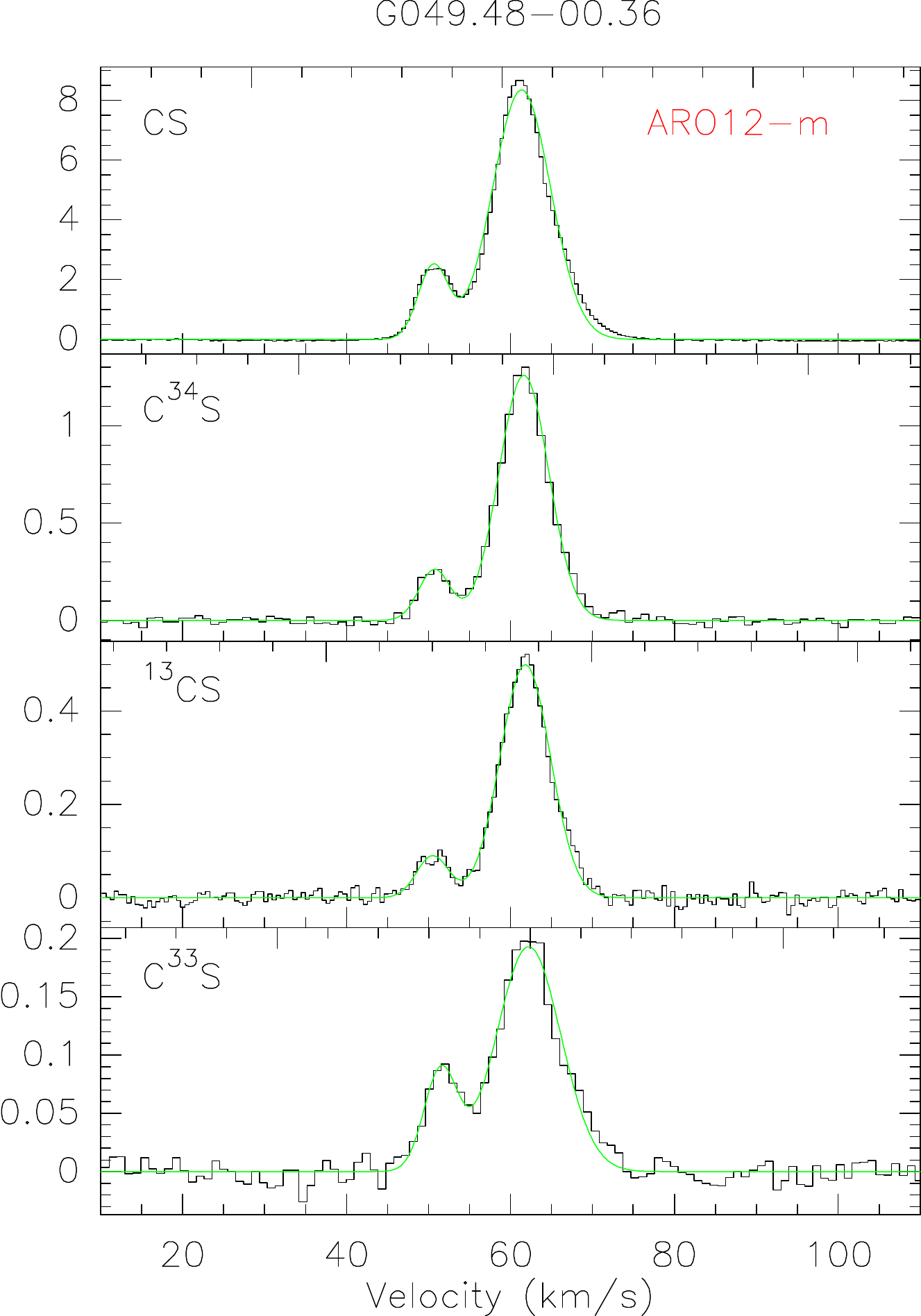}
  \includegraphics[width=110.3pt]{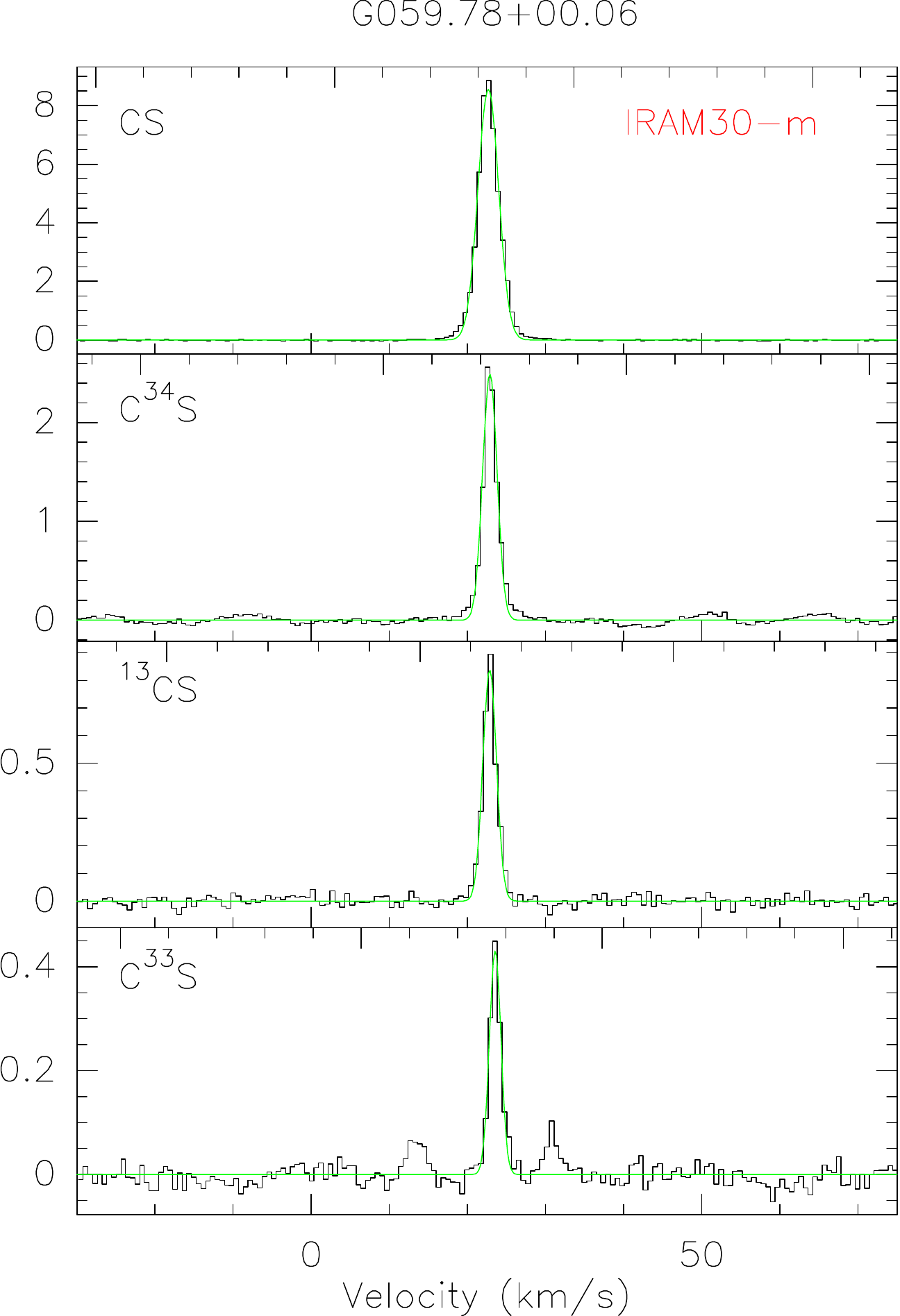}
  \includegraphics[width=113pt]{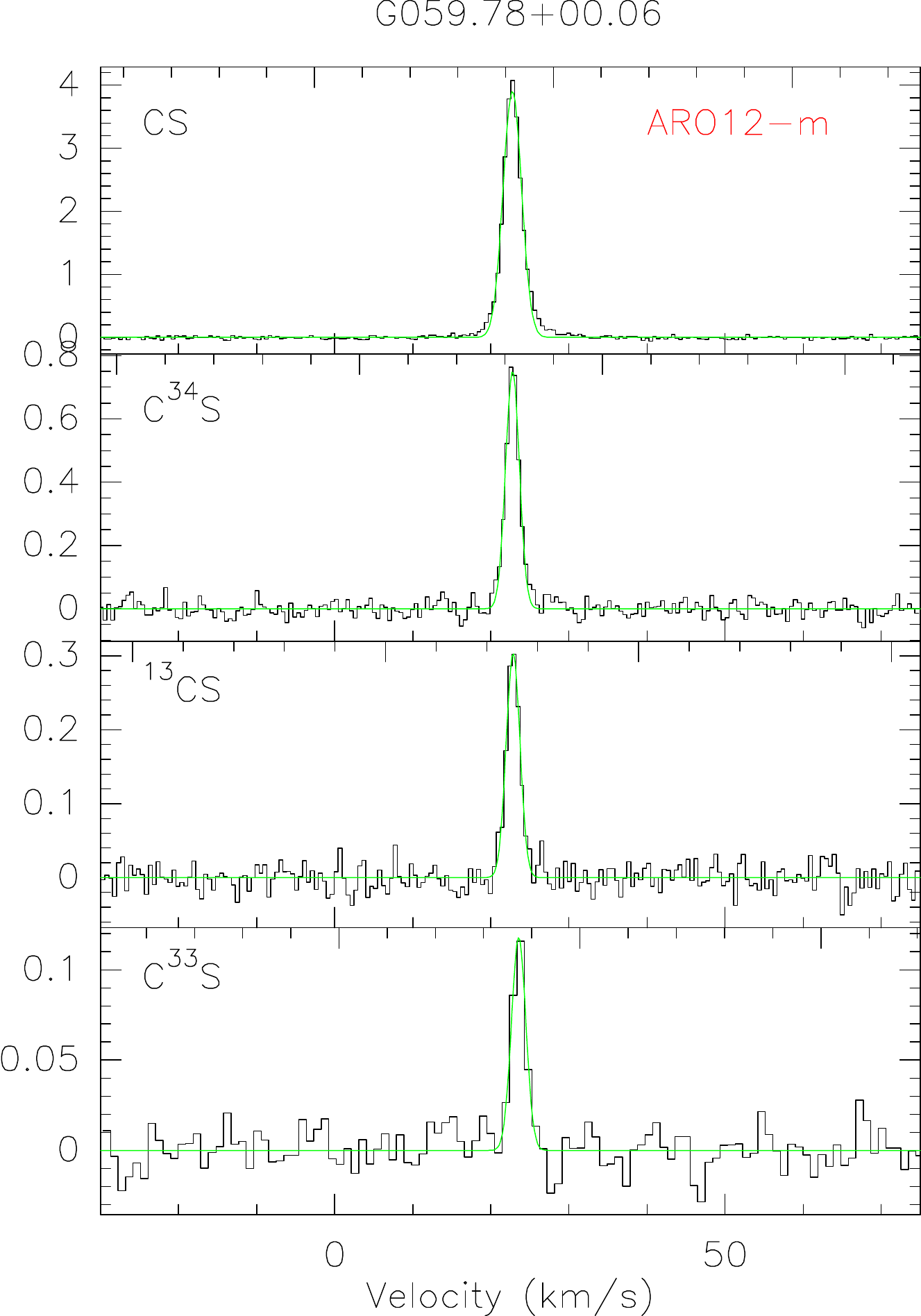}
  \includegraphics[width=110.5pt]{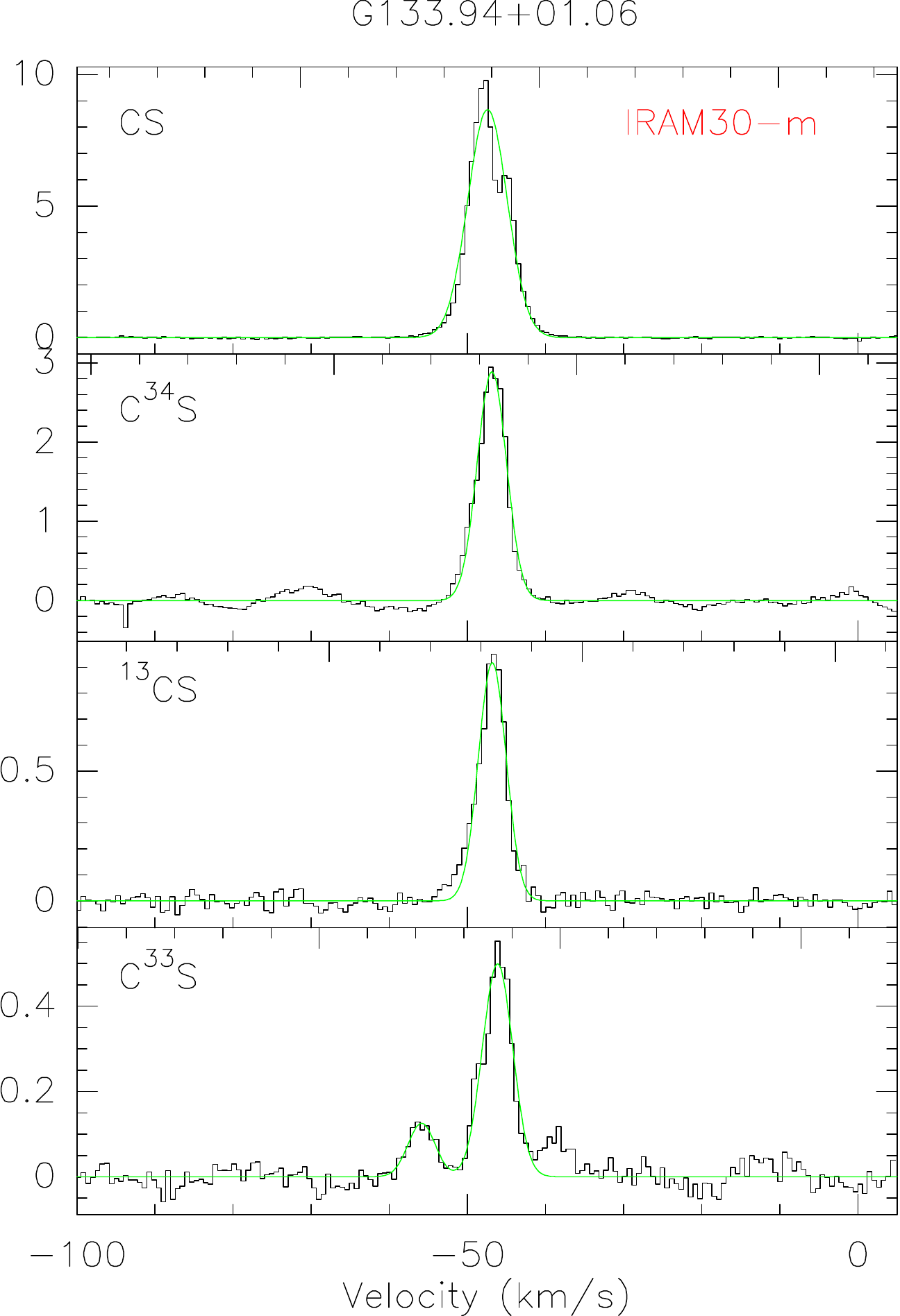}
  \includegraphics[width=110.5pt]{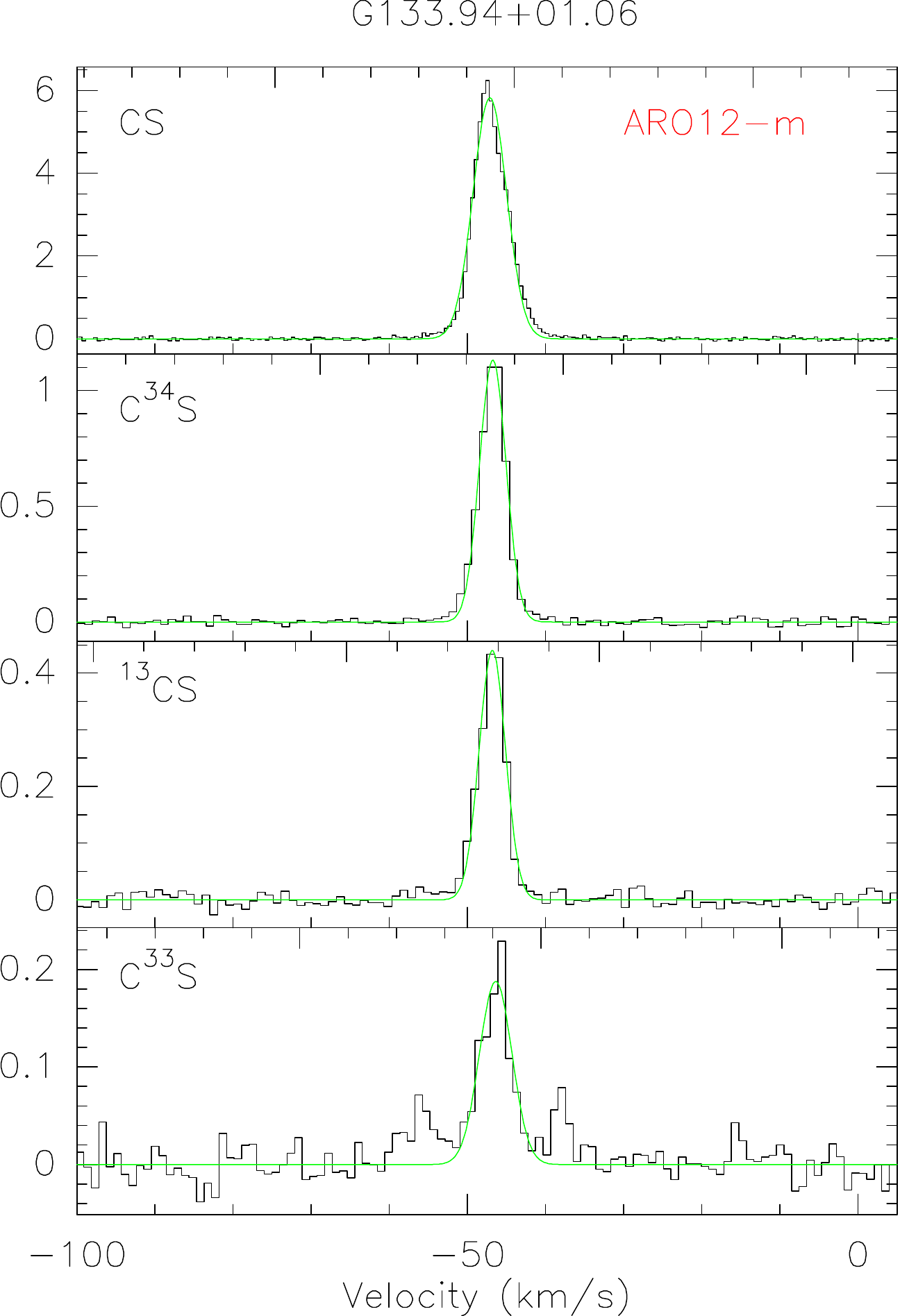}
 \caption{ IRAM 30-m and ARO 12-m $^{12}$C$^{32}$S, $^{13}$C$^{32}$S, $^{12}$C$^{33}$S and $^{12}$C$^{34}$S spectra of those six sources where we obtained data from both telescopes.}
\end{figure*}

\begin{figure*}
\gridline{\fig{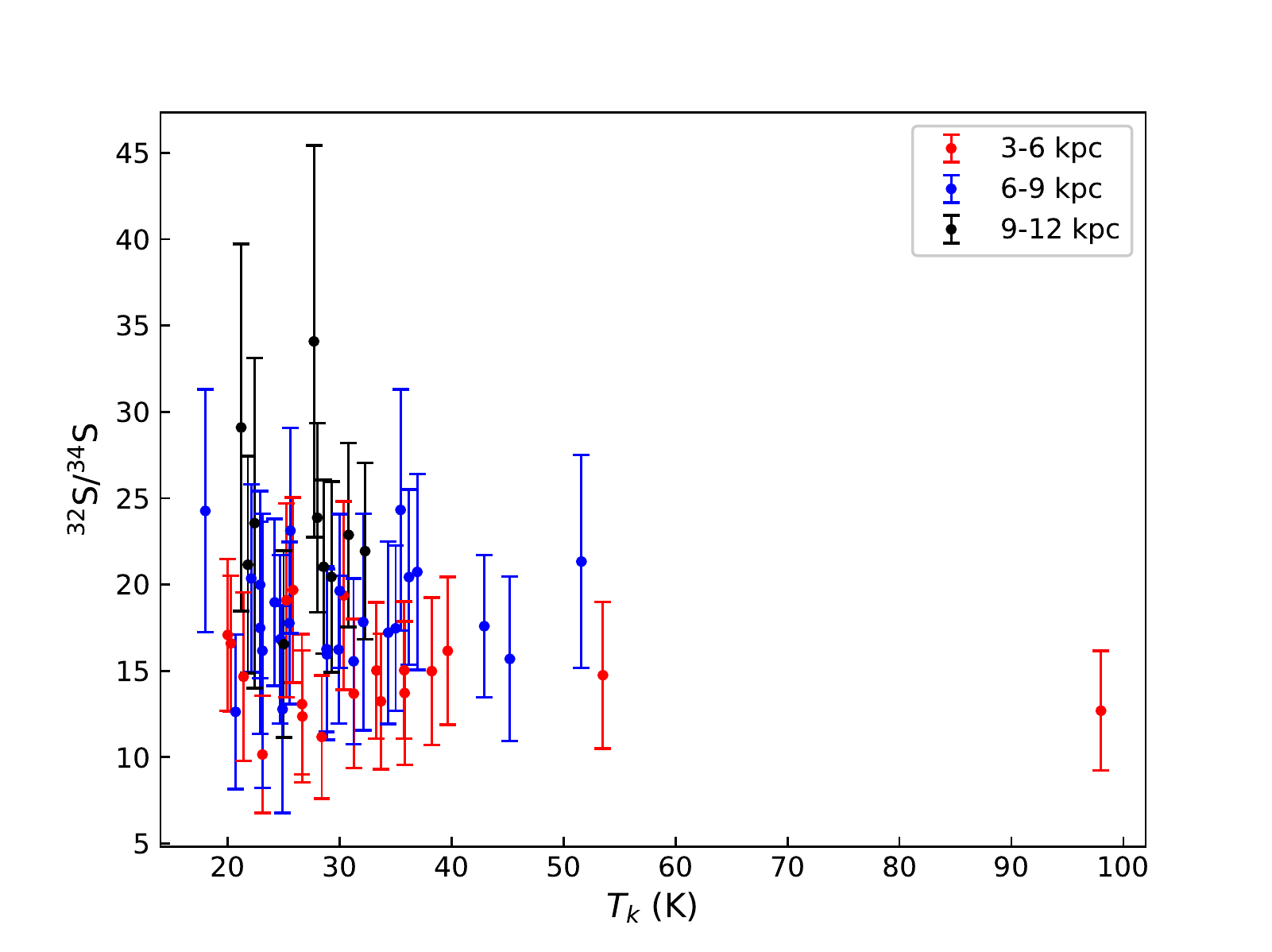}{0.5\textwidth}{(a)}
          \fig{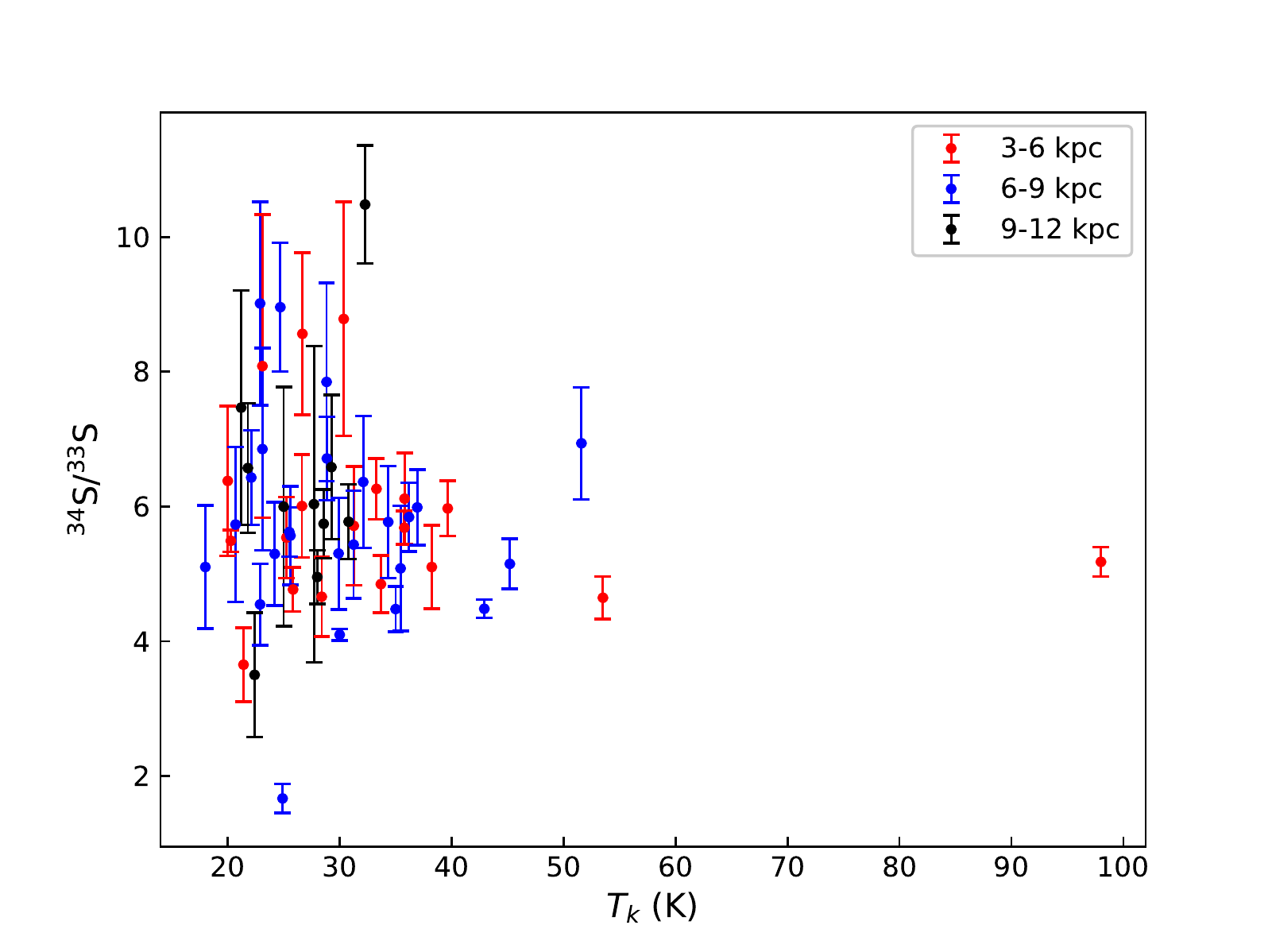}{0.5\textwidth}{(b)}
          }
\caption{(a) $^{32}$S$/$$^{34}$S and (b) $^{34}$S$/$$^{33}$S ratios plotted against gas kinetic temperature.
Red, blue and black dots represent sources from different radial galactocentric bins (for 3 - 6, 6 - 9 and 9 - 12 kpc, respectively). No correlation is found between gas kinetic temperature and sulfur isotope ratios. }
\end{figure*}

\begin{figure*}
\gridline{\fig{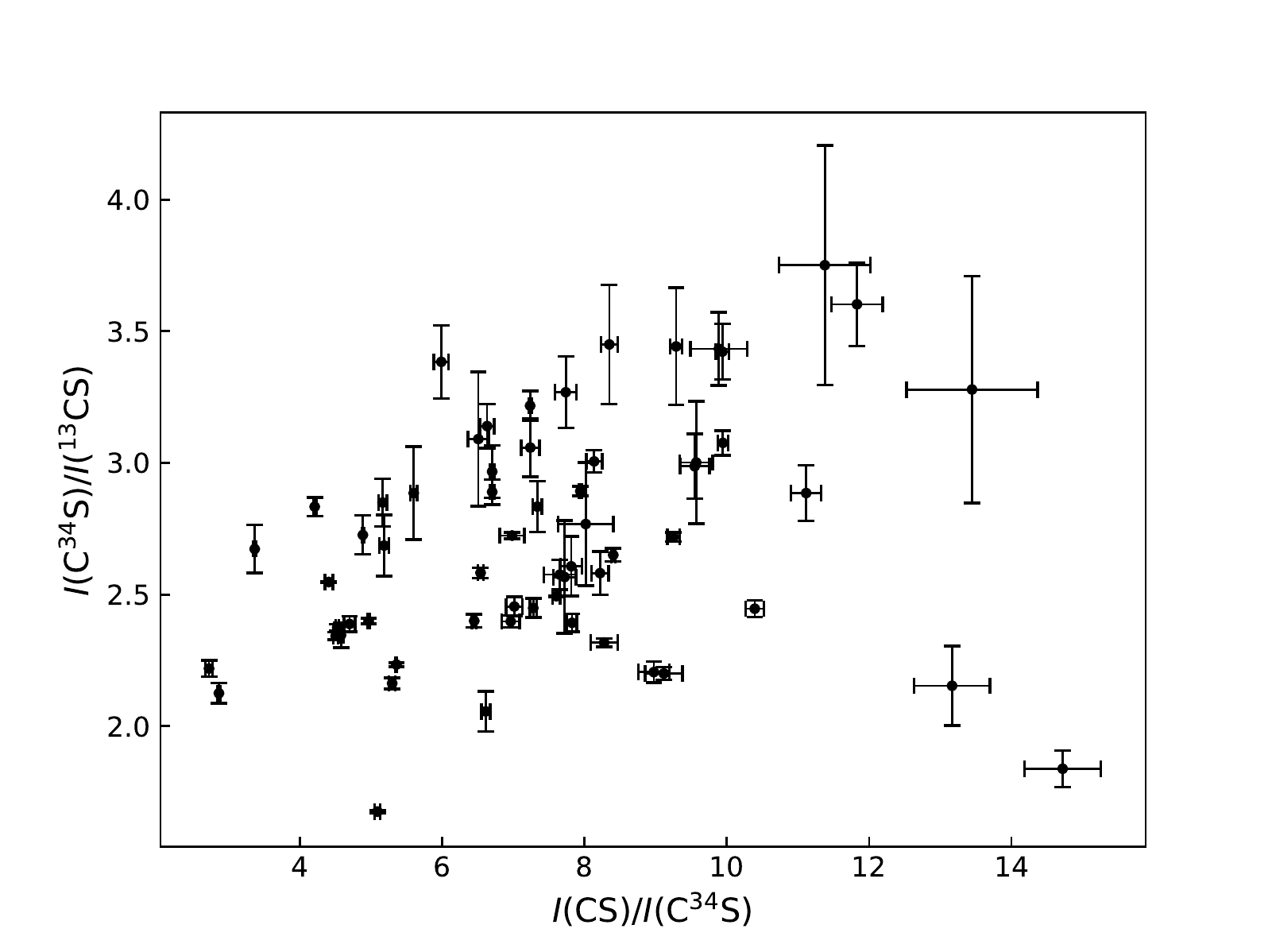}{0.5\textwidth}{(a)}
          \fig{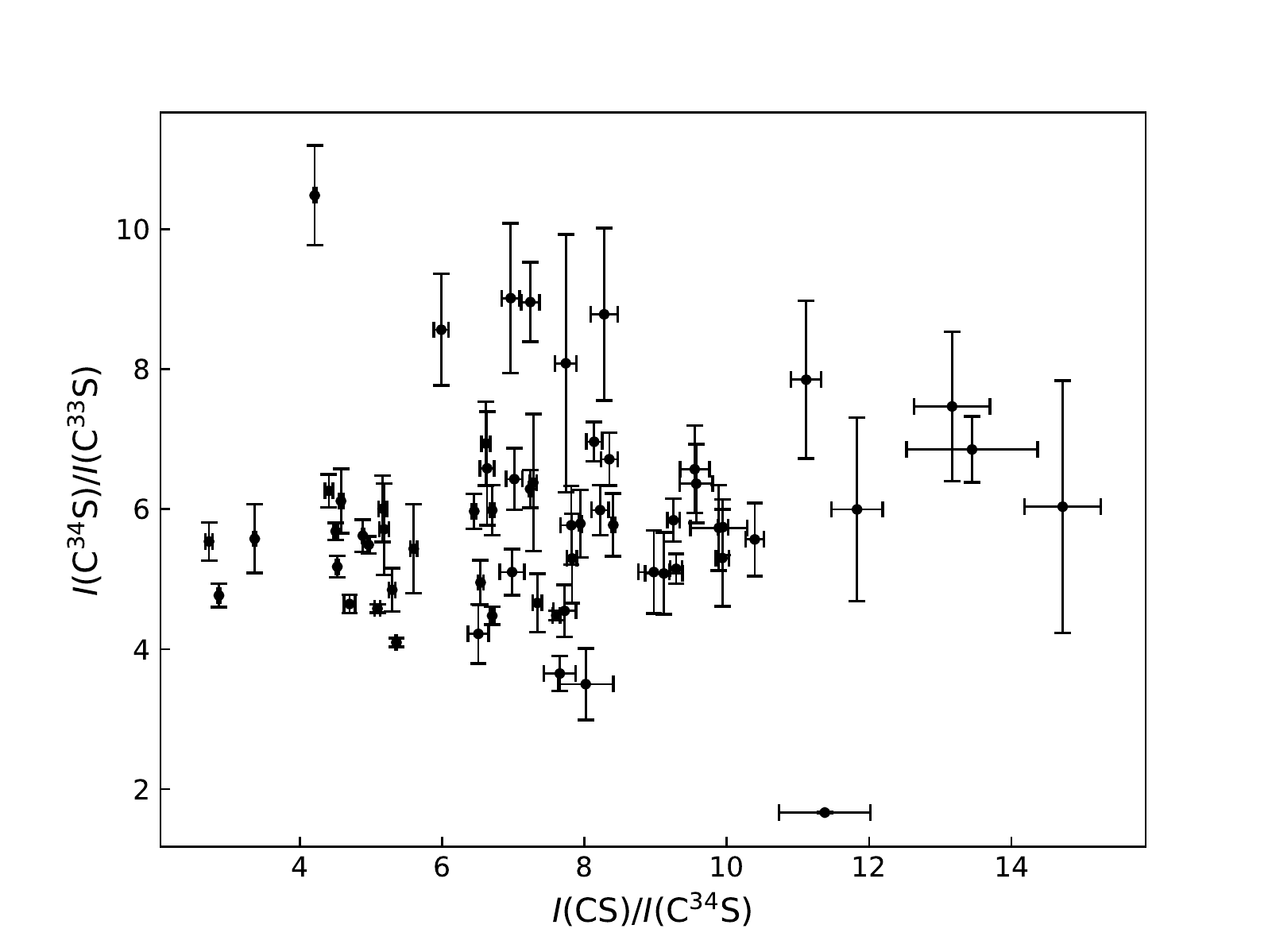}{0.5\textwidth}{(b)}
          }
\caption{(a) $I$(C$^{34}$S)/$I$($^{13}$CS) and (b) $I$(C$^{34}$S)/$I$(C$^{33}$S) as a function of $I$(CS)/$I$(C$^{34}$S). No significant correlation is found between these ratios, indicating that saturation of C$^{34}$S and other even rarer CS isotopologues is not relevant for our sample.}
\end{figure*}

\begin{figure*}
\gridline{\fig{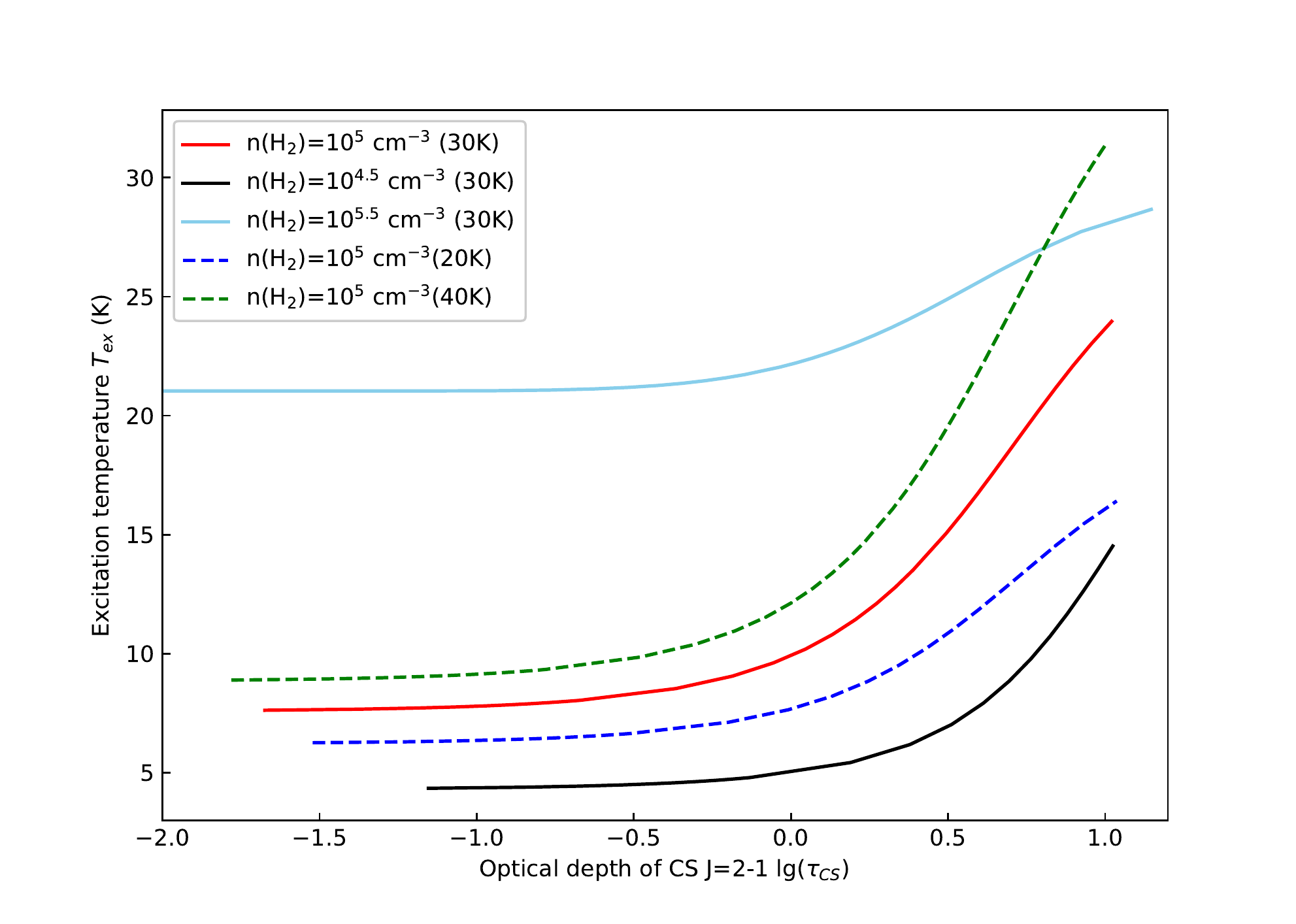}{0.8\textwidth}{}
          }
\caption{\textbf{The excitation temperature is plotted against CS $J$ = 2 - 1 optical depth, using different kinetic temperatures ($T_{\rm k}$) and particle densities ($n$($\rm H_{2}$)). The plot shows that the excitation temperatures are only slightly enhanced with increasing $\tau_{\rm CS}$ as long as $\tau_{\rm CS}$ is less than unity. However, the situation drastically changes under optically thick conditions ($\tau_{\rm CS}$ $\gg$ 1).} }
\end{figure*}

\begin{figure*}
\gridline{\fig{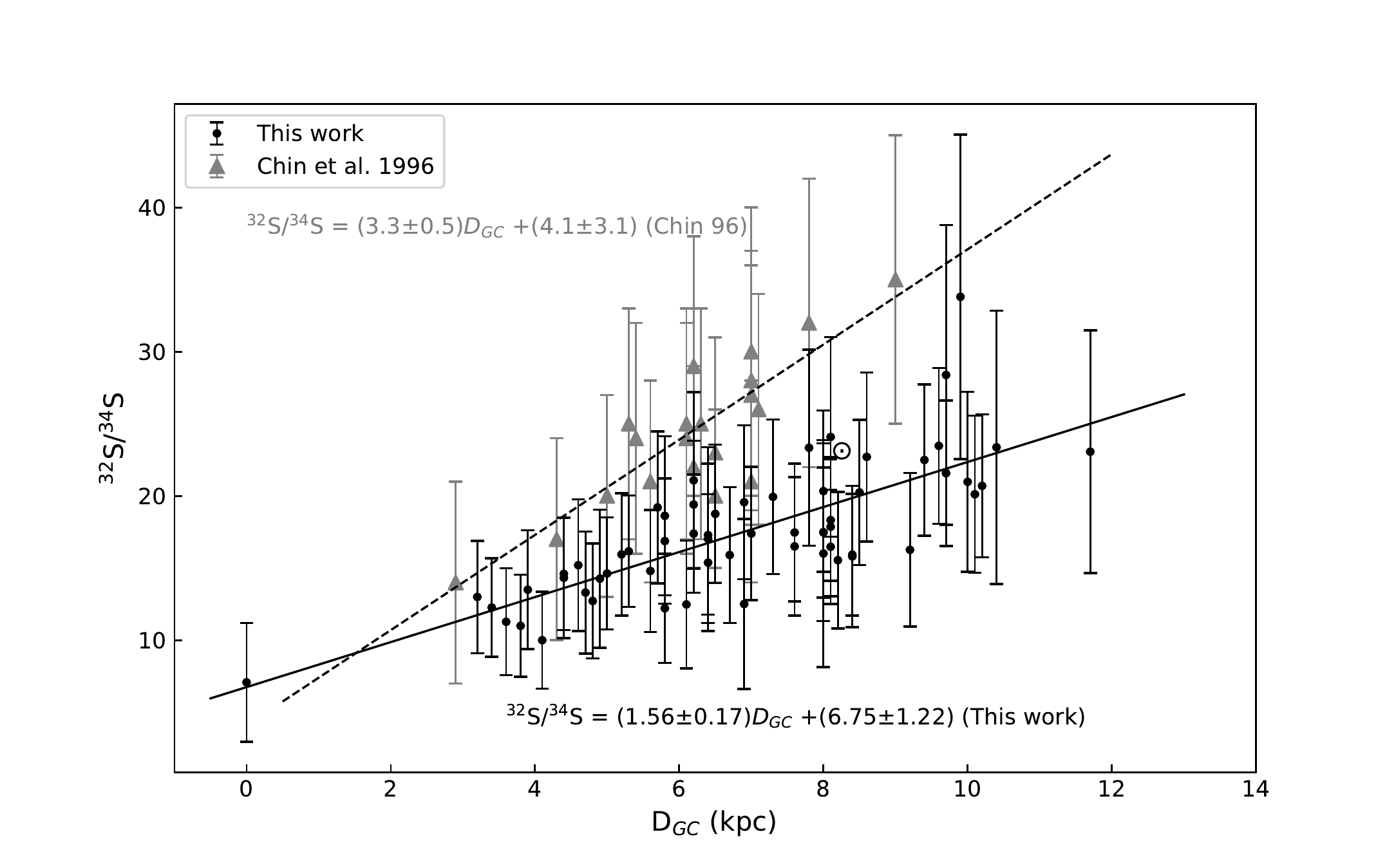}{0.8\textwidth}{(a)}
          }
\gridline{\fig{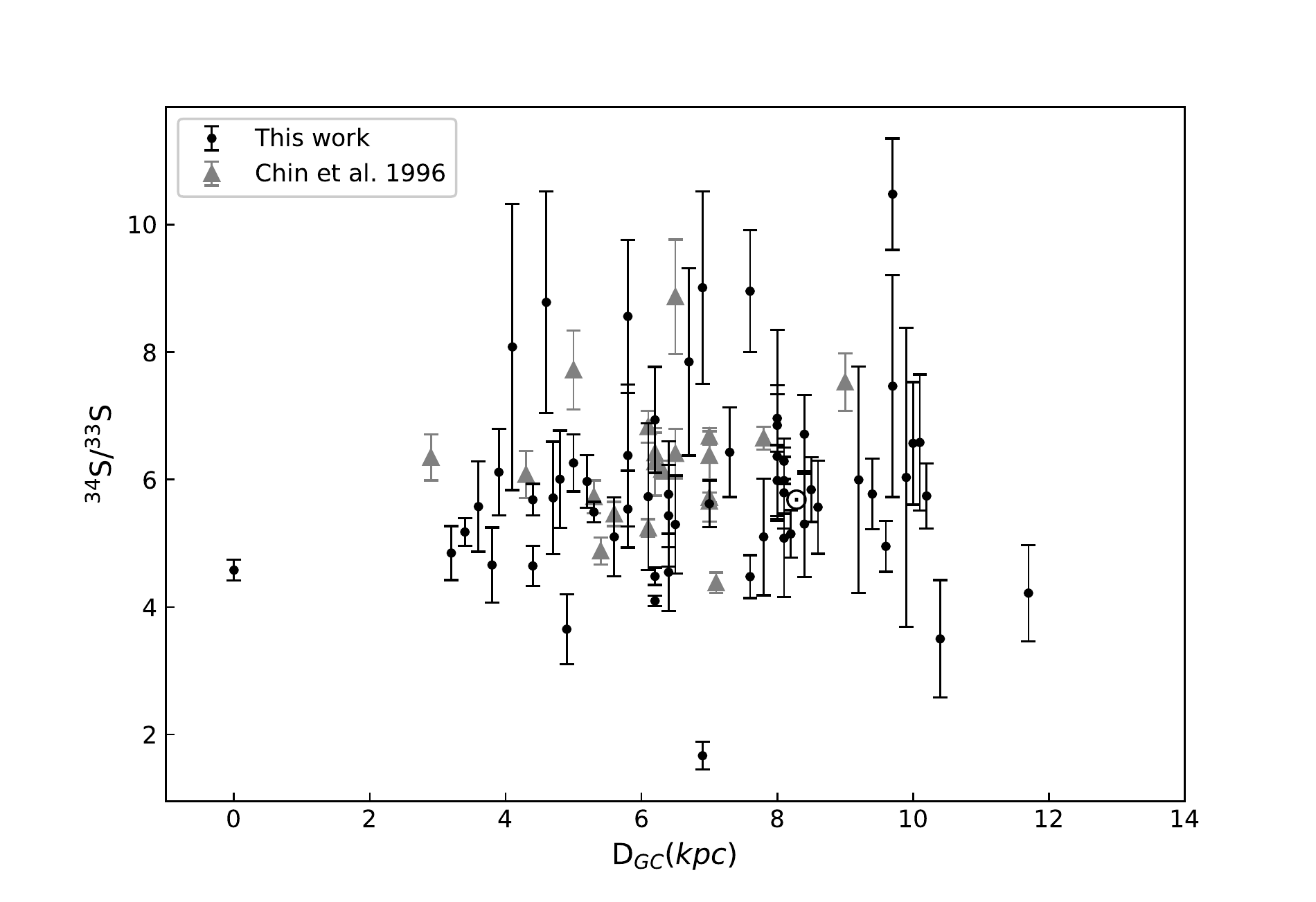}{0.8\textwidth}{(b)}
          }
\caption{(a) $^{32}$S$/$$^{34}$S and (b) $^{34}$S$/$$^{33}$S are plotted as a function of distance to the Galactic center $D_{\rm GC}$(kpc). The solar system value is also given as $\odot$ with the assumption that the solar distance to the Galactic center is 8.122 kpc \citep{2018A&A...615L..15G}. The black dots are the values obtained from the ARO 12-m. The grey triangles represent the values measured by \citet{1996A&A...305..960C}. A linear fit of the ratios from our work is plotted as a black solid line. The black dashed line was fitted from \citet{1996A&A...305..960C} in (a). As shown in (b), there is no apparent $^{34}$S$/$$^{33}$S gradient as a function of galactocentric distance.}
\end{figure*}

\begin{figure*}
\gridline{\fig{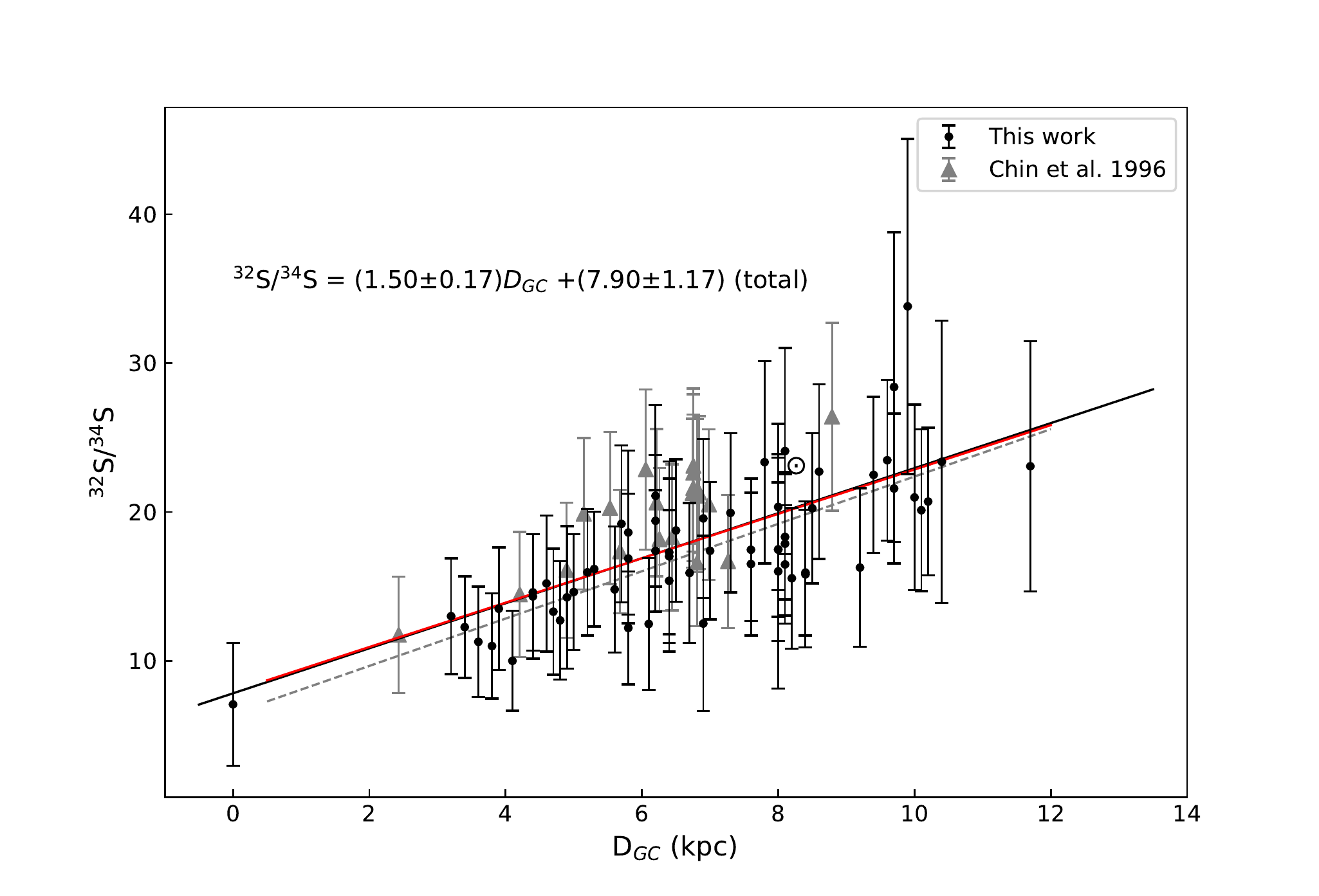}{0.8\textwidth}{}
          }
\caption{ \textbf{$^{32}$S$/$$^{34}$S ratios obtained from \citet{1996A&A...305..960C} and our work with recent $^{12}$C$/$$^{13}$C ratios from \citet{2019ApJ...877..154Y} are plotted as a function of distance to the Galactic center, $D_{\rm GC}$(kpc). A linear fit of the $^{32}$S$/$$^{34}$S ratios to all data is plotted as a black solid line. The red solid line was fitted including all data points except the Galactic center value. The black dashed line represents the fitting results of our own data excluding the Galactic center value.}  }
\end{figure*}

\begin{figure*}
\gridline{\fig{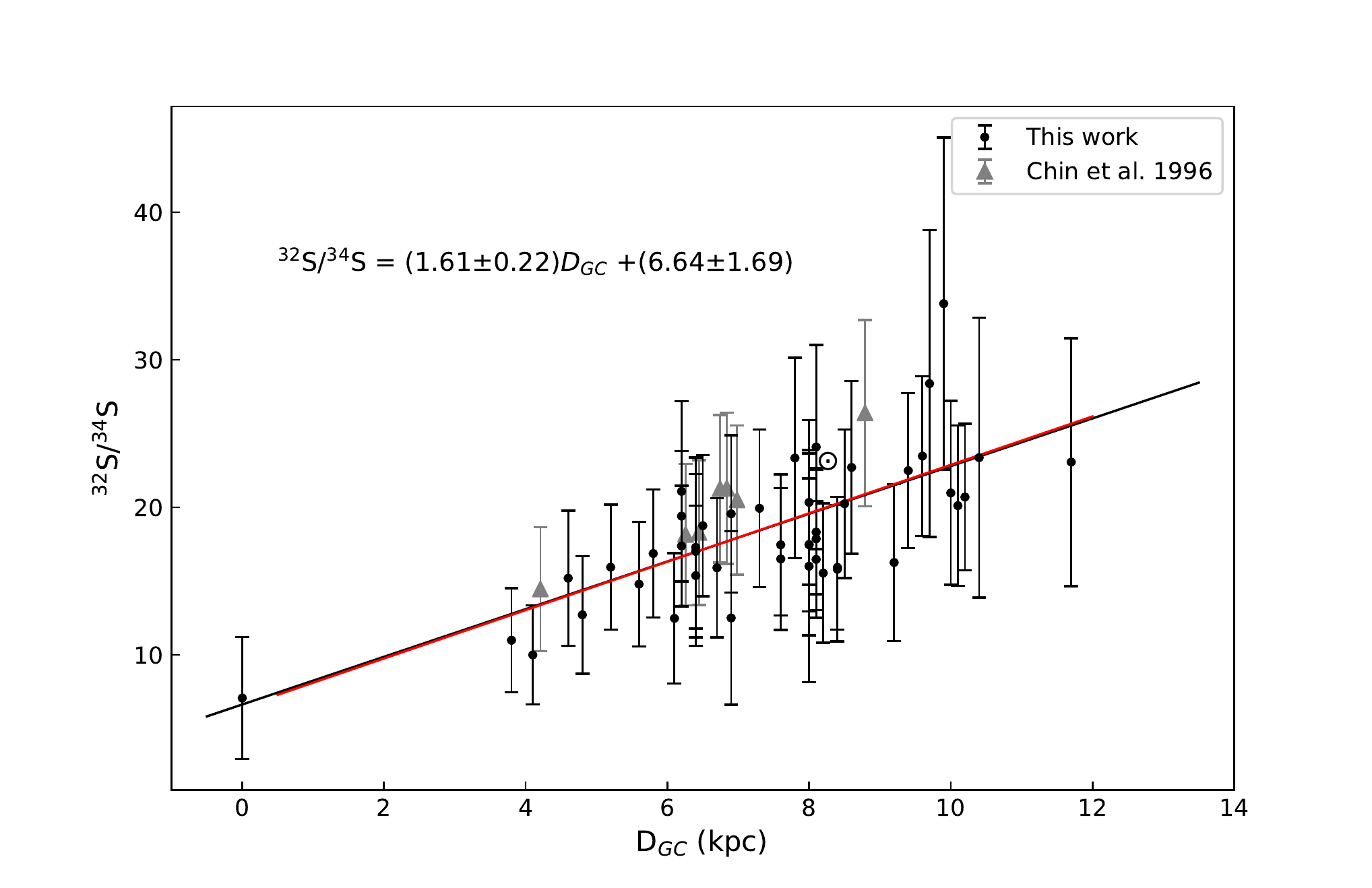}{0.8\textwidth}{}
          }
\caption{ $^{32}$S$/$$^{34}$S ratios obtained from \citet{1996A&A...305..960C} and our work with $\tau$(C$^{34}$S) $<$ 0.25 are plotted against the distance to the Galactic center, $D_{\rm GC}$(kpc). A linear fit of the $^{32}$S$/$$^{34}$S ratios toward these sources is plotted as a black solid line. The red solid line was fitted from all data points without the Galactic center values.}
\end{figure*}

\begin{figure*}
\gridline{\fig{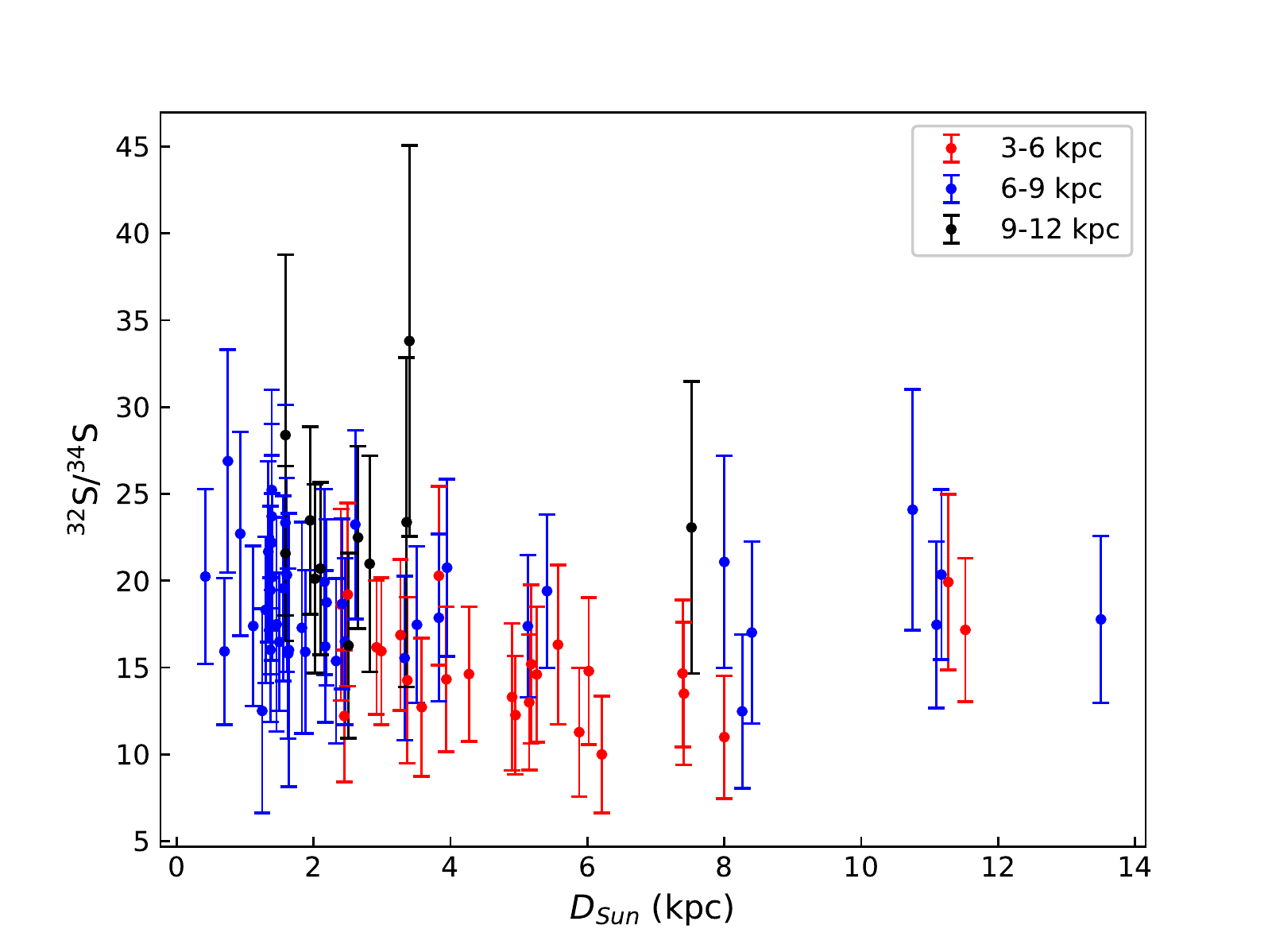}{0.5\textwidth}{(a)}
          \fig{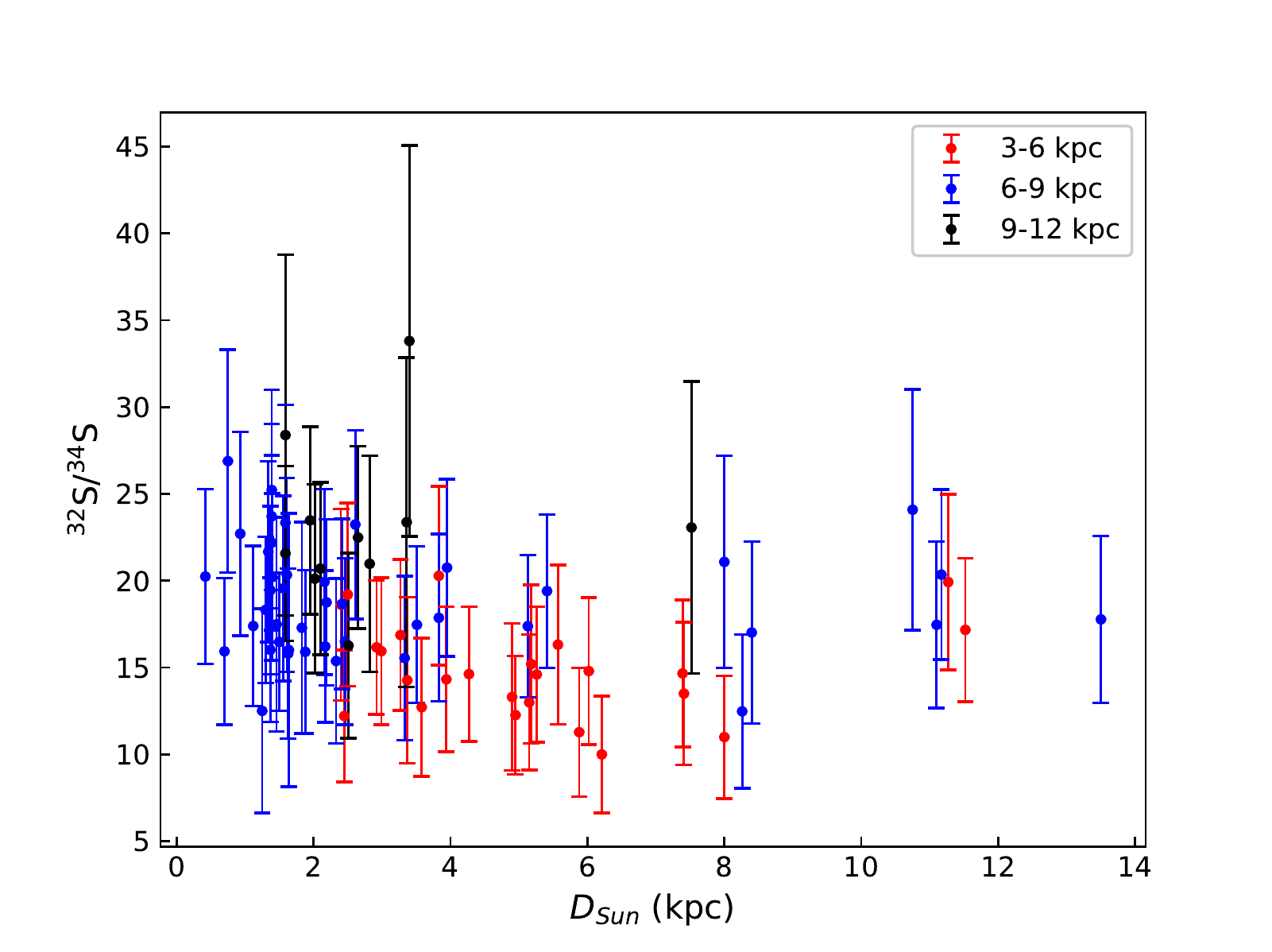}{0.5\textwidth}{(b)}
          }
\caption{(a) $^{32}$S$/$$^{34}$S and (b) $^{34}$S$/$$^{33}$S isotope ratios plotted against the distance to the Sun. Red, blue and black dots represent sources from different radial galactocentric bins (for 3 - 6, 6 - 9 and 9 - 12 kpc, respectively). No significant correlation is found between isotope ratios and the distance to the Sun.}
\end{figure*}

\begin{figure*}
\gridline{\fig{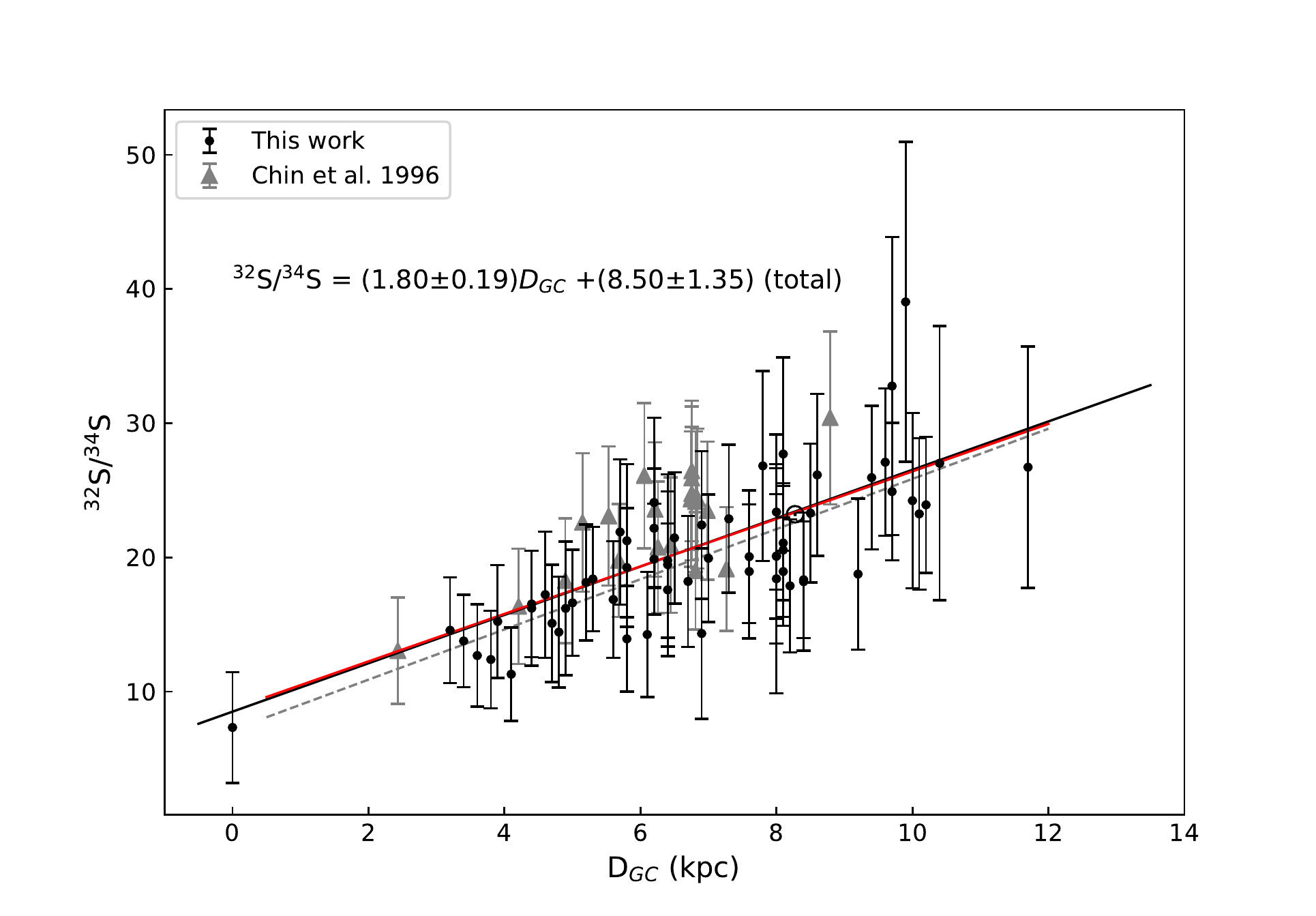}{0.502\textwidth}{(a)}
          \fig{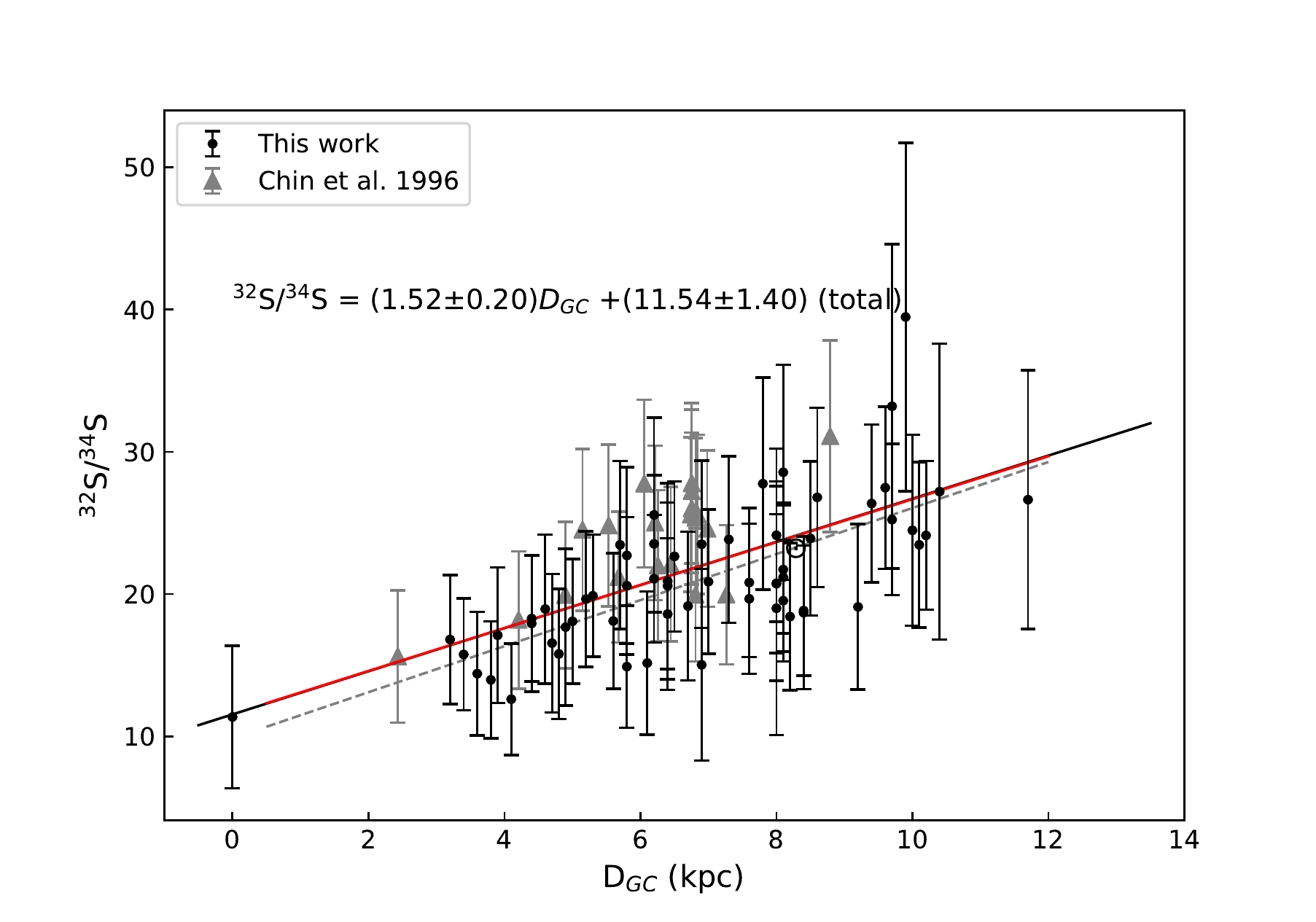}{0.5\textwidth}{(b)}
          }
\gridline{\fig{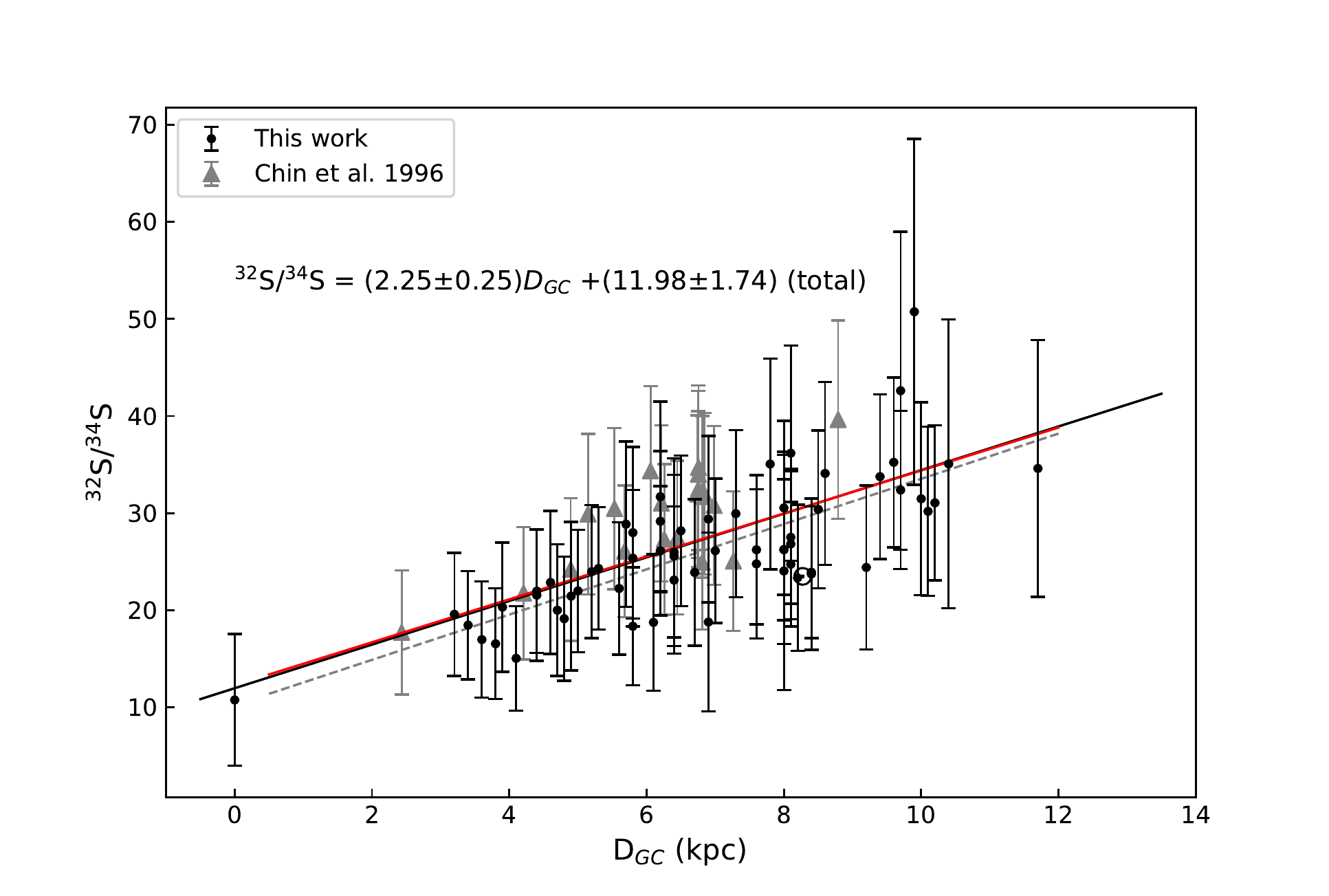}{0.5\textwidth}{(c)}
          }
\caption{\textbf{$^{32}$S$/$$^{34}$S ratios obtained from \citet{1996A&A...305..960C} and our work with different sets of $^{12}$C$/$$^{13}$C values, taken from \citet{2005ApJ...634.1126M} are plotted as a function of distance to the Galactic center, $D_{\rm GC}$(kpc). A linear fit of the $^{32}$S$/$$^{34}$S ratios to all data is plotted as a black solid line. The red solid line was fitted from all data points without the Galactic center value. The black dashed line represents the fitting result based entirely on our data excluding the Galactic center value. Plots (a), (b) and (c) show results with $^{12}$C$/$$^{13}$C values derived from old CN, C$^{18}$O and H$_2$CO data, respectively \citep{2005ApJ...634.1126M}.}}
\end{figure*}


\begin{thebibliography}{}
\expandafter\ifx\csname natexlab\endcsname\relax\def\natexlab#1{#1}\fi
\providecommand{\url}[1]{\href{#1}{#1}}
\providecommand{\dodoi}[1]{doi:~\href{http://doi.org/#1}{\nolinkurl{#1}}}
\providecommand{\doeprint}[1]{\href{http://ascl.net/#1}{\nolinkurl{http://ascl.net/#1}}}
\providecommand{\doarXiv}[1]{\href{https://arxiv.org/abs/#1}{\nolinkurl{https://arxiv.org/abs/#1}}}

\end{thebibliography}


\begin{thebibliography}{}
\bibitem[Anders \& Grevesse(1989)]{1989GeCoA..53..197A} Anders, E. \& Grevesse, N. 1989, Cosmochem. Acta 53, 197
\bibitem[Bogey et al.(1981)]{1981Bogey...81...256B} Bogey, M., Demuynck, C., Destombes, J.L., 1981, Chem. Phys. Lett. 81, 256
\bibitem[Calahan et al.(2018)]{2018ApJ...862...63C} Calahan, J. K., Shirley, Y. L., Svoboda, B. E., et al. 2018, ApJ, 862, 63
\bibitem[Chin et al.(1996)]{1996A&A...305..960C} Chin, Y.-N., Henkel, C., Whiteoak, J.B., Langer, N., Churchwell, E.B. 1996, A\&A 305, 960
\bibitem[Chiappini \& Mateucci(2001)]{2001ApJ...554.1044C} Chiappini, C. and Matteucci, F. and Romano, D. 2001. ApJ, 554, 1044
\bibitem[Dunham et al.(2010)]{2010ApJ...717.1157D} Dunham, M. K., Rosolowsky, E., Evans, II, N. J., et al. 2010, ApJ, 717, 1157
\bibitem[Dunham et al.(2011)]{2011ApJ...741..110D} Dunham, M. K., Rosolowsky, E., Evans, II, N. J., Cyganowski, C., \& Urquhart, J. S. 2011, ApJ, 741, 110
\bibitem[Frerking et al.(1980)]{1980ApJ...240...65F} Frerking, M.A., Wilson, R.W., Linke, R.A., Wannier, P.G. 1980, ApJ 240, 65
\bibitem[Gravity Collaboration et al.(2018)]{2018A&A...615L..15G} Gravity Collaboration, Abuter R., Amorim A., et al. 2018, A\&A, 615, L15
\bibitem[Harju et al.(1993)]{1993A&AS...98...51H} Harju, J., Walmsley, C. M., \& Wouterloot, J. G. A. 1993, A\&AS, 98, 51
\bibitem[Hill et al.(2010)]{2010MNRAS.402.2682H} Hill, T., Longmore, S. N., Pinte, C., et al. 2010, MNRAS, 402, 2682
\bibitem[Hughes et al.(2008)]{2008MNRAS.390.1710H} Hughes G. L., Gibson B. K., Carigi L., S\'{a}nchez-Bl\'{a}zquez P., Chavez J. M.,
Lambert D. L., 2008, MNRAS, 390, 1710
\bibitem[Linke \& Goldsmith(1980)]{1980APJ...235.437L} Linke, R.A., Goldsmith, P.F., 1980, APJ 235, 437
\bibitem[Lique et al.(2006)]{2006A&A...451.1125L} Lique, F. Spielfiedel, A. Cernicharo, J. 2006, A\&A 451, 1125L
\bibitem[Milam et al.(2005)]{2005ApJ...634.1126M} Milam, S. N., Savage, C., Brewster, M. A., Ziurys, L. M., \& Wyckoff, S. 2005,
ApJ, 634, 1126
\bibitem[Mauersberger et al.(1996)]{1996A&A...313L...1M} Mauersberger, R., Henkel, C., Langer, N.m Chin, Y.-N., 1996, A\&A 313, L1
\bibitem[Mauersberger et al.(2004)]{2004A&A...426..219M} Mauersberger, R., Ott, U., Henkel, C., Cernicharo, J., Gallino, R. 2004, A\&A 426, 219
\bibitem[Molinari et al.(1996)]{1996A&A...308..573M} Molinari, S., Brand, J., Cesaroni, R., \& Palla, F. 1996, A\&A, 308, 573
\bibitem[Ott et al.(2014)]{2014ApJ...785...55O} Ott, J., Weiß, A., Staveley-Smith, L., Henkel, C., \& Meier, D. S. 2014, ApJ, 785,
55
\bibitem[Reid et al.(2014)]{2014ApJ...783..130R} Reid, M. J., Menten, K. M., Brunthaler, A., et al. 2014, ApJ, 783, 130
\bibitem[Reid et al.(2016)]{2016ApJ...823..77R} Reid, M. J., Dame, T. M., Menten, K. M., et al. 2016, ApJ, 823, 77
\bibitem[Reid et al.(2019)]{2019ApJ...885..131R} Reid, M. J., Menten, K. M., Brunthaler, A., et al. 2019, ApJ, 885, 131
\bibitem[Roman-Duval et al.(2009)]{2009ApJ...699.1153R} Roman-Duval, J., Jackson, J. M., Heyer, M., et al. 2009, ApJ, 699, 1153
\bibitem[Schreyer et al.(1996)]{1996A&A...306..267S} Schreyer K., Henning T., Koempe C., Harjunpaeae P., 1996, A\&A, 306, 267
\bibitem[Svoboda et al.(2016)]{2016ApJ...822...59S} Svoboda, B. E., Shirley, Y. L., Battersby, C., et al. 2016, ApJ, 822,
59
\bibitem[Tieftrunk et al.(1998)]{1998A&A...336..991T} Tieftrunk, A. R., Megeath, S. T., Wilson, T. L., \& Rayner, J. T. 1998, A\&A,
336, 991
\bibitem[Thielemann \& Arnett(1985)]{1985ApJ...295..604T} Thielemann, F.-K. \& Arnett, W.D. 1985, ApJ 295, 604
\bibitem[Urquhart et al.(2011)]{2011MNRAS.418.1689U} Urquhart, J. S., Morgan, L. K., Figura, C. C., et al. 2011, MNRAS, 418, 1689
\bibitem[Van der Tak et al.(2007)]{2007A&A...468..627V} Van der Tak, F.F.S., Black, J.H., Schier, F.L., Jansen, D.J., van Dishoeck, E.F. 2007, A\&A, 468, 627-635
\bibitem[Woosley \& Weaver(1995)]{1995ApJS..101..181W} Woosley, S.E. \& Weaver, T.A. 1995, ApJS 101, 18
\bibitem[Wilson \& Matteucci(1992)]{1992A&ARv...4....1W} Wilson, T.L. \& Rood, R.T. 1994, ARA\&A 32, 191
\bibitem[Wilson \& Rood(1994)]{1994ARA&A..32..191W} Wilson, T.L. \& Matteucci, F. 1992, A\&AR 4, 1
\bibitem[Wienen et al.(2012)]{2012A&A...544A.146W} Wienen, M., Wyrowski, F., Schuller, F., et al. 2012, A\&A, 544, A146
\bibitem[Xu et al.(2006)]{2006Sci...311..54X} Xu, Y., Reid, M. J., Zheng, W. W., \& Menten, K. M. 2006, Sci, 311, 54
\bibitem[Yan et al.(2019)]{2019ApJ...877..154Y} Yan, Y. T., Zhang, J.S., Henkel, C., et al. 2019, APJ, 877, 154Y
\bibitem[Zhang et al.(2018)]{2018Natur.558..260Z}Zhang, Z.-Y., Romano, D., Ivison, R. J., et al. 2018, Natur, 558, 260
\bibitem[Zhang et al.(2017)]{2017A&A...606A..74Z} Zhang X.-Y., Zhu Q.-F., Li J., Chen X., Wang J.-Z., Zhang J.-S., 2017, A\&A, 606, A74

\end{thebibliography}
\end{document}